\documentclass{ws-rmp}

\usepackage{amsmath}%
\usepackage{amsfonts}%
\usepackage{amssymb}%
\usepackage{graphicx}
\usepackage{amsbsy,amscd}
\numberwithin{equation}{section}

\newcommand{\alg}[1]{\mathfrak{#1}}
\newcommand{\yang}[1]{\mathcal{Y}(\mathfrak{#1})}
\newcommand{\dyang}[1]{\mathcal{DY}(\mathfrak{#1})}
\newcommand{\DY}[1]{\mathcal{DY}(\mathfrak{#1})}
\newcommand{\cdoub}[1]{\mathcal{D}(\mathfrak{#1})}
\newcommand{\casiten}{t}
\newcommand{\casi}{\mathcal{C}}
\newcommand{\nln}{\nonumber\\}
\newcommand{\nn}{\nonumber}
\newcommand{\half}{\frac{1}{2}}
\newcommand{\quarter}{\frac{1}{4}}
\newcommand{\nl}[1][0pt]{\nonumber\\[#1]&\hspace{-4\arraycolsep}&\mathord{}}

\newcommand{\sun}{\mathfrak{su}(n)}
\newcommand{\psu}{\mathfrak{psu}(2|2)}
\newcommand{\sutt}{\mathfrak{su}(2|2)}
\newcommand{\uone}{\mathfrak{u}(1)}
\newcommand{\psucentral}{\mathfrak{psu}(2|2)\ltimes\alg{u}(1)^3}
\newcommand{\pslcentral}{\mathfrak{psl}(2|2)\ltimes\mathbb{C}^3}
\newcommand{\psltt}{\alg{psl}(2|2)}
\newcommand{\utt}{\alg{u}(2|2)}
\newcommand{\lutt}{\alg{u}(2|2)[u,u^{-1}]}

\newcommand{\yutt}{\mathcal{Y}(\mathfrak{u}(2|2))}

\newcommand{\ypsucentral}{\mathcal{Y}(\mathfrak{psu}(2|2)\ltimes\mathbb{R}^3)}
\newcommand{\ypslcentral}{\mathcal{Y}(\mathfrak{psl}(2|2)\ltimes\mathbb{C}^3)}
\newcommand{\sconf}{\mathfrak{psu}(2,2|4)}
\newcommand{\sln}{\alg{sl}(n)}
\newcommand{\slm}{\alg{sl}(m)}
\newcommand{\slnm}{\alg{sl}(n|m)}
\newcommand{\glnm}{\alg{gl}(n|m)}
\newcommand{\slnn}{\alg{sl}(n|n)}
\newcommand{\pslnn}{\alg{psl}(n|n)}
\newcommand{\glnn}{\alg{gl}(n|n)}
\newcommand{\yglnm}{\mathcal{Y}(\mathfrak{gl}(n|m))}

\newcommand{\lalg}[1]{\mathfrak{#1}[[u,u^{-1}]]}
\newcommand{\palg}[1]{\mathfrak{#1}[u]}
\newcommand{\nalg}[1]{\mathfrak{#1}[u^{-1}]u^{-1}}

\newcommand{\sprod}[2]{\left(#1,#2\right)}

\newcommand{\scomm}[2]{[#1,#2\}}
\newcommand{\acomm}[2]{\{#1,#2\}}
\newcommand{\sacomm}[2]{\{#1,#2]}
\newcommand{\comm}[2]{[#1,#2]}

\newcommand{\scons}{f^{ab}{}_{c}}
\newcommand{\sconsb}[2]{f^{#1}{}_{#2}}
\newcommand{\bcons}{\gamma_{ab}{}^{c}}
\newcommand{\tr}{\mathop{\mathrm{tr}}}
\newcommand{\str}{\mathop{\mathrm{str}}}

\newcommand{\copro}{\Delta}
\newcommand{\cobra}{\delta}
\newcommand{\coprocl}{\delta}
\newcommand{\coproop}{\Delta^{op}}

\newcommand{\cybe}[1]{[#1_{12},#1_{13}]+[#1_{12},#1_{23}]+[#1_{13},#1_{23}]}

\newcommand{\state}[1]{\mathopen{|}#1\mathclose{\rangle}}

\newcommand{\cartnn}{A^{\alg{gl}(n|n)}}

\newcommand{\qnum}[1]{\frac{q^{#1} - q^{-#1}}{q - q^{-1}}}
\newcommand{\qnumb}[1]{[#1]_q}

\newcommand{\gen}[1]{\mathfrak{#1}}
\newcommand{\geny}[1]{\widehat{\mathfrak{#1}}}
\newcommand{\genY}[1]{\widehat{\mathfrak{#1}}}
\newcommand{\csgh}[1]{\mathfrak{H}_{#1}}
\newcommand{\csgp}[1]{\mathfrak{E}^+_{#1}}
\newcommand{\csgm}[1]{\mathfrak{E}^-_{#1}}

\newcommand{\gfhp}[2]{\mathfrak{H}^+_{#1}(\lambda_{#2})}
\newcommand{\gfhm}[2]{\mathfrak{H}^-_{#1}(\lambda_{#2})}

\newcommand{\cwgp}[1]{\mathfrak{E}^+_{#1}}
\newcommand{\cwgm}[1]{\mathfrak{E}^-_{#1}}
\newcommand{\charge}{\mathfrak{Q}}
\newcommand{\gR}[2]{\mathfrak{R}^{#1}{}_{#2}}
\newcommand{\gL}[2]{\mathfrak{L}^{#1}{}_{#2}}
\newcommand{\gQ}[2]{\mathfrak{Q}^{#1}{}_{#2}}
\newcommand{\gS}[2]{\mathfrak{S}^{#1}{}_{#2}}
\newcommand{\gB}[2]{\mathfrak{B}^{#1}{}_{#2}}
\newcommand{\fR}{\mathfrak{R}}
\newcommand{\fL}{\mathfrak{L}}
\newcommand{\fS}{\mathfrak{S}}
\newcommand{\fQ}{\mathfrak{Q}}
\newcommand{\fP}{\mathfrak{P}}
\newcommand{\fK}{\mathfrak{K}}
\newcommand{\gC}{\mathfrak{C}}
\newcommand{\gCt}[2]{\mathfrak{C}^{#1}{}_{#2}}
\newcommand{\gP}{\mathfrak{P}}
\newcommand{\gK}{\mathfrak{K}}
\newcommand{\cwgen}[1]{\mathfrak{#1}}
\newcommand{\csgen}[1]{\mathfrak{#1}}
\newcommand{\smat}{\mathcal{S}}
\newcommand{\rmat}{\mathcal{R}}
\newcommand{\fmat}{\mathcal{F}}
\newcommand{\crmat}{r}
\newcommand{\ident}{\mathcal{I}}
\newcommand{\idm}{\text{Id}}
\newcommand{\perm}{\mathcal{P}}
\newcommand{\ferm}{\mathcal{F}}
\newcommand{\eip}{\mathcal{U}}
\newcommand{\gat}[1]{\tilde{\gamma}_{#1}}
\newcommand{\xp}[1]{x^{+}_{#1}}
\newcommand{\xm}[1]{x^{-}_{#1}}
\newcommand{\xpm}[1]{x^{\pm}_{#1}}

\newcommand{\lrbrk}[1]{\left(#1\right)}
\newcommand{\bigbrk}[1]{\bigl(#1\bigr)}

\def\[{\begin{equation}}
\def\]{\end{equation}}
\def\<{\begin{eqnarray}}
\def\>{\end{eqnarray}}
\newcommand{\beq}{\begin{equation}}
\newcommand{\eeq}{\end{equation}}
\newcommand{\bea}{\begin{eqnarray}}
\newcommand{\eea}{\end{eqnarray}}
\newcommand{\bal}{\begin{align}}
\newcommand{\eal}{\end{align}}

\newcommand{\earel}[1]{\mathrel{}&\hspace{-2\arraycolsep}#1\hspace{-2\arraycolsep}&\mathrel{}}
\newcommand{\eq}{\earel{=}}

\newcommand{\sym}{$\mathcal{N}=4$ Super Yang-Mills theory }
\newcommand{\symng}{$\mathcal{N}=4$ Super Yang-Mills theory}
\newcommand{\ads}{$AdS_5\times S^5\ $}

\newcommand{\gym}{g_{YM}}

\begin{document}

\markboth{Fabian Spill}
{Yangians in Integrable Field Theories, Spin Chains and Gauge-String Dualities}

%
\catchline{}{}{}{}{}
%

\title{Yangians in Integrable Field Theories, Spin Chains and Gauge-String Dualities}

\author{Fabian Spill}

\address{Theoretical Physics, Department of Physics\\
 Imperial College London, London SW7 2AZ, United Kingdom\\ 
\email{fabian.spill@gmail.com}
}

\maketitle

\begin{history}
\received{(Day Month Year)}
\revised{(Day Month Year)}
\end{history}

\begin{abstract}
In the following paper, which is based on the authors PhD thesis submitted to Imperial College London, we explore the applicability of Yangian symmetry to various integrable models, in particular, in relation with S-matrices. One of the main themes in this work is that, after a careful study of the mathematics of the symmetry algebras one finds that in an integrable model, one can directly reconstruct S-matrices just from the algebra. It has been known for a long time that S-matrices in integrable models are fixed by symmetry. However, Lie algebra symmetry, the Yang-Baxter equation, crossing and unitarity, which are what constrains the S-matrix in integrable models, are often taken to be separate, independent properties of the S-matrix. Here, we construct scattering matrices purely from the Yangian, showing that the Yangian is the right algebraic object to unify all required symmetries of many integrable models. In particular, we reconstruct the S-matrix of the principal chiral field, and, up to a CDD factor, of other integrable field theories with $\sun$ symmetry. Furthermore, we study the AdS/CFT correspondence, which is also believed to be integrable in the planar limit. We reconstruct the S-matrices at weak and at strong coupling from the Yangian or its classical limit.\\
We give a pedagogical introduction into the subject, presenting a unified perspective of Yangians and their applications in physics. This paper should hence be accessible to mathematicians who would like to explore the application of algebraic objects to physics as well as to physicists interested in a deeper understanding of the mathematical origin of physical quantities. 
\end{abstract}



\section{Introduction}

In this paper we will investigate the applicability of Yangian symmetry to integrable models. We give a unified, self-contained introduction to the mathematics of Yangians and their universal R-matrices and explore recently discovered applications to physical systems. In physics, obtaining exact analytic expressions for physical quantities is usually at best difficult, and in most cases impossible. Instead, one is restricted to the use of perturbative techniques or to numerical investigations. Indeed, besides improvement of algorithms, numerics has greatly benefited from the advancement of computer technology in the last decades. To obtain a result to the desired accuracy, one often just requires the use of a faster computer. However, for certain questions, it is just not sufficient to simply get better numerics. Consider, for instance, QCD. QCD is a quantum field theory developed to explain phenomena of the theory of strong interactions. It is asymptotically free \cite{Gross:1973id,Politzer:1973fx,Gross:1973ju,Gross:1974cs}, which implies that perturbation theory at high energies is effective in the investigation of the physics. However, the number of Feynman diagrams increases significantly with the order of perturbation theory, and there is no hope for a general all loop answer. At low energies, much less is understood, as the theory is strongly coupled and confining. Indeed, quarks and gluons never appear as individual particles in nature, but only as colour singlets. As perturbation in the coupling does not work in this regime, one usually restricts the investigation to lattice discretisation. After decades of improved methods and dramatically advanced computer technology, still no exact masses of all the mesons and baryons have been derived\footnote{For a review including more recent successes we refer to \cite{Jansen:2008vs}.}, and confinement is yet to be understood better. On the perturbative side, computer power seems to have been used at its maximum, and the derivation of three loop Feynman diagrams was already quite involved \cite{Moch:2004pa,Vogt:2004mw}.\\
Besides trying to improve numerics to get more precise quantitative results, a way to understand better a complicated physical system such as QCD is to simplify a model. One can switch off certain degrees of freedom, or integrate them out. Indeed, perturbation theory itself works in this way, as one throws away terms which are suppressed by some power in a small quantity. Then, the quantitative result obtained should be close to the real physical result. Occasionally, one might simplify the theory under investigation too much to obtain correct numbers, but the simplified theory is still good enough to share qualitative features with the original theory. This is for instance the case in lattice simulations of QCD. These might lead to the correct formation of mesons and baryons, but their masses are not necessarily correct. In QCD, one can think of some other simplifications. One might want to study the system without fundamental fermions, i.e. the pure $SU(3)$ Yang-Mills theory. Another possibility is to generalise QCD by considering any $SU(N)$ as the gauge group. In particular, it was shown in \cite{'tHooft:1974jz} that the theory simplifies dramatically in the large $N$ limit, if one also sends the coupling constant $\gym$ to zero, but keeps the 't Hooft coupling $\lambda = \gym^2 N$ fixed. Indeed, in this limit, only planar Feynman diagrams survive. The remaining theory is still a complicated field theory which depends on the coupling constant $\lambda$. One can then study additional non-planar corrections, i.e. perturbations in $\frac{1}{N}$. It was noticed in \cite{'tHooft:1974jz} that the perturbation theory in $\frac{1}{N}$ is similar to the topological expansion of string theory.\\

String theory was originally developed in the 1960s as a theory of the strong interactions (see the reviews \cite{Becker:2007zj, DiVecchia:2007we, Schwarz:2007yc}). The original idea was that as between two oppositely charged quarks a flux tube is formed, and the potential grows linearly with the separation distance, the dynamics can be described by a string. However, the quantisation of such a model leads to the appearance of a spin two particle in the spectrum, which is not observed in this regime. Furthermore, consistency required the theory to be defined in $26$ space-time dimensions, and the spectrum contained tachyons. With the subsequent discovery of QCD in the 1970s, string theory became less popular, despite the interesting claim that the spin-two particle should actually be related to the graviton. Hence, string theory was conjectured to be a theory of quantum gravity. String theory became popular again with the discovery that the supersymmetric extension of strings can be anomaly free if it is defined in $10$ dimensions \cite{Green:1984sg}. It was hence considered a more serious candidate for a theory of quantum gravity. Indeed, having $10$ instead of $4$ dimensions was now considered as a virtue rather than a flaw, as one was trying to use $6$ of the $10$ dimensions as internal dimensions in a Kaluza-Klein compactification of the full theory, leading to the gauge theories of nature. Hence, string theory was considered to be able to describe both gauge theories and quantum gravity, and is considered as a candidate for the theory of everything. \\

To date, it is still not quite clear how exactly one should obtain the standard model from string theory. For the purpose of scientific interest of this work, it does not matter whether or not string theory will ultimately turn out to be the theory of everything or not. What is important is that in the decades of development of string theory, lots of interesting and useful mathematical structures have been found. In this sense, string theory has a similar status as quantum field theory. Quantum field theory provides a useful framework which can be applied to many problems in physics, especially in particle physics and in condensed matter theory. Likewise, string theory can be applied to different problems, or it can be seen as the fundamental, unified theory of all particles and forces. The most prominent of the new applications of string theory is via a new connection between gauge and string theory, in the line of thought of \cite{'tHooft:1974jz} discussed above. Such a precise connection was found in the late 1990s in \cite{Maldacena:1998re,Gubser:1998bc,Witten:1998qj}, and claims the exact duality of string theory on Anti-de Sitter spaces in $d+1$ dimensions, with conformal field theories in $d$ dimensions. As string theory is consistent only in $10$ dimensions, the $d+1$ dimensional Anti-de Sitter space has to be accompanied by a $9-d$ dimensional internal manifold\footnote{There is an analogous correspondence for M-theory in eleven dimensions, which we will not discuss in this paper.}. This connection of string theory and conformal field theory is known as the AdS/CFT correspondence. \\

In this paper we will be only concerned with the best understood example of the AdS/CFT correspondence, which is between \sym in four dimensions with $SU(N)$ gauge group, and string theory on an \ads space. Shortly after the discovery of the AdS/CFT correspondence \cite{Maldacena:1998re}, only comparatively simpler physical quantities such as BPS operators and supergravity states could be compared \cite{Gubser:1998bc,Witten:1998qj}. The reason is that the correspondence is in fact a strong-weak duality. This means weakly coupled gauge theory is related to strongly coupled string theory, and vice versa. As we argued above, we can often understand physical systems well if we have a small parameter in the theory, which we can use for perturbation theory. Now the problem is that if physical quantities in the gauge theory are perturbatively calculated in a series in one parameter, the corresponding dual string theory quantity will have the inverse parameter as the natural expansion parameter. This makes a perturbative comparison of physical quantities hard to achieve. Indeed, the quantities compared in \cite{Gubser:1998bc,Witten:1998qj} are independent of the coupling. \\

Furthermore, when the AdS/CFT correspondence was formulated, not even an action describing the motion of strings on \ads was known. In general, it is a hard task to explicitly describe the motion of strings on curved space-times. The action was later fixed in \cite{Metsaev:1998it}. An attempt in computations beyond the planar limit of the AdS/CFT correspondence started with a proposal of a new limit of string and gauge theory in \cite{Berenstein:2002jq}. In this limit long operators with large R-charge were compared to strings spinning fast around a big circle in the five sphere of the \ads space. These operators or string states were allowed a small number of excitations of other charges. Hence, these operators are not protected under supersymmetry. In the dual string theory, the corresponding limit amounts to taking a plane wave limit. Such limit was also found in \cite{Blau:2002dy}, and it was shown that string theory is exactly solvable in that limit \cite{Metsaev:2001bj,Metsaev:2002re}. Related limits of fast spinning strings were found in \cite{Gubser:2002tv,Frolov:2003qc}, see also the reviews \cite{Tseytlin:2003ii,Plefka:2005bk}. It turned out later that the precise limit proposed in \cite{Berenstein:2002jq} was ill-defined (see e.g. the discussion in \cite{Beisert:2006ez}). Nevertheless, \cite{Berenstein:2002jq} can be seen as a starting point for developments in understanding the non-trivial parts of the AdS/CFT correspondence, which we will now summarize. \\

The next crucial development towards a precise understanding of the AdS/CFT correspondence was the observation in \cite{Minahan:2002ve} that in certain sectors at one loop in the 't Hooft coupling $\lambda$ of \symng, the dilatation generator, which describes the behaviour of the theory under scaling transformations, behaves like an integrable spin chain Hamiltonian. Subsequently, this observation was generalised to the whole one loop \sym \cite{Beisert:2003yb}, and integrability was conjectured to survive for the whole AdS/CFT correspondence \cite{Beisert:2003tq} in the 't Hooft limit. Indeed, on the dual string side, a Lax connection was found \cite{Bena:2003wd}, indicating the classical integrability of the string sigma model. This generalises earlier work about the classical integrability of the bosonic part of the sigma model \cite{Mandal:2002fs}. Consequently, non-local conserved charges were found on the dual gauge side, and shown to form a Yangian algebra \cite{Dolan:2003uh,Dolan:2004ps}.\\

Before continuing the discussion of the AdS/CFT correspondence, we should clarify what we actually mean by an integrable system. If we have a mechanical system with a finite number of degrees of freedom, there is a mathematically well defined notion of integrablility (see e.g. \cite{Arnold1995}). Namely, a system is called integrable, if it possesses as many independent conserved charges as it has degrees of freedom. Then, the theorem of Liouville-Arnold implies that the phase space factorises into (projective) tori. In suitable coordinates on these tori, the motion of the system is just linear. Integrability in the AdS/CFT correspondence arises on both sides of the correspondence in a completely different fashion. On the gauge side, we are dealing with the dilatation operator, which acts as an integrable Hamiltonian on a spin chain of arbitrary length. On the string side, we are dealing with an (at least classically) integrable sigma model. In both cases, we have an arbitrary number of degrees of freedom. Furthermore, we are dealing with quantum systems. Hence, we should discuss the meaning of integrability in those cases.\\

The way integrability is manifest in most known integrable systems with infinitely many degrees of freedom is in terms of the appearance of a certain quantum group, which acts as their algebra of symmetries. Quantum groups are Hopf algebras, which are extensions of ordinary algebras combined with a so-called coalgebra structure. This coalgebra structure consists mainly of a coproduct, which defines the action of the symmetry generators on tensor products. Indeed, ordinary Lie algebras act on tensor products, or multi-particle states, just as the sum on the individual states. This implies that the quantum numbers on multi-particle states should simply add up. For many integrable systems, new kinds of non-local charges appear. Non-locality is realised by the fact that the symmetry generators act non-trivially on a multi-particle state, i.e. not simply as the sum of the actions on each individual particle. In field theories, Lie algebra generators are usually realised as integrals over local conserved currents. In the case of the two dimensional non-linear sigma models, classical non-local conserved charges were constructed as double integrals over the product of two currents at different points in space \cite{Luscher:1977rq}. For the $O(n)$ sigma model, these charges were argued to survive quantisation \cite{Luscher:1977uq}. These non-local charges are the first in a series of infinitely many non-local conserved charges. In \cite{Luscher:1977uq}, it was also argued that this implies the factorisation of the scattering matrix into two-particle S-matrices, and the order of the two particle scattering processes does not matter. Furthermore, there can be no particle production or annihilation. This implies that the so-called Yang-Baxter equation holds. As one has a relativistic theory, one can derive a crossing equation for the two particle S-matrix, and the S-matrix should also be unitary and have the right analytic properties. Having established these properties imply that the two particle S-matrix is completely fixed up to a simple CDD factor\footnote{Castillejo-Dalitz-Dyson (CDD) factors appeared first in the study of Low's scattering equation \cite{Castillejo:1955ed}.}, which satisfies the homogeneous crossing and unitarity equations. This implies that the CDD factor is a trigonometric function, and it can be fixed if one knows about the particle content of the theory, and hence about the pole structure of the S-matrix. This line of thought was proposed in the 1970s, see e.g. \cite{Karowski:1977th,Zamolodchikov:1978xm}. The two particle S-matrix allows one to reconstruct all S-matrices by the property of factorisability. Often in the literature, the higher conserved charges implying factorisability of the S-matrix are local charges \cite{Iagolnitzer:1977sw,Shankar:1977cm, Parke:1980ki}, which can be used to disentangle a multi-particle scattering process. Indeed, it seems for most known cases of integrable models, local and non-local conserved charges coexist. Furthermore, their origin can often be traced back to a quantum groups. The arguments in \cite{Luscher:1977uq} for the factorisation are however given by using non-local charges.\\

As the conserved charges imply the factorisation of the S-matrix, this factorisation and the absence of particle production is in turn often taken as a definition of quantum integrability. In this paper we will give a reason why this makes sense. We will show that the scattering matrices for at least some integrable models also follow directly from the underlying infinite dimensional quantum group symmetry. If one takes for instance the two dimensional principal chiral field with $\sun\times\sun$ symmetry, similar arguments as for the $O(n)$ sigma models imply the factorisability of the S-matrix. The two particle S-matrix has been derived and used to construct so-called Bethe equations \cite{Polyakov:1983tt,Wiegmann:1984pw,Wiegmann:1984ec}, which are periodicity conditions on the wave function. It is well known that the principal chiral field also has conserved non-local charges, similar in nature to the charges of the $O(n)$ model \cite{Luscher:1977uq}. These charges are related to the Yangian \cite{MacKay:1992rc}. The matrix part of the scattering matrix is proportional to two copies of the standard Yang R-matrix, which is simply of the form $\rmat\propto\idm+\frac{1}{u}\perm$. Here, $\perm$ is the permutation operator, and $u=u_1-u_2$ is the difference of rapidities of the scattered particles. This matrix is well known to be invariant under the Yangian. However, it is not clear at all that the complicated scalar factor of the S-matrix should come directly from the Yangian. Furthermore, even though the conserved charges have, in many cases, been rigorously constructed in the classical theories, their survival in the quantum theory is usually not easy to see. We refer the reader to the discussion in \cite{Abdalla:1986xb,Evans:1999mj} about conserved charges of the principal chiral field.\\

Let us have a look at the mathematical side of Yangians. They are infinite dimensional symmetry algebras extending traditional Lie algebras. Indeed, they can be considered as deformations of the universal enveloping algebra of $\palg{g}$, which is the algebra of polynomials in $u$ with values in a simple Lie algebra $\alg{g}$. The original definition of Yangians was given in \cite{Drinfeld:1985rx}, where the algebraic structure behind rational solutions to the Yang-Baxter equation was investigated. It is named after C.N. Yang, who found the first rational R-matrix with $\sun$ invariance as given above. If one thinks of scattering problems, the parameter $u$ will be related to the rapidity of the particles. An important property of Yangians is that they allow for a so-called quantum double construction \cite{Khoroshkin:1994uk}. On the classical level, this ``doubles'' the polynomial algebra $\palg{g}$ to the loop algebra $\lalg{g}$. The importance of this construction lies in the fact that these quantum doubles have a universal R-matrix, which is an R-matrix defined in terms of the abstract generators of the Yangian Double. It satisfies an abstract Yang-Baxter equation, and is inverted by the action of the antipode map. Upon specifying a representation, this universal R-matrix should automatically lead to crossing invariant solutions of the Yang-Baxter equation. One problem is that simple evaluation representations, which enable one to evaluate the universal R-matrix on a representation, do not exist for Yangians of algebras other than $\slnm$. Furthermore, the universal R-matrix is defined as an infinite product, as there are infinitely many generators of the Yangian. There are ordering issues, and, to our knowledge, such an explicit evaluation of the whole R-matrix has not been done so far for algebras of rank greater than one. Another problem of the construction of \cite{Khoroshkin:1994uk} is that the resulting S-matrices on representations are not unitary. Consequently, the interest in universal R-matrices in the physics literature has been somewhat limited.\\

Another way to construct the Yangian is via monodromy matrices. These are the same monodromy matrices underlying the algebraic Bethe ansatz, or quantum inverse scattering method (see e.g. \cite{Faddeev:1990qg,Faddeev:1996iy}), which is a method to find the spectrum of integrable models. Indeed, consider e.g. integrable models with $\sun$ invariance. The monodromy matrices are usually thought of as $n\times n$ matrices, where each entry of the matrix corresponds to an abstract generator. For a realisation on a spin chain, the abstract part becomes a big matrix acting on the Hilbert space of the spin chain. Additionally, the monodromy matrix depends on a spectral parameter, and the expansion in this parameter around infinity leads, at lowest order, to the usual Lie algebra generators, whereas the higher orders correspond to the non-local charges discussed before. Furthermore, the trace of the monodromy matrix also contains the local conserved charges, if one expands about certain special poles. The commutation relations of the Yangian are defined via the famous RTT relations. These impose relations for the entries of the monodromy matrix, which are essentially the Yangian generators. RTT relations are derived from the Yang-Baxter equation, as for lattice models the monodromy matrix can be constructed as a product of R-matrices. Indeed, the RTT relations were known before the formal, mathematical definition of the Yangian given in \cite{Drinfeld:1985rx}.\\

Yangians also play an important role in integrable spin chain models. Indeed, the simple Heisenberg XXX spin chain with nearest-neighbour interaction and its generalisation with $\sun$ symmetry are invariant under Yangian charges, at least in the infinite length limit, when boundary terms can be neglected. The spectrum can be computed with the algebraic Bethe ansatz (see \cite{Faddeev:1996iy}), where the fundamental object is the $\sun$ monodromy matrix, which, as we argued before, is the defining object of the Yangian. Furthermore, one can also obtain Bethe equations by considering excitations over a ferromagnetic vacuum, and the resulting S-matrix is invariant under the Yangian of $\alg{su}(n-1)$. Its scalar factor in this case is trivial. Interestingly, if one scatters spinons defined over the antiferromagnetic vacuum, one gets an S-matrix invariant under the Yangian of $\sun$, which now has a crossing invariant dressing factor resembling the structure as found in other relativistic integrable models with $\sun$ symmetry (see also \cite{Faddeev:1996iy}).\\

Indeed, the Yangian seems to be the central algebraic object underlying a range of integrable models\footnote{Furthermore, many integrable models without Yangian symmetry have q-deformed symmetry. Indeed, trigonometric S-matrices are related to quantum affine algebras, which in some sense can be considered as deformations of the Yangian. q-deformed quantum groups appeared at the same time as Yangians \cite{Drinfeld:1985rx,Jimbo:1985zk}}. In particular, the Heisenberg XXX spin chain, which appears in one-loop \symng, is invariant under a Yangian, and the charges on the string side of the AdS/CFT correspondence are also related to a Yangian. Hence, it is natural to expect that the Yangian plays an important role in the AdS/CFT correspondence itself. Indeed, also at higher loops in the gauge theory, Yangian charges have been found in \cite{Agarwal:2004sz,Zwiebel:2006cb}, even though just in restricted subsectors. A full proof of quantum integrability is currently beyond reach. Indeed, quantum integrability is hard to show even in standard relativistic integrable systems. However, in many classically integrable models, one could proceed by assuming that the conserved charges survive quantisation. Provided that quantum conserved charges exist, one can conjecture the factorisation of the S-matrix, and derive the two particle S-matrix. The same procedure was proposed for the AdS/CFT correspondence \cite{Staudacher:2004tk}, and all-loop Bethe equations, conjectured to describe the spectrum of long operators, were written down in \cite{Beisert:2005fw} without a proof of quantum integrability. These Bethe equations were later derived from the asymptotic S-matrix, which was obtained purely from the Lie algebra symmetry \cite{Beisert:2005tm}. This S-matrix scatters magnons on the spin chain underlying \sym, or world sheet excitations in the light-cone gauged fixed string theory \cite{Klose:2006zd}. Interestingly, the underlying Lie algebra of the S-matrix is the centrally extended $\alg{psu}(2|2)$ algebra, and the central charges already encode the momentum dependence of the underlying particles. $\psu$ is the only simple Lie superalgebra which allows for more than one non-trivial central charge, and the existence of those charges modifies the behaviours of tensor products of representations. In particular, the tensor product of two fundamental representations is generically irreducible, which explains why the S-matrix is fixed without referring to a higher symmetry such as the Yangian. However, the S-matrix was later shown to be still invariant under a Yangian based on centrally extended $\alg{psu}(2|2)$ symmetry \cite{Beisert:2007ds}. \\

An important piece of the S-matrix and the Bethe equations which is not fixed by Lie algebra symmetries is the scalar prefactor. As string theory in the light cone gauge is not relativistically invariant, the S-matrix is not of difference form. This is in contrast to standard S-matrices related to the Yangian and caused by the fact that the underlying Lie algebra is centrally extended. However, crossing seems an intrinsic algebraic feature of quantum groups. Indeed, in \cite{Janik:2006dc}, this fact was used, without knowing the underlying Hopf algebra, to derive a crossing equation for the AdS/CFT S-matrix. Furthermore, on the classical string theory side, the Bethe ansatz was conjectured to contain a complicated dressing phase \cite{Arutyunov:2004vx}, based on earlier investigations of finite gap solutions in \cite{Kazakov:2004qf}. The one-loop corrections to this phase were subsequently investigated in \cite{Beisert:2005cw,Hernandez:2006tk} and analytically continued to weak coupling, where the phase is trivial at the leading few orders in perturbation theory. Finally, the all loop result of the phase was conjectured in a strong coupling asymptotic expansion in \cite{Beisert:2006ib}, and its weak coupling continuation in \cite{Beisert:2006ez}. Later, the crossing equation was solved explicitly and shown to reproduce the conjectured phases \cite{Volin:2009uv}.\\

Having included the correct dressing phase, the asymptotic Bethe equations seem to correctly describe the whole spectrum of long operators in the gauge theory, or fast spinning strings in \ads. This has been confirmed in many tests based on semi-classical, fast spinning strings of the type introduced in \cite{Gubser:2002tv,Frolov:2002av}. However, the asymptotic Bethe equations seem to break down when the interaction range of the Hamiltonian becomes longer than the spin chain the Hamiltonian is acting on, as was confirmed in \cite{Kotikov:2007cy}. Luscher corrections can be used to obtain the first corrections to the anomalous dimension of short operators, such as the Konishi operator \cite{Bajnok:2008bm}. These corrections were first introduced on the strong coupling side \cite{Janik:2007wt}, where finite-size corrections to certain string excitations dual to the magnons on the gauge side were obtained. These excitations on the string side are called giant magnons, and were initially introduced in \cite{Hofman:2006xt}. In general, the order from which finite size corrections should play a role is not as clear as from the weak coupling side. Indeed, in \cite{Rej:2009dk}, the strong coupling expansion of the Bethe equation for the Konishi operator was investigated, and the first terms agree with the physical expectations. Higher terms violate the analytic structure and depend logarithmically on the coupling constant. Interestingly, the Luscher corrections still depend on the asymptotic S-matrix, which now has to depend on different representations, namely on those describing bound state particles in the mirror theory. The mirror theory is related to the original theory by Wick rotation. The reason for the appearance of the mirror theory is as follows. To define an S-matrix, one needs to define scattering states. This requires one to separate in and out state, so one should define the theory on an infinite line. If the theory is defined on a cylinder of finite length, one can Wick rotate, and consider the theory with finite periodic time, or, analogously, finite temperature, and take instead the Wick rotated time coordinate, which is now spatial, to infinity. Hence, one can define a scattering theory for the Wick rotated theory. This trick was introduced in \cite{Zamolodchikov:1989cf} to derive spectral equation for the original theory. These equations are called Thermodynamic Bethe ansatz equations, or TBA equations. The main problem for the applicability of this trick to the AdS/CFT correspondence is that string theory in the light cone gauge is not Lorentz invariant, hence the Wick rotated mirror theory looks quite different from the original theory. Nevertheless, the mirror theory was investigated in \cite{Arutyunov:2007tc}. Later, another important ingredient towards the TBA equations, the string hypothesis, was written down in \cite{Arutyunov:2009zu}. Ultimately, TBA equations and their associated Y-systems were conjectured in \cite{Gromov:2009tv,Gromov:2009bc,Arutyunov:2009ur,Bombardelli:2009ns}. In principle, TBA equations only describe the ground state energy, but analytic continuation often leads to excited state TBA equations \cite{Dorey:1996re}, which are supposed to describe the whole spectrum of any operator. \\

In the present state, the TBA equations are fairly difficult to evaluate for specific states, even just numerically. Indeed, a goal would be to find an integral equation, similar to the Destri-deVega equation describing the spectrum of sine-Gordon theory \cite{Destri:1997yz}, or the integral equation describing the spectrum of the $SU(2)\times SU(2)$ principal chiral field \cite{Gromov:2008gj}. What we should note is that the bound state S-matrices underlying the TBA equations are still invariant under the same Yangian of the centrally extended $\alg{psu}(2|2)$ symmetry \cite{Arutyunov:2009mi}. Only the representation is different to the fundamental S-matrix, as we are now dealing with bound states in the mirror theory. Hence, even if the definition of Yangian charges on field theories on a finite cylinder is not well-defined, the Yangian is still present in the corresponding mirror theory on the infinite line.\\

Finally, we note that on the \sym side, the S-matrix we are considering is not directly related to the gluon amplitudes in four space-time dimensions. What we scatter here are magnons on the spin chain. In this sense, the S-matrix we are investigating has the sole purpose to find the right spectrum of the single-trace local operators. Excitingly, it was found that the tree level space-time gluon scattering amplitudes in \sym are also invariant under a Yangian \cite{Drummond:2009fd}. A further investigation of the Yangian might shed light to the behaviour of the amplitudes at higher loops as well. We refer the reader to \cite{Drummond:2010qh,Drummond:2010uq,Beisert:2010gn,Beisert:2010jq} for more information on Yangians in scattering amplitudes, and \cite{Alday:2008yw,Alday:2009zza} for general references on scattering amplitudes.

\subsection*{Outline}

We have seen that Yangians appear in many integrable systems, such as the XXX Heisenberg spin chain, relativistic integrable field theories and the AdS/CFT correspondence in the large $N$ limit. Hence, it is well worth to gain a better understanding of Yangians. In this paper we will introduce mathematical features of Yangians in chapter \ref{ch:yangians}. Besides recalling some definitions and features from the literature, we also include some new results concerning a modification of the universal R-matrix of Yangians, such that its representation satisfies the physically important property of unitarity and crossing. Furthermore, we obtain R-matrices for all superalgebras of type $\slnm$, generalising previous work on R-matrices based on simple Lie algebras. These results are based on recent work with Adam Rej \cite{Rej:2010mu}. Chapter \ref{ch:yangians} also focuses particularly on the centrally extended $\psu$ algebra, which is the symmetry algebra appearing in the light cone string theory. It has some mathematically distinct features, making it special amongst all simple superalgebras. We also include some results found in collaboration with Alessandro Torrielli \cite{Spill:2008tp} concerning the Yangian in Drinfeld's second realisation, which is suitable for the construction of the universal R-matrix as well as for the study of the representation theory. As many physical results discussed in the later chapters can be derived directly from the Yangian, chapter \ref{ch:yangians} plays a major role in this paper.\\

Chapter \ref{ch:integrablemodels} reviews some well known integrable systems, such as the principal chiral field, or $\sun$ symmetric Heisenberg spin chains. We do not include new physical results, but propose a new powerful derivation of the scattering matrix of the principal chiral field and other relativistic integrable field theories. Namely, we argue that one does not need to separate ordinary Lie algebra symmetry from the additional requirements of the Yang-Baxter equation and crossing and unitarity. Instead, the Yangian provides a unified algebraic framework, which directly leads to the full scattering matrix, up to a possible CDD factor. Consequently, we derive the S-matrix of the principal chiral field directly from the universal R-matrix of the underlying Yangian.\\

In chapter \ref{ch:adscft}, we review some features of the AdS/CFT correspondence, focusing on its integrability properties. \\

Chapter \ref{ch:smatrix} deals particularly with the S-matrix arising in the light-cone gauge of string theory, or in the scattering problem of magnons of the spin chain on the Yang-Mills side of the AdS/CFT correspondence. We include a derivation of the S-matrix at strong coupling purely from classical Yangian symmetries, as found in the publication \cite{Beisert:2007ty} in collaboration with Niklas Beisert. Furthermore, we derive the weak coupling S-matrix also from the Yangian. These results have been published in \cite{Spill:2008yr}. As we have established the mathematical machinery of universal R-matrices in chapter \ref{ch:yangians}, we can obtain this S-matrix in a straightforward fashion. Finally, we speculate about the form of the all-loop universal R-matrix.

\section{Yangians of Lie Superalgebras}\label{ch:yangians}

In this chapter we will assemble some features of Yangians and their classical
limits. Yangians are quantum groups which enlarge Lie algebras to
infinite dimensional symmetry algebras. They play a prominent role in many integrable
models. In particular, they are often related to the infinitely many conserved
charges of an integrable model. One way to define them is through the monodromy matrices and the RTT relations. This was the way Yangians historically appeared in the investigation of the quantum inverse scattering problem, see the reviews \cite{Faddeev:1990qg,Faddeev:1996iy}. The mathematics of the Yangian in this realisation is reviewed in \cite{Molev:2002}. From the perspective of S-matrices, Yangians are
closely related to rational solutions of the Yang-Baxter equation.
They are mathematically defined as deformations of the algebra of
polynomials with values in a Lie algebra. So before discussing Yangians in section \ref{sec:yangians}, we will start by describing these
polynomial algebras and their related loop and Kac-Moody algebras in section
\ref{sec:lieandloop}. In particular, we will show how to construct the loop algebra as a classical double of the polynomial algebra. This double automatically contains a classical r-matrix, which satisfies the classical Yang-Baxter equation. In section \ref{sec:yangians} we will then construct a Yangian double, which contains a quantum R-matrix satisfying the quantum Yang-Baxter equation. Furthermore, the Yangian R-matrix satisfies the crossing equation, and can be chosen such that it leads to unitary R-matrices on representations. In particular, we directly recover the integrable and crossing-invariant $\sun$ S-matrices. This is a new result and can be found in the preprint \cite{Rej:2010mu}, written in collaboration with Adam Rej. Furthermore, the analysis done in this paper is valid for Yangians based on Lie superalgebras, whereas previous investigations were only true for simple Lie algebras.\\

The focus in this work lies on Lie algebras of type $\slnm$ and their Yangians, as the physical applications we are dealing with in the later chapters have this symmetry. We will also discuss the peculiarities of the special series $\slnn$ and the algebra $\psltt$. They have the unique property that they allow for a non-trivial central extension. These central charges will be of greatest importance for the physics in relation with the S-matrix of the AdS/CFT correspondence, as discussed in chapter \ref{ch:smatrix}. In particular, $\psltt$ allows for three independent central charges. This distinguishes it from all other simple Lie superalgebras. We contribute some new results towards the Yangian of $\psltt$ in Drinfeld's second realisation, which is important for the construction of the universal R-matrix and the study of the representation theory (see e.g. the recent studies of long representations of $\psucentral$ \cite{Arutyunov:2009pw}). These results were mainly published in our paper \cite{Spill:2008tp} with Alessandro Torrielli, but we have changed the conventions and also included some new results.

\subsection{Lie Algebras and Lie Bialgebras}\label{sec:lieandloop}

We start by recalling important definitions for Lie algebras and Lie superalgebras in section \ref{sec:liealgebras}, and for loop algebras in section \ref{sec:loopalg}. General references for this part are \cite{Fuchs:1997jv,Kac:1977em,Kac:1977qb}. In section \ref{sec:liebialgebras}, we define Lie bialgebras, and focus on the example of loop algebras, which can be constructed as a classical double from the polynomial algebra. One can find more background information on Lie bialgebras as well as classical r-matrices in the textbook \cite{Chari:1994pz}.

\subsubsection{Lie Superalgebras}\label{sec:liealgebras}

A Lie superalgebra $\alg{g}$ is a vector space equipped with a Lie superbracket
or supercommutator. Let us denote a basis by

\[
\gen{J}^a,\quad a=1,\dots , \textit{dim}(\alg{g}),
\]
then we denote the supercommutator by

\begin{equation}\label{def:liecomm}
\scomm{\gen{J}^a}{\gen{J}^b}=\scons\gen{J}^c ,
\end{equation}
where $\scons$ are the structure constants of $\alg{g}$. We denote by

\beq 
|a| \equiv |\gen{J}^a| = \left\{ \begin{array}{ccc}0&,\quad a \text{ bosonic}&  \\
1&,\quad a \text{ fermionic}&  \end{array} \right.
\eeq
the degree of the generator $\gen{J}^a$. If $\gen{J}_{UEA}^a$ is the element in the universal enveloping algebra $\alg{U(g)}$ corresponding to $\gen{J}^a$, then the supercommutator is given by 

\begin{equation}
\scomm{\gen{J}_{UEA}^a}{\gen{J}_{UEA}^b} = \gen{J}_{UEA}^a\gen{J}_{UEA}^b - (-1)^{|a||b|}\gen{J}_{UEA}^b\gen{J}_{UEA}^a.
\end{equation}
Indeed, whenever we will use a product of the form $\gen{J}^a\gen{J}^b$, it is understood to be in the universal enveloping algebra, so we will henceforth drop the label distinguishing elements from a Lie superalgebra to the corresponding elements in the universal enveloping algebra.

An important concept is that of a non-degenerate invariant supersymmetric
bilinear form. We will always assume its existence for the Lie superalgebras we
are dealing with in this paper, or the existence of an extension of the algebra
which has such form. We will denote this form by

\begin{equation}\label{def:killform}
\kappa^{ab}=\sprod{\gen{J}^a}{\gen{J}^b}.
\end{equation}
For simple Lie algebras this form is just the usual Killing form (up to an
overall factor), and in general on representations one has, up to rescaling,
\[
\kappa^{ab}=\str(\gen{J}_{r}^a\gen{J}_{r}^b),
\]
where $\str$ denotes the supertrace over the corresponding representation space $V_r$. $\gen{J}^a_r$ denotes a representation of $\gen{J}^a$, and we will often drop the subscript as long as this is unambiguous. We also have to assign a degree to the vectors we are representing. Let e.g. the representation space be $n+m$ dimensional, such that the first $n$ elements are bosonic, and the last $m$ are fermionic. Then the generators of the Lie superalgebra can be schematically written in block form
\[
\begin{pmatrix}
A & B \\ 
C & D
\end{pmatrix}, 
\]
such that $A$ is an $n\times n$ matrix acting only on the $n$ bosons, and $D$ is an $m\times m$ matrix acting only on the fermions. As these two blocks to not change the statistics of bosons or fermions they are acting on, they are themselves of bosonic nature, i.e. they constitute the degree $0$ generators of $\alg{g}$. $C$ and $D$ swap bosons and fermions, and are henceforth the fermionic, or degree $1$ generators of $\alg{g}$. The supertrace in this block form is given by
\[
\str\begin{pmatrix}
A & B \\ 
C & D
\end{pmatrix} = \tr(A) -\tr(D).
\]
As the Killing form, i.e. the supertrace
over the adjoint representation, vanishes in the case of $\alg{sl}(n|n)$\footnote{$\alg{sl}(n|n)$ itself is not simple, but the algebra obtained by projecting the centre out is simple. These details will be discussed in more detail in later chapters.}, which
is of importance for the physics dealt with in this work, we will define the
invariant form as the supertrace over the fundamental representation. A further property is that the form acts on the bosonic and fermionic blocks separately, i.e. 

\[
\kappa^{ab} = 0
\]
if $|a|=0$ and $|b| =1$, or the other way round.
Supersymmetry means that the form satisfies

\begin{equation}
\kappa^{ab}=(-1)^{|a||b|}\kappa^{ba} = (-1)^{|a|}\kappa^{ba},
\end{equation}
where the last equation holds because of the equation above. Invariance is the property that

\[
\sprod{\gen{J}^a}{\scomm{\gen{J}^b}{ \gen{J}^c}}=\sprod{\scomm{\gen{J}^a}{ \gen{J}^b}}{\gen{J}^c}.
\]
{\bf Chevalley-Serre basis}

The Chevalley-Serre basis allows for a unified description of contragredient Lie superalgebras\footnote{These are basic, classical Lie superalgebras, i.e. simple Lie superalgebras where the bosonic part acts on the fermionic as a completely reducible representation, and where the algebra admits a non-degenerate invariant bilinear form.} by encoding their
commutation relations with their Cartan matrix $A$, or with their corresponding
Dynkin diagram. It is also useful for
applications in physics, as it directly involves the important Cartan
subalgebra, the maximal subalgebra of
commuting generators.

Let $A$ be the $r\times r$ Cartan matrix, $\csgh{i}$, $\csgp{i}$ and $\csgm{i}$
be the Cartan generator and 
positive and negative simple root generators, respectively. Then these
generators satisfy

\<\label{eq:chevserre}
\comm{\csgh{i}}{\csgp{j}}=A_{ij}\csgp{j}\nln
\comm{\csgh{i}}{\csgm{j}}=-A_{ij}\csgm{j}\nln
\scomm{\csgp{i}}{\csgm{j}}=\delta_{ij}\csgh{i}.
\>
These $3r$ generators form, in general, not a linear basis of the Lie
superalgebra. When commuting positive simple 
root generators with each other one will get, in general, new positive root
generators. For contragredient Lie superalgebras,
one does only get a finite number of additional positive root generators, and a
corresponding number of negative roots. Note that for bosonic simple Lie algebras, simplicity of the algebra requires the Cartan matrix to be non-degenerate. We will see that this condition of non-degeneracy is not necessary in the case of Lie superalgebras. In particular, the important series of $\pslnn$ has a degenerate Cartan-matrix.
How to precisely construct the additional generators is encoded in the Serre
relations

\[
\scomm{\csgp{i}}{\scomm{\csgp{i}}{\scomm{...}{\scomm{\csgp{i}}{\csgp{j}}}}}=0,
\]
where one applies the commutator $|A_{ij}|+1$ times. In the case of simple Lie
algebras, these Serre relations together 
with the Chevalley-Serre relations completely determine the Lie algebra
structure. Hence, all information is encoded in
the Cartan matrix. For superalgebras, however, one usually needs some additional
Serre relations. We will later spell 
them out in the case of $\glnm$.

Let us note that the Cartan Matrix can usually be derived from the inner product
in the root space. This is also related to 
the invariant form $\kappa$. Indeed, if $\alpha_i$, $i=1,\dots r$ are the simple
roots corresponding to the root generators 
$\csgp{i}$, and the root product is denoted by $\sprod{\alpha_i}{\alpha_j}$,
then the Cartan matrix can be written as follows:
\<
\tilde{A}_{ij} = 2\frac{\sprod{\alpha_i}{\alpha_j}}{\sprod{\alpha_i}{\alpha_i}}, \text{if }\sprod{\alpha_i}{\alpha_i}\neq 0\nln
\tilde{A}_{ij} = \frac{\sprod{\alpha_i}{\alpha_j}}{\sprod{\alpha_i}{\alpha'_i}}, \text{if }\sprod{\alpha_i}{\alpha_i}= 0.
\>
Here, $\alpha'_i$ is chosen such that $\sprod{\alpha_i}{\alpha'_i}$ is minimal. Note that for simple Lie algebras $\sprod{\alpha_i}{\alpha_i}\neq 0$, so one can always normalise the diagonal elements to $2$. In this paper, we will usually work with the symmetric Cartan matrix, which is simply given by

\[
A_{ij} = \sprod{\alpha_i}{\alpha_j}.
\]
It can be obtained from $\tilde{A}_{ij}$ by multiplication with a diagonal matrix.\\

The whole Cartan-Weyl basis, which is a particularly nice form of a linear basis
of a simple Lie algebra, is parametrised by a set of all roots $\beta_k$. To each
such root belongs
a root generator $\cwgp{\beta_i}$ if $\beta_i$ is positive, and a second generator
$\cwgm{\beta_i}$ if
it is negative. Furthermore, there are Cartan generators $\cwgen{H}_\beta$, which
together form a vector space dual
to the root space, i.e. one has
$\cwgen{H}_\beta(\gamma)=\gamma(\cwgen{H}_\beta)=\sprod{\beta}{\gamma}$.
The number of linearly independent Cartan generators is by definition the rank $r$ of the Lie
superalgebra, and, upon rescaling, one can obtain the Chevalley-Serre generators $\csgen{H_i}$.
For contragredient Lie superalgebras, the root space can be degenerate, but nevertheless the Cartan-Weyl 
basis and the root space are useful concepts there.

\subsubsection{Loop Algebras}\label{sec:loopalg}

Let $\alg{g}$ be a Lie superalgebra with a linear basis $\gen{J}^a$ as before.

The algebra $\palg{g}$ is defined as the algebra of polynomials in $u$ with
values in $\alg{g}$, i.e. the elements are linear combinations of the monomials

\[
\gen{J}^a_n = u^n\gen{J}^a_0 ,\quad \quad a=1,\dots , \textit{dim}(\alg{g}),
\quad n=0,\dots,\infty,
\]
such that the generators of zeroth degree form the basis of the Lie superalgebra
$\alg{g}$, i.e. $\gen{J}^a_0=\gen{J}^a$. 

Likewise, we define the loop algebra $\lalg{g}$ to be the algebra of all Laurent
Series in $u$ with values in $\alg{g}$, i.e. it is generated by 

\[
\gen{J}^a_n = u^n\gen{J}^a_0 ,\quad \quad a=1,\dots , \textit{dim}(\alg{g}),
\quad n=-\infty,\dots,\infty.
\]
In this paper, we will always work with completed algebras allowing for linear
combinations of infinitely many generators.

In terms of the structure constants of the underlying Lie algebra \eqref{def:liecomm} the
commutation relations of the loop and polynomial algebras read

\begin{equation}
\scomm{\gen{J}^a_n}{\gen{J}^b_m}=\scons\gen{J}^c_{n+m} .
\end{equation}
From this relation it is clear that $\palg{g}$ is a subalgebra of $\lalg{g}$. Likewise, one can define a closed subalgebra $\nalg{g}$ which consists of generators of purely negative degree.

We also introduce an invariant form on the loop algebra: We extend the bilinear invariant 
form \eqref{def:killform} of the Lie superalgebra $\alg{g}$ to loop algebras $\lalg{g}$ by putting

\begin{equation}
\sprod{\gen{J}^a_n}{\gen{J}^b_m}=\textit{Res}(\str(\gen{J}^a_n
\gen{J}^b_m))=\kappa^{ab}\delta_{n,-m-1}. 
\label{eq:innerprod}
\end{equation}
It is straightforward to see that this product is again invariant. Furthermore, if one considers the generators of $\lalg{g}$ as functions of $u$ with values in $\alg{g}$, this inner product can be seen as taking the residue over $u$. This also means that this product splits $\lalg{g}$ in such a way that 
\<
\sprod{\palg{g}}{\palg{g}} = 0,\nln
\sprod{\nalg{g}}{\nalg{g}} = 0,\nln
\sprod{\palg{g}}{\nalg{g}} \neq 0.
\>
This way of splitting the loop algebra will be important for the construction of the classical r-matrix.

Note that if we want to describe the loop algebra by roots, we can take the set of roots $\beta_i$ of the underlying Lie superalgebra, and add one imaginary root $\delta$. If, as before, one introduces a Chevalley-Serre basis for the Lie superalgebra, i.e. generators $\csgh{i}$, $\csgp{i}$ and $\csgm{i}$, then the generators $\csgp{i,k}$ and $\csgm{i,k}$ correspond to the roots $\alpha_i + k \delta$ and $-\alpha_i + k \delta$, respectively. Likewise, one can add $k \delta$ to the Cartan generators $\csgh{i}$, denoting them by $\csgh{i,k}$. We will refer to all $\csgh{i,k}$ as Cartan generators, as they commute with each other. However, as they are now related to the imaginary root vector $\delta$, $\csgh{i,k}$ are sometimes called imaginary roots in the literature.

Finally, we would like to mention that loop algebras have a central extension and an external derivation, leading to the definition of affine Kac-Moody algebras. We will not deal with these extensions in this paper. However, it is interesting to note that the mathematics leading to this central extension, as well as the structure of the derivation, is quite similar to the central extension of the simple Lie superalgebras $\pslnn$ to $\slnn$, which we will discuss in section \ref{sec:pslcentral}.

\subsubsection{Lie Bialgebras}\label{sec:liebialgebras}

A Lie bialgebra $\alg{g}$ is a Lie superalgebra with an additional
structure $\cobra:\alg{g}\rightarrow\alg{g}\wedge\alg{g}$ called the cobracket, which has to satisfy the
so-called cocycle condition 

\begin{equation}
\cobra(\scomm{\gen{J}^a}{\gen{J}^b}) =
\scomm{\cobra\gen{J}^a}{\gen{J}^b}+\scomm{\gen{J}^a}{\cobra\gen{J}^b}
\label{eq:cocycle}
\end{equation}
and is skew-symmetric as well as linear. Here, by
$\scomm{\cobra\gen{J}^a}{\gen{J}^b}$ we mean the adjoint action of $\gen{J}^b$
on $\cobra\gen{J}^a$, i.e.
$\scomm{\cobra\gen{J}^a}{\gen{J}^b}=\scomm{\cobra\gen{J}^a}{\gen{J}^b\otimes
1+1\otimes\gen{J}^b}$. The cocycle condition just means that the cobracket acts
as a derivation on the Lie bracket. Furthermore, the skew-symmetric, or anti-supersymmetric tensor product is defined as
\[
 \gen{J}^a\wedge\gen{J}^b = \gen{J}^a\otimes\gen{J}^b-(-1)^{|a||b|}\gen{J}^b\otimes\gen{J}^a,
\]
i.e. in case both generators are fermionic, the tensor product is in fact symmetric.\\

The cobracket will become the coproduct upon
quantisation, i.e. the structure equipping Hopf algebras with an action on
tensor products of representations. Furthermore, the cobracket is the dual to
the normal Lie bracket on the dual Lie algebra. That is, if, as before, $\scons$
are the structure constants of the Lie algebra, and the cobracket has structure
constants $\bcons$, i.e.

\begin{equation}
\cobra(\gen{J}^c)=\bcons \gen{J}^a\otimes\gen{J}^b, 
\end{equation}
then the dual Lie superalgebra $\alg{g^*}$ with generators $\gen{J}_a$ has
commutations relations 

\begin{equation}
\scomm{\gen{J}_a}{\gen{J}_b}=\bcons\gen{J}_c,
\end{equation}
The dual basis is fixed by the usual requirement $\sprod{ \gen{J}_a}{\gen{J}^b}=\delta_a^b$. 
Then this means that the structure constants of the commutator of the dual $\alg{g^*}$ are the same as the structure
constants of the cobracket of $\alg{g}$. Likewise,  $\alg{g^*}$ can also be promoted to a Lie
bialgebra with the cobracket

\begin{equation}
\cobra(\gen{J}_c)=-\scons \gen{J}_a\otimes\gen{J}_b. 
\end{equation}

As we assume the existence of an non-degenerate form $\kappa^{ab}$ we just have 

\[
\gen{J}_a = \kappa_{ab} \gen{J}^b,
\]
with $\kappa_{ab}\kappa^{bc}=\delta_a^c$.

{\bf Quasitriangular Lie bialgebras}\label{sec:quasitribi}

In this paper we are only interested in a particular subclass of bialgebras,
namely those where the cobracket is generated by an r-matrix. These Lie
bialgebras are called coboundary. 

A classical r-matrix $\crmat$ is an element in $\alg{g}\otimes\alg{g}$ such that

\begin{equation}
\cobra(\gen{J}^a) = \scomm{\gen{J}^a}{\crmat}.
\label{eq:rcobra}
\end{equation}
For generic $\crmat$ the above equation defines a cobracket precisely if
$\crmat_{12} + \crmat_{21}$ and $\cybe{r}$ are invariant under the adjoint
action of $\alg{g}$. Here, we introduced the notation $\crmat_{kl}$ which means
that the first tensor factor in $\crmat$ lives in the $k$'th tensor product
under consideration, and the second in the $l$'th. That is, if $\crmat =
\sum_i{a_i\otimes b_i}$, for some elements $a_i, b_i$ of the Lie algebra, then
$\crmat_{21} = \sum_i{b_i\otimes a_i}$. If one is working on higher tensor
products one should think of the missing coefficients as identities, i.e.
$\crmat_{12} = \sum_i{a_i\otimes b_i\otimes 1}$ if we consider triple tensor products.

Now in what follows we want to consider r-matrices which satisfy the classical
Yang-Baxter equation

\begin{equation}
\cybe{r}=0 .
\label{eq:cybe}
\end{equation}
If the classical Yang-Baxter equation holds we call the Lie bialgebra
quasi-triangular.

Before proceeding, let us look at the example of the polynomial algebra
$\palg{g}$. Then one can see that

\begin{equation}
\crmat = \frac{t}{u_1-u_2},
\label{eq:rationalr}
\end{equation}
with $t=\kappa_{ab}\gen{J}^a\otimes\gen{J}^b$ being the quadratic Casimir acting
on the tensor product, is a classical r-matrix, even though it is not an element of $\palg{g}\otimes\palg{g}$. To show this we note that

\<
\comm{t_{12}}{t_{13}+t_{23}}=\comm{t_{12}}{\kappa_{ab}(\gen{J}^a\otimes
1\otimes\gen{J}^b+1\otimes\gen{J}^a\otimes
\gen{J}^b)}=\kappa_{ab}\comm{t_{12}}{\gen{J}^a}\otimes \gen{J}^b=0.\nln
\>
Likewise, $\comm{t_{23}}{t_{12}+t_{13}}=0$. 

But then

\<\label{eq:casicomm}
&&\comm{\frac{t_{12}}{u_1-u_2}}{\frac{t_{13}}{u_1-u_3}}+\comm{\frac{t_{13}}{
u_1-u_3}}{\frac{t_{23}}{u_2-u_3}}\nonumber\\&=&\comm{t_{12}}{t_{23}}\left(\frac{
1}{u_2-u_1}\frac{1}{u_1-u_3}+\frac{1}{u_1-u_3}\frac{1}{u_3-u_2}
\right)\nonumber\\
&=&-\comm{\frac{t_{12}}{u_1-u_2}}{\frac{t_{23}}{u_2-u_3}},
\>
so the classical Yang-Baxter equation holds. Note that the polynomial algebra acts on different tensor factors with a different spectral parameter $u_i$. As \eqref{eq:rationalr} is itself not an element of $\palg{g}\otimes\palg{g}$, it is henceforth more natural to consider it as an r-matrix of the loop algebra $\lalg{g}$. We will investigate this in the next section. Here, we would like to mention that \eqref{eq:rationalr} still defines a consistent cobracket on $\palg{g}$ via \eqref{eq:rcobra}, which is given by
\[
\cobra{(\gen{J}^a_k)} = \scomm{\gen{J}^a_k}{\crmat} = \frac{1}{2}\sum_{l=0}^{k-1}f^a_{cd}\gen{J}^c_l\wedge\gen{J}^d_{k-l-1}.
\]
Mathematically, one is interested in classifying the solutions to the classical
Yang-Baxter equation. Indeed, the r-matrix $\frac{t}{u_1-u_2}$ is in fact the
simplest solution of the classical Yang-Baxter equation in the case where $\crmat$ is a rational
function of two spectral parameters $u_1$ and $u_2$. Other cases involve spectral parameters which are periodic or elliptic functions of the spectral parameter. If the underlying Lie algebra is simple, the solution to the classical Yang-Baxter equation have been classified in \cite{Belavin:1982}.

\subsubsection{The Classical Double}\label{sec:classdouble}

Given a Lie bialgebra $\alg{g}$ with structure constants $\scons$, $\bcons$ as
before, the classical double is a construction to automatically obtain a
quasitriangular Lie bialgebra. Indeed, if $\gen{J}^a$ is a basis of $\alg{g}$,
and $\gen{J}_a$ is its dual basis in $\alg{g^*}$, then the double of $\alg{g}$
is the vector space $\cdoub{g}=\alg{g}+\alg{g^*}$ with the classical r-matrix 

\begin{equation}
\crmat := \sum_a{\gen{J}^a\otimes\gen{J}_a}.
\label{eq:rdouble}
\end{equation}
By plugging this formula into \eqref{eq:rcobra} one can see that this r-matrix equips the double with a cobracket such that,
if we restrict to the subspace $\alg{g}\subset\alg{g}+\alg{g^*}$, we recover the
initial bialgebra structure defined by the structure constants $\scons$, $\bcons$. Likewise, $\alg{g^*}$ is embedded into $\cdoub{g}$ provided that one swaps the tensor factors of the cobracket. Furthermore, the Yang-Baxter equation automatically holds for r-matrices of the form \eqref{eq:rdouble}. As the r-matrix is embedded into $\cdoub{g}\otimes \cdoub{g}$ as
\[
\crmat\in\alg{g}\otimes\alg{g}^*,
\]
we can also consider $\crmat$ as the trivial operator mapping $\alg{g}$ to $\alg{g}$.
Let us see how this construction works for the case where $\alg{g}$ equals the
polynomial algebra $\palg{g}$. We have established the invariant product in the
loop algebra \ref{eq:innerprod}, and can clearly see that generators with
negative degree are paired with non-negative degree generators. Indeed, the dual
generator for $\gen{J}^a_n, n\geq 0$ is $\gen{J}_{a,-n-1} =
\kappa_{ab}\gen{J}^b_{-n-1}$, so the loop algebra $\lalg{g}$ can be seen as a
double of the polynomial algebra $\palg{g}$. Furthermore, the classical r-matrix
is given by

\begin{equation}
\crmat = -\sum_{n=0}^\infty\sum_{a=1}^{\textit{dim}(\alg{g})}{\gen{J}^a_n\otimes\gen{J}_{a,-n-1}}.
\label{def:crmatpos}
\end{equation}
Taking into account that $\gen{J}^a_n = u^n \gen{J}^a$, we get

\begin{equation}
\crmat =\sum_{n=0}{u_1^n \gen{J}^a\otimes u_2^{-n-1}\gen{J}_{a}} =
\frac{t}{u_1-u_2}.
\label{def:crmateval}
\end{equation}
Hence we have derived the classical r-matrix for the polynomial and loop
algebras introduced before. As it appeared via the construction of the double,
it automatically satisfies the classical Yang-Baxter equation, confirming the explicit calculation \eqref{eq:casicomm}.\\

An often overlooked fact in the literature is the fact that there is an ambiguity how to construct the loop algebra as a double from the polynomial algebra. Indeed, one could equally well define the double of $\nalg{g}$, which would yield the classical r-matrix

\begin{equation}
\crmat = \sum_{n=0}^\infty\sum_{a=1}^{\textit{dim}(\alg{g})}{\gen{J}^a_{-n-1}\otimes\gen{J}_{a}^n}.
\label{def:crmatneg}
\end{equation}
On evaluation representations, \eqref{def:crmatneg} would be completely equivalent to \eqref{def:crmatpos}. Indeed, both choices differ just by the invariant element

\begin{equation}
\sum_{n=-\infty}^\infty\sum_{a=1}^{\textit{dim}(\alg{g})}{\gen{J}^a_{-n-1}\otimes\gen{J}_{a}^n}.
\end{equation}
The classical double construction is well-known to lead to a quasi-triangular Lie bialgebra. However, in physics one would like to have a unitary r-matrix, which, at the classical level means that

\[
\crmat(u_1-u_2) = - \crmat(u_2 -u_1).
\]
This is realised on representations for both choices \eqref{def:crmatpos} and \eqref{def:crmatneg}, as the invariant element evaluates to $0$. One can construct a unitary r-matrix, i.e. where the classical r-matrix satisfies

\[
\crmat_{21} = -\crmat_{12},
\]
by choosing the antisymmetric combination

\begin{equation}
\crmat = \half\sum_{n=0}^\infty\sum_{a=1}^{\textit{dim}(\alg{g})}{\gen{J}^a_{-n-1}\wedge\gen{J}_{a}^n}.
\label{def:crmattriangular}
\end{equation}
This r-matrix will still define a Lie bialgebra, but not satisfy the classical Yang-Baxter equation. However, it automatically guarantees unitarity on representations. The analogous case of the Yangian will be discussed in section \ref{sec:universalR}.

\subsection{Yangians}\label{sec:yangians}

The Yangian $\yang{g}$ based on a Lie superalgebra $\alg{g}$ is a deformation of
the polynomial algebra $\palg{g}$, or, more precisely, a deformation of the universal enveloping algebra of $\palg{g}$.
It is a Quantum Group, by which we mean a quasitriangular Hopf Algebra. Hopf
Algebras are appearing naturally in many areas of physics, in particular, in
integrable systems. They equip a symmetry algebra with additional structures such
as the coproduct, which defines the action of the symmetry generators on multiparticle states. Furthermore, they usually posses an
antipode, which is often associated with antiparticle representations. As the coproduct and antipode are rather trivial for ordinary Lie superalgebras, Hopf algebras are not widely studied for systems with only Lie algebra symmetry. The situation changes for most integrable models in two dimensions, which often have a more complicated symmetry structure than just Lie algebras and are indeed often Quantum Groups. 
Quasitriangularity means that the Hopf algebra is equipped with a universal
R-matrix $\rmat$. Such an R-matrix can often be used to define an integrable
model by means of the RTT formalism, where both R and T here are representations
of the same universal R-matrix $\rmat$. Furthermore, the RTT relations can also be used to define the Yangian (see the review \cite{Molev:2002} on Yangians in this realisation). Indeed, there are several equivalent ways to define the Yangian, and it depends on the problem one is studying which approach is the best. Universal R-matrices also give rise to
S-matrices. Quasitriangularity is basically the property which guarantees that the
Yang-Baxter equation as well as a crossing equation hold. Even though it was known before that S-matrices are somehow related to the universal R-matrix, a direct derivation of scattering matrices just from the universal R-matrix is, to our knowledge, a new application of universal R-matrices, and published in the paper \cite{Rej:2010mu} and in this paper for the first time. \\

We begin by giving the definition of Hopf algebras and Quantum Groups in section \ref{sec:quantumgroups}. For further references we refer the reader to the textbooks \cite{Chari:1994pz,Kassel:1995xr}. Then we define the Yangian in the first realisation in section \ref{sec:drinf1} following \cite{Drinfeld:1985rx}, and in section \ref{sec:drinf2} we define the Yangian in Drinfeld's second realisation, following \cite{Drinfeld:1987sy}. The references mentioned deal with Yangians of ordinary Lie algebras, whereas generalisations of the Yangian to superalgebras have been studied e.g. in \cite{Gow:2007th,Gow:2007aa}. The second realisation is suitable for the construction of the Yangian double, which is a quantisation of the classical double of the polynomial algebra, as encountered in section \ref{sec:classdouble}. The Yangian double for simple Lie algebras has been worked out in \cite{Khoroshkin:1994uk} and was generalised to superalgebras in \cite{Stukopin:2005aa}. We will develop our own approach to the Yangian double of superalgebras, as published in \cite{Rej:2010mu}. Only in this approach does the universal R-matrix lead to unitary, crossing invariant S-matrices on representations. This is of crucial importance for physical applications, and indeed, the resulting R-matrices for $\sun$ algebras will be shown to describe the scattering in $\sun$ integrable field theories in section \ref{sec:fieldtheory}. Furthermore, the results of \cite{Stukopin:2005aa} are not valid for $\slnn$, and we are not sure about the conventions used in this paper. Hence, we present the results independently of \cite{Stukopin:2005aa}.

\subsubsection{Quantum Groups}\label{sec:quantumgroups}

Let us first describe the mathematical notion of what we mean by a Quantum Group. We mean a quasi-triangular Hopf-Algebra which possesses a universal R-matrix, which satisfy some conditions which imply the Yang-Baxter as well as the crossing equations. Let us start by describing a Hopf Algebra. It is, first of all, an associative algebra, i.e. a vector space which posses a multiplication, and also an identity element. In this paper, we are mainly considering complex algebras, but we can also choose to work over the real numbers. It is useful to write the multiplication as a map 

\[
\mu: \alg{A}\otimes \alg{A}\rightarrow\alg{A},
\]
where $\alg{A}$ is the Hopf algebra we are considering. Note that generically the multiplication map is just defined on the direct product of two copies of the algebra, but we extend it linearly to the tensor product. Then we can write the law of associativity as the following commuting diagram:

\begin{equation*}
\begin{CD}
\alg{A}\otimes \alg{A}\otimes\alg{A} @> id\otimes\mu>>\alg{A}\otimes\alg{A}\\
@V\mu\otimes id VV	@VV\mu V\\
\alg{A}\otimes\alg{A} @>\mu >> \alg{A}	
\end{CD}
\end{equation*}
Furthermore, we also define the identity of the algebra as a map

\[
\eta: \mathbb{C}\rightarrow \alg{A},
\]
such that the following two diagrams commute:
\begin{equation*}
\begin{CD}
\alg{A}\otimes \mathbb{C} @> id\otimes\eta>>\alg{A}\otimes\alg{A}\\
@V\cong VV	@VV\mu V\\
\alg{A} @>id>> \alg{A}	
\end{CD}
\end{equation*}
\begin{equation*}
\begin{CD}
\mathbb{C}\otimes \alg{A}@> \eta\otimes id>>\alg{A}\otimes\alg{A}\\
@V\cong VV	@VV\mu V\\
\alg{A} @>id>> \alg{A}	
\end{CD}
\end{equation*}
Defining an associative multiplication and the identity in such way might look more complicated than necessary, but the advantage is that it allows one to straightforwardly define the remaining properties of the Hopf algebra. In particular, a Hopf algebra is also a coalgebra, that is a vector space which posses a comultiplication map

\[
\Delta:\alg{A}\rightarrow\alg{A}\otimes\alg{A}
\]
which is defined by the commuting diagram

\begin{equation*}
\begin{CD}
\alg{A}\otimes \alg{A}\otimes\alg{A} @< id\otimes\Delta <<\alg{A}\otimes\alg{A}\\
@A\Delta\otimes id AA	@AA\Delta A\\
\alg{A}\otimes\alg{A} @<\Delta << \alg{A}	
\end{CD}
\end{equation*}
We see that this diagram is identical to the defining diagram of the multiplication map $\mu$, up to the fact that all arrows are reversed. This explains the origin of the name coproduct. Likewise, the coalgebra possesses a counit, i.e. a reversed identity, or unit map

\[
\epsilon: \alg{A}\rightarrow \mathbb{C},
\]
which is defined via the following commuting diagram:

\begin{equation*}
\begin{CD}
\alg{A}\otimes \mathbb{C} @< id\otimes\epsilon<<\alg{A}\otimes\alg{A}\\
@A\cong AA	@AA\Delta A\\
\alg{A} @<id<< \alg{A}	
\end{CD}
\end{equation*}
\begin{equation*}
\begin{CD}
\mathbb{C}\otimes \alg{A}@< \epsilon\otimes id<<\alg{A}\otimes\alg{A}\\
@A\cong AA	@AA\Delta A\\
\alg{A} @<id<< \alg{A}	
\end{CD}
\end{equation*}
Of course, a Hopf algebra is not merely an algebra and a coalgebra at the same time, but all defined structures should be compatible in a certain way. In particular, we demand that the coproduct and the counit are algebra homomorphisms, i.e. equations like

\[
\Delta(\gen{X}\gen{Y})= \Delta(\gen{X})\Delta(\gen{Y})
\]
hold for all $\gen{X}, \gen{Y} \in \alg{A}$. 

Furthermore, a Hopf algebra posses an antipode map

\[
S:\alg{A}\rightarrow\alg{A},
\]
which is an antihomomorphism with respect to the algebra structure, i.e.

\[
S(\gen{X}\gen{Y})= (-1)^{|\gen{X}||\gen{Y}|}S(\gen{Y})S(\gen{X}).
\]
The importance of a coproduct $\Delta$ is the fact that it is usually used to define tensor products of representations of $\alg{A}$, i.e. if
\[
\pi_{1,2}:\alg{A}\rightarrow End(V_{1,2}),
\]
are two representations of $\alg{A}$ on two vector spaces $V_{1,2}$, then $\alg{A}$ acts on the tensor product $V_1\otimes V_2$ as
\[
\pi_1\otimes\pi_2 \copro(\gen{X}),
\]
for all $\gen{X}\in\alg{A}$. If one is interested in defining tensor products of more than two representation spaces, one can apply the coproduct several times. Coassociativity guarantees that the order in which the coproducts are applied does not matter.

Let us look at one of the simplest examples of a Hopf algebra, the universal enveloping algebra of a Lie superalgebra $\alg{g}$. The supercommutator of $\alg{g}$ is this language is simply realised in the usual way,

\[
\scomm{\gen{X}}{\gen{Y}} = \mu(\gen{X},\gen{Y}) - (-1)^{|\gen{X}||\gen{Y}|}\mu(\gen{Y},\gen{X}),
\]
for all $\gen{X},\gen{Y} \in \alg{g}$ as embedded into the enveloping algebra. Likewise, the coproduct is realised by

\[
\Delta\gen{X} =\gen{X}\otimes \idm + \idm\otimes\gen{X},
\]
where we have to observe that the tensor product is graded. This means that when multiplying two elements $A_1\otimes B_1$ and $A_2\otimes B_2$ in the tensor product, we have to observe the rule
\[
(A_1\otimes B_1) (A_2\otimes B_2) = (-1)^{|B_1||A_2|}A_1 A_2\otimes B_1B_2 .
\]
Likewise, as the coproduct enables us to act on tensor products $v_1\otimes v_2$, where $v_1,v_2$ are vectors of some unspecified representation space, then we have to observe that

\[
(\gen{X}\otimes \idm + \idm\otimes\gen{X})(v_1\otimes v_2) = (\gen{X}v_1)\otimes v_2 + (-1)^{|X||v_1|}v_1\otimes (\gen{X}v_2).
\]
An equivalent way to realise the presence of fermions, which is sometimes more advantageous for implementation on a computer, is to work with the usual tensor product, i.e. $(A_1\otimes B_1) (A_2\otimes B_2) = A_1 A_2\otimes B_1B_2$,
 and instead introducing the Fermi operator

\[
\ferm \gen{X} \ferm = (-1)^{|\gen{X}|}\gen{X},
\]
i.e. it commutes with all bosonic generators, and one picks up a minus sign when commuting it with a fermion.
Then the coproduct is realised as

\[
\Delta\gen{X} =\gen{X}\otimes \idm + \ferm^{|\gen{X}|}\otimes\gen{X}.
\]
The general rule is that the Fermi operator appears on the left-hand side of a fermionic factor in a tensor product.

 The coproduct defines, as mentioned before, the action of the Hopf algebra on tensor products of representations. The simple coproduct for the universal enveloping algebra represents the intuition that a Lie algebra generator acts on a tensor product like a sum over the actions on each factor in the tensor product. When we have a superalgebra, the grading of the tensor product, or equivalently, the Fermi generator, guarantees that we pick up a minus sign whenever we pull one fermion over another fermion.

We finally mention that the antipode for the enveloping algebra is simply given by

\[
S\gen{X} = -\gen{X},
\]
if we work with the graded tensor product. For the corresponding Lie group, this means that the antipode maps an element to its inverse.
This is related to the fact that the antipode is often used to define antiparticle representations. Let the original representation $\pi$ defining the action of the generators on a vector space $V$ correspond to the particle representation. Then the antiparticle representation correspond to the dual, or contragredient representation $\bar{\pi}$, which is defined by taking a minus sign and supertranspose of the original representation $\pi$. The generalisation for Hopf algebras is to define an antiparticle representation by taking the supertranspose of the antipode, i.e.

\[
\bar{\pi}(\gen{X}) = \pi(S(\gen{X}))^{st}.
\]
We will see later that Yangians, which also contain a copy of the universal enveloping algebra of the underlying Lie algebra, have a more complicated coproduct, which is why we introduced these structures. Furthermore, for Yangians and other Quantum Groups, the important property that $S^2 = \text{Id}$, which holds for Lie algebras, is not true any longer. 

{\bf Quasitriangular Hopf Algebras}

Let us now come to the important concept of a quasi-triangular Hopf algebra. This is a Hopf algebra $\alg{A}$, with a universal R-matrix $\rmat$ which satisfies the axioms

\<\label{def:quasitriang}
(\copro \otimes \idm)(\rmat) = \rmat_{13}\rmat_{23}\nln
(\idm \otimes\copro)(\rmat) = \rmat_{13}\rmat_{12},
\>
as well as the property of almost cocommutativity

\[\label{def:cocomm}
\coproop(X)\rmat = \rmat \copro(X).
\]
Here, $\coproop$ denotes the opposite coproduct 

\[
\coproop = \perm \copro.
\]
$\perm$ is the permutation operator, flipping the two factors of the tensor product.
Hence, almost cocommutativity is behind the idea that the R-matrix can be used to define an S-matrix (scattering matrix) on representations, 
\[
\smat = \perm \rmat,
\]
which maps the tensor product $V_1\otimes V_2$ of two representation spaces of the Hopf algebra $\alg{A}$ to $V_2\otimes V_1$. Indeed, if the Hopf algebra has a generic, complicated coproduct, it is by no means clear that $V_1\otimes V_2$ and $V_2\otimes V_1$ are isomorphic as Hopf algebra modules.
The S-matrix establishes such isomorphism, hence mathematically, the S-matrix is simply an intertwiner of two representation spaces.

From the quasi-triangularity properties it follows that $\rmat$ satisfies the Yang-Baxter equation
\[
\rmat_{23}\rmat_{13}\rmat_{12} =\rmat_{12}\rmat_{13}\rmat_{23}
\]
as well as the crossing equation

\[
(S\otimes \idm)(\rmat) = \rmat^{-1}.
\]
Indeed, the Yang-Baxter equation follows from permuting the first two tensor factors in the first equation of \eqref{def:quasitriang},

\<
&&\rmat_{23}\rmat_{13}=(\coproop \otimes \idm)(\rmat) \nln
&=& \rmat_{12}(\copro \otimes \idm)(\rmat) \rmat_{12}^{-1} = \rmat_{12}\rmat_{13}\rmat_{23}\rmat_{12}^{-1}.
\>
Taking $\rmat_{12}^{-1}$ to the other side, we get

\[
\rmat_{23}\rmat_{13}\rmat_{12} =\rmat_{12}\rmat_{13}\rmat_{23}.
\] 
Let us also show how the crossing equation follows. The antipode of a Hopf Algebra satisfies the equation

\[
\mu (S\otimes \idm)\copro(X) = \eta(\epsilon(X)), X\in\alg{A},
\]
with $\epsilon$ being the counit and $\eta$ the unit map, i.e. it maps the complex number $1$ to the identity element in the Hopf algebra, which we will often just call $1$ as well. The counit satisfies

\[
(\epsilon\otimes \idm) \copro X = X,
\]
so we have 

\<
&&(\epsilon\otimes \idm\otimes \idm) (\copro\otimes\idm)(\rmat) =\rmat\nln
&=&(\epsilon\otimes \idm\otimes \idm)\rmat_{13}\rmat_{23}\nln
&=&((\epsilon\otimes \idm\otimes \idm)\rmat_{13})\rmat.
\>
Hence,

\[
(\epsilon\otimes \idm\otimes \idm)\rmat_{13} = \idm\otimes\idm,
\]
and as the second tensor factor plays no role, we have

\[
(\epsilon\otimes \idm)\rmat = \idm.
\]
Now we write the R-matrix as some sum over unspecified elements of the algebra, $\rmat=\sum r_1\otimes r_2$. Then

\<
&&(\mu (S\otimes \idm)\copro(r_1))\otimes r_2 = (\eta\otimes\idm) (\epsilon(r_1)\otimes r_2)= \eta(1)\otimes\idm = \idm\otimes\idm\nln
&=&(\mu (S\otimes \idm)\otimes \idm) \rmat_{13}\rmat_{23} = (\mu \otimes \idm)(S(r_1)\otimes \idm \otimes r_2)(\idm\otimes r_1' \otimes r_2')\nln
&=& (S(r_1) \otimes r_2)(r_1' \otimes r_2') =  (S\otimes \idm)(\rmat) \rmat .
\>
Hence,

\[
(S\otimes \idm)(\rmat) =\rmat^{-1}.
\]

\subsubsection{Yangians in Drinfeld's first Realisation}\label{sec:drinf1}

Consider a Lie superalgebra $\alg{g}$ with basis $\gen{J}^a$. We introduce an
additional set of generators $\geny{J}^a$, which have the following commutation
relations:

\<\label{def:yangian1}
\scomm{\gen{J}^a}{\gen{J}^b} = \scons\gen{J}^c,\\
\scomm{\gen{J}^a}{\genY{J}^b} = \scons\genY{J}^c.
\>
\<\label{def:yang1serre}
\scomm{\genY{J}^{a}}{\scomm{\genY{J}^b}{\gen{J}^{c}}} + 
\scomm{\genY{J}^{b}}{\scomm{\genY{J}^c}{\gen{J}^{a}}} +
\scomm{\genY{J}^{c}}{\scomm{\genY{J}^a}{\gen{J}^{b}}}
\nln = \frac{\hbar^2}{4}
\sconsb{ag}{d}\sconsb{bh}{e}\sconsb{ck}{f}f_{ghk}
\gen{J}^{\{d}\gen{J}^e\gen{J}^{f]}.
\>
The resulting algebra is infinite dimensional, as the commutator of $\geny{J}^a$ with $\geny{J}^b$ will, in general, not be a linear combination of the generators $\gen{J}^a$ or $\geny{J}^a$. Indeed, the relation \eqref{def:yang1serre} tells one about the behaviour of the commutators of higher order, and is henceforth referred to as a Serre relation for the Yangian. This is in analogy to the case of simple Lie algebra, whose full linear basis could be constructed from a minimal set of Chevalley-Serre generators by observing compatibility with the Serre relations. Note also that to raise or lower indices, we use the bilinear form $\kappa$ introduced in \eqref{def:killform}, which needs to be non-degenerate for these purposes.

In the important case if the Yangian is of type $\glnm$ there exists an evaluation
representation such that
\[
\genY{J}^{a}=u \gen{J}^{a}.
\]
Even though this looks like the loop algebra, or polynomial algebra, we note
that the coproduct is different. It is given by
\<\label{eq:coproductyang}
\copro(\gen{J}^a)&=& \gen{J}^a\otimes 1+1\otimes \gen{J}^a\nln
\copro(\geny{J}^a)&=& \geny{J}^a\otimes 1+1\otimes \geny{J}^a
+\hbar \frac{1}{2}\comm{\gen{J}^a\otimes 1}{\casiten} = \geny{J}^a\otimes 1+1\otimes \geny{J}^a
+\hbar \frac{1}{2}  f^a_{bc}\gen{J}^b\otimes\gen{J}^c,\nln
\>
where $\casiten$ is the quadratic Casimir of the underlying Lie algebra $\alg{g}$, defined on the tensor product $\alg{g}\otimes\alg{g}$.
Hence, we see that the universal enveloping algebra of $\alg{g}$ is embedded into the Yangian as a Hopf subalgebra.
Furthermore, we recover the polynomial algebra $\palg{g}$ in the limit $\hbar
\rightarrow 0$. In this sense, $\yang{g}$ is considered as a deformation of the universal enveloping algebra of the polynomial algebra $\palg{g}$, and $\hbar$ is the corresponding deformation parameter. Even though $\hbar$ does not necessarily correspond to the Planck constant, such deformation is often referred to as a quantisation, or a deformation quantisation. If $\hbar\neq 0$, then we can actually scale $\hbar$ away, i.e. without loss of generality set to $\hbar=1$. 

{\bf The Antipode for the Yangian}

The action of the antipode for the Yangian is most easily calculated using Drinfeld's first realisation, as we have a general formula for the coproduct in this case, valid for any generator $\geny{J}^a$. We apply the formula 

\<
0&=&\eta(\epsilon(\geny{J}^a))=\mu (S\otimes \idm)\copro(\geny{J}^a)\nln 
&=&\mu(S(\geny{J}^a)\otimes 1+1\otimes \geny{J}^a
-\hbar \frac{1}{2}  f^a_{bc}\gen{J}^b\otimes\gen{J}^c)\nln
&=&S(\geny{J}^a)+ \geny{J}^a
-\hbar \frac{1}{2}  f^a_{bc}\gen{J}^b\gen{J}^c,
\>
and get
\<\label{eq:antiyangian}
S(\geny{J}^a) = - \geny{J}^a+\hbar \frac{1}{2}  f^a_{bc}\gen{J}^b\gen{J}^c.
\>
Note that if the underlying Lie superalgebra is simple, the above formula for the antipode can be simplified to
\<
S(\geny{J}^a) = - \geny{J}^a+\hbar \frac{1}{4}c \gen{J}^a,
\>
where $c$ is the eigenvalue of the quadratic Casimir on the adjoint representation. This eigenvalue in turn is proportional to the dual Coxeter number.

\subsubsection{Yangians in Drinfeld's second Realisation}\label{sec:drinf2}

Yangians are particularly interesting as they lead to rational solutions of the
Yang-Baxter equation. As in 
the classical case of Lie bialgebras, where we could construct a classical
r-matrix based on the 
double of the polynomial algebra $\palg{g}$, which turned out to be the loop
algebra $\lalg{g}$, we will now show that
there is a universal R-matrix associated to the double of $\yang{g}$. As
$\yang{g}$ is a quantisation of $\palg{g}$, the double will be
a quantisation of $\lalg{g}$.

To write down a universal R-matrix we first need an appropriate linear basis, such that
for each generator in the Yangian we know its dual
generator. Then, similarly to the case of the classical double, the universal R-matrix will be the canonical element
\[
\rmat = \sum \gen{X}\otimes\gen{X}^*,
\]
where the sum is taken over the whole Yangian with unspecified basis $\gen{X}$, and its corresponding dual basis $\gen{X}^*$.

The first realisation of the Yangian, as given in the previous section, is not suitable for the construction of the universal R-matrix. This is due to the fact that only finitely many generators $\gen{J}^a$ and $\geny{J}^a$ are realised explicitly in this basis. The infinitely many other generators are in principle constructed by observing the Serre relations, but such construction is not possible in an explicit form. This was one of the main reasons why a new realisation of the Yangian was given in \cite{Drinfeld:1987sy}. This realisation defines the Yangian in a Chevalley-Serre type basis, and is called Drinfeld's second realisation of the Yangian. This realisation is isomorphic to the realisation given in the previous section.\\

One introduces generators $\csgh{i,n}$, $\csgp{i,n}$, $\csgm{i,n}$,
$i=1,\dots,$rank$(\alg{g})$ such that again the degree zero generators are identified
with the generator of the underlying Lie superalgebra. With the help of the symmetric Cartan
matrix $A$, the Yangian is defined as follows

\<
\label{def:drinf2}
\comm{\csgh{i,m}}{\csgh{j,n}}=0,\quad [\csgh{i,0},\csgp{j,m}]=A_{ij}
\,\csgp{j,m},\nonumber\\
\comm{\csgh{i,0}}{\csgm{j,m}}=- A_{ij} \csgm{j,m},\quad
\scomm{\csgp{i,m}}{\csgm{j,n}}=\delta_{i,j}\, \csgh{j,n+m},\nonumber\\
\comm{\csgh{i,m+1}}{\csgp{j,n}}-[\csgh{i,m},\csgp{j,n+1}] = \frac{1}{2} A_{ij}
\{\csgh{i,m},\csgp{j,n}\},\nonumber\\
\comm{\csgh{i,m+1}}{\csgm{j,n}}-[\csgh{i,m},\csgm{j,n+1}] = - \frac{1}{2} A_{ij}
\{\csgh{i,m},\csgm{j,n}\},\nonumber\\
\scomm{\csgp{i,m+1}}{\csgp{j,n}}-\scomm{\csgp{i,m}}{\csgp{j,n+1}} =
\frac{1}{2} A_{ij} \sacomm{\csgp{i,m}}{\csgp{j,n}},\nonumber\\
\scomm{\csgm{i,m+1}}{\csgm{j,n}}-\scomm{\csgm{i,m}}{\csgm{j,n+1}} =
-\frac{1}{2} A_{ij} \sacomm{\csgm{i,m}}{\csgm{j,n}},\nonumber\\
 Sym_{\{k\}}
[\csgp{i,k_1},[\csgp{i,k_2},\dots [\csgp{i,k_{n_{ij}}},
\csgp{j,l}\}\dots\}\}=0,\nln
Sym_{\{k\}}
[\csgm{i,k_1},[\csgm{i,k_2},\dots [\csgm{i,k_{n_{ij}}},
\csgm{j,l}\}\dots\}\}=0,\nln
i\neq j,\, \, \, \, n_{ij}=1+|A_{ij}|.
\>
Here, the superanticommutator is defined as $\sacomm{A}{B}:=A B + (-1)^{|A||B|}B
A$. Again, as in the case of the Chevalley-Serre basis of Lie superalgebras, one
needs to add extra Serre relations if we have a proper Lie superalgebra, i.e.
with non-zero fermionic part.

The advantage of this presentation of the Yangian is that, unlike in the case of
the first realisation, we have explicit relations for the generators of arbitrary degree $n$,
$\gen{J}^a_n, n\geq 0$. As mentioned before, the generators of degree $0$ are identified with
the 
generators of the underlying Lie superalgebra $\alg{g}$. The generators of
degree $1$
are in principle associated with the generators $\geny{J}^a$ of the Yangian in
the first realisation,
but not in a direct way. In fact, one derives that to match the defining relations
in the first realisation
one needs the following isomorphism:

\begin{align}
\label{def:isom}
&\gen{H}_{i,0}=\gen{H}_i,\quad \gen{E}^+_{i,0}=\gen{E}^+_i,\quad
\gen{E}^-_{i,0}=\gen{E}^-_i,\nonumber\\
&\gen{H}_{i,1}=\hat{\gen{H}}_i-v_i,\quad \gen{E}^+_{i,1}=\hat{\gen{E}}^+_i-w_i,\quad
\gen{E}^-_{i,1}=\hat{\gen{E}}^-_i-z_i,
\end{align}
where

\<\label{def:special}
v_i &=&
\frac{1}{4}\sum_{\beta}\sprod{\alpha_i}{\beta}\sacomm{\cwgen{E}^-_{\beta}}{
\cwgen{E}^+_{\beta}} - \frac{1}{2}\csgen{H}_{i}^2, \nonumber\\ 
w_i &=& \frac{1}{4}\sum_{\beta}(-1)^{\beta
i}\sacomm{\cwgen{E}^-_{\beta}}{\scomm{\csgen{E}^+_{i}}{\cwgen{E}^+_{\beta}}} -
\frac{1}{4}\acomm{\csgen{H}_{i}}{\csgen{E}^+_{i}}, \nonumber\\
z_i &=&
-\frac{1}{4}\sum_{\beta}\sacomm{\scomm{\csgen{E}^-_{i}}{\cwgen{E}^-_{\beta}}}{
\cwgen{E}^+_{\beta}} - \frac{1}{4}\acomm{\csgen{H}_{i}}{\csgen{E}^-_{i}}.
\>
The sums over $\beta$ goes over all positive roots.

Note that even though the second realisation explicitly contains generator of arbitrary degree, we have not established a full linear basis of the Yangian yet. First of all, we need to add root generators corresponding to the non-simple roots of the underlying Lie superalgebra. This can be done in a similar fashion as for the underlying Lie algebra, but is still quite non-trivial. We give an explicit realisation in the case $\yang{\slnm}$. Furthermore, the Yangian contains arbitrary powers of the generators, i.e. elements of the form $(\csgp{\beta_1,k_1})^{n_1}\dots(\csgp{\beta_r,k_r})^{n_r}(\csgh{1,l_1})^{m_1}\dots(\csgh{s,l_s})^{m_s}(\csgm{\beta_1,x_1})^{p_1}\dots(\csgm{\beta_t,x_t})^{p_t}$.
All such elements form a linear basis provided one avoids double counting. This is guaranteed if one orders the roots, and then sticks to the block form above, which reads schematically $EHF$, i.e. one starts with the ordered positive roots, then the Cartan generators and then the ordered negative roots.

\subsubsection{The Yangian Double}

We will now define the double $\dyang{g}$ of the Yangian $\yang{g}$, which
naturally contains a universal
R-matrix. We have established the Yangian in Drinfeld's second realisation,
which contains already generators of arbitrary positive degree.
We will first construct a linear basis of the whole Yangian, and then appropriate
dual generators. However, unlike in the classical limit,
it turns out that the explicit description of the dual generators is more
involved here.

First, we note that the generators in the second realisation, $\csgh{i,n}$,
$\csgp{i,n}$, $\csgm{i,n}$ form only a Chevalley-Serre type basis of the Yangian.
To get a linear basis, as in the case of the underlying Lie superalgebra, we
have to construct a corresponding 
Cartan-Weyl basis. Indeed, such generators are simply of the form
$\cwgp{\beta,n}$, $\cwgm{\beta,n}$, where $\cwgp{\beta,0}$, $\cwgm{\beta,0}$
correspond to the generators of $\alg{g}$, i.e. $\beta$ is a positive root vector. Unlike for $\alg{g}$, for the Yangian one cannot 
construct them explicitly. However, we know that there should be generators of exactly this form as 
the Yangian is a deformation of the enveloping algebra of $\palg{g}$, hence there 
should be a one-to-one correspondence between the appropriate generators. The obstruction can be seen from the defining relations \eqref{def:drinf2}. Unlike for $\palg{g}$, commutators of generators of degree $n$ and $m$ are not of homogeneous degree $n+m$, but contain generators of lower order. Mathematically speaking, the degree $n$ does not provide a grading on the Yangian, but only a filtration. If one has a root vector $\beta + n\delta$, where $\delta$ is the imaginary root vector, then there is not a unique way to construct the corresponding root generator. \\

Furthermore, as we know that the classical double of $\palg{g}$ is the loop algebra 
$\lalg{g}$, the double Yangian $\dyang{g}$ should now also contain generators of the form 
$\csgh{i,n}$, $\csgp{i,n}$, $\csgm{i,n}$, $n<0$, as well as corresponding Cartan-Weyl generators.
Now we have two choices: we could simply proceed by setting e.g. $\csgp{i,n}$ to be dual to $\csgm{i,-n-1}$,
just as in the case of the polynomial algebra. Then one could derive the algebra and coalgebra relations from 
the requirement that dual product imposes a non-degenerate invariant form which also respects the algebra and coalgebra relations. However, it turns out that then the generators of negative degree would not have a nice behaviour, at least for those generators associated with the Cartan subalgebra. 
Instead, one would like them to be as closely related to the generators of the loop algebra. In particular, on representations, one would like to have a simple evaluation type representation. So we will follow the approach of \cite{Khoroshkin:1994uk} to construct the quantum double. Here the generators of negative degree behave similarly to the generators of positive degree, but the dual product on the Cartan subalgebra is non-trivial.

The inner product is given by

\<
\sprod{\csgen{E}^+_{i,k}}{\csgen{E}^-_{j,l}}=(-1)^{|i|}\sprod{\csgen{E}^-_{j,l}}{\csgen{E}^+_{i,k}}=-\delta_{ij}\delta_{k,-l-1}\nonumber\\
\sprod{\csgen{H}_{i,k}}{\csgen{H}_{j,-l-1}}=-A_{ij}\left(\frac{A_{ij}}{2}\right)^{k-l}{k\choose l} \quad \text{for} \quad k\geq l ,
\>
with all other products vanishing. 
Here, we will not list the coproduct and the product relations for the dual generators, as we will not use them. They have the same structure as in \cite{Khoroshkin:1994uk}.

To construct the universal R-matrix, one needs to construct a dual basis with respect to the inner product. To do this, it is useful to introduce generating functions for the generators as follows

\<
 \csgen{E}^+_i(\lambda):=\sum_{k=0}^{\infty}\csgen{E}^+_{i,k}\lambda^{-k-1}\,, \quad(\csgen{E}^+)^*_i(\lambda):=-\sum_{k=-1}^{-\infty}\csgen{E}^-_{i,k}\lambda^{-k-1}\,, \nln
\csgen{E}^-_i(\lambda):=\sum_{k=0}^{\infty}\csgen{E}^-_{i,k}\lambda^{-k-1}\,, \quad(\csgen{E}^-)^*_i(\lambda):=-\sum_{k=-1}^{-\infty}\csgen{E}^+_{i,k}\lambda^{-k-1}\,,
\>

\<
 \csgen{H}^+_i(\lambda):=1+\sum_{k=0}^{\infty}\csgen{H}_{i,k}\lambda^{-k-1}, \quad\csgen{H}^-_i(\lambda):=1-\sum_{k=-1}^{-\infty}\csgen{H}_{i,k}\lambda^{-k-1}.
\>
The parameter $\lambda$ is the formal parameter of the generating functions. It should be noted that on evaluation representations with spectral parameter $u$ the generating function will depend on the difference $u-\lambda$, so effectively $\lambda$  may be interpreted as the spectral parameter. 

The dual of the function $\gen{J}(\lambda_1) = \sum_{k=0}^{\infty}\gen{J}_k \lambda^{-k-1}_1$ is defined as the function $\gen{J}^*(\lambda_2)= -\sum_{k=-1}^{-\infty}\gen{J^*}_k \lambda^{-k-1}_2$ such that
\<
\sprod{\gen{J}(\lambda_1)}{\gen{J}^*(\lambda_2)} = \frac{1}{\lambda_1-\lambda_2}\,.
\>
This is equivalent to introducing the generator $\gen{J}^*_l$ dual to $\gen{J}_k$ in the sense of

\[
\sprod{\gen{J}_k}{\gen{J}^*_{-l-1}}=-\delta_{k,l}.
\]
According to this definition the root generators are already written in terms of a dual basis. What remains to be found is the dual basis for the Cartan generators. Note that here the superscripts $\pm$  indicate the expansion of $\csgen{H}^\pm_i(\lambda)$ at $\lambda=0$ and $\lambda=\infty$ respectively. On evaluation representations one finds that $\csgen{H}^+_i(\lambda)$ and $\csgen{H}^-_i(\lambda)$ represent formally the same function. Their scalar product is given by \cite{Khoroshkin:1994uk}
\<
 \sprod{\gfhp{i}{1}}{\gfhm{j}{2}} = \frac{\lambda_1-\lambda_2 + \frac{A_{ij}}{2}}{\lambda_1-\lambda_2 -\frac{A_{ij}}{2}}.
\>
It turns out to be useful to consider the formal logarithms $\log(\csgen{H}^\pm_i(\lambda))$ due to the following property 
\<
\sprod{\log(\gfhp{i}{1})}{\log(\gfhm{j}{2})} = \log\frac{\lambda_1-\lambda_2 + \frac{A_{ij}}{2}}{\lambda_1-\lambda_2 -\frac{A_{ij}}{2}}\,.
\>
Therefore, 
\<
\sprod{\frac{d}{d\lambda_1}\log(\gfhp{i}{1})}{\log(\gfhm{j}{2})} = \frac{1}{\lambda_1-\lambda_2 +\frac{A_{ij}}{2}}-\frac{1}{\lambda_1-\lambda_2 -\frac{A_{ij}}{2}}\,.\nln
\>
If one introduces the shift operator
\[
 T f(\lambda_2)= f(\lambda_2+1),
\]
then the above formula may be written as
\[\label{eq:diagform}
\sprod{\frac{d}{d\lambda_1}\log(\gfhp{i}{1})}{\log(\gfhm{j}{2})} = (T^{-A_{jk}/2}-T^{A_{jk}/2})\frac{\delta_{ik}}{\lambda_1-\lambda_2}\,.
\]
This is a matrix equation and to complete the task of the diagonalisation one needs to invert the 
operator 
\[\label{def:dop}
D_{ij} = T^{-A_{ij}/2}-T^{A_{ij}/2}\,.
\]
Note that on evaluation representations $T$ effectively shifts the spectral parameter $u_2$ due to the aforementioned fact that the Drinfeld currents depend on the difference $\lambda-u$. For the sake of the following discussion it is useful to introduce the q-deformed symmetric Cartan matrix, i.e. we replace each number $x$ by its q-number
\beq
x \to [x]_q = \frac{q^x-q^{-x}}{q-q^{-1}}\,.
\eeq
Hence, the q-deformed Cartan matrix takes the following form 
\[ \label{eq:Aqdeformiert}
 A(q)_{ij} = \qnumb{(\alpha_i,\alpha_j)} = \qnum{(\alpha_i,\alpha_j)}.
\]
The $D_{ij}$ operator is then related to the q-deformed Cartan matrix through
\[\label{eq:dascartan}
D_{ij} = -(T^{1/2} - T^{-1/2})A(T^{1/2})_{ij}\,.
 \]

\subsubsection{The Universal R-Matrix}\label{sec:universalR}

In the previous section we have established the Yangian double $\DY{\alg{\slnm}}$ by
generalising the analysis of \cite{Khoroshkin:1994uk} for the case of simple Lie algebras. The universal R-matrix can now be easily stated with the help of the diagonalised form \eqref{eq:diagform} since it is simply the canonical element of the Yangian double, i.e. the sum over all elements of the Yangian $\mathcal{Y}({\alg{\slnm}})$ tensor its appropriate dual. One should stress that the Yangian consists not only of the Chevalley-Serre generators \eqref{eq:chevserre} and their commutators, but also of all monomials of the corresponding generators. Schematically, the dual product decomposes as follows:
\<
\sprod{E^+ H E^- }{(E^+)^{*}(H)^{*}(E^-)^{*}} = \sprod{E^+}{(E^+)^*}\sprod{H}{H^*}\sprod{E^-}{(E^-)^*}\nln
\>
Here, $E^+, E^-$ and $H$ stand for generators of the Yangian corresponding to the positive root, negative root or Cartan generators of the underlying Lie algebra, and $ ^*$ donates the dual with respect to the inner product. This decomposition works as in the case for simple Lie algebras \cite{Khoroshkin:1994uk}. Hence, the universal R-matrix has the quasi-triangular structure
\beq\label{eq:rsplit}
R = R_+ R_H R_-\,.
\eeq
The positive and negative root parts are given in terms of ordered products
\bea\label{def:Rpm}
R_+ &=& \prod_{\beta,k\geq 0}^\rightarrow \exp(-(-1)^{|\beta|}\mathcal{F}^{|\gamma|}\cwgp{\beta+k\delta}\otimes \cwgm{\beta-(k+1)\delta})\,,\nln
R_- &=& \prod_{\beta,k\geq 0}^\leftarrow \exp(-\mathcal{F}^{|\beta|}\cwgm{\beta+k\delta}\otimes \cwgp{\beta-(k+1)\delta}).
\eea
Here, $\mathcal{F}$ is the usual Fermi-number generator. The product is only taken over positive roots $\beta \in \slnm$. The symbol $\delta$ denotes the imaginary root. An important feature of (\ref{def:Rpm}) is that the product in $R_+$ is taken in a specified order, whereas for $R_-$ the reverse ordering is applied. Such ordering can be defined inductively. Let two roots $\gamma_1, \gamma_3$ be already ordered as $\gamma_1<\gamma_3$. Then we say $\gamma_1<\gamma_2<\gamma_3$ if we can write $\scomm{\cwgp{\gamma_1}}{\cwgp{\gamma_3}}=\cwgp{\gamma_2}$. This procedure for the root ordering was introduced for Yangians based on simple Lie algebras in \cite{Khoroshkin:1994uk}. We also refer to \cite{Rej:2010mu} for the explicit ordering of roots of $\slnm$.

The Cartan part of the universal R-matrix is the significantly more complicated
\[\label{def:RH}
R_H = \prod_{i,j}\exp\left(\sum_{t=0}^\infty\left(\left(\frac{d}{d\lambda_1}\log(\gfhp{i}{1})\right)_{t}\otimes\left( D^{-1}_{ij}\log(\csgh{j}^-(\lambda_2)\right)_{-(t+1)}\right)\right)\,.
\]
The subscripts $t$ and $-(t+1)$ denote the respective coefficients of the expansion of the generating functions in $\lambda_1 \gg 1$, $\lambda_2 \ll 1$. Tensoring them together is thus equivalent to taking the residue at $\lambda_1 = \lambda_2$.  In the case of integer-valued Cartan matrices we find that the inverse of the q-Cartan matrix is also q-integer valued up to an overall constant q-number $\qnumb{l}$.

The concrete definition of the inverse $D^{-1}_{ij}$ on the set of functions of the spectral parameter determines the scalar part of the R-matrix. According to \eqref{eq:dascartan}, the operator $D$ may be expressed solely through the translation operator $T$, for which the action on functions of spectral parameter is well-defined. Clearly, one may define $D^{-1}_{ij}$ by expanding it in a power series around either $T=0$ or $T=\infty$. It turns out, however, that both expansions \textit{do not} result in the same scalar part $R_H$. We would like to argue that the guiding principle should be unitarity. Indeed, the Yangian Double is not triangular, i.e. the equation 
\[
R_{12} R_{21} = \idm
\]
does not hold. Only a balanced expansion in power series in $T$ and $T^{-1}$ will lead to unitary dressing factors. It follows immediately from \eqref{eq:Aqdeformiert} that $A(q^{-1}) = A(q)$ so that the operator $D_{ij}$ defined in \eqref{eq:dascartan} satisfies $D_{ij}(q) = - D_{ij}(q^{-1})$. The same must hold for its inverse $D^{-1}_{ij} (q)$ thus one may write
\beq \label{eq:Dunitary}
D^{-1}_{ij} (q) = \tfrac{1}{2} \left(D^{-1}_{ij}(q)-D^{-1}_{ij}(q^{-1}) \right)\,.
\eeq
We propose to expand the first term at $q=0$ and the second one at $q=\infty$ and setting  $q \to T^{1/2}$ in the expansions. Subsequently, the principal branch of the square root should be applied.  We conjecture that the resulting R-matrix contains a unitary dressing factor satisfying the corresponding crossing equation. Moreover, in all cases studied in what follows the dressing factor found in this way is a \textit{meromorphic} function up to a square root of a CDD factor. This suggest that this may be a general feature of this procedure. Please note also that analytic properties of a given solution to crossing and unitarity equations are dictated by the concrete physical model and cannot be determined with help of the universal R-matrix.

Generic Cartan matrices of superalgebras, in particular those corresponding to the $\glnn$ algebra where the external $\uone$ automorphism is rescaled and shifted by the identity, have non-integer elements and the aforementioned prescription needs to be applied. Integer-valued Cartan matrices allow for further simplification since there exists a matrix $C(q)$
\beq
 A(q) C(q) = \qnumb{l} \textit{Id}\,,
\eeq
such that its elements are \text{polynomials} in $q$ and $q^{-1}$. Here, $l$ is assumed to take minimal value for which such $C(q)$ exists.
The inverse of \eqref{eq:dascartan} may now be written as
\bea\label{eq:dinv}
D^{-1} =C(T^{1/2})\frac{1}{T^{-l/2}-T^{l/2}}\,.
\eea
When expanded in the vicinity of $T=\infty^{\mp 1}$, one finds
\<\label{RHpos}
&&R^{\pm}_H =\nln
 &&\prod_{i,j,k}\exp\left(\pm\left(\frac{d \log(\gfhp{i}{1})}{d\lambda_1}\right)_{t}\otimes\left( C_{ij}(T^{1/2})\log(\csgh{j}^-(\lambda_2\pm(k+1/2)l)\right)_{-t-1}\right)\,.\nln
\>
For $\sln$ algebras $l=n$ and the above formulae reduce to the one proposed in \cite{Khoroshkin:1994uk}. The scalar part leading to a unitary dressing factor may be formally written as
\beq
R_{H} = \sqrt{ \frac{R^{+}_H}{R^{-}_H}}
\eeq
This formula, however, remains also valid for the supersymmetric counterpart $\slnm$ with $l = n-m$ and $n \neq m$. Clearly, the case of $n=m$ is special and the definition of $C(q)$ becomes redundant for non-canonical choice of the extension parameters. Thus the matrix $C(q)$ is convenient for classification purposes only, but becomes ill-defined in the general case of real-valued Cartan matrices.

\subsection{$\slnm$ Lie Superalgebras and their Extensions}\label{sec:glnm}

Of particular importance for the physics we are dealing with in this paper are
Lie superalgebras of type $\slnm$ or $\glnm$, as well as their real forms. This series
includes the Lie algebras of type $\alg{sl}(n)$ in the special case $\alg{sl}(n|0)$. We will define $\slnm$ for $n\neq m$ in section \ref{sec:slnmalgebra}, and treat the special case $n=m$ separately in section \ref{sec:slnn}. Due to its importance for later applications in physics, and as it has some remarkable mathematical features, we devote the additional section \ref{sec:pslcentral} to the study of the centrally extended $\pslcentral$ algebra. We refer the reader to the references \cite{Kac:1977em,Kac:1977qb} for the general theory of Lie superalgebras, and \cite{Beisert:2006qh} for background on the centrally extended $\pslcentral$ algebra.

\subsubsection{$\slnm$}\label{sec:slnmalgebra}

Let us give the definition of Lie superalgebras of type $\glnm$ in terms of their $n+m$
dimensional fundamental representation. The representation space is spanned by
$n$ bosons and $m$ fermions

\begin{eqnarray}\label{eq:funrepglnm}
&&\state{\phi^a},\quad a=1,\dots n,\nonumber\\
&&\state{\psi^\alpha},\quad \alpha=1,\dots m.
\end{eqnarray}
They can be written as $n+m$ dimensional column vectors such that
$\state{\phi^a}$ is the vector with a $1$ in the $a$'th row and $0$ otherwise,
whereas $\state{\psi^\alpha}$ has a $1$ in the $n+\alpha$'th row. 

We will generally denote bosons by lower case Latin characters $a,b,c,\dots$, and fermionic indices by Greek characters $\alpha, \beta, \gamma, \dots$. For some purposes we will combine the bosonic and fermionic indices into upper case Latin characters $I, J, \dots$. Using those indices, $\glnm$ is simply the Lie algebra consisting of all $(n+m)\times(n+m)$ dimensional matrices, where a useful base is given by the matrices

\[
E_{IJ},\quad I,J=1,\dots, n+m ,
\]
which have a $1$ at the $J$'th row and the $I$'th column and $0$ otherwise. Furthermore, the generators $E_{IJ},\quad I,J\leq n$ as well as $E_{IJ},\quad I,J > n$ are bosonic, whereas $E_{IJ},\quad I\leq n, J>n$ and $E_{IJ},\quad J\leq n, I > n$ are fermionic. We will use the notation $|IJ|=0,1$ to characterise whether $E_{IJ}$ is bosonic or fermionic. Hence, the commutation relations read

\[\label{eq:commglnm}
\scomm{E_{IJ}}{E_{KL}} = \delta_{JK}E_{IL} - (-1)^{|IJ||KL|}\delta_{IL}E_{JK}.
\]
The non-degenerate bilinear form in this basis is simply given by
\[
\sprod{E_{IJ}}{E_{KL}}=\str(E_{IJ}E_{KL})=\delta_{JK}\str(E_{IL}) =\delta_{JK}\delta_{IL}(-1)^{I}.
\]
Then, the quadratic Casimir is given by 
\[\label{def:casiglnm}
\casi = \sum_{I,J=1}^{n+m}(-1)^{|J|}E_{IJ}E_{JI} = \sum_{I=1}^{n+m}\sum_{J=1}^{n}E_{IJ}E_{JI} - \sum_{I=1}^{n+m}\sum_{J=n+1}^{n+m}E_{IJ}E_{JI}.
\]
On the tensor product, this Casimir 
\[\label{def:casiglnmten}
\casiten = \sum_{I,J=1}^{n+m}(-1)^{|J|}E_{IJ}\otimes E_{JI}
\]
will reduce to the graded permutation operator $\perm$.
It is often useful to write the matrices of $\glnm$ in the following block form:

\[
\begin{pmatrix}
\gR{a}{b}&\gS{a}{\alpha}\\
\gQ{\alpha}{a}&\gL{\alpha}{\beta}
\end{pmatrix}.
\]
Here, $\gR{a}{b}$ forms a basis of $\alg{gl}(n)$, $\gL{\alpha}{\beta}$ forms a
basis of $\alg{gl}(m)$, and $\gQ{\alpha}{a}$ and $\gS{a}{\alpha}$ are fermionic
generators forming $n\times m$ or $m\times n$ blocks, respectively. The advantage of this notation is that we have clearly separated bosons and fermions. Latin and
Greek indices indicate the action on bosons or fermions of the basis \eqref{eq:funrepglnm}, and in matrix form, a
generator $\gen{X}^x{}_y$ means that the appropriate generator has a $1$ in the
$x$'th column and the $y$'th row of the appropriate bosonic or fermionic block.
For instance, $\gR{a}{b}$ has a $1$ in the element where column $a$ and row $b$
intersect, and $0$ otherwise. Hence, the translation to the indices $I, J$ used above is given by 
\<\label{def:funrepindices}
a=I&\quad& \text{for } 1\leq I\leq n\nln
\alpha+n=I&\quad& \text{ for } n< I\leq n+m.
\>
A basis for the $n+m$ dimensional representation space \eqref{eq:funrepglnm} is then given by vectors $V^I$, $I=1,\dots n+m$, with
\<
V^I = \phi^a,\quad I=a,\nln
V^I = \psi^\alpha,\quad I=\alpha + n. 
\>
The degree of an index is denoted by $|I|$, so the short-hand notation of the degree of a generator $E_{IJ}$, which was denoted by $|IJ|$, translates to $|IJ|=|I| + |J|$.\\

In the $a, \alpha$ convention it is manifest that the fermionic generators $\gQ{\alpha}{a}$, $
\gS{a}{\alpha}$ form fundamental and antifundamental representations of the bosonic $\alg{gl}(n)$ and $\alg{gl}(m)$ subalgebras. This is realised by the natural commutation relations
\<
\comm{\gL{\alpha}{\beta}}{\gQ{\gamma}{a}}&=&\delta_\beta^\gamma\gQ{\alpha}{a}\nonumber\\
\comm{\gL{\alpha}{\beta}}{\gS{a}{\gamma}}&=&-\delta_\alpha^\gamma\gS{a}{\beta}\nonumber\\
\comm{\gR{a}{b}}{\gQ{\gamma}{c}}&=&-\delta_a^c\gQ{\gamma}{b}\nonumber\\
\comm{\gR{a}{b}}{\gS{c}{\gamma}}&=&\delta_b^c\gS{a}{\gamma} .
\>
The standard $\alg{gl}(n)$ and $\alg{gl}(m)$ commutation relations are given by
\<\label{eq:sln}
\comm{\gR{a}{b}}{\gR{c}{d}} &=& \delta_b^c \gR{a}{d}-\delta_d^a
\gR{c}{b}\nonumber\\
\comm{\gL{\alpha}{\beta}}{\gL{\gamma}{\delta}} &=& \delta_\beta^\gamma
\gL{\alpha}{\delta}-\delta_\delta^\alpha \gL{\gamma}{\beta}.
\>

Of course, these are special cases of \eqref{eq:commglnm} with $m=0$ or $n=0$, respectively. Another advantage of dividing the generators into blocks is that their action on the fundamental representation \eqref{eq:funrepglnm} is realised in a canonical fashion. The non-vanishing actions are given by

\<
\gR{a}{b}\state{\phi^c}=\delta_b^c\state{\phi^a},\nln
\gL{\alpha}{\beta}\state{\psi^\gamma}=\delta_\beta^\gamma\state{\psi^\alpha},\nln
\gQ{\alpha}{b}\state{\phi^c}=\delta_b^c\state{\psi^\alpha},\nln
\gS{a}{\beta}\state{\psi^\gamma}=\delta_\beta^\gamma\state{\phi^a}.
\>
The quadratic Casimir in this realisation is given by
\[
\casi = \sum \gR{a}{b}\gR{b}{a} - \gL{\alpha}{\beta}\gL{\beta}{\alpha} + \gQ{\alpha}{a}\gS{a}{\alpha}-\gS{a}{\alpha}\gQ{\alpha}{a}.
\]
It is important to note that the algebras $\glnm$ are not simple. This can be easily seen, as they have a
central element 

\[
\sum_a{\gR{a}{a}} +\sum_\alpha{\gL{\alpha}{\alpha}}= \sum_I E_{II}.
\]
In the case $n\neq m$, one can factor it out by removing all matrices with non-vanishing supertrace 

\[
\str \left(\begin{pmatrix}
R&S\\
Q&L
\end{pmatrix}\right) = \tr(R)-\tr(L)=0.
\]
As all the diagonal elements $\gR{a}{a}$, $\gL{\alpha}{\alpha}$ introduced
earlier have supertrace $1$ or $-1$, we change their fundamental representation
towards

\<
\gR{a}{a} = 
\begin{pmatrix}
0&\hdots& 0&\hdots &\\
\vdots&\ddots & &\\
0 & & 1 & \\
\vdots & & & \ddots
\end{pmatrix} - \frac{1}{n}\begin{pmatrix}
1&0&\hdots & &\\
0&\ddots & &\\
\vdots & &  & \\
 & & & 1
\end{pmatrix}
\>
and likewise 

\<
\gL{\alpha}{\alpha} = 
\begin{pmatrix}
0&\hdots& 0&\hdots &\\
\vdots&\ddots & &\\
0 & & 1 & \\
\vdots & & & \ddots
\end{pmatrix} - \frac{1}{m}\begin{pmatrix}
1&0&\hdots & &\\
0&\ddots & &\\
\vdots & &  & \\
 & & & 1
\end{pmatrix}.
\>
As before, $\gL{\alpha}{\alpha}$ and $\gR{a}{b}$ are written as $m\times m$ or
$n\times n$ matrices, which are understood to be embedded block-wise into the
superalgebra, and no summation convention was used. Using the basis vectors $\state{\phi^a}$, $\state{\psi^a}$ as before, we get 

\<\label{def:funrepslnm}
\gR{a}{b}\state{\phi^c}=\delta_b^c\state{\phi^a}-\frac{1}{n}\delta_b^a\state{\phi^c},\nln
\gL{\alpha}{\beta}\state{\psi^\gamma}=\delta_\beta^\alpha\state{\psi^\gamma}-\frac{1}{m}\delta_\beta^\gamma\state{\psi^\alpha}.
\>
Note that those
shifts do not alter the $\alg{sl}(n)$ or $\alg{sl}(m)$ commutation relations \eqref{eq:sln}.

However, the action on the fermions is slightly modified, we now get

\<
\comm{\gL{\alpha}{\beta}}{\gQ{\gamma}{a}}&=&\delta_\beta^\gamma\gQ{\alpha}{a} -
\frac{1}{m}\delta^{\alpha}_{\beta}\gQ{\gamma}{a}\nonumber\\
\comm{\gL{\alpha}{\beta}}{\gS{a}{\gamma}}&=&-\delta_\alpha^\gamma\gS{a}{\beta} +
\frac{1}{m}\delta^{\alpha}_{\beta}\gS{a}{\gamma}\nonumber\\
\comm{\gR{a}{b}}{\gQ{\gamma}{c}}&=&-\delta_a^c\gQ{\gamma}{b} +
\frac{1}{n}\delta^{a}_{b}\gQ{\gamma}{c}\nonumber\\
\comm{\gR{a}{b}}{\gS{c}{\gamma}}&=&\delta_b^c\gS{a}{\gamma} -
\frac{1}{n}\delta^{a}_{b}\gS{c}{\gamma}
\>

This does not mean that the fermions form a different representation with respect to the bosonic subalgebras than before. Indeed, as we have imposed the constraint $\sum_a \gR{a}{a}=0$, $\sum_\alpha \gL{\alpha}{\alpha}=0$, it is more natural to consider $n-1$ independent generators $\gR{a}{a}-\gR{a+1}{a+1}$, $a=1,\dots n-1$, and likewise $\gL{\alpha}{\alpha}- \gL{\alpha+1}{\alpha+1}$, $\alpha = 1,\dots m-1$. These combinations are supertraceless even in the original $\glnm$ basis, and we will later use them for the Chevalley-Serre basis.

Demanding that the supertrace vanishes removes only one
diagonal generator from $\glnm$, but now we have imposed two constraints reducing $\alg{gl}(n)$ and $\alg{gl}(m)$ to $\alg{sl}(n)$ and $\alg{sl}(m)$, respectively. Hence $\alg{sl}(n|m)$ contains one more $\alg{u}(1)$ generator
$\gen{C}$. Indeed, if we commute two fermions, we get

\[\label{eq:slnmfermi}
\acomm{\gQ{\alpha}{a}}{\gS{b}{\beta}}=\delta_a^b\gL{\alpha}{\beta}+
\delta_\beta^\alpha\gR{b}{a}+ \delta_\beta^\alpha\delta_a^b\gen{C}.
\]
Here, we have put 

\<\label{def:centralc}
\gen{C}\state{\phi^a}&=&\frac{1}{n}\state{\phi^a}\nonumber\\
\gen{C}\state{\psi^\alpha}&=&\frac{1}{m}\state{\psi^\alpha}.
\>
To go from $\glnm$ to $\alg{sl}(n|m)$ in the $E_{IJ}$ basis we can simply set

\[\label{eq:gltosl}
E_{IJ} \rightarrow  E_{IJ} - (-1)^{|I|}\frac{1}{n-m}\delta_{IJ}\sum_K E_{KK}.
\]
Note that this removes only exactly one $\alg{u}(1)$ generator. However, the simple bosonic $\sln$ and $\slm$ subalgebras are not manifestly realised, as the bosonic blocks are not traceless. Only the supertrace over the whole space vanishes. Also, this shift apparently does not work for $n=m$. Indeed, this is not an artefact of the basis, but reminiscent of the fact that $\alg{sl}(n|n)$ is indeed mathematically different from the general series $\glnm$. We will treat this case separately in the next subsection. For now, we would like to give a choice of the Chevalley-Serre basis. We will work with the distinguished Dynkin diagram of figure \ref{fig:dynkindist}.

\begin{figure}\centering
\setlength{\unitlength}{1pt}%
\small\thicklines%
\begin{picture}(260,20)(-10,-10)
\put(  5,00){\circle{15}}%
\put(  3.5,12){1}%
\put(  12,00){\line(1,0){21}}%
\put( 40,00){\ldots}%
\put( 57,00){\line(1,0){21}}%
\put( 85,00){\circle{15}}%
\put( 80.5,12){n-1}%
\put( 92,00){\line(1,0){21}}%
\put(120,00){\circle{15}}%
\put(117.5,12){n}%
\put(127,00){\line(1,0){21}}%
\put(155,00){\circle{15}}%
\put(149,12){n+1}%
\put(162,00){\line(1,0){21}}%
\put(190,00){\ldots}%
\put(207,00){\line(1,0){21}}%
\put(235,00){\circle{15}}%
\put(226,12){n+m-1}%
\put( 115,-5){\line(1, 1){10}}%
\put( 115, 5){\line(1,-1){10}}%
\end{picture}
\caption{The distinguished Dynkin diagram of $\slnm$.}\label{fig:dynkindist}
\end{figure}
Then the corresponding symmetric Cartan matrix reads

\[
 A = \begin{pmatrix}
2 &-1&0&\dots & &  & & & & \\
-1&2&-1&\dots &\vdots & & &0 & & \\
0&\dots & \ddots & -1&0& & & & & \\
\vdots&\dots&-1&2 &-1 & & & & \\
0&\dots& 0&-1& 0& 1 & 0 &\dots  & & \\
& & & &1 &-2 &1&0&\dots&\\
& & & &0 &1&-2&1&\dots&\\
& & 0& &\vdots & & & \ddots&1&\\
& & & &0 &\dots& &1&-2
\end{pmatrix} .
\]
This defines the commutation relation \eqref{eq:chevserre} of the generators $\csgh{I}$, $\csgp{I}$ and $\csgm{I}$, where $I=1,\dots, n+m-1$. In this distinguished basis, the roots $\csgp{n}$ and $\csgm{n}$ are fermionic, and all other generators are bosonic.
To make contact with our previously used bases we see that a choice for such Chevalley-Serre basis is given by

\<
&&\csgh{I} = E_{I,I} - E_{I+1,I+1},\quad 1\leq I < n\nln
&&\csgh{n} = E_{n,n} + E_{n+1,n+1},\nln
&&\csgh{I} = -E_{I,I} + E_{I+1,I+1},\quad n< I \leq n+m\nln
&&\csgp{I} = E_{I,I+1},\nln
&&\csgm{I} = E_{I+1,I} ,\quad 1\leq I \leq n\nln
&&\csgm{I} = -E_{I+1,I} ,\quad n< I \leq n+m .\nln
\>
Note that in this basis is left invariant by the shift \eqref{eq:gltosl}. Put differently, all Chevalley-Serre generators already have vanishing supertrace. Indeed, building a complete linear basis by commuting the simple root generators, one always gets the algebra $\slnm$, as the supercommutator of generators with vanishing supertrace again has vanishing supertrace. In the case $n\neq m$, the remaining generator $\ident = \sum_I E_{I,I}$ turning $\slnm$ into $\glnm$ is hence just a trivial $\alg{u}(1)$ charge not connected in any sense to the other $\slnm$ generators. This also means this $\alg{u}(1)$ charge has vanishing Killing form with each of the other generators of $\slnm$. Indeed, the derived algebra of $\slnm$ is again $\slnm$, i.e. $\scomm{\slnm}{\slnm}=\slnm$. Hence, for each $\gen{J}^1\in\slnm$, we can find some $\gen{J}^2,\gen{J}^3\in\slnm$ such that $\scomm{\gen{J}^2}{\gen{J}^3} = \gen{J}^1$. Then we have for the Killing form
\[
\sprod{\gen{J}^1}{\ident} = \sprod{\scomm{\gen{J}^2}{\gen{J}^3}}{\ident} = \sprod{\gen{J}^2}{\scomm{\gen{J}^3}{\ident}} =0.
\]
It follows that $\glnm$ is a trivial central extension of the simple Lie superalgebra $\slnm$, just as in the case of the (non-super) Lie algebras $\alg{gl}(n)$, which are trivial central extensions of $\alg{sl}(n)$. Again we emphasise that for $n=m$ the situation will be different, and we will investigate this case in the next section.\\

Let us, for completeness, also spell out the Chevalley-Serre generators in the basis $\gL{\alpha}{\alpha}$, $\gR{a}{b}$,$\gQ{\alpha}{a}$, $\gS{b}{\beta}$. We put

\<\label{def:repchevserre}
&&\csgh{a} = \gR{a}{a} - \gR{a+1}{a+1},\quad 1\leq a < n,\nln
&&\csgh{n} = \gR{n}{n} + \gL{1}{1} + \gC,\nln
&&\csgh{\alpha + n} = -\gL{\alpha}{\alpha} + \gL{\alpha+1}{\alpha+1},\quad \alpha \leq m\nln
&&\csgp{a} =  \gR{a}{a+1},\quad 1\leq a < n,\nln
&&\csgp{n} =  \gS{n}{1},\nln
&&\csgp{\alpha+n} =  \gL{\alpha}{\alpha+1},\quad 1\leq \alpha < m,\nln
&&\csgm{a} =  \gR{a+1}{a},\quad 1\leq a < n,\nln
&&\csgm{n} = \gQ{1}{n} ,\nln
&&\csgm{\alpha+n} =  -\gL{\alpha}{\alpha+1},\quad 1\leq \alpha < m,
\>

\subsubsection{$\alg{sl}(n|n)$}\label{sec:slnn}

Within the family of Lie superalgebras  $\slnm$, the case $n=m$ is special in several aspects. The $\alg{u}(1)$ generator
$\gen{C}$, as introduced in \eqref{eq:slnmfermi}, becomes central, as can be seen directly from the representation \eqref{def:centralc}. This means that removing generators with
non-vanishing supertrace from $\alg{gl}(n|m)$ is not sufficient to obtain a
simple Lie superalgebra. Put differently, $\alg{sl}(n|n)$ is the only subseries of
the special linear superalgebras $\alg{sl}(n|m)$ where the identity matrix is
part of the algebra and is identical to $\gen{C}$. To obtain a simple Lie superalgebra we project the identity
out and consider

\begin{equation}
\alg{psl}(n|n) = \alg{sl}(n|n)/<\gen{C}>.
\end{equation}
Let us stress the important fact that the central charge we projected out is a
nontrivial central charge, i.e. $\alg{sl}(n|n)$ is a non-trivial central
extension of $\alg{psl}(n|n)$. This is different e.g. to the case of
$\alg{gl}(n)$, which is a trivial central extension of $\alg{sl}(n)$. Indeed, this
 means we cannot find a basis such that $\gen{C}$ decouples from the other generators, i.e.
 does not appear on the right-hand side of Lie brackets of $\alg{sl}(n|n)$ generators. Phrased differently, 
 if we think of $\alg{psl}(n|n)$ as a subalgebra of $\alg{sl}(n|n)$, we have
 $\comm{\alg{psl}(n|n)}{\alg{psl}(n|n)}=\alg{sl}(n|n)$ as sets.
 Irreducible highest weight representation are specified by the eigenvalues of the Cartan generators, or the Dynkin labels. The generator $\gen{C}$ can have, in general, arbitrary eigenvalues, whereas the eigenvalues of the other Cartan generators belonging to the two $\alg{sl(n)}$ subalgebras are restricted to be integers. However, in the important case of atypical or short representations, to which the fundamental representation belongs, the eigenvalue of $\gen{C}$ is constrained by the eigenvalues of the other Cartan generators.

Note that for $\alg{psl}(n|n)$ there is no $2n$-dimensional fundamental
representation. This can be understood as follows: All $2n \times
2n$-dimensional supermatrices, i.e. $\alg{gl}(n|n)$, have $2n$ Cartan generators,
which correspond to the $2n$ diagonal matrix elements. Imposing the two
conditions of factoring out the identity as well as demanding vanishing
supertrace leaves one with the $ 2\times (n-1)$ Cartan generators of the two
bosonic subalgebras $\alg{sl}(n)$. However, when commuting a positive fermionic
root generator with its dual negative root, one will always end up with a
generator which is not a direct sum of the Cartan generators of the
$\alg{sl}(n)$'s. One can also explain the lack of existence of the fundamental representation for $\alg{psl}(n|n)$ by considering the above mentioned constraint for the Dynkin lables on the fundamental representation of $\alg{sl}(n|n)$, which implies that the central element has a non-zero eigenvalue, and can henceforth not be dropped. Hence, for $\alg{psl}(n|n)$, it is often useful to work with
projective representations, which are obtained from representations of $\alg{sl}(n|n)$ such that each matrix is just defined modulo
the central element. 

We can write the partition of $\alg{gl}(n|n)$ into its simple part $\pslnn$ and two
$\alg{u}(1)$ charges as

\[\label{eq:glnnsplit}
\alg{gl}(n|n) = \alg{u}(1)\ltimes\alg{psl}(n|n)\ltimes\alg{u}(1),
\]
where $A \ltimes B$ indicates a non-direct product such that $B$
is an ideal in $A \ltimes B$. The $\uone$ on the right-hand side corresponds to the central charge $\gen{C}$, which completes $\pslnn$ to $\slnn$.  Furthermore, the $\uone$ on the left-hand corresponds to the generator which is removed by requiring the vanishing of the supertrace. It acts as an external derivation on $\slnn$, and will be denoted by $\gen{H}_{2n}$.The action on the remaining Chevalley-Serre generators is given as follows:
\<\label{def:commh2n}
\comm{\gen{H}_{2n}}{\gen{H}_j} &=& 0, \nonumber\\
\comm{\gen{H}_{2n}}{\gen{E}_n} &=& \gen{E}_n,\nonumber  \\
\comm{\gen{H}_{2n}}{\gen{F}_n} &=& -\gen{F}_n,\nonumber  \\
\comm{\gen{H}_{2n}}{\gen{E}_j} &=& \comm{\gen{H}_{2n}}{\gen{F}_j} = 0,\quad j\neq n .
\>
Due to this diagonal action it can henceforth be considered as an additional Cartan generator. In the literature, this generator is sometimes called outer automorphism, as each derivation of a Lie superalgebra also defines an automorphism via the adjoint action of its exponential. Note that as any multiple of $\gen{H}_{2n}$ can be used to define an automorphism, we have a one dimensional vector space of external derivations, and a one-parameter family of external automorphisms. This is in contrast with all other series of simple Lie superalgebras, which have only a discrete set of external automorphisms.\\

Another curious point to note is that for the simple algebra $\pslnn$ the definition of its rank is somewhat ambiguous. Indeed, the Chevalley-Serre basis consists of $3\times (2 n+1)-1$ generators, as \eqref{def:repchevserre} for generic $\slnm$, but as mentioned, with $\gen{C}=0$. This constraint removes one Cartan generator, but one is still left with $2n+1$ positive and $2n+1$ negative simple root generators. Furthermore, a consequence is that the Cartan matrix is degenerate for $\pslnn$, but also for the central extension $\slnn$. However, if one extends the algebra further to $\glnn$, one obtains an extended Cartan-matrix 
\bea\label{def:cartnn}
\nonumber && \cartnn = \begin{pmatrix}
2 &-1&0&\dots & \dots &\dots  & \dots & \dots & \dots &0 \\
-1&2&-1&0 &\dots & \dots & \dots & \dots & \dots & \vdots\\
0&\ddots & \ddots & -1& 0& \dots & \dots & \dots & \dots &\vdots \\
\vdots&\dots&-1&2 &-1 & \ddots & \dots &\dots &\dots & 0 \\
0&\dots& 0&-1&0&1&0&\dots  &\dots &1 \\ \vdots
& \dots & \dots & \dots &1 &-2 &1&0&\dots&0 \\ \vdots
& \dots & \dots & \dots &0 &1&-2&1&\ddots&\vdots\\ \vdots
& \dots & \dots & \dots &\vdots & \dots & \ddots& \ddots&1&\vdots\\ \vdots
& \dots & \dots & \dots &0 &\dots& \dots &1&-2&0
\\ 0
&\dots &\dots&0 &1&0&\dots &\dots&0&0
\end{pmatrix}.
.\\
\eea
This matrix describes the action of a generalised Chevalley-Serre basis which includes the generators of $\slnn$ and the additional Cartan generator $\csgen{H}_{2n}$, defined via \eqref{def:commh2n}. It is non-degenerate.\\

As $\csgen{H}_{2n}$ does not appear on the right-hand side of any commutator of generators of $\glnn$, one is free to rescale $\csgen{H}_{2n}$ by a constant $\mu$, without changing the structure of the algebra. The only difference will be that the commutators with $\gen{E}_n$, $\gen{F}_n$, which now read

\<\label{def:commh2nmu}
\comm{\gen{H}_{2n}}{\gen{E}_n} &=& \mu\gen{E}_n,\nonumber  \\
\comm{\gen{H}_{2n}}{\gen{F}_n} &=& -\mu\gen{F}_n.
\>
Furthermore, one can add a multiple of the central element to $\csgen{H}_{2n}$, as this will not change any commutators at all. The shifted and rescaled generator $\csgen{H}_{2n}$ is then represented by

\[\label{def:generalauto}
\csgen{H}_{2n} = \frac{\mu}{2}\left(\sum_{I=1}^{n}E_{I,I} - \sum_{I=n+1}^{2n}E_{I,I}\right) + \frac{\lambda}{2 n}\sum_{I=1}^{2n}E_{I,I}\,.
\]
The rescaling and the shift modify the Cartan matrix, which now reads

\bea\label{def:cartnnmula}
\nonumber && \cartnn = \begin{pmatrix}
2 &-1&0&\dots & \dots &\dots  & \dots & \dots & \dots &0 \\
-1&2&-1&0 &\dots & \dots & \dots & \dots & \dots & \vdots\\
0&\ddots & \ddots & -1& 0& \dots & \dots & \dots & \dots &\vdots \\
\vdots&\dots&-1&2 &-1 & \ddots & \dots &\dots &\dots & 0 \\
0&\dots& 0&-1&0&1&0&\dots  &\dots &\mu \\ \vdots
& \dots & \dots & \dots &1 &-2 &1&0&\dots&0 \\ \vdots
& \dots & \dots & \dots &0 &1&-2&1&\ddots&\vdots\\ \vdots
& \dots & \dots & \dots &\vdots & \dots & \ddots& \ddots&1&\vdots\\ \vdots
& \dots & \dots & \dots &0 &\dots& \dots &1&-2&0
\\ 0
&\dots &\dots&0 &\mu&0&\dots &\dots&0&\lambda\mu
\end{pmatrix} 
.\\
\eea

\subsubsection{$\pslcentral$}\label{sec:pslcentral}

As we have seen in the last section, $\alg{psl}(n|n)$ is a special series of Lie
superalgebras in the sense that it allows for a non-trivial central extension,
and also has a continuous one-parameter family of outer automorphism. The case $n=2$ is even more
special, as it allows for in total three non-trivial central charges, as well as
three independent outer automorphisms forming an $\alg{sl}(2)$ algebra. This
feature crucially modifies the representation theory and allows physical models
with such symmetry to have special features. In particular, we will see in
chapter \ref{ch:smatrix} that all three additional central charges are required in the AdS/CFT
correspondence, so that the central charge corresponding to energy eigenvalues can have non-trivial values on representations. 

In this chapter we only want to describe the mathematical features of the
centrally extended $\pslcentral$. We start by describing the $\alg{sl}(2)$ outer
automorphism, which will, upon application on $\psltt$, automatically lead
to the central extension.

Using the same conventions as in the previous two sections, $\psltt$ consists of two $\alg{sl}(2)$'s generated by $\gR{a}{b}$ and $\gL{\alpha}{\beta}$, with all indices running from $1$ to $2$ now. The indices at the fermionic generators $\gS{a}{\alpha}$,$\gQ{\alpha}{a}$ are arranged, as before, in such a way that they transform in fundamental and antifundamental representation of the two bosonic subalgebras. As these are now simply $\alg{sl}(2)$ algebras, the fundamental and antifundamental representation are actually isomorphic. It is convenient to raise all indices of the fermions, and combine the fermions corresponding to positive roots with the negative roots into one doublet $\gen{Q}^{a\beta \mathfrak{c}}$ as follows:
\[
\gen{Q}^{a\beta 1}=\varepsilon^{ac}\gen{Q}^\beta{}_c,
\quad
\gen{Q}^{a\beta 2}=\varepsilon^{\beta\gamma}\gen{S}^a{}_\gamma.
\]
Hence, the new index $\mathfrak{c}=1,2$ appears in a similar fashion as the Latin and Greek indices. These did correspond to representation indices of the $\alg{sl}(2)$ algebras generated by $\gR{a}{b}$ and $\gL{\alpha}{\beta}$. Interestingly, one can define $3$ external generators $\gB{\mathfrak{a}}{\mathfrak{b}}$, $\mathfrak{a},\mathfrak{b}= 1,2$, with $\gB{1}{1}+\gB{2}{2}=0$, which act on $\alg{psl}(2|2)$ as follows: 
\<\label{def:sl2auto}
\comm{\gen{B}^{\mathfrak{a}}{}_{\mathfrak{b}}}{\gen{Q}^{c\delta\mathfrak{e}}}
&=& \delta_{\mathfrak{b}}^{\mathfrak{e}}\gen{Q}^{c\delta\mathfrak{a}}
-\frac{1}{2}\delta_{\mathfrak{b}}^{\mathfrak{a}}\gen{Q}^{c\delta\mathfrak{e}}\nln
\comm{\gen{B}^{\mathfrak{a}}{}_{\mathfrak{b}}}{\gR{a}{b}} &=& \comm{\gen{B}^{\mathfrak{a}}{}_{\mathfrak{b}}}{\gL{\alpha}{\beta}} = 0.
\>
Here, we combined the supercharges $\alg{Q}^\alpha{}_b,\alg{S}^a{}_\beta$
into one doublet of generators.
Indeed, one can convince oneself that the
$\gen{B}^{\mathfrak{a}}{}_{\mathfrak{b}}$'s generate an $\alg{sl}(2)$ algebra

\<
\comm{\gen{B}^{\mathfrak{a}}{}_{\mathfrak{b}}}{\gen{B}^{\mathfrak{c}}{}_{
\mathfrak{d}}}
= \delta_{\mathfrak{b}}^{\mathfrak{c}}\gen{B}^{\mathfrak{a}}{}_{\mathfrak{d}}
   -\delta^{\mathfrak{a}}_{\mathfrak{d}}\gen{B}^{\mathfrak{c}}{}_{\mathfrak{b}}.
\>
The global transformations they generate are given by

\<
\gen{Q}^{a\beta 1} \rightarrow a \gen{Q}^{a\beta 1} + b\gen{Q}^{a\beta 2}\nln
\gen{Q}^{a\beta 2} \rightarrow c \gen{Q}^{a\beta 1} + d\gen{Q}^{a\beta 2},
\>
with four complex numbers $a,b,c,d$ which, for compatibility with the commutation
relations, have to satisfy the condition

\[
ad-bc=1 .
\]
Then, all previous $\alg{psl}(2|2)$ relations are satisfied as before, apart from the appearance
of three additional charges $\gC^{\mathfrak{a}}_{\mathfrak{b}}$, which are central with respect to
$\alg{psl}(2|2)$:

\<\label{def:commfermicentral}
\acomm{\gen{Q}^{111}}{\gen{Q}^{221}} =
\gC^1_2\nln
\acomm{\gen{Q}^{112}}{\gen{Q}^{222}} =
-\gC^2_1
\>
We can combine the full commutation relations for all fermionic generators as

\<\label{def:commfermi}
\acomm{\gen{Q}^{a\alpha\mathfrak{a}}}{\gen{Q}^{b\beta\mathfrak{b}}} =
\epsilon^{ab}\epsilon^{\alpha\gamma}\epsilon^{\mathfrak{a}\mathfrak{b}}\gL{\beta}{\gamma}+\epsilon^{ac}\epsilon^{\alpha\beta}\epsilon^{\mathfrak{a}\mathfrak{b}}\gR{b}{c} + \epsilon^{ab}\epsilon^{\alpha\beta}\epsilon^{\mathfrak{a}\mathfrak{c}}\gC^\mathfrak{b}_\mathfrak{c}.
\>
If we denote the three independent central charges $\gen{C},\gen{P},\gen{K}$,
such that

\[
\gen{C}^1{}_1=-\gen{C}^2{}_2=\gen{C},
\quad
\gen{C}^1{}_2=\gen{P},
\quad
\gen{C}^2{}_1=-\gen{K},
\]
we see that $\gC$ appears precisely in the way as in \eqref{eq:slnmfermi}. $\gC$ is
central as for all $\glnn$, but the crucial difference in the case $n=2$ is that it is accompanied by $2$ further central charges. Likewise, the generator $\gB{1}{1}$ is quite similar to the generator $\gen{H}_{2n}$ introduces for $\glnn$, but again accompanied by two further external derivations. Note that if one would set $\gen{P}=\gen{K}=0$, and does not consider the action of $\gB{1}{2}$ and $\gB{2}{1}$, then one would indeed recover the algebra $\alg{gl}(2|2)$. However, if one considers the fully centrally extended algebra $\pslcentral$, then it is quite clear that the $\alg{sl}(2)$ automorphisms $\gB{\mathfrak{a}}{\mathfrak{b}}$ cannot act on the fundamental representation. This is because $\gen{B}^{\mathfrak{a}}{}_{\mathfrak{b}}$ act on the central charges
as

\<\label{def:autooncentral}
\comm{\gen{B}^{\mathfrak{a}}{}_{\mathfrak{b}}}{\gen{C}^{\mathfrak{c}}{}_{\mathfrak{d}}}
= \delta_{\mathfrak{b}}^{\mathfrak{c}}\gen{C}^{\mathfrak{a}}{}_{\mathfrak{d}}
   -\delta^{\mathfrak{a}}_{\mathfrak{d}}\gen{C}^{\mathfrak{c}}{}_{\mathfrak{b}},
\>
which can be derived from consistency with the commutation relations \eqref{def:sl2auto}. Note that the same consistency requires the eigenvalues of the
central charges to be related to the parameters of the $\alg{sl}(2)$ rotations as

\<
\gC &=& \frac{ad + bc}{2}\nln
\gP &=& ab\nln
\gK &=& cd.
\>
Furthermore, the fundamental representation of the fermionic generators needs to be modified to

\[\label{eq:fermigencentralrep}
\begin{array}[b]{rclcrcl}
\gen{Q}^\alpha{}_a\state{\phi^b}\eq a\,\delta^b_a\state{\psi^\alpha},&&
\gen{Q}^\alpha{}_a\state{\psi^\beta}\eq b\,\varepsilon^{\alpha\beta}\varepsilon_{ab}\state{\phi^b},\\[3pt]
\gen{S}^a{}_\alpha\state{\phi^b}\eq c\,\varepsilon^{ab}\varepsilon_{\alpha\beta}\state{\psi^\beta},&&
\gen{S}^a{}_\alpha\state{\psi^\beta}\eq d\,\delta^\beta_\alpha\state{\phi^a}.
\end{array}\]
As the automorphisms do not act on the bosonic part of the theory, their action remains unchanged. For completeness we give their representation:

\[\label{eq:bosegencentralrep}
\gen{R}^a{}_b\state{\phi^c}=\delta^c_b\state{\phi^a}
  -\half \delta^a_b\state{\phi^c},
\qquad
\gen{L}^\alpha{}_\beta\state{\psi^\gamma}=\delta^\gamma_\beta\state{\psi^\alpha}
  -\half \delta^\alpha_\beta\state{\psi^\gamma}.
\]
Note that for generic parameters $a,b,c,d$, the fundamental representation does not allow for a triangular decomposition, which means that there is no basis such that the positive (negative) roots can be represented in the upper (lower) triangular block. The fundamental representation is also not of highest weight type. Instead, it is a cyclical representation. \\

For some purposes, it is useful to consider the maximally extended algebra

\[
\alg{sl}(2)\ltimes\pslcentral,
\]
where indeed $\alg{sl}(2)$ acts as an outer automorphism on the ideal
$\pslcentral$. The problem of considering this big algebra is that the
$\alg{sl}(2)$ automorphism cannot be represented on the four dimensional
fundamental representation, and it is hence of limited interest for the applications
dealt with in this paper. The automorphisms are also not compatible with many other finite-dimensional representations. However, the physics of the AdS/CFT correspondence requires the existence of the central charges. This renders the invariant form to be degenerate, for the same reason as $\slnn$ has a degenerate form. For generic $n$ we only have one central charge, and the degeneracy could be lifted by considering the extension by the sole external derivation $\gen{H}_{2n}$, giving the algebra $\glnn$. It is easy to see that extending $\pslcentral$ to $\alg{sl}(2)\ltimes\pslcentral$ lifts this degeneracy, but also obstructs the construction of representations. This is a certain dilemma, as e.g. the construction of the R-matrix for the corresponding Yangian, as done in section \ref{sec:drinf2}, requires the existence of a non-degenerate form, whereas physics needs representations. We will overcome this dilemma, at least for the classical r-matrix, in section \ref{sec:adsbialgebra}.\\

\subsection{Yangian of $\slnm$}\label{sec:slnmyangians}

As we have seen in section \ref{sec:drinf1}, the Yangian in Drinfeld's first realisation can be defined
for any Lie superalgebra with non-degenerate invariant bilinear form. Hence, we can straightforwardly use the definitions \eqref{def:yangian1} for the case of the simple Lie superalgebras $\alg{psl}(n|m)$, but also their extension
$\glnm$. This will be discussed in section \ref{sec:yangglnm}. As we will see, this is also true when $n=m$, as well as for the case of 
$\alg{sl}(2)\ltimes\pslcentral$. However, we recall that $\slnn$ and $\pslcentral$ do not posses a non-degenerate form. Consequently, the $\alg{sl}(2)$ automorphisms are necessary to define a consistent Yangian structure. This is potentially problematic, as the automorphisms are, in most cases, incompatible with the $\pslcentral$ representations. If the automorphisms would appear in the coproduct of the $\ypslcentral$ generators, it could indeed be problematic to define tensor products of $\ypslcentral$. However, it turns out that the automorphisms do not appear explicitly in the coproducts of $\ypslcentral$. Hence, one can define tensor
products of $\ypslcentral$, despite the fact that the automorphisms cannot
be represented on most representation spaces of $\pslcentral$. Furthermore, it is consistent to define the Yangian of $\slnn$ and $\pslcentral$ in Drinfeld's second realisation. the particular case of $\pslcentral$ will be investigated in section \ref{sec:yangpslcentral}, with the results mainly based on \cite{Spill:2008tp}. Before introducing the Yangian, we will briefly discuss its classical counterpart, the Lie bialgebra structure corresponding to the loop algebra of $\glnm$, in section \ref{sec:glnmbialgebra}. Section \ref{sec:funR} deals with the evaluation of the universal R-matrix on the fundamental representation of the Yangian of $\glnm$, as investigated in \cite{Rej:2010mu}.

\subsubsection{The Lie Bialgebra of $\glnm$}\label{sec:glnmbialgebra}

Before coming to the actual Yangian, let us briefly state the classical Lie bialgebra of $\glnm$, or, rather, of $\lalg{\glnm}$. If we use the basis generators $E_{IJ}$, as introduced in \eqref{eq:commglnm}, then the classical r-matrix, derived from the classical double construction as shown in section \ref{sec:classdouble}, reads

\[\label{def:crmatglnm}
\crmat = \sum_{k,J}(-1)^{J}E_{IJ,k}\otimes E_{JI,-k-1} = \sum_{J}(-1)^{J}\frac{E_{IJ}\otimes E_{JI}}{u_1 - u_2} = \frac{\perm}{u_1 - u_2}.
\] 
The resulting cobrackets of the generators $E_{JI,k}$ of the polynomial algebra read explicitly

\[\label{def:cobraglnm}
\cobra(E_{KL,k}) = \sum_{l=0}^{k-1}\left((-1)^{|J|}E_{KJ,l+k}\otimes E_{JL,-l-1}\right)
\]
Hence, the classical r-matrix is simply the graded permutation operator, divided by the difference of the spectral parameters of the associated representation spaces. This is in line with the observation that the quadratic Casimir of $\glnm$ reduces to the permutation operator on the tensor product of two fundamental representations \eqref{def:casiglnmten}. For the centrally extended $\psucentral$, we are again facing the dilemma of having no Casimir. However, the extended algebra $\alg{sl}(2)\ltimes\psucentral$ has a non-degenerate form with associated quadratic Casimir

\[\label{def:casicentral}
\casiten =  \gR{a}{b}\otimes\gR{b}{a} - \gL{\alpha}{\beta}\otimes\gL{\beta}{\alpha} + \gQ{\alpha}{a}\otimes\gS{a}{\alpha}-\gS{a}{\alpha}\otimes\gQ{\alpha}{a} + \gB{\mathfrak{a}}{\mathfrak{b}}\otimes\gCt{\mathfrak{b}}{\mathfrak{a}}+ \gCt{\mathfrak{a}}{\mathfrak{b}}\otimes\gB{\mathfrak{b}}{\mathfrak{a}}.
\]
Interestingly, the resulting cobrackets of the generators of $\pslcentral$, which are spelled out in table \ref{tab:cobrapslcentral}, do not include any automorphism $\gB{\mathfrak{b}}{\mathfrak{a}}$. This will be crucial when constructing the Yangian of $\pslcentral$ in section \ref{sec:yangpslcentral}. However, the classical r-matrix related to the Casimir \eqref{def:casicentral}, and we will study the solution to this problem in section \ref{sec:adsbialgebra}.
\begin{table}\centering
\<
\coprocl(\gCt{\mathfrak{b}}{\mathfrak{a}})_n\eq \sum_{k=0}^{n-1}(\gCt{\mathfrak{b}}{\mathfrak{c}})_{k}\wedge(\gCt{\mathfrak{c}}{\mathfrak{a}})_{n-k-1}
\nln
\coprocl(\gen{R}_n)^a{}_b\eq
+\sum_{k=0}^{n-1}
(\gen{R}_k)^a{}_c\wedge(\gen{R}_{n-1-k})^c{}_b
\nl
-\sum_{k=0}^{n-1}
\Bigl[
(\gen{S}_k)^a{}_\gamma\wedge(\gen{Q}_{n-1-k})^\gamma{}_b
-\half\delta^a_b\,(\gen{S}_k)^d{}_\gamma\wedge(\gen{Q}_{n-1-k})^\gamma{}_d
\Bigr]
\nln
\coprocl(\gen{L}_n)^\alpha{}_\beta\eq
-\sum_{k=0}^{n-1}
(\gen{L}_k)^\alpha{}_\gamma\wedge(\gen{L}_{n-1-k})^\gamma{}_\beta
\nl
+\sum_{k=0}^{n-1}
\Bigl[
(\gen{Q}_k)^\alpha{}_c\wedge(\gen{S}_{n-1-k})^c{}_\beta
-\half\delta^\alpha_\beta\,(\gen{Q}_k)^\delta{}_c\wedge(\gen{S}_{n-1-k})^c{}_\delta
\Bigr]
\nln
\coprocl(\gen{Q}_{n})^\alpha{}_b\eq
-\sum_{k=0}^{n-1}
(\gen{L}_k)^\alpha{}_\gamma\wedge(\gen{Q}_{n-1-k})^\gamma{}_b
-\sum_{k=0}^{n-1}
(\gen{R}_k)^c{}_b\wedge(\gen{Q}_{n-1-k})^\alpha{}_c
\nl
-\sum_{k=0}^{n-1}
\gen{C}_{k-1}\wedge(\gen{Q}_{n-k-1})^\alpha{}_b
+\sum_{k=0}^{n-1}
\varepsilon^{\alpha\gamma}\varepsilon_{bd}\gen{P}_{k}\wedge(\gen{S}_{n-1-k})^d{}_\gamma
\nl
\nln
\coprocl(\gen{S}_n)^a{}_\beta\eq
+\sum_{k=0}^{n-1}
(\gen{R}_k)^a{}_c\wedge(\gen{S}_{n-1-k})^c{}_\beta
+\sum_{k=0}^{n-1}
(\gen{L}_k)^\gamma{}_\beta\wedge(\gen{S}_{n-1-k})^a{}_\gamma
\nl
+\sum_{k=0}^{n-1}
\gen{C}_{k}\wedge(\gen{S}_{n-k-1})^a{}_\beta
+\sum_{k=0}^{n-1}
\varepsilon^{ac}\varepsilon_{\beta\delta}\gen{K}_{k}\wedge(\gen{Q}_{n-1-k})^\delta{}_c
\>
\caption{Cobrackets of $\pslcentral$.}
\label{tab:cobrapslcentral}
\end{table}

\medskip

\subsubsection{$\yglnm$}\label{sec:yangglnm}

Let us start by considering the fundamental representation of $\glnm$ with
generic generators $\gen{J}^a$. Then the corresponding first level Yangian
generators $\geny{J}^a$ are represented as

\[
\geny{J}^a = u \gen{J}^a, 
\]

where $u$ is some complex number called the spectral parameter. As shown in
\cite{Chari:1994pz}, for $\alg{sl}(n)$ this satisfies all defining
relations of the Yangian, whereas for other simple Lie algebras such simple evaluation
representation does not work. The proof carries over to $\glnm$.
 We have introduced the Yangian as a deformation of the polynomial
algebra $\palg{g}$, but from the perspective of the fundamental representation
there is actually no difference to the Yangian. The crucial difference comes
into play when considering tensor products. The action on tensor products is
given by the coproduct \eqref{eq:coproductyang}. Now, if $u_i$ labels the spectral
parameter corresponding to tensor factor $i$, the coproduct of the Yangian
generators in the first realisation on two representation spaces reads

\<
\copro(\gen{J}^a)&=& \gen{J}^a\otimes 1+1\otimes \gen{J}^a\\
\copro(\geny{J}^a)&=& \geny{J}^a\otimes 1+1\otimes \geny{J}^a
+ \frac{1}{2}  f^a_{bc}\gen{J}^b\otimes\gen{J}^c \nln
&=&u_1\gen{J}^a\otimes 1+1\otimes u_2\gen{J}^a
+ \frac{1}{2} f^a_{bc}\gen{J}^b\otimes\gen{J}^c.
\>
Note that here we put $\hbar = 1$. 
Introducing the basic matrices $E_{IJ}$, as in \eqref{eq:commglnm}, and using the quadratic Casimir \eqref{def:casiglnm}, we get

\<
\copro \hat{E}_{KL} &=& \hat{E}_{KL}\otimes1+ 1\otimes\hat{E}_{KL} + \frac{1}{2}\scomm{E_{KL}\otimes 1}{\sum_{IJ} (-1)^{|J|}E_{IJ}\otimes E_{JI}} \nln
&=& \hat{E}_{KL}\otimes1+ 1\otimes\hat{E}_{KL} + \nln
&&\frac{1}{2}\sum_J \left((-1)^{|J|}E_{KJ}\otimes E_{JL} -(-1)^{|K|+|KL||JK|)}E_{JL}\otimes E_{KJ}\right).\nln
\>
Note that one can quickly convince oneself that this is a quantisation of the cobrackets \eqref{def:cobraglnm}, i.e. $\delta \sim \Delta -\Delta^{op}$.

With this information one can already derive the R-matrix on evaluation representations. Indeed, the r-matrix is defined as the intertwiner
\[\label{eq:yangintertwiner}
R\copro(\gen{X}) = \coproop(\gen{X}) R
\]
for all generators of the Yangian $\gen{X}$.
On the fundamental representation we get the well-known Yang R-matrix
\[
\rmat = 1 + \frac{1}{u_1-u_2}\perm,
\]
where $\perm$ is, as before, the graded permutation operator, which permutes the two representations. Note that if $u\sim \frac{1}{\hbar}$, then this quantum R-matrix is expanded as 

\[
\rmat = 1 + \hbar \crmat,
\]
with the classical r-matrix of \eqref{def:crmatglnm}.\\

Of course, one cannot fix the scalar factor of the R-matrix by solving equation \eqref{eq:yangintertwiner}. In this paper, we are mainly interested in a different, more powerful approach to obtain R-matrices from Yangians, namely via universal R-matrices, as studied in section \ref{sec:universalR} . Hence, we should study the fundamental evaluation representation of the Yangian in the second realization. Indeed, if we take the Chevalley-Serre basis \eqref{eq:chevserre} of $\glnm$, and consider the defining relation \eqref{def:drinf2}, we can straightforwardly derive the evaluation representation in this case. Motivated by the form of the isomorphism between the first and the second realisation \eqref{def:isom} we make the ansatz

\<
\csgh{i,k} = (u + a_i)^k \csgh{i,0}\nln
\csgp{i,k} = (u + a_i)^k \csgp{i,0}\nln
\csgm{i,k} = (u + a_i)^k \csgm{i,0}.
\>
Take e.g. the relation 

\<
\comm{\csgh{i,k+1}}{\csgp{j,l}}-[\csgh{i,k},\csgp{j,l+1}] = \frac{1}{2} A_{ij}
\{\csgh{i,k},\csgp{j,l}\}.
\>
Plugging in our ansatz we get 

\<
\rightarrow (a_i - a_j)A_{ij}\csgp{j,0} = \frac{1}{2} A_{ij}
\{\csgh{i,0},\csgp{j,0}\}.
\>
For $\glnm$ this gives us enough conditions to fix all but one $a_i$. Indeed, if we put $a_1=1/2$, we get

\<
a_i = \frac{i}{2}, \quad i\leq n\nln
a_i = \frac{2n-i}{2}, \quad n<i .
\>
One can check that all other defining relations \eqref{def:drinf2} are also satisfied with these parameters. We have also checked explicitly that the isomorphism \eqref{def:isom} is compatible with this choice, up to an overall shift in $u$.

\subsubsection{The R-matrix on the Fundamental Representation}\label{sec:funR}

In this section we will systematically evaluate the universal R-matrix on the 
fundamental representation of $\glnm$ and its Yangian in Drinfeld's second realization, as studied in the last section. The first step towards the explicit evaluation of the Cartan part \eqref{def:RH} of the universal R-matrix is to invert the operator $D$ given in (\ref{def:dop}).\\

The factors $R_+$ and $R_-$ are fairly easy to evaluate due to the nilpotence of the roots
\bea \label{REfun}
R_+&=&\prod^{\frac{1}{2} (m+n-1) (m+n)}_{k=1} \mbox{exp}(\frac{1}{u}(-1)^{|\beta_k|}\mathcal{F}^{|\beta_k|}E^{+}_{\beta_k}\otimes E^{-}_{\beta_k})\nln
&=&\prod^{\frac{1}{2} (m+n-1) (m+n)}_{k=1} \left(1+\frac{1}{u}(-1)^{|\beta_k|}\mathcal{F}^{|\beta_k|}E^{+}_{\beta_k}\otimes E^{-}_{\beta_k}\right) \,,
\eea
\bea \label{RFfun}
R_-&=&\prod^{\frac{1}{2} (m+n-1) (m+n)}_{k=1} \mbox{exp}(\frac{1}{u}\mathcal{F}^{|\beta_k|}E^{-}_{\beta_k}\otimes E^{+}_{\beta_k})\nln
&=&\prod^{\frac{1}{2} (m+n-1) (m+n)}_{k=1} \left(1+\frac{1}{u}\mathcal{F}^{|\beta_k|}E^{-}_{\beta_k}\otimes E^{+}_{\beta_k}\right)\,.
\eea
Here, $u=u_1-u_2$ is the difference of the spectral parameters on both factors of the tensor product.
The product in $R_+, R_-$ is taken in a particular order, as defined in \cite{Rej:2010mu}. On the fundamental representation the individual blocks commute, however. Thus what remains is the evaluation of the \eqref{def:RH}.  Let us first show that the factor \eqref{def:RH} is convergent. Indeed, each element of $R_H=\prod_{ij} (R_H)_{ij}$ is of the form

\[
(R_H)_{ij} \sim \prod_{n=0}^{\infty} \frac{a n + b}{a n + b + h_i}\frac{a n + b+h_i - h_j}{a n + b - h_j}\,.
\]
Using the following product representation of the Gamma function
\beq
 \Gamma(z) = \lim_{M\rightarrow\infty}\frac{1}{z}e^{-z(\sum_{k=1}^M 1/k - \log M)} \prod_{n=1}^{M} \frac{1}{1+z/n}e^{z/n}\,,
\eeq
one easily finds
\beq
(R_H)_{ij} \sim  \prod_{n=0}^{\infty} \frac{a n + b}{a n + b + h_i}\frac{a n + b+h_i - h_j}{a n + b - h_j}= \frac{\Gamma(\frac{b+h_i}{a})\Gamma(\frac{b- h_j}{a})}
{\Gamma(\frac{b+h_i - h_j}{a})\Gamma(\frac{b }{a})}\,.
\eeq
The matrix \eqref{def:RH} is diagonal since the Cartan algebra elements are diagonal. Using the prescription \eqref{eq:Dunitary} one finds
\beq \label{R0mix}
(R_H)_{11; 11}\equiv R_0 (u)= \left\{ \begin{array}{ccc}\sqrt{h(u)} \frac{ \Gamma \left(\frac{1-u}{n-m}\right) \Gamma
   \left(\frac{u}{n-m}\right)}{\Gamma
   \left(-\frac{u}{n-m}\right) \Gamma
   \left(\frac{u+1}{n-m}\right)}, \quad n \neq m, \\ \\
\frac{u+\frac{1}{2}}{u-\frac{1}{2}}, \quad n=m.  \end{array} \right.
\eeq
The function $h(u)$ is a simple ratio of trigonometric functions and, as follows from the discussion in the following section, may be dropped being solely a CDD factor. Surprisingly, this factor coincides for $m=0$ with the $u(n)$ dressing factor found in \cite{Berg:1977dp}. Moreover, for $m >0$ and $m \neq n$ it is identical to the $u(N)$ dressing factor with $N=n-m$. 

Evaluating the remaining elements of $R_H$ and combining them with the formulas \eqref{REfun} and \eqref{RFfun} one finds the following compact result for the R-matrix
\bea\label{eq:funR}
R &=& R_0(u) \left(\frac{u}{u+1} + \frac{1}{u+1}\perm \right)\,.\nonumber\\
\eea
Here, $\perm$ denotes the graded permutation operator
\[
\perm V_i\otimes V_j = (-1)^{|i||j|}V_j\otimes V_i.
\]
The matrix part of this R-matrix is the supersymmetric version of the Yang's R-matrix. It does not depend on whether the expansion is taken for $T \ll1$ or $T \gg 1$ in the Cartan part. This is expected, as this is the only solution to the rational Yang-Baxter equation on those representation spaces.

\subsubsection{Yangian of $\pslcentral$}\label{sec:yangpslcentral}

The principal definition of the Yangian works for the algebra
$\alg{sl}(2)\ltimes\psucentral$, 
as this has a non-degenerate invariant form. As argued before, we are particularly
interested
in the fundamental evaluation representation, on which the automorphisms do not
act.

Considering the generic coproduct of a Yangian generator,

\<
\copro(\geny{J}^a)&=& \geny{J}^a\otimes 1+1\otimes \geny{J}^a
+ \frac{1}{2}  f^a_{bc}\gen{J}^b\otimes\gen{J}^c,
\>
we notice that in principle the automorphisms can appear on the right-hand side
even for generators of 
$\pslcentral$. This would be a problem as then one could not act with
this coproduct on a tensor product of two representations, if the automorphisms
do not act on the individual representations. Luckily, it turns out that the
automorphisms do not appear on the 
right-hand side of coproducts of $\pslcentral$ generators. This can be seen by
recalling that 
$\psucentral$ forms an ideal in $\alg{sl}(2)\ltimes\psucentral$. Hence, the
non-trivial part of the coproduct

\[
 \comm{t}{\gen{J}^a\otimes 1} =
\comm{t_{\alg{psl}(2|2)}+\gB{x}{y}\otimes\gCt{y}{x}+\gCt{x}{y}\otimes\gB{y}{x}}{
\gen{J}^a\otimes 1} 
\]
is also in $\psucentral\otimes\psucentral$ for all elements $\gen{J}^a$ in $\pslcentral$.

Having established this fact, we indeed find that tensor products of representations of
$\psucentral$ can
be lifted to representations of the Yangian. 

\subsubsection{Drinfeld's Second Realisation of $\mathcal{Y}(\pslcentral)$}\label{sec:drinf2central}

In \cite{Spill:2008tp} it was shown that one can also define the Yangian of the centrally extended $\pslcentral$ in a consistent way in Drinfeld's second realisation. This is non-trivial, as the generic isomorphism between first and second realisation, valid for simple Lie algebras, is not working directly. It is however this realisation which is used for the construction of of the universal R-matrix. In this section, we give the details of \cite{Spill:2008tp}, but also include some new results on how to include one of the outer automorphisms into the algebra. Such automorphisms were already necessary to be included in the classical r-matrix (see section \ref{sec:adsbialgebra}), to lift the degeneracy of the invariant form. So one expects that it also plays a role for the quantum R-matrix. The first crucial remark is that the second realisation of the Yangian seems incompatible with the distinguished Dynkin diagram. Indeed, one can show that the some defining relations from \eqref{def:drinf2}, like 

\<
\comm{\csgh{i,m+1}}{\csgp{j,n}}-[\csgh{i,m},\csgp{j,n+1}] = \frac{1}{2} A_{ij}
\{\csgh{i,m},\csgp{j,n}\},\nonumber
\>
are not possible to satisfy for the fermionic root $\csgp{j,2}$, if one assumes simple evaluation type representations

\[
\csgh{i,m} = (u+a_i)^m\csgh{i,0}, \quad \gen{E}^\pm_{i,m} = (u+a_i)^m\gen{E}^\pm_{i,0}.
\]
One could cure this problem by allowing for more complicated representations, or introduce additional Fermi number generators. We are not sure if such choices are particularly natural, and will henceforth work with the fermionic Dynkin diagram, where such problems do not arise. This is probably related to the fact that all simple roots are fermionic and hence treated on an equal footing. We note that also the Bethe equations of the AdS/CFT correspondence, which derive from the centrally extended S-matrix, cannot be written in all Dynkin bases \cite{Beisert:2005fw}, so this might be related to the problems we face here. \\

Let us introduce the purely fermionic Chevalley-Serre basis. The extended Cartan matrix is given by
\[
A=\left(
\begin{array}{cccc}
 0 & 1 & 0 & 0 \\
 1 & 0 & -1 & -1 \\
 0 & -1 & 0 & 2 \\
 0 & -1 & 2 & 0
\end{array}
\right),
\]
and the corresponding Chevalley-Serre generators, satisfying the standard relations \eqref{eq:chevserre}, are given in terms of the generators used in section \ref{sec:pslcentral}, by

\<\label{def:chevserrexxx}
&&\csgh{1} = - \gR{1}{1}+\gL{1}{1}-\gC,\nln
&&\csgh{2} = -\gR{1}{1} - \gL{1}{1} + \gC,\nln
&&\csgh{3} = \gR{1}{1} - \gL{1}{1} - \gC,\nln
&&\csgp{1} =  \gQ{2}{1},\nln
&&\csgp{2} =  \gS{2}{2},\nln
&&\csgp{3} =  \gQ{1}{2},\nln
&&\csgm{1} =  -\gS{1}{2},\nln
&&\csgm{2} =  \gQ{2}{2},\nln
&&\csgm{3} =  -\gS{2}{1}.
\>
The fourth Cartan generator can be realised by 
\[
\csgh{4} = \gR{1}{1} + \gL{1}{1} - \gB{1}{1},
\]
which would satisfy the standard Chevalley-Serre relations, but does not act on the fundamental representation due to the involvement of the automorphism $\gB{1}{1}$. However, note the following observation. If one represents the generator $\gB{1}{1}$ as 

\[\label{def:representauto}
\gB{1}{1} = \frac{1}{4\gC} \left(E_{11} + E_{22} - E_{33} -E_{44}\right),
\] 
then $\csgh{4}$ correctly produces the Cartan matrix from the invariant product on the fundamental representation, which is given by 
\[
A_{ij} = \str(\csgh{i}\csgh{j}).
\]
However, $\csgh{4}$ will then not act diagonally on the roots. As $\gB{1}{1}$ is an external automorphism, we might just redefine the abstract action on the roots in such a way that is compatible with \eqref{def:representauto}.\\

Let us now give the definition of the Yangian in Drinfeld's second realisation. For $\pslcentral$, we find that the Chevalley-Serre generators $\csgh{i,k}$, $\csgp{i,k}$ and $\csgm{i,k}$, where as usual $k=0$ corresponds to the Lie generators \eqref{def:chevserrexxx}, satisfy the standard relations \eqref{def:drinf2}. However, the isomorphism to the first realisation \eqref{def:isom} does not exactly the standard form. Instead, the special elements \eqref{def:special} appearing in \eqref{def:isom} need to be modified to

\begin{eqnarray}
v_1 &=& - \frac{1}{2} \gen{H}_{2,0}^2 + \frac{1}{4} (\fR^2{}_1 \fR^1{}_2 +  \fR^1{}_2 \fR^2{}_1 -  \fL^2{}_1 \fL^1{}_2 -  \fL^1{}_2 \fL^2{}_1 \nonumber\\
&&+ \fQ^1{}_1 \fS^1{}_1 + \fQ^2{}_2 \fS^2{}_2 -  \fS^1{}_1 \fQ^1{}_1  - \fS^2{}_2 \fQ^2{}_2) -  \half\gen{P}\gen{K}  ,\nonumber\\ 
v_2 &=& - \frac{1}{2} \gen{H}_{1,0}^2 +  \frac{1}{4} (\fR^2{}_1 \fR^1{}_2 +  \fR^1{}_2 \fR^2{}_1 +  \fL^2{}_1 \fL^1{}_2 +  \fL^1{}_2 \fL^2{}_1 \nonumber\\
&&+ \fQ^2{}_1 \fS^1{}_2 - \fQ^1{}_2 \fS^2{}_1 -  \fS^1{}_2 \fQ^2{}_1  + \fS^2{}_1 \fQ^1{}_2) + \half\gen{P}\gen{K} ,\nonumber\\
v_3 &=& - \frac{1}{2} \gen{H}_{3,0}^2 + \frac{1}{4} (- \fR^2{}_1 \fR^1{}_2 -  \fR^1{}_2 \fR^2{}_1 +  \fL^2{}_1 \fL^1{}_2 +  \fL^1{}_2 \fL^2{}_1 \nonumber\\
&&- \fQ^1{}_1 \fS^1{}_1 - \fQ^2{}_2 \fS^2{}_2 +  \fS^1{}_1 \fQ^1{}_1  + \fS^2{}_2 \fQ^2{}_2) -  \half\gen{P}\gen{K},\nonumber  
\end{eqnarray}
\begin{eqnarray}
\label{def:specialcentral}
w_1 &=& - \frac{1}{4} (\gen{E}^+_{2,0} \gen{H}_{2,0} + \gen{H}_{2,0} \gen{E}^+_{2,0}) +\nln
&& \frac{1}{4} (\fS^1{}_1 \fL^1{}_2 +  \fL^1{}_2 \fS^1{}_1 -  \fS^2{}_2 \fR^1{}_2 -  \fR^1{}_2 \fS^2{}_2 - 2 \fQ^1{}_2 \fK),\nonumber\\
w_2 &=& - \frac{1}{4} (\gen{E}^+_{1,0} \gen{H}_{1,0} + \gen{H}_{1,0} \gen{E}^+_{1,0}) + \nln
&&\frac{1}{4} (\fQ^2{}_1 \fR^1{}_2 +  \fR^1{}_2 \fQ^2{}_1 +  \fQ^1{}_2 \fL^2{}_1 +  \fL^2{}_1 \fQ^1{}_2 + 2 \fS^1{}_1 \fP) +  \half\gen{S}_1^1\gen{P},\nonumber\\
w_3 &=& - \frac{1}{4} (\gen{E}^+_{3,0} \gen{H}_{3,0} + \gen{H}_{3,0} \gen{E}^+_{3,0}) +\nln
&& \frac{1}{4} (\fS^1{}_1 \fR^2{}_1 +  \fR^2{}_1 \fS^1{}_1 -  \fS^2{}_2 \fL^2{}_1 -  \fL^2{}_1 \fS^2{}_2 - 2 \fQ^2{}_1 \fK),\nonumber\\
z_1 &=& - \frac{1}{4} (\gen{E}^-_{2,0} \gen{H}_{2,0} + \gen{H}_{2,0} \gen{E}^-_{2,0}) + \nln
&&\frac{1}{4} (\fQ^1{}_1 \fL^2{}_1 +  \fL^2{}_1 \fQ^1{}_1 -  \fQ^2{}_2 \fR^2{}_1 -  \fR^2{}_1 \fQ^2{}_2 - 2 \fS^2{}_1 \fP),\nonumber\\
z_2 &=& - \frac{1}{4} (\gen{E}^-_{1,0} \gen{H}_{1,0} + \gen{H}_{1,0} \gen{E}^-_{1,0}) + \nln
&&\frac{1}{4} (\fS^1{}_2 \fR^2{}_1 +  \fR^2{}_1 \fS^1{}_2 + \fS^2{}_1 \fL^1{}_2 +  \fL^1{}_2 \fS^2{}_1 + 2 \fQ^1{}_1 \fK)+ \half\gen{Q}_1^1\gen{K},\nonumber\\
z_3 &=& - \frac{1}{4} (\gen{E}^-_{3,0} \gen{H}_{3,0} + \gen{H}_{3,0} \gen{E}^-_{3,0}) +\nln
&& \frac{1}{4} (\fQ^1{}_1 \fR^1{}_2 +  \fR^1{}_2 \fQ^1{}_1 -  \fQ^2{}_2 \fL^1{}_2 -  \fL^1{}_2 \fQ^2{}_2 - 2 \fS^1{}_2 \fP).\nonumber\\
\end{eqnarray}
Note the appearance of the central charges $\gC, \fK, \fP$, which is certainly not expected from the general form \eqref{def:special}. Indeed, the difference to \eqref{def:special}, when naively plugging in the generators, is given by

\<
(v_1)_{\text{standard}} - v_1 &=& + \half\gen{P}\gen{K},\\
(v_2)_{\text{standard}} - v_2 &=& - \half\gen{P}\gen{K},\\
(v_3)_{\text{standard}} - v_3 &=& + \half\gen{P}\gen{K},\\
(w_2)_{\text{standard}} - w_2 &=& - \half\gen{S}_1^1\gen{P},\\
(z_2)_{\text{standard}} - z_2 &=& - \half\gen{Q}_1^1\gen{K}.
\>
Here, we have labelled the standard special elements appearing in \eqref{def:special} by the subscript standard. We can now evaluate these special elements on the fundamental representation, or alternatively make the same ansatz as for $\slnm$,
\<
\csgh{i,k} = (u + a_i)^k \csgh{i,0}\nln
\csgp{i,k} = (u + a_i)^k \csgp{i,0}\nln
\csgm{i,k} = (u + a_i)^k \csgm{i,0}.
\>
Then we find the following interesting result:
\<
a_1 = 0,\nln
a_2 = \gC,\nln
a_3 = 0
\>
That means the spectral parameter corresponding to the second set of Chevalley-Serre generators is shifted by the eigenvalue of the central charge $\gC$. As $\gC$ can have continuous values in the centrally extended case, it is not a shift by half-integers, as in the case of $\glnm$ studied in section \ref{sec:yangglnm}. It is quite interesting that the rapidity, which parametrises the momentum, is shifted by the energy. Such combination also appears in the all-loop S-matrix, which can be seen as an indication that this S-matrix can be derived from a Yangian double.
\section{Integrable Models}\label{ch:integrablemodels}
In this chapter we will review several integrable models. In particular, we will discuss relativistic integrable field theories in section \ref{sec:fieldtheory} and integrable spin chains in section \ref{sec:xxxspinchain}. We will focus on discussing models with Yangian symmetries. Indeed, we will argue that the Yangian is a central object in the investigation of those theories. On the example of the principal chiral field, we will see that the Yangian leads, by direct evaluation of the universal R-matrix derived in section \ref{sec:drinf2}, to the exact S-matrix of this model. This offers a new perspective on the S-matrix formalism in integrable models. Previously S-matrices have usually been obtained by first using the invariance with respect to the Lie algebra symmetry, which combined with the Yang-Baxter equation yields the matrix structure of the S-matrix. Alternatively to the Yang-Baxter equation, one can use Yangians or other Quantum Groups to fix the matrix structure of the S-matrix \cite{Bernard:1990jw,Bernard:1992mu}. This is because the Yang-Baxter equation can be used as a defining relation for the Yangian, as reviewed in \cite{Molev:2002}. The scalar factor of the S-matrix is then fixed by exploiting crossing symmetry, as well as imposing unitarity and analyticity. Often, crossing, unitarity and Lie algebra symmetry are considered as independent concepts, whereas e.g. in \cite{Bernard:1992mu} it has been already pointed out that Quantum Groups can include all those symmetries. Our approach goes further, as we show that, in the case that the underlying Quantum Group is a Yangian, we can reconstruct the full S-matrix including a crossing symmetric, unitary prefactor, directly from the universal R-matrix. As we have extensively studied the Yangian in the last chapter, the derivation of the S-matrix is straightforward. This demonstrates the usefulness of our approach to S-matrices.\\

 It is well known that there are different links between spin chain models and integrable field theories. The existence of the same symmetry in two models can be seen as a first evidence for a link of the models. We will not elaborate on the details of how integrable field theories arise as limits of spin chains, and refer the reader to the review \cite{Volin:2010cq}. In \cite{Volin:2010cq}, the same integrable models are discussed as in this work, but from the perspective of functional equations rather than the symmetry algebras. Interestingly, the techniques used to solve the crossing equations are quite similar to the techniques of constructing the universal R-matrix, as studied in chapter \ref{ch:yangians}.\\ 
 
 The S-matrices are the fundamental object in constructing the Bethe equations, which were proposed in \cite{Bethe:1931hc} for the Heisenberg Hamiltonian. Bethe equations for symmetries of higher rank were investigated in \cite{Yang:1967bm,Sutherland:1975vr}. We will discuss the ideas of the Bethe ansatz only briefly, and refer the reader to the reviews \cite{Faddeev:1996iy,Nepomechie:1998jf,Beisert:2004ry}.

\subsection{Integrable Field Theories}\label{sec:fieldtheory}

In this section we would like to discuss some features of integrable relativistic field theories in $1+1$ dimensions. A crucial object of these theories is the S-matrix, which can be derived provided some quite general properties hold, namely, if the S-matrix is factorisable and satisfies the Yang-Baxter equation, it is crossing invariant and it is unitary. These properties are closely related to the abstract algebraic properties of quasi-triangular Hopf algebras, which we discussed before. We discuss these properties of S-matrices in section \ref{sec:smatrixfieldtheories}. A review on integrable field theories and their S-matrices can be found in \cite{Dorey:1996gd}. In section \ref{sec:lax} we introduce another important object in integrable models, the Lax connection. Section \ref{sec:pcf} deals with the Principal Chiral Field, which possesses, at least at the level of the classical theory, conserved Yangian charges.\footnote{There are actually ambiguities in the classical theory related to the appearance of non-ultra local terms.} We then reproduce its complete S-matrix by exploiting this Yangian symmetry\footnote{We assume there is no quantum anomaly, as argued in \cite{Luscher:1977uq, Abdalla:1986xb}}. More information on Yangians in integrable field theories can be found in the review \cite{MacKay:2004tc}.

\subsubsection{The S-matrix of Integrable Field Theories}\label{sec:smatrixfieldtheories}

Let us investigate a generic field theory defined in $1+1$ dimensions. It is often useful to work with light cone momenta
\[
p_i = p_i^0 + p_i^1,\quad \bar{p}_i =p_i^0 - p_i^1,
\]
where the index $i$ denotes particle $i$ with mass $m_i$. For simplicity we will take all masses to be equal. The light cone momenta can then be written in terms of rapidities $\theta_i$ as
\[
p_i = m e^{\theta_i} \quad \bar{p}_i = m e^{-\theta_i},
\]
where $\theta_i$ is a positive (negative) real number if the particle moves in positive (negative) direction along the real line. The transformation
\[
\theta_i \rightarrow - \theta_i
\]
transforms a particle moving in positive direction to one moving in negative direction. This can also be seen by expressing the original Lorentzian energy and momentum in terms of the rapidity,
\<
p^0_i &=& \frac{1}{2}(p_i+\bar{p}_i) = m \cosh(\theta_i)\nln
p^1_i &=& \frac{1}{2}(p_i-\bar{p}_i) = m \sinh(\theta_i).
\>
Furthermore, 
\[\label{def:thetacross}
\theta_i \rightarrow i \pi - \theta_i
\]
leads to a change in sign of the energy, which is why this transformation corresponds to an antiparticle transformation.

We would now like to investigate the scattering of particles. The S-matrix connects an in-state in the infinite past with an out-state in the infinite future. Scattering of all incoming particles can only take place if the particle to the furthest on the left is the fastest, and the one to the right is the slowest. Hence, we sort our in-state as
\[
\state{A_{a_1}(\theta_1)A_{a_2}(\theta_2)\dots A_{a_n}(\theta_n)}_{in}
\]
such that 
\[
\theta_1>\theta_2>\dots >\theta_n.
\]
Here, $\state{A_{a_i}(\theta_i)}$ denotes a particle of type ${a_i}$ with rapidity $\theta_i$\footnote{To be precise, the particle should be represented by a wave packet such that the dominant contribution comes from a momentum corresponding to the rapidity $\theta_i$.}. The S-matrix is now simply the operator
\[
S^{a_1,\dots,a_n}_{b_1,\dots,b_n}\state{A_{a_1}(\theta_1)\dots A_{a_n}(\theta_n)}_{in}=\state{A_{b_m}(\theta'_m)\dots A_{b_1}(\theta'_1)}_{out},
\]
which maps an in-state to an out-state. It depends on the rapidities of the scattered particles as well as on the type of particles it scatters.
Note that the order of rapidities of the particles of the out-state is reversed compared to the in-state, i.e.
\[
\theta'_1<\theta'_2<\dots <\theta'_m.
\]
Generically, the particle labels $a_k, b_l$ of the in and out state will correspond to a representation of some symmetry algebra, in many cases a simple Lie algebra. The classical way to look at it is to say that a particle carries a Lie algebra index, and consider the rapidity as something additional, independent from the Lie algebra. We would like to advertise that Quantum Groups are a more natural way unifying the matrix structure with the rapidity dependence. Indeed, in the case of Yangian symmetry, the rapidity $\theta$ is related to the spectral parameter $u$ of the Yangian. Hence, one can consider a particle described by a state $\state{A_{a_i}(\theta_i)}$ as living on the representation of the Yangian, and the rapidity comes in naturally.\\

Before investigating the precise symmetry structures for certain models, let us continue with some fairly general arguments for integrable relativistic field theories. Consider the total energy and momentum, which in the light cone are just given by
\<
P\state{A_{a_1}(\theta_1)A_{a_2}(\theta_2)\dots A_{a_n}(\theta_n)} &=& \sum_{k=1}^{n}m e^{\theta_k}\state{A_{a_1}(\theta_1)A_{a_2}(\theta_2)\dots A_{a_n}(\theta_n)}\nln
\bar{P}\state{A_{a_1}(\theta_1)A_{a_2}(\theta_2)\dots A_{a_n}(\theta_n)}&=& \sum_{k=1}^{n}m e^{-\theta_k}\state{A_{a_1}(\theta_1)A_{a_2}(\theta_2)\dots A_{a_n}(\theta_n)}.\nln
\>
These are of course conserved in a scattering process. We remark that in Hopf algebra language this simply means that energy and momentum act with the trivial coproduct on multi-particle states.

In a scattering process, energy and momentum conservation certainly restrict the S-matrix, but the effects become much more dramatic if one has an infinite set of higher spin\footnote{Spin refers as usual to representations of the Lorentz group, here in $1+1$ dimensions. Higher spins hence correspond to higher dimensional representations.} conserved currents. Consider charges with spin $s$ acting as
\[
\charge_s\state{A_{a_1}(\theta_1)} = e^{s\theta_1}\state{A_{a_1}(\theta_1)}.
\]
If infinitely many of them are conserved, then the scattering process of $n$ to $m$ particles,
\[
\state{A_{a_1}(\theta_1)A_{a_2}(\theta_2)\dots A_{a_n}(\theta_n)}_{in}\rightarrow\state{A_{b_1}(\theta'_1)A_{b_2}(\theta'_2)\dots A_{b_m}(\theta'_m)}_{out},
\]
leads to infinitely many equations
\[
e^{s\theta_1} + e^{s\theta_2} + \dots + e^{s\theta_n} = e^{s\theta'_1} + e^{s\theta'_2} + \dots + e^{s\theta'_m}.
\]
These equations can generally only be satisfied if $n=m$, i.e no particle production or annihilation takes place. Furthermore, the momenta can only be permuted, so the set of outgoing equals the set of incoming momenta. This would already greatly simplify the S-matrix structure. However, even more is true \cite{Shankar:1977cm}. As the charges $\charge_s$ roughly act as $p^s$, the action on a particle described by
\[
\psi(x) \propto \int_{-\infty}^\infty dp e^{-a^2(p-p_1)^2 + i p (x-x_1)},
\]
i.e. a wave packet with momentum centred at $p_1$, is given by
\[
e^{i \charge_3}\psi(x)\propto \int_{-\infty}^\infty dp e^{i p^3 - a^2(p-p_1)^2 + i p (x-x_1)},
\]
so if one expands the argument in the exponent about $p_1$, one gets a shift of $x_1$ by something depending on $p_1$. As the different incoming particles are assumed to have different momenta, we can hence use the higher spin conserved charges to separate the $n \rightarrow n$ scattering into $2\rightarrow 2$ particle scattering processes\footnote{In \cite{Parke:1980ki} it was shown that indeed the existence of two conserved charges of higher spin is sufficient for the factoristaion, whereas in \cite{Shankar:1977cm} infinitely many conserved charges are assumed to exist. However, in most known cases, if one has one additional conserved charge, one also finds infinitely many others. We believe that the reason for this is that we usually have Quantum Groups like the Yangian as the symmetry algebra in integrable models. The corresponding monodromy matrix generates both local and non-local conserved charges. The non-local charges are not commuting and form a Yangian, and indeed, one extra Yangian charge is usually sufficient to generate, together with the underlying Lie algebra, the whole Yangian.}. Hence, the $n\rightarrow n$ particle S-matrix factorises into a product of 2-particle scattering matrices, i.e.
\[
S(\theta_1,\dots,\theta_n))=\prod S(\theta_k,\theta_l).
\]
Here, the two particle S-matrix $S(\theta_k,\theta_l)$ is understood to act only on particles $k$ and $l$.

 Even more, one can argue that the order of the two particle scattering matrices composing the $n$ particle scattering process does not matter. This leads to the important Yang-Baxter equation, which is hence a necessary equation to hold for the theory to be integrable.
\[ 
S(\theta_1,\theta_2)S(\theta_1,\theta_3)S(\theta_2,\theta_3)=S(\theta_2,\theta_3)S(\theta_1,\theta_3)S(\theta_1,\theta_2).
\]
The Yang-Baxter equation itself just states that the order of the two-particle scattering processes within a factorised process of scattering three particles does not matter. As we will see in the case of the S-matrix of the AdS/CFT correspondence in chapter \ref{ch:smatrix}, the existence of an S-matrix which satisfies the Yang-Baxter equation is often seen as a strong hint for integrability, even though the existence of conserved charges at the quantum level was not shown there yet. Indeed, the Yang-Baxter equation is not a sufficient condition for the integrability of the theory. This is clear if one considers e.g. a theory without flavour, such that the S-matrix is just a scalar. This could be for instance $\phi^4$ theory in $1+1$ dimensions. This theory is not integrable. Interestingly, it can be made integrable by adding terms of higher order in the field, ultimately leading to the sinh-Gordon theory, see \cite{Dorey:1996gd}.

\begin{figure}[htb]\centering
\setlength{\unitlength}{8cm}%
\small\thicklines%
\begin{picture}(1, 1)
  \put(0.5, 0.5){\vector(1, 1){0.25}}
  \put(0.5, 0.5){\vector(-1, 1){0.25}}
  \put(0.75, 0.75){\line(1, 1){0.15}}
  \put(0.65, 0.5){$\smat_{12}(p_1,p_2)$}
  \put(0.25, 0.75){\line(-1, 1){0.15}}
  \put(0.25, 0.25){\line(1, 1){0.25}}
  \put(0.10, 0.10){\vector(1, 1){0.15}}
  \put(0, 0.05){$p_1,a_1$}
  \put(0, 0.95){$p_2,b_2$}
  \put(0.95,0.05){$p_2,a_2$}
  \put(0.95, 0.95){$p_1,b_1$}
  \put(0.75, 0.25){\line(-1, 1){0.25}}
  \put(0.90, 0.10){\vector(-1, 1){0.15}}
\end{picture}
\caption{Two particle scattering process.}\label{pic:smat2}
\end{figure}
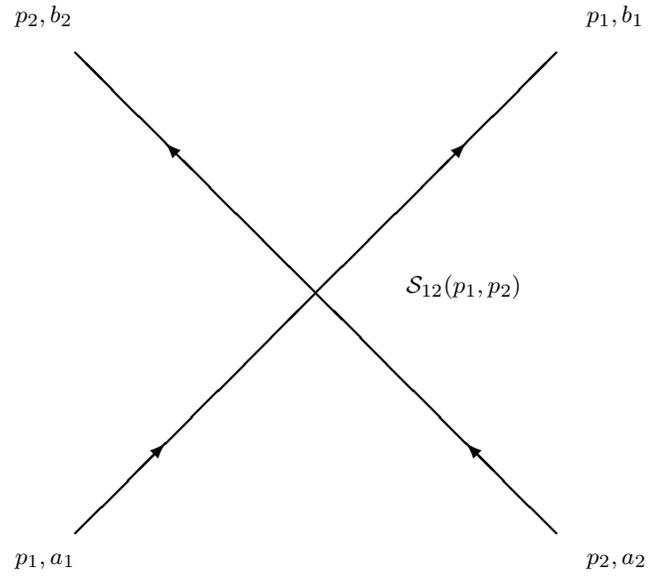

Due to the factorisation property of S-matrices in integrable field theories, one of the main remaining problems is to find the two particle S-matrix. In general, if we scatter two particles $1$ and $2$, the S-matrix will depend on the two rapidities $\theta_1, \theta_2$. This situation is depicted in figure \ref{pic:smat2}.

As we argued before, we have $p_3=p_2$ and $p_1 = p_4$, so the momenta are only permuted.
However, if we have a relativistic theory, then relativistic invariance implies also that the two particle S-matrix depends only on the difference  of the two rapidities. Indeed, the S-matrix should depend only on invariant quantities, and one can express the invariant Mandelstam variable 
\[
s = (p_{1}^\mu + p_{2}^\mu)^2
\]
as
\<
\frac{s}{m^2} &=& \cosh^2(\theta_1)-\sinh^2(\theta_1)+\cosh^2(\theta_2)-\sinh^2(\theta_2) +\nln
&& 2\cosh(\theta_1)\cosh(\theta_2) - 2\sinh(\theta_1)\sinh(\theta_1)\nln
&=& 2+2\cosh(\theta_1-\theta_2),\nln
\>
i.e. $s$ depends only on $\theta = \theta_1 - \theta_2$. 

\begin{figure}[htb]\centering
\setlength{\unitlength}{4cm}%
\small\thicklines%
\begin{picture}(1, 2)
    \qbezier(.5, .5)(0,1)(0.5, 1.5)
    \qbezier(.5, .5)(1,1)(0.5, 1.5)
    \put(0.25, 0.25){\vector(1, 1){0.1}}
    \put(0.75, 0.25){\vector(-1, 1){0.1}}
    \put(0.5, 0.5){\line(1, -1){0.15}}
    \put(0.5, 0.5){\line(-1, -1){0.15}}
    \put(0.5, 1.5){\vector(1, 1){0.15}}
    \put(0.5, 1.5){\vector(-1, 1){0.15}}
    \put(0.65, 1.65){\line(1, 1){0.1}}
    \put(0.35, 1.65){\line(-1, 1){0.1}}
    \put(0.15, 1.80){$p_1$}
    \put(0.1, 0.95){$p_2$}
    \put(0.80,0.2){$p_2$}
    \put(0.80, 0.95){$p_1$}
    \put(0.80,1.8){$p_2$}
    \put(0.1, 0.2){$p_1$}
    \put(0.65, 0.5){$\smat(p_1,p_2)$}
    \put(0.65, 1.5){$\smat(p_2,p_1)$}
    \put(0.25,1){\vector(0,1){0.01}}
    \put(0.75,1){\vector(0,1){0.01}}
\end{picture}
\caption{Unitarity of the S-matrix.}\label{pic:unitarity}
\end{figure}
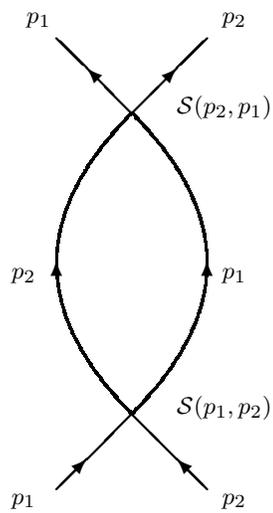

An important issue is that if we apply the S-matrix to a two particle state, where the difference of rapidities of the two particles is $\theta = \theta_1 - \theta_2$, and then apply the S-matrix again, the initial and final states are indistinguishable. As the momenta got permuted after the first application of the S-matrix, this means that the second S-matrix depends on the relative rapidity $-\theta = \theta_2 - \theta_1$. This situation is depicted in figure \eqref{pic:unitarity}.

In terms of rapidity depending S-matrices we have
\[
S(\theta)S(-\theta)=\idm ,
\]
or with indices
\[
S(\theta)_{a_1 a_2}^{b_1 b_2}S(-\theta)_{b_1 b_2}^{c_1 c_2}=\delta_{a_1}^{c_1}\delta_{a_1}^{c_1}.
\]
This property of the S-matrix is called unitarity. \\

\begin{figure}[htb]\centering
\setlength{\unitlength}{8cm}%
\small\thicklines%
\begin{picture}(1, 1)
  \put(0.5, 0.5){\vector(1, 1){0.25}}
  \put(0.5, 0.5){\vector(-1, 1){0.25}}
  \put(0.75, 0.75){\line(1, 1){0.15}}
  \put(0.65, 0.5){$\bar{\smat}_{12}(p_1,-p_2)$}
  \put(0.25, 0.75){\line(-1, 1){0.15}}
  \put(0.25, 0.25){\line(1, 1){0.25}}
  \put(0.10, 0.10){\vector(1, 1){0.15}}
  \put(0, 0.05){$p_1,a_1$}
  \put(0, 0.95){$-p_2,\bar{b}_2$}
  \put(0.95,0.05){$-p_2,\bar{a}_2$}
  \put(0.95, 0.95){$p_1,b_1$}
  \put(0.75, 0.25){\line(-1, 1){0.25}}
  \put(0.90, 0.10){\vector(-1, 1){0.15}}
\end{picture}
\caption{The crossed S-matrix.}\label{pic:crossing}
\end{figure}
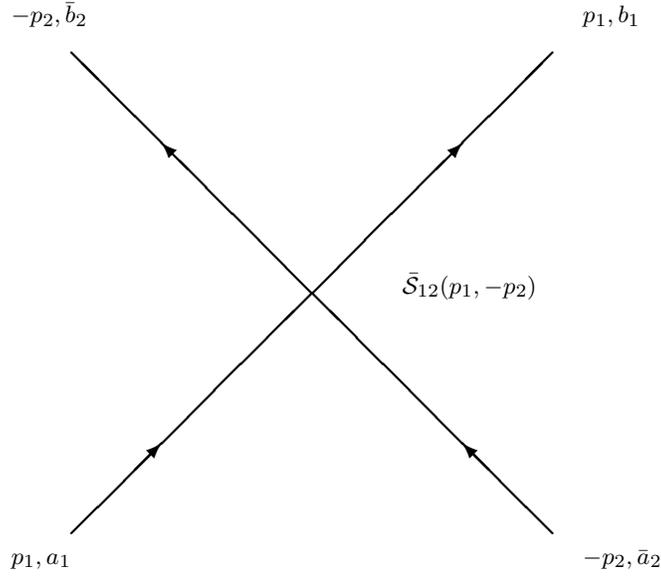

As this S-matrix acts like
\[
S(\theta_2-\theta_1 )^{a_1,a_2}_{b_1,b_2}\state{A_{a_1}(\theta_1)A_{a_2}(\theta_2)}_{in} = \state{A_{b_2}(\theta_2)A_{b_1}(\theta_1)}_{out},
\]
it is considered as the forward scattering matrix $a_1, a_2\rightarrow b_1, b_2$. If we consider the crossed S-matrix $a_1, \bar{b}_1\rightarrow \bar{b}_2, a_2$, we see that the antiparticle $\bar{b}_1$ has momentum $-p_2$. The situation is shown in figure \ref{pic:crossing}.

The invariant in this channel is the Mandelstam variable
\[
t=(p_1-p_3)^2 = (p_1-p_2)^2,
\]
which can be expressed in term of the crossed rapidity \eqref{def:thetacross} as
\[
\frac{t}{m^2}=2 + 2\cosh(i \pi - \theta).
\] 
We should note that, as after all both $s$ and $t$ depend only on $\theta$, they are not really independent variables. This is of course expected in $1+1$ dimensions. Cross-channel unitarity reads
\[
S(i \pi - \theta)_{a_1 \bar{a}_2}^{\bar{b}_1 b_2}S(i \pi + \theta)_{\bar{b}_1 b_2}^{c_1 \bar{c}_2}=\delta_{a_1}^{c_1}\delta_{\bar{a}_2}^{\bar{c}_2}.
\]
Now, crossing invariance precisely means that the t-channel matrix should equal the s-channel matrix, i.e. in components we have
\[
S(i \pi - \theta)_{a_1 \bar{a}_2}^{\bar{b}_1 b_2} = S(\theta )^{a_1,a_2}_{b_1,b_2}.
\]
This is the crossing equation.\\

If one has a relativistic integrable model, one can now use the Yang-Baxter equation, combined with crossing and unitarity, to derive the two-particle S-matrix. This will generically completely fix the S-matrix up to one undetermined factor $X(\theta)$, which satisfies
\<
X(\theta)X(-\theta)=1,\nln
X(\theta)X(i\pi -\theta)=1.
\>
Such factor is called a CDD factor. It cannot be fixed by the symmetries directly, but is instead fixed by assumptions on the pole structure of the S-matrix. Certain poles in the S-matrix correspond to bound states of the fundamental particles, so knowledge about the particle content of the theory can be used to fix the CDD factor.\\

Having established the two particle S-matrix, we can compose multi-particle S-matrices of two particle S-matrices using factorizability. Then, one can use the Bethe ansatz to diagonalise the scattering process. If we consider the field theory on a large cylinder, periodic boundary conditions allow one to solve for the momenta, or rapidities. Combined with the dispersion relation, this allows one to find the asymtotic energy spectrum of the theory on a large cylinder.\\

We would like to comment on this general approach to find the S-matrix for integrable models. We have argued that crossing, unitarity and the Yang-Baxter equation are quite general features of an S-matrix in integrable field theories. Additionally, the S-matrix should be invariant under the symmetry algebra, which is usually just taken to be a Lie algebra. Then, crossing, unitarity and the Yang-Baxter equation are somewhat disconnected from the Lie algebra symmetry. We would now like to advertise the point of view that quantum groups offer a unified approach to S-matrices of integrable field theories. We have seen in section \ref{sec:quantumgroups} that quasi-triangular Hopf algebras automatically lead to S-matrices which satisfy crossing and the Yang-Baxter equation\footnote{Unitarity should be related to triangularity. We have given a prescription to yield unitary S-matrices from the Yangian Double, despite that it is only quasi-triangular.}. In particular, Yangians extend ordinary Lie algebra symmetry, and provide rational S-matrices. Hence, with this logic, all we have to do is to show our model has a quantum group symmetry, and all desired properties for the S-matrix follow directly. We will follow this logic by studying precise examples in section \ref{sec:pcf} for the principal chiral field.

\subsubsection{Lax Connection and Conserved Charges}\label{sec:lax}

Another important concept in integrable models is the Lax pair and the monodromy matrix. Indeed, the trace of the monodromy matrix, which is called the transfer matrix, is the object which generates all the conserved local charges, rendering a model integrable. To start with, the equations of motion of the integrable field theory under investigation must have the property that they can be written in terms of the flatness condition for the connection $(\partial_0 + L_0,\partial_1 + L_1)$, i.e.
\[
\comm{\partial_0 + L_0}{\partial_1 + L_1}=0
\]
must imply the equations of motion. The precise form of the Lax matrices $L_0,L_1$ will depend on the model. Generically, they will depend on a spectral parameter $\lambda$. This will be important, as this means we have a one-parameter family of flat connections. \\
Let us consider the monodromy matrix associated to the Lax connection. It is defined by
\[
T(t,\lambda) = \mathcal P \exp\left(-\int_0^{2\pi} L_1(x,\lambda)dx\right),
\]
where $\mathcal P$ is the path-ordering operator. Its time derivative is given by
\<
&&\partial_t T(t,\lambda) = \partial_t \lim\prod_{k}^\leftarrow (1-L_1(x_k,\lambda)dx)\nln
&=&-\lim\sum_r\prod_{k=n}^{r+1}(1-L_1(x_k,\lambda)dx)\partial_tL_1(x_r,\lambda)dx\prod_{k=r-1}^{1}(1-L_1(x_k,\lambda)dx)\nln
&=&-\lim\sum_r\prod_{k=n}^{r+1}(1-L_1(x_k,\lambda)dx)(\partial_x L_0(x_r,\lambda)+\comm{L_1(x_r,\lambda)}{L_0(x_r,\lambda)})dx\nln
&&\prod_{k=r-1}^{1}(1-L_1(x_k,\lambda)dx)\nln
&=& L_0(2\pi,\lambda)\exp\left(-\int_0^{2\pi} L_1(y,\lambda)dy\right)-\exp\left(-\int_0^{2\pi} L_1(y,\lambda)dy\right) L_0(0,\lambda)\nln
&=& \comm{ L_0(0,\lambda)}{T(t,\lambda)}.\nln
\>
Here, we have used that the theory is defined on a cylinder with periodicity  $2 \pi$, and suppressed the time dependence for $L$. As we find that the time derivative of $T(t,\lambda)$ is a commutator, this implies that the trace of $T(t,\lambda)$ is time independent. Furthermore, even $\partial_t \tr(T^n(t,\lambda) = 0$ is true. If we expand those quantities in $\lambda$, we get infinitely many conserved charges. The precise form of the expansion will influence the nature of the conserved charges. Indeed, a priori it is not clear which charges, if any, can be written as integrals of local densities. Such charges were used in the arguments to show factorisability of the S-matrix. Indeed, local charges can be obtained by expanding about the poles, see \cite{Babelon:2003book}. In \cite{Luscher:1977uq}, also non-local charges, which are obtained from the expansion about a point traditionally assumed to be $\lambda = \infty $, were shown to lead to the absence of particle production and the factorisation of the S-matrix.

\subsubsection{Principal Chiral Field}\label{sec:pcf}

In this section we study the $1+1$ dimensional Principal Chiral Field with $SU(n)\times SU(n)$ symmetry. This is one of the simplest integrable field theories one can write down. We will show in particular its relation to the Yangian, culminating in the reproduction of the complete S-matrix just from Yangian symmetry. We will only work on the level of the classical theory and Poisson brackets, relying on arguments already given \cite{Luscher:1977uq, Abdalla:1986xb} that there will be no quantum anomalies upon quantization.

The action of the theory is given by
\[
S = \frac{1}{2 \lambda}\int d^2x \tr(G^{-1}\partial_\mu G)(G^{-1}\partial^\mu G),
\]
where $G$ takes values in the $SU(n)$ group, and $\lambda$ is a coupling constant, which plays no role for the investigation of the classical theory in this section.

The Principal Chiral Field possesses local as well as non-local conserved charges, and the latter form a Yangian. All these charges have been investigated in \cite{Goldschmidt:1980wq,Evans:1999mj}. Let us start by describing the usual Noether charges. $SU(n)$ acts globally from the left and from the right in the usual way,
\[
G\rightarrow U_L G U_R^{-1},
\]
with $U_L\in SU(n)_L$ and $U_R\in SU(n)_R$. This transformation is apparently a symmetry of the Lagrangian, and the corresponding conserved currents are
\<
j_L^\mu = \frac{1}{\lambda}\partial^\mu G G^{-1}\nln
j_R^\mu = -\frac{1}{\lambda}G^{-1}\partial^\mu G .
\>
The conservation equation
\[
\partial^\mu j_\mu =0
\]
implies, as usual, that the associated charges
\[
\charge^{a}_0 = \int_{-\infty}^{\infty} j_0^a(x) dx 
\]
are constant in time. Here, we have made the Lie algebra index $a$ explicit. Now one can convince oneself that the charge
\[\label{def:nonlocalcharge}
\charge^{a}_1 = \int_{-\infty}^{\infty} j_1^a(x) dx - \frac{\lambda}{2}\scons \int_{-\infty}^{\infty} j_0^b(x)\int_{-\infty}^{x} j_0^c(y) dy dx 
\]
is also conserved. This follows from the fact that the current satisfies the flatness condition
\[\label{def:pcfflatcurrent}
\partial_\mu j_\nu - \partial_\nu j_\mu -\lambda\comm{j_\mu}{j_\nu}=0.
\]
Note that the index $0,1$ at the currents corresponds to the specification of the space-time index $\mu$, whereas the $0,1$ at the charges $\charge$ corresponds to a numeration of infinitely many such charges. In particular, the charges $\charge_0$ Poisson commute to a $\sun$ Lie algebra, whereas the charges $\charge_1$ correspond to the additional Yangian generators of $\yang{\sun}$, as we introduced them in \eqref{def:yangian1} (we ignore complications due to the non-ultra local terms as discussed below and in the references \cite{Maillet:1985fn,Maillet:1985ec,Maillet:1985ek,MacKay:2004tc}. Indeed, it was shown that all defining relations for the Yangian are satisfied (see \cite{MacKay:2004tc}). Having in mind that the Lie Group acts from the left and the right, there are also two corresponding copies of the Yangian.

The Poisson brackets of the currents \cite{Maillet:1985fn,Maillet:1985ec,Maillet:1985ek} are given by
\<
\acomm{j_0^a(x)}{j_0^b(y)}=f^{abc}{j_0^c(x)}\delta(x-y),\nln
\acomm{j_0^a(x)}{j_1^b(y)}=f^{abc}{j_1^c(x)}\delta(x-y)+\frac{1}{\lambda}\delta^{ab}\delta'(x-y) .
\>
These hold again for left and right $SU(n)$ separately. The derivative of the delta function, which renders the Poisson brackets non-ultra local, complicates the investigation of the commutation relations. We will not discuss the consequences of these terms in detail, as we will now merely state the local conserved charges, but not derive their Poisson brackets. Further details can be found in the above mentioned references. It is useful to work in light cone coordinates $x_\pm =\frac{1}{2}(t\pm x)$, where the flatness condition takes the form
\[
\partial_-j_+ = \partial_+j_- = -\lambda\comm{j_+}{j_-}.
\]
Then it is easy to see that the non-vanishing components of the energy-momentum tensor,
\[
T_{\pm \pm} = - \frac{\lambda}{2} \tr(j_\pm j_\pm),
\]
are conserved:
\[
\partial_+ T_{--}\ =\partial_- T_{++} = 0.
\]
Furthermore, their powers $T_{\pm\pm}^n$, which are charges of higher spin, also satisfy
\[
\partial_+ T_{--}^n\ =\partial_- T_{++}^n = 0
\]
and are hence also conserved.

Let us now choose an orthonormal basis of $\sun$ with respect to the Killing form, i.e. $\sprod{\gen{J}^{a}}{\gen{J}^{b}}=-\delta^{ab}$. 
Furthermore, consider the tensors $d_{a_1,\dots ,a_m}$ associated to Casimir operators $\mathcal C$ of $\sun$ via
\[
\mathcal C = d_{a_1,\dots ,a_m} \gen{J}^{a_1}\dots\gen{J}^{a_m}.
\]
They can be chosen to be totally symmetric. These details can be found in any Lie algebra book, see e.g. \cite{Fuchs:1997jv}.
In terms of $d$ invariance of the Casimir means
\[
d_{c (a_1,\dots ,a_{m-1}}f_{a_m)bc} = 0,
\]
where the brackets $()$ denote symmetrisation of the appropriate indices. 

Then one finds that
\<
\partial_\pm d_{a_1,\dots ,a_m}j_\mp^{a_1}j_\mp^{a_2}\dots j_\mp^{a_m} &\propto& d_{a_1,\dots ,a_m} \sum_{k=1}^mj_\mp^{a_1}\dots j_\mp^{a_{k-1}}f^{a_k}_{bc}j_\mp^{b}j_\pm^{c}\dots j_\mp^{a_m}\nln
&=& 0,
\>
using the flatness condition for the current. 

The corresponding conserved charges are of higher spins, and hence should imply the factorisation of the S-matrix. Furthermore, it was shown in \cite{Evans:1999mj} that these local charges commute with the Yangian charges, and, by taking certain combinations of the conserved charges, one could construct a basis of Poisson commuting local charges.

As we have seen in section \ref{sec:fieldtheory}, if one has established the factorisation of the S-matrix, the usual way to derive the S-matrix in a relativistic, integrable quantum field theory is solve for the two particle S-matrix. This is usually done by solving the Yang-Baxter equation, imposing unitarity as well as crossing invariance. The remaining CDD ambiguity is then fixed by investigating the particle content of the theory. All these equations are fairly complicated. We will advocate here another method to find the S-matrix, as established in our recent work \cite{Rej:2010mu}. We learned that the principal chiral field is invariant under two copies of the Yangian $\mathcal{Y}(\sun)$. Furthermore, we know that the (double) Yangian is quasi-triangular, i.e. it has a universal R-matrix, which, on representation, automatically satisfies the Yang-Baxter and crossing equation, and, upon the right choice of the diagonalisation of the Cartan part, is also unitary \eqref{R0mix}. As an S-matrix satisfying all those equations is unique up to a CDD ambiguity, the S-matrix of the principal chiral field should be of the form
\[
S(\theta)_{PCF} = X_{PCF}(\theta)S(\theta)_{\mathcal{Y}(\sun)}\otimes S(\theta)_{\mathcal{Y}(\sun)},
\]
where $S(\theta)_{\mathcal{Y}(\sun)}$ is the $\mathcal{Y}(\sun)$ invariant S-matrix, coming from a representation of the universal R-matrix of sections \ref{sec:universalR},\ref{sec:funR}. $X_{PCF}$ is a CDD factor, which is fixed by making the diagonalisation of the Cartan part of the Yangian compatible with the particle spectrum of the theory. We recall the result for the $\mathcal{Y}(\sun)$ S-matrix in the antisymmetric expansion of the shift operator \eqref{R0mix},
\beq\label{def:Smatpcf}
S(u)_{\mathcal{Y}(\sun)}= \sqrt{\frac
{\Gamma(\frac{u}{n-m})\Gamma(1+\frac{u}{n-m})}{\Gamma(\frac{1+u}{n-m})\Gamma(1-\frac{1-u}{n-m})}\frac
{\Gamma(\frac{1-u}{n-m})\Gamma(1-\frac{1+u}{n-m})}{\Gamma(\frac{-u}{n-m})\Gamma(1+\frac{-u}{n-m})}}\left(\frac{1}{u+1}\idm +\frac{u}{u+1}\perm\right).
\eeq
 If we put $\theta = \frac{u N}{2 \pi i}$, we recover the precise result from \cite{Abdalla:1984iq,Wiegmann:1984ec},
\<\label{def:Smatpcfliterature}
S(\theta)_{PCF} &=& \frac{\sinh(\theta/2 +\frac{i\pi}{N})}{\sinh(\theta/2 -\frac{i\pi}{N})}\left(\frac{\Gamma(1-\frac{\theta}{2\pi i})\Gamma(\frac{1}{N}+\frac{\theta}{2\pi i})}{\Gamma(1+\frac{\theta}{2\pi i})\Gamma(\frac{1}{N}-\frac{\theta}{2\pi i})}\right)^2\nln
&&\left(P^+ + \frac{\theta + 2 \pi i/N}{\theta - 2 \pi i/N}P^-\right)\otimes\left(P^+ + \frac{\theta + 2 \pi i/N}{\theta - 2 \pi i/N}P^-\right),\nln
\>
provided that we put the CDD factor to
\[
 X_{PCF}(\theta)=1.
\]
Note that, as here the square of the $\mathcal{Y}(\sun)$ S-matrix is involved, this also resolves the square root ambiguity encountered in \eqref{R0mix}. Furthermore, we have used the operators $P^+, P^-$ projecting on the symmetric and antisymmetric part of the tensor products. They are defined as usual by
\[
P^\pm = \frac{1}{2}(\idm \pm \perm).
\]   
This S-matrix leads to the Bethe Ansatz equations
\[
e^{i p_k L} = \prod_{l\neq k}^M \smat_{kl}(\theta_k -\theta_l),
\]
describing the phase shift of a particle with momentum $p_k$, when it lives on the cylinder with radius $L$ together with $M-1$ other particles. The intuition is that if particle $p_k$ goes once around the cylinder, it will scatter the $M-1$ other particles and acquire a phase shift $\prod_{l\neq k}^M \smat_{kl}(-\theta_k +\theta_l)$ from the scattering, and $e^{i p_k L}$ from the free propagation, and the total phase shift should equal one. We will not deal with the precise derivation or solution of the Bethe Ansatz in this paper, and refer the reader to \cite{Polyakov:1983tt,Wiegmann:1984pw,Wiegmann:1984ec} for the solution at finite $n$, and \cite{Fateev:1994ai,Fateev:1994dp} for the solution for large $n$. We also refer to the review \cite{Volin:2010cq}.\\

\paragraph*{Lax connection for the principal chiral field}
Let us now construct the Lax connection for the principal chiral field. We had argued that the current $j$ is flat \eqref{def:pcfflatcurrent}. Let us rescale this current by $\frac{1}{\lambda}$ for convenience, and write it as a one-form
\[
 j = j_\mu dx^\mu = j_0 dt + j_1 dx = j^0 dt - j^1 dx.
\]
The flatness condition reads 
\[
 d j + j\wedge j =0.
\]
Then the Hodge dual is
\[
 *j = j_\mu dx^\mu = j_1 dt + j_0 dx 
\]
where we have used the Lorentzian signature $(+,-)$. It follows that
\[
 d*j = 0
\]
is equivalent to the equation of motion $\partial_\mu j^\mu =0$.

Then we consider the connection
\[
L = \alpha j + \beta *j,
\]
with $\alpha, \beta$ some constants,
and get
\[
 dL + L \wedge L = d j + \alpha^2 j\wedge j + \beta^2 *j\wedge *j = (\alpha^2 - \alpha -\beta^2)j\wedge j .
\]
As a consequence, the current is flat precisely if 
\[
 (\alpha^2 - \alpha -\beta^2) =0.
\]
This implies that we have constructed a one parameter family of flat connections. A parametrisation usually found in the literature is given by introducing a parameter $u$ such that
\[
\alpha=\frac{1}{2}(1\pm\cosh(u)),\quad\beta=\frac{1}{2}\sinh(u).
\]
Then we denote the families by $L(u)$, and construct the monodromies in the same way as in \ref{sec:lax}.

\subsection{Spin Chain Models}\label{sec:xxxspinchain}

In this section, we will discuss one of the simplest spin-chain models, a generalisation of the Heisenberg XXX spin chain to $\sun$ symmetry. We give its definition in section \ref{sec:xxxdef}. In section \ref{sec:yangianxxx}, we deal with the symmetry of the XXX spin chain, and show that the $\sun$ symmetry is enlarged to a Yangian by using the Yangian in Drinfeld's first realisation. The Hamiltonian can also be derived directly from the R-matrix via the RTT relations, which makes the symmetries explicit. This construction is reviewed in \cite{Faddeev:1990qg,Beisert:2004ry}. The definition of Yangians via RTT relations is reviewed in \cite{Molev:2002}.
We discuss the Bethe ansatz for the spin chain in its simple coordinate form in section \ref{sec:BAxxx}. References on Bethe ansatze can be found in \cite{Faddeev:1996iy,Karbach:1998ba1,Karbach:1998ba2,Beisert:2004ry}.

\subsubsection{XXX Spin Chains} \label{sec:xxxdef}

A spin chain is a one-dimensional lattice of $L$ sites, where at each site one has a spin-degree of freedom. In this paper, we will only be interested in closed spin chains, i.e. we impose periodic boundary conditions on the lattice. However, it is also natural to consider open spin chains. Furthermore, we will deal with spin chains where the spins form a vector space which on which $\sun$ acts in the fundamental representation. Let us denote the basis of the representation space by
\[
\state{\phi^a},\quad a=1,\dots n.
\]
The Heisenberg XXX spin chain is given by $L$ such spins interacting with the simple nearest neighbour Hamiltonian
\<\label{def:heisenham}
H = \lambda \sum_{k=1}^L H_{k,k+1},\nln
H_{k,k+1} = \ident-\perm_{k,k+1}.
\>
For the closed spin chain, the $L+1$'th site is identified with the first one.
The permutation operator flips two neighbouring sites, 
\[
\perm\state{XY}=\state{YX}.
\]
If one generalises this Hamiltonian to superalgebras, one should simply substitute the permutation operator with the graded permutation operator, which picks up a minus sign whenever two fermions are interchanged. If the constant $\lambda$ is positive, the Hamiltonian has a ferromagnetic vacuum, whereas if $\lambda$ is negative, the ground state is antiferromagnetic.

Now as usual in physics the task is to find the eigenvalues of this Hamiltonian. Despite the fact that this Hamiltonian is extraordinarily simple, its diagonalisation is, for fixed $n$, but arbitrary $L$, not straightforward. This is because the Hamiltonian as a matrix grows with $n^L$. It is only due to the fact that this Hamiltonian is integrable that we can find a more efficient method to diagonalise it, which is the Bethe ansatz. Crucial is the existence of higher conserved charges. We will discuss non-local charges in the next section.

\subsubsection{The Yangian of the XXX Spin Chain}\label{sec:yangianxxx}

The Hamiltonian \eqref{def:heisenham} is invariant under $\sun$. We choose the same notation as in section \ref{sec:glnm} and note that
\[
\perm_{k,k+1} = \sum_{IJ}E^k_{IJ}\otimes E^{k+1}_{JI}.
\]
Here, $E^k_{IJ}$ denotes the action of $E_{IJ}$ on the k'th factor of the tensor product, i.e.
\[
 E^k_{IJ} = \idm\otimes\dots\otimes E_{IJ}\otimes\dots\otimes\idm
\]
But the action of $E_{IJ}$ on the tensor product of $L$ fundamental representations is just given by the standard coproduct $\sum_{k=1}^LE^k_{IJ}$, and we get
\<
&&\scomm{\sum_{k=1}^L E^k_{IJ}}{\sum_{l=1}^L E^l_{ST}\otimes E^{l+1}_{TS}} = \sum_{l,k=1}^L(\delta_{kl}(\delta_{JS}E^k_{IT}-\delta_{IT}E^k_{SJ})\otimes E^{k+1}_{TS}\nln
& &+\delta_{k(l+1)}E^{k}_{ST}\otimes(\delta_{JT}E^{k+1}_{IS}-\delta_{IS}E^{k+1}_{TJ}))\nln
&=& \sum_{k=1}^L((E^k_{IT}\otimes E^{k+1}_{TJ}-E^k_{SJ}\otimes E^{k+1}_{IS})+E^{k}_{SJ}\otimes E^{k+1}_{IS}-E^{k}_{IT}\otimes E^{k+1}_{TJ}))\nln
&=&0\nln
\>
Hence, the Heisenberg XXX Hamiltonian is invariant under the Lie algebra $\sun$. We now proceed to the Yangian, and recall the coproduct of the generators $\hat{E}_{KL}$ in the first realisation \eqref{eq:coproductyang}:
\<
\copro \hat{E}_{KL} &=& \hat{E}_{KL}\otimes1+ 1\otimes\hat{E}_{KL} + \frac{1}{2}\scomm{E_{KL}\otimes 1}{\sum_{IJ} E_{IJ}\otimes E_{JI}} \nln
&=& \hat{E}_{KL}\otimes1+ 1\otimes\hat{E}_{KL} + \frac{1}{2}\sum_J \left(E_{KJ}\otimes E_{JL} -E_{JL}\otimes E_{KJ}\right)\nln
&=& \hat{E}_{KL}\otimes1+ 1\otimes\hat{E}_{KL} + \frac{1}{2}\sum_J \left(E_{KJ}\wedge E_{JL}\right) .
\>
We work here on the simplest evaluation representation, namely the one were the spectral parameter vanishes, i.e. $\hat{E}^s_{KL}=0$. 
Then, the coproduct on the $L$-times tensor product is simply given by
\<
\copro^{L-1} \hat{E}_{KL} =  \frac{1}{2}\sum_{s=1}^{L-1}\sum_{t=s+1}^L\sum_J \left(E^s_{KJ}\wedge E^t_{JL} \right).\nln
\>
Taking tensor products of representations does not give information whether the chain is closed or open. This is a property of the corresponding Hamiltonian. Let us consider for the moment the open chain. Commuting the Yangian generator with the Heisenberg Hamiltonian leads to boundary terms. This can be seen from 

\[
 \half\comm{\perm}{E_{KJ}\wedge E_{JL}} = E_{KL}\wedge\idm ,
\]
and arguing that the cross-terms cancel out. Then one gets
\<
 \comm{\sum_k\perm_{k,k+1}}{\frac{1}{2}\sum_{s=1}^{L-1}\sum_{t=s+1}^L\sum_J \left(E^s_{KJ}\wedge E^t_{JL} \right)} = E^1_{KL}-E^L_{KL}.
\>
These terms can be neglected upon working on asymptotic states, as one can interpret them as discretized total derivatives. Such terms correspond to boundary terms in the dual field theory and are hence expected to vanish. In the case of a finite, periodic spin chain the Yangian charges do not commute with the Hamiltonian, as the boundary terms will remain. Having established the first Yangian charges, one can use them to commute them to the whole Yangian algebra. The monodromy matrix is then the formal power series consisting of all Yangian charges. Conversely, one can derive the whole Hamiltonian from the transfer matrix. We will not pursue this derivation, and refer the reader to the reviews. The transfer matrix and Yangian are closely related to the algebraic Bethe ansatz. Instead, we will present the physical ideas of the coordinate Bethe Ansatz in the next subsection.

\subsubsection{Bethe Ansatz for the $\sun$ Spin Chain}\label{sec:BAxxx}

We now present the ideas of the coordinate Bethe Ansatz, which is a method to calculate the spectrum of the Heisenberg spin chain. The idea is to consider a vacuum state, where all the spins at each site point into the same direction. Then one excites such vacuum state by flipping a spin, and, by taking linear combinations of such states with one spin flipped, one calculates the eigenstate of the Hamiltonian and the corresponding energy eigenvalue, which yields the dispersion relation. Such an excitation will be called a magnon. Then one goes on and considers several magnons and calculates the scattering matrix, which describes the scattering of two magnons. As the model is integrable, the S-matrix describing the scattering of multiple magnons factorises into the product of two-particle S-matrices, in the same way as it was happening for integrable field theories. Finally, imposing appropriate boundary conditions for the wave function leads to the Bethe equations, which one can solve for the momenta, and, via the dispersion relation, for the energies of the magnons. Let us start by describing the vacuum of the spin chain. The ferromagnetic vacuum of \eqref{def:heisenham} with length $L$ is just the state
\[\label{def:ferrovac}
 \state{0} = \state{\phi^n,\dots,\phi^n},
\]
where we have put $L$ $\phi^n$'s into the state. Of course, we could have chosen any other of the $n$ $\phi^i$'s as the ground state. Note also that for our discussions, it does not matter whether the constant $\lambda$ in \eqref{def:heisenham} is positive or not. In the case of an antiferromagnetic choice $\lambda<0$, \eqref{def:ferrovac} is not a true vacuum of the system, but still a suitable state to construct the Bethe Ansatz. One can imagine that one just diagonalises the negative Hamiltonian in that case. We now proceed to consider states where single sites have a ``spin flipped'' compared to the vacuum, i.e. at one site we have a different vector than $\phi^n$. We denote a state where at the $k'th$ site we have a vector $\phi^a$, with $a=1,\dots,n-1$ and $\phi^n$'s at the other sites, by
\[
 \state{\phi^a_k} = \state{\phi^n,\dots,\phi^n,\phi^a,\phi^n,\dots,\phi^n}.
\]
Such states are not eigenstates of the Hamiltonian. However, direct calculation shows that the plane wave
\[
 \state{\phi^a(p)} = \sum_k e^{ipk}\state{\phi^a_k}
\]
is an eigenstate with the energy 
\[\label{def:dispersionxxx}
 E(p) = \lambda (2 -2\cos(p)) = 4\lambda \sin^2(\frac{p}{2}),
\]
provided the momentum is satisfying the periodicity requirement for a standing wave, 
\[
e^{ipL} = 1.
\]
As we would like to find all eigenstates of the Hamiltonian, the way to proceed is to consider states with two spin flips now. Direct calculations shows that the state of two plane waves,
\[
 \sum_{k,l} e^{ip_1k+ip_2 l}\state{\phi^a_k\phi^b_l},
\]
is not an eigenstate of the Hamiltonian. In some sense it is almost an eigenstate, up to the terms in the above sum where $k=l\pm 1$. The idea is to consider a linear combination of this plane wave state with two spin flips as above with a plane wave state where the two spin flips are exchanged, or scattered. This leads to the ansatz
\[
 \state{\phi^{a,b}(p_1,p_2)} = \sum_{k<l} e^{ip_1k+ip_2 l}\state{\phi^a_k\phi^b_l} + \smat^{a b}_{c d}(p_1,p_2)\sum_{k<l} e^{ip_2k+ip_1 l}\state{\phi^c_k\phi^d_l},
\]
where $\smat^{a b}_{c d}(p_1,p_2)$ is the S-matrix. For the above state to be an eigenstate one has to put
\[
 \smat^{a b}_{c d}(p_1,p_2) = \frac{e^{-ip_1-i p_2}+1}{e^{-ip_1}-2e^{ip_2-ip_1}+e^{ip_2}}\delta^a_c\delta^b_d - \frac{1-e^{ip_1}-e^{ip_2}+e^{ip_1+ip_2}}{1-2e^{ip_2}+e^{ip_1+ip_2}}\delta^a_d\delta^b_c.
\]
Note that $\delta^a_c\delta^b_d$ is simply the $(n-1)\times (n-1)$ dimensional identity matrix, whereas $\delta^a_d\delta^b_c$  is the $(n-1)\times (n-1)$ dimensional permutation operator.

Furthermore, the energy of this state is simply the sum of the energies of the two magnons, 
\<
  H \state{\phi^a(p_1)\phi^b(p_2)} &=& (E(p_1) + E(p_2)) \state{\phi^a(p_1)\phi^b(p_2)}\nln
  E(p_i) &=& 4\lambda \sin^2(\frac{p_i}{2}).
\>
Let us do introduce the new variable
\[\label{def:uofp}
 u_i = \frac{1}{2}\cot{\frac{p_i}{2}}.
\]
Then the S-matrix takes the simple rational form
\[\label{def:smatu}
 \smat(u_1-u_2) = \frac{u_1-u_2}{u_1-u_2+i}\left(\idm - \frac{i}{u_1-u_2}\perm\right) .
\]
Note that this is precisely the S-matrix we derived from $\mathcal{Y}(\alg{su}(n-1))$ Yangian symmetry \eqref{eq:funR}, but here the S-matrix has no complicated scalar prefactor as in \eqref{eq:funR}. This is of course not unexpected. The choice of the vacuum state \eqref{def:ferrovac} breaks the Lie algebra symmetry from $\sun$ to $\alg{su}(n-1)$. However, we discussed that the whole chain has $\mathcal{Y}(\alg{su}(n))$ symmetry. Indeed, the vacuum also breaks this higher symmetry to $\mathcal{Y}(\alg{su}(n-1))$. However, as the eigenstate basis is now basically Fourier transformed, i.e. we have momentum eigenstates, the representation of the Yangian now contains the evaluation parameter $u$, which equals precisely \eqref{def:uofp}. Henceforth, it also automatically satisfies the Yang-Baxter equation, and is furthermore unitary, i.e.
\[
\smat_{12}(u_1-u_2)\smat_{21}(u_2-u_1)=\idm.
\]
As we have solved the two-magnon problem, we still have to diagonalise all further states with more than two magnons. This is where we can use that the system is integrable. Indeed, the idea is to start with a plane wave state consisting of $M$ spin-flips with flavour $a_1,\dots,a_M$ propagating with corresponding momenta $p_1, \dots, p_M$, and consider linear combinations of such states were the $k$ momenta are permuted. The coefficients in those linear combinations are then taken to be S-matrices which precisely describe the permutation of the momenta. If one permutes e.g. $l$ of the $M$ momenta, the appropriate S-matrix is the $l$ particle S-matrix. This $l$ particle S-matrix factorises, by integrability, into two particle S-matrices. In particular, let us denote the $M$ magnon eigenstate by
\[\label{def:Meigenstate}
  \state{\phi^{a_1,\dots,a_M}(p_1,\dots,p_M)} = \sum_\pi\sum_{l_1,\dots,l_k} e^{ip_1\pi(1)+\dots+ip_M \pi(M)}S_\pi\state{\phi^{a_1}_{l_1},\dots,\phi^{a_M}_{l_M}}.
\]
Here, $\smat_\pi$ is an S-matrix which realises the permutation $\pi$. If $\pi$ is decomposed into transpositions 
\[
 \pi = (i_1,j_1)\dots(i_r,j_r),
\]
then
\[
 \smat_{\pi} = \smat_{i_1,j_1}\dots\smat_{i_r,j_r}.
\]
By the Yang-Baxter equation, the particular decomposition of the permutation into transpositions does not matter. One can now convince oneself that the states \eqref{def:Meigenstate} are eigenstates of the Hamiltonian in the infinite length limit, and the eigenvalue is again simply the sum of the eigenvalues of the single plane wave states, i.e.
\[
   H\state{\phi^{a_1,\dots,a_M}(p_1,\dots,p_M)} = (E(p_1)+\dots+E(p_M))  \state{\phi^{a_1,\dots,a_M}(p_1,\dots,p_M)}.
\]
Hence, we have constructed eigenstates of the Hamiltonian for an arbitrary number of magnons\footnote{We refer the reader to \cite{Kirillov:1985ab,Faddeev:1984aa} for discussions about the completeness of the eigenstates.}.\\
 As in this work, we are interested with closed spin chains, an important issue are boundary conditions. If one has $M$ magnons as before, and we consider the scattering of magnon $l$ with all the $M-1$ other magnons, we came back to the starting position in the periodic chain. Hence, the total phase shift of the magnon $l$ is given by
\[
 \prod_{k\neq l}^M S_{kl}(p_k,p_l)\state{\phi^{a_1,\dots,a_M}(p_1,\dots,p_M)} = e^{ip_lL}\state{\phi^{a_1,\dots,a_M}(p_1,\dots,p_M)}.
\]
In the case that the initial spin chain had just $\alg{su}(2)$ symmetry, the S-matrix just has $\alg{u}(1)$ as residual symmetry, and is hence a scalar, i.e. \eqref{def:smatu} reduces to 
\[
 \smat(u_1-u_2) = \frac{u_1-u_2-i}{u_1-u_2+i}.
\]
If we invert \eqref{def:uofp}, getting 
\[
 e^{ip}=\frac{u+i/2}{u-i/2},
\]
we can write the Bethe equations as
\[\label{def:su2be}
 \left(\frac{u_l+i/2}{u_l-i/2}\right)^L = \prod_{k\neq l}^M\frac{u_k-u_l-i}{u_k-u_l+i}.
\]
These are algebraic equations for the spectral parameters $u_k$, which are related to the momenta. After solving for the momenta, we get the energy eigenvalues just by plugging into the dispersion relation \ref{def:dispersionxxx}.
If the original symmetry algebra is $\sun$, with $n>2$, the idea is to stepwise reduce the rank of the algebra, till one arrives at  $\alg{su}(2)$ Bethe equations \eqref{def:su2be}. We will not study the details of the nested Bethe Ansatz in this paper, and refer the reader to the literature \cite{Yang:1967bm,Sutherland:1975vr}.
\section{The AdS/CFT Correspondence}\label{ch:adscft}

The AdS/CFT correspondence, discovered in the late 1990s \cite{Maldacena:1998re,Gubser:1998bc,Witten:1998qj}, relates, in its most general form, conformal field theories in a flat $d$ dimensional space to string theory on a ten dimensional space. This ten dimensional space is the direct product of a $d+1$ dimensional Anti-de Sitter space, denoted by $AdS_{d+1}$, and a $9-d$ dimensional internal manifold. These theories are dual to each other in the sense that if one considers the full theory, one gets field theory in a perturbative expansion of the coupling constant, whereas the natural expansion parameter of string theory is the inverse coupling. One can consider the field theory as living on the boundary of the AdS space. This is why the AdS/CFT correspondence is often seen as a realisation of the holographic principle \cite{'tHooft:1993gx,Susskind:1994vu}. In this paper we will be only concerned with the best tested case of correspondences, namely between $\mathcal{N}=4$ Super Yang-Mills theory with $SU(N)$ gauge group in four dimensions and IIB string theory defined on an \ads space. Moreover, we will only deal with the planar, large $N$ limit, as the theories seem to become integrable only in this limit. Integrability allows for novel all-loop tests of the AdS/CFT correspondence, which indicate that, at least in the planar limit, the involved gauge and string theories are physically exactly equivalent. This seems rather surprising, as from naively looking at the action and the particle content, gauge theory in four and string theory in ten dimensions seem to have little in common. However, already in the seventies \cite{'tHooft:1974jz} it was realised that a certain expansion of gauge theory, where the expansion parameter involves the coupling constant as well as the rank of the gauge group, $N$, resembles the perturbative expansion encountered in string theory. The AdS/CFT correspondence makes such relation of gauge and string theories precise and involves a much stronger claim, namely, of the exact duality of the involved theories.\\

The reason why the AdS/CFT correspondence works well for the case of large $N$ \ads is the large amount of supersymmetry and subsequently its integrability. Classical integrability was argued to exist for the bosonic \cite{Mandal:2002fs} and the full supersymmetric sigma model \cite{Bena:2003wd} and on the dual gauge side in \cite{Minahan:2002ve,Beisert:2003tq,Bena:2003wd}, whereas quantum integrability is currently only conjectured. Under this assumption the exact S-matrix \cite{Beisert:2005tm} and the asymptotic Bethe \cite{Beisert:2005fw} and TBA equations \cite{Gromov:2009tv,Gromov:2009bc,Arutyunov:2009ur,Bombardelli:2009ns} could be written down. We will discuss the S-matrix in chapter \ref{ch:smatrix}. Recently, $\mathcal{N}=6$ Chern-Simons Theory in three dimensions was related to IIA string theory on $AdS_4\times \mathbb{CP}^3$ \cite{Aharony:2008ug}. These theories allow for a planar limit as well. Again, there is strong evidence that in the planar limit these theories are exactly equivalent. This is because there is strong evidence that these theories are also integrable \cite{Minahan:2008hf,Arutyunov:2008if,Stefanski:2008ik}. The AdS/CFT correspondence is also used in many other examples. One can try to break symmetries and still learn many things about the involved theories. As the correspondence is generically a strong/weak duality, AdS/CFT is a unique tool which allows to use perturbation theory at one side of the correspondence to understand the strong coupling behaviour on the other side. This can be useful e.g. when one wants to study quantum gravity and uses field theory to do so. Likewise, one might be interested in field theory questions and can use string theory via the correspondence. Ultimately, one might be interested in obtaining a string theory dual to QCD, but as both supersymmetry and conformal symmetry are broken, and QCD contains additional fundamental fermions, this might still be a long way to go.

We begin this chapter by stating some features of string theory on \ads in section \ref{sec:stringsads}, and continue in section \ref{sec:n4sym} to introduce the dual gauge theory. Section \ref{sec:intadscft} shows how integrable structures appear in the AdS/CFT correspondence.

As in this section we do not present new results, we refer the reader to the reviews \cite{Aharony:1999ti, Becker:2007zj} for some background on the AdS/CFT correspondence, and \cite{Green:1987sp,Polchinski:1998rq,Polchinski:1998rr, Becker:2007zj} for some general background on string theory. Integrability in the AdS/CFT correspondence has been reviewed in \cite{Plefka:2005bk,Dorey:2009zz,Arutyunov:2009ga,Rej:2009je,Volin:2010cq}

\subsection{String theory on \ads}\label{sec:stringsads}

String theory on \ads has $2$ independent parameters: The string slope $\alpha'$, which is related to the tension $T=\frac{1}{2\pi \alpha'}$ of an individual string, and the string coupling $g_s$, which controls the interaction of several strings with each other. The relation with the joint radius $R$ of $AdS_5$ and $S^5$ is given by $R^4 = 4 \pi g_s N \alpha'^2$. $N$ is an integer which will turn out to be identical to the rank of the gauge group of the dual gauge theory. On the string side, it is related to the flux of the five form field. We are interested in a limit where $g_s\rightarrow 0$, so the strings are not interacting. To keep the radius finite we will send $N\rightarrow \infty$, such that 

\[
 \lambda = g_s N
\]
remains finite. $\lambda$ will correspond to the 't Hooft coupling on the gauge theory side, and the limit is called the 't Hooft, or planar limit. \\

In the 't Hooft limit string theory is described by a non-linear sigma model \cite{Metsaev:1998it}. The bosonic part of the action takes the usual form
\[
S = \frac{1}{\alpha'}\int_0^{2 \pi} d\sigma\int_{-\infty}^\infty d\tau G_{MN} \gamma^{ab}\partial_aX^M\partial_bX^N .
\]
It describes the embedding of a world-sheet cylinder with radius $2 \pi$ and metric $\gamma^{ab}$ into the \ads target space with metric $G_{MN}$. The radius $R$ of the Anti de-Sitter space is the same as the radius of the sphere, which is required for consistency upon adding the fermionic part to the action. In general, a $d$ dimensional sphere can be written as a coset 
\[
S^d =\frac{SO(d+1)}{SO(d)},
\]
whereas a $d$ dimensional AdS space is considered to be the coset
\[
AdS_d =\frac{SO(d-1,2)}{SO(d-1,1)}.
\]
Written in this way the global symmetry algebras are manifestly given to be $\alg{so}(d+1)$ in the case $S^d$, and $\alg{so}(d-1,2)$ for $AdS_{d}$. As the bosonic part of the \ads sigma model is a direct product space of $S^5$ and $AdS_{5}$, the bosonic symmetry algebra of \ads is just the direct sum  $\alg{so}(4,2)\oplus\alg{so}(6)$. Using the well-known isomorphisms
\[
\alg{so}(4,2)\cong \alg{su}(2,2), \quad \alg{so}(6)\cong \alg{su}(4),
\]
the elements of $\alg{so}(4,2)\oplus\alg{so}(6)\cong \alg{su}(4)\oplus\alg{su}(2,2)$ can be though of in block form as
\[\label{eq:bossymm}
\begin{pmatrix}
B_1 &0  \\ 
 0&B_2 
\end{pmatrix} ,
\]
with $B_1\in\alg{su}(4)$, and $B_2\in\alg{su}(2,2)$. Now there is a unique simple Lie superalgebra which has $\alg{su}(4)\oplus\alg{su}(2,2)$ as the bosonic subalgebra. It is the algebra $\sconf$, which is a real form of $\alg{psl}(4|4)$, i.e. part of the algebras $\alg{psl}(n|n)$ we have studied in section \ref{sec:glnm}. If we consider \eqref{eq:bossymm}, then getting $\sconf$ is quite natural if we think of filling the zero-blocks in \eqref{eq:bossymm} with as many fermionic elements as possible. Recall that on the fundamental representation the anticommutator of fermions will automatically lead to a central extension, i.e. to the algebra $\alg{su}(2,2|4)$, just as we have discussed in \ref{sec:slnn}. We will briefly discuss the supersymmetric extension of the action in section \ref{sec:adsintegrability}, but will first proceed by introducing the dual gauge theory in the next section.

\subsection{$\mathcal{N}=4$ Super Yang-Mills Theory}\label{sec:n4sym}

The other side of the AdS/CFT correspondence is $\mathcal{N}=4$ Super Yang-Mills theory, which is the four dimensional supersymmetric Yang-Mills theory with $SU(N)$ gauge group and maximal supersymmetry. The Lagrangian is given by
\<
 L=\tr\left(\frac{1}{4}F^{\mu\nu}F_{\mu\nu}+\frac{1}{2}D^\mu\phi^nD_\mu\phi_n-\frac{1}{4}\gym^2\comm{\phi^n}{\phi^m}\comm{\phi_n}{\phi_m}+\right.\nln
 \left. \dot{\psi}^a_{\dot{\alpha}}\sigma^{\dot{\alpha}\beta}_\mu D^\mu\psi_{\beta a}- \frac{1}{2}i \gym\psi_{\alpha a}\sigma^{ab}_m\epsilon^{\alpha\beta}\comm{\phi^m}{\psi_{\beta b}}-\frac{1}{2}i \gym\dot{\psi}^a_{\dot{\alpha}}\sigma_{ab}^m\epsilon^{\dot{\alpha}\dot{\beta}}\comm{\phi_m}{\dot{\psi}_{\dot{\beta}}^b}\right).\nln
\>
All fields transform in the adjoint of the $SU(N)$ gauge group. Let us describe the fundamental fields appearing in the Lagrangian. First of all, we have the gauge field $A_\mu$, with $\mu=0,1,2,3$ being the usual Lorentz index. The gauge field appears in the Lagrangian in form of the usual gauge invariant objects, the covariant derivative

\[
 D_\mu = \partial_\mu - i \gym A_\mu,
\]
as well as the field strength

\[
 F^{\mu\nu} = \frac{i}{\gym}\comm{D_\mu}{D_\nu} = \partial_\mu A_\nu - \partial_\nu A_\mu - i \gym \comm{A_\mu}{A_\nu}.
\]
Furthermore, the Lagrangian contains $6$ scalar fields $\phi^n$, $n=1,\dots,6$, which transform in the fundamental representation of $\alg{so}(6)$. $\alg{so}(6)$ is called the R-symmetry. Finally, there are the fermionic fields $\psi_{\alpha a},\dot{\psi}^a_{\dot{\alpha}}$ which transform both under the R and Lorentz symmetries. The Greek indices $\alpha, \dot{\alpha}$ run from $1$ to $2$ and correspond to the usual splitting of the complexified Lorentz algebra

\[
 \alg{so}(1,3)=\alg{su}(2)\oplus\alg{su}(2),
\]
which is due to the fact that the Lorentz algebra is not simple. We will not make explicit use of the reality conditions in this paper, and hence refer from now on to the complexified algebra as $\alg{su}(2)$. Then $\alpha, \dot{\alpha}$ correspond to the $2$ dimensional representations of the two $\alg{su}(2)$'s. The indices $a,b = 1,\dots,4$, correspond to the spinor representation of the $\alg{so}(6)$ R-symmetry, which makes use of the isomorphism

\[
 \alg{so}(6)=\alg{su}(4).
\]
The matrices $\sigma_{ab}^m$ and $\sigma^{\dot{\alpha}\beta}_\mu$ are the chiral gamma matrices of $\alg{su}(4)$ and $\alg{su}(2)\oplus\alg{su}(2)$, respectively. They allow one to write e.g. a Lorentz $4$-vector $V^\mu$ in terms of the left/right $\alg{su}(2)$'s, i.e.

\[\label{def:lorentzsplit}
 V^{\dot{\alpha}\beta} = \sigma^{\dot{\alpha}\beta}_\mu V^\mu .
\]
The way the Lagrangian is written shows that the $\alg{so}(1,3)$ Lorentz symmetry as well as the $\alg{so}(6)$ R-symmetry are manifest. Let us briefly describe our notations for the whole symmetry algebra. We will work with the isomorphisms to the appropriate $\sun$ algebras, so we denote a basis of the R-symmetry by

\[
 \gen{R}_a^b,\quad a,b=1,\dots,4 ,
\]
which satisfy the usual $\alg{su}(4)$ relations 

\[
 \comm{\gen{R}_a^b}{\gen{R}_c^d} = \delta_c^b\gen{R}_a^d-\delta_a^d\gen{R}_c^b,
\]
just as outlined in section \ref{sec:glnm} for the general case $\slnm$. As before, we have traceless matrices, which implies $\sum \gR{a}{a}=0$. The Lorentz generators are split into two sets of $\alg{su}(2)$ generators denoted by

\<
 \gen{L}_\alpha^\beta,\quad \alpha,\beta=1,2, \nln
 \gen{L}_{\dot{\alpha}}^{\dot{\beta}},\quad \dot{\alpha},\dot{\beta}=1,2,
\>
which again satisfy the usual $\alg{su}(2)$ relations. Note that for each of the $\alg{su}(2)$ algebras, we again impose the trace condition, e.g. $\sum_\alpha\gen{L}_\alpha^\alpha = \sum_{\dot{\alpha}}\gen{L}_{\dot{\alpha}}^{\dot{\alpha}}=0$.\\

Let us comment on the further symmetries. First of all, the bosonic symmetry is enhanced. Indeed, the Lorentz algebra is enlarged to the Poincare algebra, as expected. Furthermore, it was argued in \cite{Sohnius:1981sn} that the theory is conformal even in the quantum case, and the beta function vanishes exactly. The conformal algebra in four dimensions is

\[
  \alg{so}(2,4).
\]
The additional generator of $\alg{so}(2,4)$, which are not in $\alg{so}(1,3)$, are the $4$ translation generators $\gen{P}_\mu$, the special conformal generators $\gen{K}_\mu$, and the dilatation generator $\gen{D}$. Again, we will often make use of the isomorphism $\alg{so}(2,4)= \alg{su}(2,2)$, and we will rewrite the Lorentz indices in the same way as in \eqref{def:lorentzsplit}. Then the commutator of translation and boost reads

\<\label{eq:confcomm}
\comm{\alg{K}^{\alpha \dot\beta}}{\gen{P}_{\dot\gamma\delta}}= 
  \delta_{\dot\gamma}^{\dot\beta} \alg{L}^{\alpha}{}_{\delta}
  +\delta_\gamma^\alpha \dot{\alg{L}}^{\dot\beta}{}_{\dot\delta}
  +\delta_\gamma^\alpha\delta_{\dot\delta}^{\dot\beta} \gen{D}
\>
Furthermore, the theory is supersymmetric and contains four sets of supercharges. We will denote the supercharges by $\gen{Q}^b{}_\beta$, $\dot{\gen{Q}}_{\dot\alpha a}$, $\gen{S}^\alpha{}_a$ and $\dot{\gen{S}}^{a\dot\alpha}$, with the same indices as before. The indices manifestly realise the action of the bosonic generators. The nontrivial commutation relations are given by

\[\label{eq:sconfacomm}
\arraycolsep0pt
\begin{array}{rclcrcl}
\comm{\gen{S}^\alpha{}_a}{\gen{P}_{\dot\beta\gamma}}= 
   \delta^\alpha_\gamma \dot{\gen{Q}}_{\dot\beta a},
&\qquad&
\comm{\alg{K}^{\alpha\dot\beta}}{\dot{\gen{Q}}_{\dot\gamma c}}=
  \delta^{\dot\beta}_{\dot\gamma} \gen{S}^\alpha{}_c,
\\[3pt]
\comm{\dot{\gen{S}}^{a\dot\alpha}}{\gen{P}_{\dot\beta\gamma}}= 
  \delta^{\dot\alpha}_{\dot\beta} \gen{Q}^a{}_{\gamma},
&&
\comm{\alg{K}^{\alpha\dot\beta}}{\gen{Q}^c{}_\gamma}=
  \delta^\alpha_\gamma \dot{\gen{S}}^{c\dot\beta},
\\[3pt]
\acomm{\dot{\gen{Q}}_{\dot\alpha a}}{\gen{Q}^b{}_\beta}= 
  \delta_a^b \gen{P}_{\dot\alpha\beta},
&&
\acomm{\dot{\gen{S}}^{a\dot\alpha}}{\gen{S}^\beta{}_b}= 
  \delta^a_b \alg{K}^{\beta\dot\alpha},
\end{array}
\]
\<
 \acomm{\gen{S}^\alpha{}_a}{\gen{Q}^b{}_\beta}=
  \delta^b_a \alg{L}^\alpha{}_\beta
  +\delta_\beta^\alpha \alg{R}^b{}_a
  +\half \delta_a^b \delta_\beta^\alpha (\gen{D}-\gen{C}),\nln
\acomm{\dot{\gen{S}}^{a\dot \alpha}}{\dot{\gen{Q}}_{\dot\beta b}}=
  \delta^a_b \dot{\alg{L}}^{\dot\alpha}{}_{\dot\beta}
  -\delta_{\dot\beta}^{\dot\alpha} \alg{R}^a{}_b
  +\half \delta^a_b \delta_{\dot\beta}^{\dot\alpha} (\gen{D}+\gen{C}).\nonumber
  \>
  Here, we have introduced the central charge $\alg{C}$, and together with all the other generators we get the algebra
  
 \[
   \sconf\ltimes \alg{u}(1) = \alg{su}(2,2|4).
 \]
The superconformal algebra 
\[
 \sconf
\]
is obtained by projecting out $\gen{C}$. It is a real form of the simple Lie superalgebra $\alg{psl}(4|4)$, and the corresponding series of Lie superalgebras were discussed in section \ref{sec:slnn}. We should note that the existence of supersymmetry of the theory is crucial for the persistence of conformal symmetry at the quantum level. \\

Now, we would briefly like to comment on the 't Hooft limit of this gauge theory. Usually, one fixes the rank $N$ of the gauge group, and remains with the coupling constant $\gym$ as the only free parameter in the theory. The idea of \cite{'tHooft:1974jz} was to consider $N$ as a parameter of the gauge theory. Physical quantities in perturbation theory can be expanded in a double series in $\gym$ and $\frac{1}{N}$. Then, one can introduce a new, effective coupling constant

\[
 \lambda = \gym^2 N,
\]
and take a limit $N\rightarrow \infty$ and $\gym\rightarrow 0$ such that $\lambda$ is fixed. In this limit only planar Feynman diagrams contribute to physical quantities, and hence, the theory dramatically simplifies.


\subsection{Integrability in AdS/CFT}\label{sec:intadscft}

In this section we study how integrability appears in the AdS/CFT correspondence. Section \ref{sec:adsintegrability} deals with the classical integrability of the string sigma model, which is shown via the existence of a Lax connection. In \symng, integrability appears in the action of the dilatation operator on single trace local operators. This is explored in section \ref{sec:integrabilitysym}.

\subsubsection{Integrability of the AdS Sigma Model}\label{sec:adsintegrability}

Consider now an element $G$ of the supergroup $SU(2,2|4)$. Then we can construct a one-form

\[
A = -G^{-1} dG, 
\]
which takes values in $\alg{su}(2,2|4)$. It satisfies the flatness condition

\[
\partial_\mu A_\nu - \partial_\nu A_\mu -\comm{A_\mu}{A_\nu}=0,
\]
similarly as for the $\sun$ principal chiral field \eqref{sec:pcf}. However, the action we construct now is different, and is written down by exploiting the $\mathbb{Z}_4$ automorphism of $\alg{su}(2,2|4)$. This automorphism imposes a grading,

\[
\alg{su}(2,2|4) = \alg{su}(2,2|4)^{(0)} + \alg{su}(2,2|4)^{(1)}+\alg{su}(2,2|4)^{(2)}+\alg{su}(2,2|4)^{(3)}
\]
 on $\alg{su}(2,2|4)$, so we can write each element as a direct sum over the graded components. In particular, we can write the one-form $A$ as
 
\[
 A=A^{(0)}+A^{(1)}+A^{(2)}+A^{(3)}.
\]
Then we define the action on the supersymmetric coset

\[\label{def:adscoset}
\frac{PSU(2,2|4)}{SO(5)\times SO(4,1)}
\]
as
\[\label{def:fermiaction}
S = -\frac{R^2}{4 \pi \alpha'}\int\left(\gamma^{\alpha\beta}\str(A_\alpha^{(2)}A_\beta^{(2)}) + \kappa\epsilon^{\alpha\beta}\str(A_\alpha^{(1)}A_\beta^{(3)})\right).
\]
Note that the $\mathbb{Z}_4$ respects the supersymmetric grading, i.e. the zeroth and second components are bosonic, whereas the first and third component are fermionic. As one can check that $\alg{su}(2,2|4)^{(0)} = \alg{so}(4,1)\oplus\alg{so}(5)$, the component $A^{(0)}$ does not appear in the action. This is in line with the intuition that in the coset \eqref{def:adscoset}, $SO(5)\times SO(4,1)$ is cancelled out.

Let us summarise some features of this action. First of all, one can explicitly construct conserved Noether currents, and show that the corresponding charges form an $\alg{su}(2,2|4)$ algebra. Then, one has a reparametrisation invariance of the world-sheet cylinder, which ultimately leads to the Virasoro constraints. Furthermore, one has a peculiar fermionic symmetry called kappa-symmetry, which fixes the real parameter $\kappa$ appearing in \eqref{def:fermiaction} to be $\kappa=\pm 1$, and, most importantly, reduces the number of fermionic degrees of freedom by one half. Finally, the Lagrangian is classically integrable. A Lax connections was constructed in \cite{Bena:2003wd}

\[
L(z) = A^{(0)} + \frac{1}{z}A^{(1)}+\frac{1}{2}(z^2 + \frac{1}{z^2})A^{(2)}+\frac{1}{2\kappa}(z^2 - \frac{1}{z^2})*A^{(2)}  +z A^{(3)}
\]
and shown to satisfy the flatness condition 

\[
\partial_\alpha L_\beta-\partial_\beta L_\alpha = \comm{L_\alpha}{L_\beta}.
\]
We have used the two dimensional Hodge operator $*$. Note that this equation holds only if kappa symmetry holds, i.e. $\kappa=\pm 1$. Hence, it seems Kappa symmetry is crucial for the integrability of the model. The flatness conditions implies that the trace of the monodromy around the closed string

\[\label{def:adsconserved}
 T(\tau,\lambda) = \mathcal P \exp\left(-\int_0^{2\pi} L_1(\tau,\sigma,z)d\sigma\right)
\]
is conserved, just as we discussed for general integrable field theories in section \ref{sec:lax}. However, it remains to be shown that those conserved charges are in involution. This was done in \cite{Magro:2008dv}, where it was conjectured that the Lax connections requires the following modification: 

\[
 L(z)\rightarrow L(z) + \frac{1}{2\sqrt{\lambda}}(1-z^4)(\mathcal{C}^{(0)}+z^{-3}\mathcal{C}^{(0)}+z^{-1}\mathcal{C}^{(3)}).
\]
Here, $\mathcal{C}^{(k)}$ are secondary constraints in the Hamiltonian analysis. We also refer the reader to \cite{Vicedo:2009sn}, where this Lax connection was derived.\\

The Poisson structure, resulting from the extended Lax connection where the constraints are not imposed, is nontrivial, containing non-local terms, similar as in the case of the principal chiral field. We believe that much work remains to be done in this line, to fully uncover the algebraic structure underlying the AdS/CFT system. In particular, no proper quantisation of the sigma model and the charges has been done yet, so a proof of quantum integrability remains to be found. Let us spell out the Poisson brackets, which have been found in \cite{Magro:2008dv}:

\<\label{def:possion}
\acomm{L_1(\sigma_1,z_1)} {L_1(\sigma_2,z_2)} = \comm{\crmat^-_{12}(z_1,z_2)}{L_1(\sigma_1,z_1)}\delta(\sigma_1-\sigma_2)\nln
+\comm{\crmat^+_{12}(z_1,z_2)}{L_2(\sigma_1,z_2)}\delta(\sigma_1-\sigma_2)-(\crmat^+_{12}(z_1,z_2)-\crmat^-_{12}(z_1,z_2))\partial_{\sigma_1}\delta(\sigma_1-\sigma_2).\nln
\>
Here, the classical r-matrix is given by
\[
\crmat_{12}(z_1,z_2) = \sum_{n=0}^3 \frac{(\Omega^n\otimes 1) \mathcal T^{\sconf}_{12}}{{e^{i \pi n/2} z_1 - z_2}},
\]
and 

\<
 \crmat_{12}^+(z_1,z_2) = -\frac{f(z_1)}{4 z_1^3}\crmat_{12}(z_1,z_2),\nln
 \crmat_{12}^-(z_1,z_2) = \frac{f(z_2)}{4 z_2^3}\crmat_{21}(z_1,z_2). 
\>
$\Omega$ is the $\mathbb Z_4$ automorphism of $\sconf$ introduced before in connection with the grading of $\sconf$. This means $\crmat_{12}(z_1,z_2)$ here is a twisted r-matrix. It is not directly related to the classical r-matrix with centrally extended $\psucentral$ invariance discussed in section \ref{sec:adsbialgebra}, which is related to the weak/strong coupling limit of the scattering matrix. Here, the r-matrix generates the Poisson structure \eqref{def:possion}. Nevertheless, there is a remarkable similarity of the algebraic structure. Using this Poisson structure it was shown in 
\cite{Magro:2008dv} that the conserved charges \eqref{def:adsconserved} are in involution. This implies that the classical sigma model is integrable. Note that this classical r-matrix also has a nice abstract algebraic formulation in terms of a Lie dialgebra \cite{Vicedo:2010qd}, in a similar way as the classical r-matrix of the $\psucentral$ scattering problem has an underlying universal r-matrix (see section \ref{sec:adsbialgebra}) related to a Lie bialgebra.

\subsubsection{Integrability in $\mathcal{N}=4$ Super Yang-Mills Theory}\label{sec:integrabilitysym}

We have established that the classical string sigma model, describing the motion of strings on an \ads backgrounds, is integrable. As the sigma-model is defined on a two dimensional world-sheet cylinder, we can consider the \ads sigma model as a generalisation of other two-dimensional integrable field theories. We learned that $\mathcal{N}=4$ Super Yang-Mills theory in the large $N$ limit is dual to the \ads string sigma model. If we assume integrability to survive quantisation, we should certainly expect integrability to appear in $\mathcal{N}=4$ Super Yang-Mills theory. In particular, we know that systems with infinitely many degrees of freedom are usually just integrable in dimensions less than three, whereas here we consider $\mathcal{N}=4$ Super Yang-Mills theory in four dimensions. It turns out that the spectral problem of finding anomalous dimensions of single-trace local operators can be seen as an effective one dimensional problem. The crucial insight of \cite{Minahan:2002ve} was to map such single-trace local operators to a spin chain. This generalised earlier observations \cite{Lipatov:1993yb,Faddeev:1994zg} of the appearance of integrable lattice models in QCD. In \cite{Minahan:2002ve}, the authors took just single trace operators composed of the $\alg{so}(6)$ scalar fields, which have the form
\[
 \tr(\phi^{a_1}\phi^{a_2}\dots\phi^{a_L}),
\]
with $a_k=1,\dots,6$. As the fields $\phi^{a}$ live in a trace, one can consider such operator as a closed spin chain of length $L$, and each $\phi^{a}$ defines a spin degree of freedom, transforming in the fundamental representation of $\alg{so}(6)$. The classical dimension of those scalar fields is just $1$, but now one is interested in calculating the quantum corrections of the dimension, i.e. the anomalous dimension. In fact, the whole dimension of an operator is the eigenvalue of the dilatation generator $\gen{D}$ of the conformal algebra, i.e. we consider the eigenvalue problem

\[
 \gen{D} \mathcal O = \Delta \mathcal O,
\]
where $ \mathcal O$ is a generic single-trace local operator, and $\Delta$ is its dimension. The dimension has an expansion

\[
 \Delta = \sum\lambda^k \Delta^{(k)},
\]
where $\Delta^{(0)}$ is simply the classical dimension. As mentioned before, we are only interested in the planar limit, so $\lambda$ denotes, as before, the 't Hooft coupling constant. Likewise, we can expand the dilatation generator itself as a power series
\[
\gen{D} = \sum\lambda^k \gen{D}^{(k)}.
\]
Now the above interpretation of a single trace-local operator as a spin chain is merely a picture, but now the dilatation generator can be seen as a Hamiltonian acting on such chain. The observation in \cite{Minahan:2002ve} was that the first correction to the classical dilatation generator, $\gen{D}^{(1)}$, is proportional to the Heisenberg XXX spin-chain Hamiltonian with nearest neighbour interaction and $\alg{so}(6)$ symmetry. As this Hamiltonian is integrable, in the same way as the spin-chains with $\sun$ symmetry, which we discussed in section \ref{sec:xxxspinchain}, one can diagonalise it with the help of the Bethe Ansatz. Soon after the discovery of one-loop integrability in the $\alg{so}(6)$ sector, integrability was argued to survive at higher loops \cite{Beisert:2003tq} in the expansion of $\gen{D}$ in $\lambda$. It turns out that one can consider each of the higher $\gen{D}^{(k)}$ as Hamiltonians, whose interaction grows linearly with $k$, i.e. $\gen{D}^{(k)}$ interacts with its $k$th nearest neighbours. Furthermore, Bethe Ansatz equations describing the whole one-loop spectrum of a spin chain with $\sconf$ symmetry have been derived in \cite{Beisert:2003yb}, and the complete one-loop Dilatation generator $\gen{D}^{(1)}$ has been written down in \cite{Beisert:2003jj}. It was argued in \cite{Dolan:2003uh,Dolan:2004ps} that the superconformal algebra at tree level is enlarged to a Yangian. As we have also established non-local charges forming a Yangian algebra at the string sigma model, it is natural to expect that there should be a Yangian for the full quantum AdS/CFT system. However, as the range of the interaction grows order by order in perturbation theory, and indeed also the Lie algebra generators themselves depend on the coupling constant, finding the Yangian even in perturbation theory is complicated. All perturbative checks performed in some subsectors to certain lower loop orders have confirmed that there are Yangian charges in each case \cite{Agarwal:2004sz,Zwiebel:2006cb}. A complete proof of integrability would require to find full all-loop charges, which can be expanded in weak coupling but also in strong coupling, giving the non-local charges of the sigma model \cite{Bena:2003wd} at leading order in strong coupling. Then there should be infinitely many related conserved local charges, which guarantee integrability in the way as argued in section \ref{sec:fieldtheory}. This has currently not been achieved. However, the picture becomes clearer if one constructs magnons states out of the spin chain, and considers the scattering problem, in the same way as we did for $\sun$ spin chains in section \ref{sec:xxxspinchain}. Likewise, there will be an associated scattering problem on the string side. This line of thought was proposed in \cite{Staudacher:2004tk}, and will be followed in the next chapter.

\section{The Magnon S-Matrix of AdS/CFT}\label{ch:smatrix}

In this chapter we would like to study the S-matrix of AdS/CFT. This S-matrix is the central object to derive spectral equations, as was advocated in \cite{Staudacher:2004tk}. It was found in its all loop-form by exploiting the centrally extended $\psucentral$ symmetry \cite{Beisert:2005tm}, up to an overall factor. This factor was later conjectured in \cite{Beisert:2006ib,Beisert:2006ez}, based on earlier observations that there has to be a non-trivial phase in the Bethe equations of classical strings \cite{Arutyunov:2004vx}. This phase was generalised to quantum strings at one-loop \cite{Beisert:2005cw, Hernandez:2006tk}, and those results are compatible with the fact that at weak coupling there is no phase at the first three loops. A derivation of the full scalar factor in an integral form \cite{Dorey:2007xn} directly from crossing symmetry \cite{Janik:2006dc} was done in \cite{Volin:2009uv,Arutyunov:2009kf}.\\

 In its weak-coupling form, the S-matrix describes the scattering of magnons of the $\sconf$ spin chain, whereas at strong coupling it describes the scattering of world-sheet excitations of the string sigma-model. The existence of crossing symmetry is not clear a priori, but it seems to hold \cite{Janik:2006dc}, and one can derive it from a Hopf Algebra which was found later in \cite{Gomez:2006va,Plefka:2006ze}. This can be seen as a confirmation for the strength of the algebraic methods advocated in this paper (see also \cite{Gomez:2007zr,Young:2007wd} for interesting discussions relating the breaking of relativistic symmetry to q-deformation of the algebra). Diagonalisation of the S-matrix leads to the all-loop Bethe Ansatz \cite{Beisert:2005fw} of AdS/CFT, which completely describes the spectrum of long operators, or corresponding string states with large rotational charges. Furthermore, the S-matrix \cite{Arutyunov:2009mi} of the so-called mirror-model \cite{Ambjorn:2005wa,Arutyunov:2007tc}, which is obtained by Wick rotation from the original theory, is, in principle, underlying the TBA equations. These TBA equations are believed to describe the full spectrum of the AdS/CFT correspondence, and they and their associated Y-system were first conjectured in \cite{Gromov:2009tv,Gromov:2009bc,Arutyunov:2009ur,Bombardelli:2009ns}. The derivation of the TBA equations relies on the string hypothesis \cite{Arutyunov:2009zu}. Hence, the S-matrix plays a central role in the spectral problem of the AdS/CFT correspondence, and we devote this chapter to the study of it.\\
 
We begin by introducing the S-matrix of the string sigma model in section \ref{sec:smatsigma}. We will not calculate the S-matrix explicitly in perturbation theory, but follow the logic to reconstruct everything from symmetry. The symmetry at this level turns out to be a Lie bialgebra, which is the classical limit of the Yangian. These results have been published in our paper with Niklas Beisert \cite{Beisert:2007ty}. In section \ref{sec:smatsym}, we derive the one-loop S-matrix appearing on the spin chain of \sym, which is a representation of the universal R-matrix studied in section \ref{sec:universalR}. We do not need to do any new calculations to obtain the desired result, as we have established the underlying mathematics in chapter \ref{ch:yangians}. This is in the line of thought followed in this paper. As we have studied the mathematics underlying the symmetries encountered in the AdS/CFT correspondence, the physics (in form of the S-matrix) follows straightforwardly. The results of section \ref{sec:smatsym} have been published in \cite{Spill:2008yr}, were an explicit derivation of the universal R-matrix of $\utt$ was done.

\subsection{The Worldsheet S-Matrix of the String Sigma Model}\label{sec:smatsigma}

In this section we would like to study the S-matrix at strong coupling, as it arises in perturbative string theory. The string action defined in section \ref{sec:adsintegrability} contains many gauge degrees of freedom. Hence, before we proceed, we need to study string theory in a suitable gauge, so we can define a physical scattering problem. A common gauge choice is the light cone gauge, which we will discuss in section \ref{sec:stringinlc}. This choice will not only break the global $\sconf$ symmetry, but also the two dimensional Lorentz symmetry. Hence, the applicability of methods similar to those encountered in relativistic field theories, as studied in section \ref{sec:fieldtheory}, is questionable. Nevertheless, we will see later that the algebraic structures encountered in the light-cone string theory are quite similar to those in other integrable field theories. In section \ref{sec:worldsheetS}, we briefly discuss how the S-matrix arises perturbatively in the light-cone string theory \cite{Klose:2006zd}. Instead of working out the details of the derivation, we construct the S-matrix from the underlying symmetry in section \ref{sec:adsbialgebra}. This derivation is based on the publication \cite{Beisert:2007ty}. The symmetry at this level is the Lie bialgebra of $\utt$. It has some peculiar mathematical features, which make this case distinct from other Lie bialgebras encountered in chapter \ref{ch:yangians}. We recall that Lie bialgebras are the classical limit of Yangians, or other Quantum Groups in general.

\subsubsection{String Theory in the Light Cone Gauge}\label{sec:stringinlc}

Just as in flat space, string theory on the \ads background has several gauge degrees of freedoms, which we need to fix. One possibility is to choose the light-cone gauge. The time-like directions live on the $AdS_5$ part of the \ads space, so we choose one direction labelled by $t$. Then a suitable choice for the space-like coordinate of the light-cone is a big circle on the $S^5$, which we call $\phi$. Hence, we take the light cone coordinates

\[
 x_\pm = \phi\pm t .
\]
The charges corresponding to the translations in $t$ and $\phi$ are given by

\<
 E = -\frac{\sqrt{\lambda}}{2\pi}\int_{-r}^r p_t d\sigma,\nln
 J = \frac{\sqrt{\lambda}}{2\pi}\int_{-r}^r p_\phi d\sigma,
\>
i.e. they are just the usual energy and angular momentum. For the moment, we will choose as the world sheet a cylinder of circumference $2r$, instead of the previous normalisation to $2\pi$. The world sheet is parametrised by the world-sheet time $\tau$ and the spatial coordinate $\sigma$. As we are interested in deriving a scattering matrix, and thus need to construct asymptotic states, we should decompactify the cylinder, i.e. send $r\rightarrow\infty$. The light cone momenta, generating translations in the light cone coordinates chosen above, are given by

\[
 p_\pm = \half(p_\phi \mp p_t).
\]
Now the uniform light-cone gauge consists of identifying $x_+$ with the world-sheet time $\tau$,

\[
 x_+ = \tau.
\]
The corresponding momentum is then simply
\[
 p_+=1.
\]
Furthermore, the total light-cone momentum is related to the length of the world-sheet by

\[
 P_+ = \frac{\sqrt{\lambda}}{2\pi}\int_{-r}^r p_+ d\sigma = \frac{\sqrt{\lambda} r}{\pi}.
\]
With $P_+$ being fixed, $P_-$ takes the role of the light-cone energy. It is the only remaining non-compact charge to be determined, as the other charges are the remaining angular momenta on the sphere and the AdS space, and will hence get quantised. In particular, $P_-$ depends on the coupling constant $\lambda$, or, more precisely, it will be a power series in $\frac{1}{\sqrt{\lambda}}$. Furthermore, due to the Virasoro constraints it is not independent of the other fields.

Choosing the light-cone gauge also breaks the $\sconf$ symmetry into generators which either commute or do not commute with $P_-$. It turns out that the residual symmetry algebra commuting with $P_-$ is the algebra

\[\label{def:residualsymm}
 \uone\ltimes\psu\oplus\psu\ltimes\uone.
\]
The $\uone$ charge on the right of this sum corresponds to the light-cone energy. Due to the direct sum structure of the two $\psu$ summands one can split the remaining $8$ bosonic light-cone fields in such a way that they carry one index from the one $\psu$, and another index corresponding to the other $\psu$. Furthermore, if kappa symmetry is fixed, one also remains with $8$ fermionic degrees of freedom. 

Let us denote the $8$ bosonic fields which are transverse to the light-cone by $Z_{\alpha\dot\alpha}$ and $Y_{a\dot a}$, where $\alpha,\dot\alpha = 1,2$ are two $\alg{su}(2)$ indices belonging to the AdS space, whereas $a,\dot a=1,2$ are again $\alg{su}(2)$ indices, which now belong to the $S^5$. One can translate to the more familiar  $\alg{so}(4)$ indices in the usual way, by multiplication with Gamma matrices. These four $\alg{su}(2)$'s are the bosonic subalgebras of $\psu\oplus\psu$. The fermions are precisely what connects the $AdS_5$ with $S^5$ subspaces, which would otherwise be independent. In the light cone gauge, the fermions carry indices

\[
 \psi_{a\dot\alpha},\quad \psi_{\alpha\dot a}.
\]
Then, the action of $\psu\oplus\psu$ is manifestly realised on all fundamental fields in the light-cone. Indeed, the fundamental fields transform in the fundamental representation with respect to each of the two $\psu$'s, as outlined in section \ref{sec:glnm}. The undotted indices correspond to the one $\psu$, whereas the dotted indices belong to the second $\psu$. \\

As we discussed in section \ref{sec:pslcentral}, one problem is that on the fundamental representation of $\psu\ltimes\uone$, the $\uone$ charge has to be fixed to $\frac{1}{2}$ due to the shortening condition. This is problematic, as this charge plays the role of the light cone Hamiltonian, so its eigenvalues should be a continuous, real valued function of the coupling constant $\lambda$. The solution to this problem on the string side was found in \cite{Arutyunov:2006ak}, confirming the previous investigations on the gauge side \cite{Beisert:2005tm}. The first observation is that physical states of the string have to satisfy the level matching condition, which means that the total momentum of all excitations propagating around the string world-sheet has to be zero. If one relaxes the level-matching condition, then both $\psu$'s undergo a central extension, just as described in section \ref{sec:pslcentral}. This so-called off-shell symmetry algebra\footnote{Off-shell in this context purely refers to the fact that the level-matching condition is relaxed. This means that the momentum into one direction of the closed string does not necessarily equal the momentum into the other direction.} is now given by

\[\label{def:residualsymmextended}
  \uone\ltimes\psu\oplus\psucentral.
\]
Note that the three central charges are shared by the two $\psu$'s. Two of them can be identified in terms of a new Hopf algebra generator, as we will discuss later. For the moment we consider all three central charges to be independent. As explained in section \ref{sec:pslcentral}, the eigenvalue of the light cone Hamiltonian can now take continuous values. We remark that the central charge corresponding to the light-cone Hamiltonian was denoted by $\gen{C}$ in section \ref{sec:pslcentral}. For further discussions it is useful to consider the mode decomposition of our fundamental fields,

\<\label{def:fourierfields}
Y_{a\dot a}(\sigma,\tau) = \frac{1}{\sqrt{2\pi}}\int dp\frac{1}{2\omega_p} (e^{ip\sigma}a_{a\dot a}(p,\tau)+e^{-ip\sigma}\epsilon^{ab}\epsilon^{\dot a\dot b}a^\dagger_{b\dot b}(p,\tau)),\nln
Z_{\alpha\dot \alpha}(\sigma,\tau) = \frac{1}{\sqrt{2\pi}}\int dp\frac{1}{2\omega_p} (e^{ip\sigma}a_{\alpha\dot \alpha}(p,\tau)+e^{-ip\sigma}\epsilon^{\alpha\beta}\epsilon^{\dot \alpha\dot\beta}a^\dagger_{\beta\dot \beta}(p,\tau)).\nln
\>
$\omega_p$ is given by $\omega_p^2 = 1 + p^2$. 
Then the operators $a^\dagger(p,\tau)$ create excitations on the world-sheet, propagating around the world-sheet with momentum $p$. We will call those excitations magnons, in line with the later discussed magnons on the Yang-Mills side of the AdS/CFT correspondence. Note that these excitations are not the giant magnons of \cite{Hofman:2006xt}, which arise in a different kinematical regime. Similar mode decompositions exist for the fermions as well as for the canonically conjugate momenta in the light-cone gauge. To study them, we should derive the light-cone action and corresponding Hamiltonian order by order in perturbation theory. We will not need the explicit form, as we only want to illustrate how the S-matrix arises directly from string theory. A better derivation of the S-matrix will be later given by using symmetry. \\

The symmetry algebra can be thought of as acting on the individual momentum eigenstates. The crucial observation is that when one has a state consisting of $M$ magnons

\[
 a^\dagger(p_1,\tau)\dots{a}^\dagger(p_M,\tau)\state{0},
\]
where $\state{0}$ is the light cone vacuum, the level matching condition only says that all momenta together should add up to zero, i.e. 
\[
 \sum_{i=1}^Mp_i = 0.
\]
Hence, one can relax the level matching condition for individual excitations, and the centrally extended algebra will act on individual magnons. Each individual magnon can have non-trivial energy depending on the coupling constant. Furthermore, we can now study the problem when two such excitations hit each other, i.e. we will discuss the scattering matrix describing the scattering of two magnons. This will be done in the next section.

\subsubsection{The Worldsheet S-Matrix}\label{sec:worldsheetS}

After fixing the symmetries of the string sigma model, we can calculate the S-matrix of worldsheet excitations in perturbation theory. The leading contribution was calculated in \cite{Klose:2006zd}, whereas the two-loop correction was obtained in \cite{Klose:2007rz}. By integrability, the scattering matrices for $M$ magnons will factorise into two-particle S-matrices, and again the order of the two-particle scattering processes will not matter. In \cite{Klose:2006zd}, it was explicitly checked that there is no $2\rightarrow 4$ particle process, which is a necessary condition for integrability, namely, that there is no particle production. Let $\smat_{\psu\times\psucentral}$ denote the whole quantum S-matrix, which is invariant under the residual light-cone symmetry ${\psu\times\psucentral}$. As in the light-cone we have $16$ physical degrees of freedom, $8$ bosonic and $8$ fermionic fields, the two-particle S-matrix is a $16^2\times 16^2$ matrix. Due to the direct sum structure of the underlying Lie algebra symmetry, as well as the integrability of the model, the two-particle S-matrix factorises with respect to the algebra as

\[\label{def:smattensor}
\smat_{\psu\times\psucentral} = \smat_{\psucentral}\otimes\smat_{\psucentral}.
\]
The S-matrix will now scatter excitations with definite momentum, as created by the creation operators appearing in \eqref{def:fourierfields}. One should consider free fields as in-state, mapping to free fields in the out-state. All sets of operators satisfying the same canonical commutation relations should be related by similarity transformations, see \cite{Arutyunov:2009ga} for details. We denote by $M,N$ and $\dot M, \dot N$ two indices belonging to the two $\psu$'s, i.e. they combine the previous $4$ indices $a,\alpha, \dot a, \dot \alpha$ belonging to the $4$ $\alg{su}(2)'s$. This is analogously to the indices used in \eqref{def:funrepindices}. The corresponding fields with momentum $p$ are denoted by $\phi_{M\dot M}(p)$. Then the S-matrix acts as

\[
 \smat_{M\dot M N \dot N}^{P \dot P Q \dot Q}\state{\phi_{P\dot P}(p_1)\phi_{Q\dot Q}(p_2)},
\]
if one uses the matrix notation $\smat_{M\dot M N \dot N}^{P \dot P Q \dot Q}$. The factorisation with respect to the algebra then means that 

\[
 \smat_{M\dot M N \dot N}^{P \dot P Q \dot Q} = \smat_{M N}^{P Q}\smat_{\dot M\dot N}^{\dot P\dot Q},
\]
up to a minus sign depending on the conventions for fermions. One can now derive the S-matrix in perturbation theory. If the Hamiltonian in the light cone is split into a free part $H_0$, and an interaction part $V$, i.e. $H= H_0 +V$, then the S-matrix is given by

\[
 \smat = \mathcal T \exp (-i \int_{-\infty}^\infty d\tau V),
\]
with $\mathcal T$ denoting the time ordering operator.\\

The leading contribution in an expansion in $\frac{1}{\sqrt{\lambda}}$ is denoted by

\[
\smat = 1 +\frac{2\pi i}{\sqrt{\lambda}}T.
\]
Hence, $T$ is given by

\[
 T=-\frac{\sqrt{\lambda}}{2\pi}\int_{-\infty}^\infty d\tau V.
\]
Here, $V$ itself should be truncated at leading order in perturbation theory. We shall not present the detailed derivation of $T$ here, and refer the reader to \cite{Klose:2006zd,Arutyunov:2009ga}. We only wanted to highlight the ideas leading to the S-matrix directly from string theory. Instead, in the spirit of this paper, we shall reproduce the classical scattering matrix from symmetry in the next section.

\subsubsection{The S-Matrix and Lie Bialgebra at Strong Coupling}\label{sec:adsbialgebra}

In this section we will reproduce the classical contribution to the scattering matrix from symmetry considerations. We have described in the previous section that in the light cone gauge the global symmetry is, upon relaxing the level matching condition, given by two copies of $\psucentral$. Furthermore, as integrability implies that the S-matrix factorises with respect to the two copies of the algebra \eqref{def:smattensor}, we only need to derive the $\psucentral$ invariant S-matrix. The non-local charges associated to the Lax connection, as studied for the sigma model before gauge fixing in section \ref{sec:adsintegrability}, seem to be related to the Yangian. This was argued in \cite{Dolan:2003uh}, where related Yangian charges on the gauge side were studied, Furthermore, in the pure spinor formalism, Yangian charges which explicitly resemble those as found in other integrable field theories \eqref{def:nonlocalcharge} were constructed \cite{Berkovits:2004jw}. For this reason, we assume that in the light cone gauge, $\psucentral$ should also be extended to the Yangian. The classical version of the Yangian (double) is the Lie bialgebra based on the loop algebra $(\psucentral)[u,u^{-1}]$. We will show that two of the central charges are dependent on the third charge and the loop parameter $u$, and can be consistently eliminated. \\

Let us briefly recall the construction of the Lie bialgebra and classical r-matrix for $\glnm$, as studied in sections \ref{sec:classdouble},\ref{sec:glnmbialgebra}. If we denote a basis of $\glnm$ by $J^a$, and the corresponding basis of the loop algebra $\lalg{gl(n|m)}$ by $J^a_n$, then we established the classical r-matrix

\[\label{eq:crmatrepeated}
 \crmat = -\sum_{n=0}^\infty\sum_{a=1}{\gen{J}^a_n\otimes\gen{J}_{a,-n-1}}.
\]
As raising and lowering the indices $a$ corresponding to the Lie algebra is done with a non-degenerate bilinear form, which does not exist for $\pslcentral$, we should in principle extend the algebra to $\alg{sl}(2)\ltimes\pslcentral$. As discussed in section \ref{sec:pslcentral}, the automorphisms have no fundamental matrix representation, so this extension cannot give a physical answer. Physics requires the r-matrix to be an ordinary $16\times 16$ matrix. Let us have a closer look at the fundamental matrix representation of $\pslcentral$. In section \ref{sec:pslcentral} we have described the eigenvalues of the three central charges $\gC, \gP, \gK$ in terms of four complex numbers $a,b,c,d$ with the constraint $ad-bc=1$. Furthermore, we argued that $\gC$ should correspond to the energy eigenvalue of the light cone Hamiltonian, which should depend on the momentum $p$ of the corresponding magnon as well as the coupling constant $\sqrt{\lambda}$. Hence, the labels $a,b,c,d$ should also depend on $p$ and $\sqrt{\lambda}$. We find that the following parametrisation for $a,b,c,d$ at first order in 
\[
 \frac{1}{g} := \frac{2\pi}{\sqrt{\lambda}}
\]
yields the correct dispersion relation \cite{Beisert:2004hm} for $\gC$:

\[\label{eq:abcd}
a = \gat{},\qquad
b = -\frac{i\alpha x}{\gat{}(x^2-1)}\,,\qquad
c = \frac{i\gat{}}{\alpha x}\,,\qquad
d = \frac{x^2}{\gat{}(x^2-1)}\,.
\]
Here, we introduced two parameters $\gat{}$, $\alpha$ which do not appear in the eigenvalue of $\gC$, and are hence not physical. They are related to rescaling the basis vectors. The spectral parameter $x$ is related to the momentum $p$ as 

\[
 p = g\frac{x}{x^2-1}.
\]
In this representation, the two central charges $\gP,\gK$ are clearly not independent. They are given by

\[
 \frac{\gP}{\alpha} = -i \frac{p}{g} = -\alpha\gK.
\]
Hence, it is useful to introduce a new generator $\gen{D}$, such that its eigenvalue is given by

\[
 \gen{D} = \frac{p}{g} = \frac{x}{x^2-1}.
\]
Then, we have

\[
 \gP = -i \alpha \gen{D},\qquad\gK = i \frac{1}{\alpha} \gen{D}.
\]
The energy eigenvalue $\gC$ in those parameters was found in \cite{Arutyunov:2006iu} and reads

\[
 \gC = \frac{1}{2}\frac{x^2+1}{x^2-1}.
\]
Let us now investigate the extension to the loop algebra, in order to construct the classical double and the classical r-matrix. We have seen in section \ref{sec:pcf} that for relativistic integrable models, the loop parameter, or, in quantised form, the parameter $u$ of the Yangian, is related to the rapidity. Here, we do not have relativistic invariance, but nevertheless, $u$ is a function of the momentum. Indeed, we find that a good parametrisation leading to the right physics is given by
\[\label{eq:uofx}
u = x + \frac{1}{x},
\]
where we choose a unitary representation for our loop algebra generators, i.e.

\[\label{eq:loopdegree}
 \gen{J}^a_n = (iu)^n \gen{J}^a_0.
\]
Hence, contrary to the case of ordinary simple Lie algebras, where the fundamental representations generically does not depend on continuous parameters, and parameter dependence is introduced via the loop algebra, in the case of the loop algebra $\lalg{\psucentral}$ the loop variable does not introduce any new independent parameter. This at least holds for the physical application we have in mind, but it is also important for the mathematical features of the loop algebra which we will discuss now. The crucial observation on the fundamental representation is that

\[\label{eq:constraintCDu}
 \gen{D} = \frac{2\gC }{u}
\]
holds. Hence, at least on the fundamental representation, one can eliminate yet another central charge, and remains with only one independent central charge $\gC$. We will now show that one can impose the constraint \eqref{eq:constraintCDu} in an abstract way in the loop algebra. In particular, adding the index $n$ for the degree of a generator as in \eqref{eq:loopdegree}, we get

\[\label{eq:constraintinloop}
  \gen{P}_n = -2 i \alpha \gC_{n-1},\qquad\gen{K}_n = 2 i \frac{1}{\alpha} \gC_{n-1}.
\]
The first observation is that as these identifications mix generators of different degree $n$, they obviously violate the generic loop algebra commutation relations $\scomm{
 \gen{J}^a_n}{\gen{J}^b_m}=\scons\gen{J}^c_{n+m}$. Instead, the fermionic generators commute like
 
 \<\label{eq:QSbi}
\acomm{(\gen{Q}_m){}^\alpha{}_b}{(\gen{S}_n){}^c{}_\delta}\eq
\delta^c_b(\gen{L}_{m+n}){}^\alpha{}_\delta
+\delta^\alpha_\delta(\gen{R}_{m+n}){}^c{}_b
+\delta^c_b\delta^\alpha_\delta (\gen{C}_{m+n}){},
\nln
\acomm{(\gen{Q}_m){}^{\alpha}{}_{b}}{(\gen{Q}_n){}^{\gamma}{}_{d}}\eq
2\alpha\varepsilon^{\alpha\gamma}\varepsilon_{bd}\gen{C}_{m+n-1},
\nln
\acomm{(\gen{S}_m){}^{a}{}_{\beta}}{(\gen{S}_n){}^{c}{}_{\delta}}\eq
-2\alpha^{-1}\varepsilon^{ac}\varepsilon_{\beta\delta}\gen{C}_{m+n-1},
\>
modifying the relations \eqref{def:commfermicentral} of $\psucentral$. Interestingly, this is compatible with the Lie algebra structure. In particular, the Jacobi identity holds. The same kind of shift in the degree in the commutation relations is not possible for loop algebras based on simple Lie superalgebras. Generically, if we have a relation of the form $\comm{J_n^1}{J_m^2} = J^3_{n+m}$, we cannot shift $J^3_{n+m}$, as it will have non-trivial commutation relations with the other generators. Hence it is quite non-trivial that this identification here does not lead to incompatibilities with the defining relations. \\

Due to the identification \eqref{eq:constraintinloop} the set of remaining generators is the same as the set of generators forming the loop algebra $\lalg{su(2|2)}$. However, the commutation relations are slightly different, as generically $\lalg{su(2|2)}$ relations would imply that the right-hand side of the second and third equation in \eqref{eq:QSbi} would be zero. Nevertheless, we end up with a well defined Lie algebra structure, which we shall call the deformed $\lalg{su(2|2)}$. This is because in some scaling limit of the loop variable $u$, we can recover the undeformed loop algebra $\lalg{su(2|2)}$.

Despite having reduced the three central charges to one we note that this deformed algebra still does not have a non-degenerate bilinear form, as needed for the construction for the classical r-matrix. We have to investigate how the external automorphism algebra $\alg{sl}(2)$ acts on $\lalg{su}(2|2)$ after the above identification \eqref{eq:constraintinloop}. The action of $\alg{sl}(2)$ on the undeformed $\pslcentral$ algebra was described in equations \eqref{def:sl2auto},\eqref{def:autooncentral}. The goal is to reduce the three $\alg{sl}(2)$ generators to just one generator, such that this remaining generator can act on the fundamental matrix representation, as required for physics. This generator should be compatible with the commutation relations, and still lift the degeneracy in the invariant product caused by the remaining central charges $\gC_n$. As on the fundamental matrix representation the central charge triplet $\gC^\mathfrak{a}_\mathfrak{b}$, before and after the reduction \eqref{eq:constraintinloop}, acts like a multiple of the identity, and henceforth commutes with all matrices, we also require that the surviving automorphism should commute with all central charges. The unique combination is given by

\[
 \gen{B}_n:=2(\gen{B}_n)^1_1 + \frac{2}{\alpha}(\gen{B}_{n-1})^1_2+ 2\alpha(\gen{B}_{n-1})^2_1.
\]
This generator acts on the fermionic generators as
\<\label{eq:BQS}
\comm{\gen{B}_m}{(\gen{Q}_n){}^\alpha{}_b}\eq
+(\gen{Q}_{m+n}){}^\alpha{}_b
-2\alpha\varepsilon^{\alpha\gamma}\varepsilon_{bd}(\gen{S}_{m+n-1}){}^d{}_\gamma,
\nln
\comm{\gen{B}_m}{(\gen{S}_n){}^a{}_\beta}\eq
-(\gen{S}_{m+n}){}^a{}_\beta
-2\alpha^{-1}\varepsilon^{ac}\varepsilon_{\beta\delta}(\gen{Q}_{m+n-1}){}^\delta{}_c.
\>
The remaining commutation relations of $\gen{B}$ are trivial, as before:

\[\label{eq:Brest}
\comm{\gen{B}_m}{\gen{C}_n}=
\comm{\gen{B}_m}{(\gen{R}_n)^a{}_b}=
\comm{\gen{B}_m}{(\gen{L}_n)^\alpha{}_\beta}=0
\]
Taking all the generators $(\gen{R}_n)^a{}_b,(\gen{L}_n)^\alpha{}_\beta,\gen{B}_n,\gen{C}_n,(\gen{S}_n){}^a{}_\beta,(\gen{Q}_n){}^\alpha{}_b$ together, we obtain the loop algebra $\lutt$. Due to the shift of degree in \eqref{eq:constraintinloop}, and consequently the modifications of the central elements and automorphisms, it is not the standard $\lutt$ loop algebra, but has deformed commutation relations as spelled out above. \\

We are now almost in a position to write down the classical r-matrix for $\lutt$. What we have to investigate is how the identifications \eqref{eq:constraintinloop} modify the invariant inner product on the loop algebra. Recall that the inner product on the loop algebra is defined by

\[\label{eq:innerprodcentral}
 \sprod{\gen{J}^a_n}{\gen{J}^b_m}=-\kappa^{ab}\delta_{n,-m-1}.
\]
On the fundamental representation, $\kappa$ is given by

\[
 \kappa^{ab} = \str(\gen{J}^a\gen{J}^b).
\]
As \eqref{eq:constraintinloop} was proclaimed via the observed identification on the fundamental representation, we argue that this identification does not change the inner product on the $\psu$ generators. This can also be checked explicitly by imposing the invariance of the inner product. Using the invariance condition $\sprod{A}{\comm{B}{C}} = \sprod{\comm{A}{B}}{C}$, we get the following inner product of the automorphism and the central element:

\[\label{def:prodCB}
  \sprod{\gen{C}_n}{\gen{B}_m} = \delta_{n,-m-1}.
\]
Furthermore, $\gen{B}$ allows for the following fundamental matrix representation:

\[\label{eq:Bclassfund}
\gen{B}\state{\phi^a}=-\frac{1}{4C}\,\state{\phi^a},
\qquad
\gen{B}\state{\psi^\alpha}=+\frac{1}{4C}\,\state{\psi^\alpha}
\]
Interestingly, the representation of $\gen{B}$ itself depends on the eigenvalue of the central element $\gC$. This is clearly required to satisfy \eqref{def:prodCB}. Furthermore, as $\gen{B}$ is an external automorphism of $\alg{su}(2|2)$, and henceforth never appears on the right-hand side of any commutation relations, we can add a multiple of $\gC$ to it without modifying any commutation relations. This is in analogy with the observations \eqref{def:generalauto} for the automorphism of $\glnn$. In principle, in the loop algebra $\lutt$, we could shift $\gen{B}_m$ by different multiples of $\gC$ for each $m$. We will not do so in this paper, as the classical r-matrix is invariant by such shifts. However, the r-matrix will have additional twists, as we will see soon. These could in principle be affected by such shifts. \\

We have now described a complete, consistent matrix representation for the generators of the deformed $\lutt$ algebra, and we also have described an inner product, which is basically unaffected by the identifications done. Furthermore, the basis generators we have chosen already form a dual basis, i.e. we have a basis $J_i\in\alg{g}[u]$ and a basis $J_j^*\in\nalg{g}$ such that $\sprod{J_i}{J_j^*} = \pm \delta_{i,j}$. The classical r-matrix \eqref{eq:crmatrepeated}, as studied in section \ref{sec:classdouble}, is written down as the canonical element in the double $\mathcal{D}(\palg{g}) = \lalg{g}$. It is important that $\alg{g}[u]$, as well as its dual $\nalg{g}$, are subbialgebras of $\lalg{g}$. This is clearly violated by the identifications we have done. Indeed, if one commutes two fermionic generators of degree $0$, we get from \eqref{eq:QSbi}

\<
\acomm{(\gen{Q}_0){}^{\alpha}{}_{b}}{(\gen{Q}_0){}^{\gamma}{}_{d}}\eq
2\alpha\varepsilon^{\alpha\gamma}\varepsilon_{bd}\gen{C}_{-1},
\nln
\acomm{(\gen{S}_0){}^{a}{}_{\beta}}{(\gen{S}_0){}^{c}{}_{\delta}}\eq
-2\alpha^{-1}\varepsilon^{ac}\varepsilon_{\beta\delta}\gen{C}_{-1}.
\>
Hence, the commutator of two elements in $\palg{g}$ is in $\nalg{g}$. However, there is a surprisingly simple solution to this problem. We can consider the two subalgebras $\alg{g}^+,\alg{g}^-$ spanned by

\<
\alg{g}^+\eq\langle \gen{R}_n, \gen{L}_n, \gen{Q}_n, \gen{S}_n, \gen{C}_{n-1}, \gen{B}_{n+1}\rangle_{n\geq 0},
\nln
\alg{g}^-\eq\langle \gen{R}_{-1-n}, \gen{L}_{-1-n}, \gen{Q}_{-1-n}, \gen{S}_{-1-n}, \gen{C}_{-2-n}, \gen{B}_{-n}\rangle_{n\geq 0}.
\>
One can immediately convince oneself that these are indeed closed subalgebras, and, importantly, they are still dual subalgebras with respect to the inner product \eqref{eq:innerprodcentral}. Hence, one can construct the classical double for $\alg{g}^+$, which will again be the loop algebra $\lutt$, but with deformed commutation relations. Furthermore, the classical r-matrix is also different, and is given as follows:

\[\label{eq:rclasspsucentral}
r
=r_{\alg{psu}(2|2)}-
\sum_{m=-1}^\infty
\gen{B}_{-1-m}\otimes\gen{C}_{m}
-
\sum_{m=+1}^\infty
\gen{C}_{-1-m}\otimes\gen{B}_{m}
\]
Here, $r_{\alg{psu}(2|2)}$ is the standard classical r-matrix for $\alg{psu}(2|2)$, given by
\<\label{eq:rclasspsu}
r_{\alg{psu}(2|2)}
\eq
+
\sum_{m=0}^\infty
(\gen{R}_{-1-m})^c{}_d\otimes(\gen{R}_{m})^d{}_c
-
\sum_{m=0}^\infty
(\gen{L}_{-1-m})^\gamma{}_\delta\otimes(\gen{L}_{m})^\delta{}_\gamma
\nl
+
\sum_{m=0}^\infty
(\gen{Q}_{-1-m})^\gamma{}_d\otimes(\gen{S}_{m})^d{}_\gamma
-
\sum_{m=0}^\infty
(\gen{S}_{-1-m})^c{}_\delta\otimes(\gen{Q}_{m})^\delta{}_c.\nln
\>
As we have constructed the classical r-matrix via the classical double construction, it follows immediately that it satisfies the classical Yang-Baxter equation

\[
 \cybe{r}=0.
\]
Details as well as an explicit check for the CYBE were performed in \cite{Beisert:2007ty}, and a proof why r-matrices obtained via a double are quasi triangular can be found in \cite{Chari:1994pz}. Furthermore, we could have equally well constructed it as a double of $\alg{g}^-$, and, just as argued in section \ref{sec:classdouble}, taking an antisymmetric combination of the two corresponding r-matrices results in a unitary r-matrix, see equation \ref{def:crmattriangular}.\\

Note that we can also subtract and readd the terms shifted in \eqref{eq:rclasspsucentral}, and obtain 

\[\label{eq:rcentralstandard}
r_{12}
=
\frac{\mathcal{T}_{\psu,12}
-\gen{B}\otimes \gen{C}
-\gen{C}\otimes \gen{B}
}{iu_1-iu_2}
-\frac{\gen{B}\otimes \gen{C}}{iu_2}
+\frac{\gen{C}\otimes \gen{B}}{iu_1}\,,
\]
with $\mathcal{T}_{\psu,12}$ being the standard Casimir of $\psu$, acting on the tensor product. The additional terms $-\gen{B}\otimes \gen{C}
-\gen{C}\otimes \gen{B}$ complete the Casimir to formally the standard Casimir of $\utt$. However, again the commutation relations are deformed with respect to the standard $\utt$ algebra. The last terms, $-\frac{\gen{B}\otimes \gen{C}}{iu_2}
+\frac{\gen{C}\otimes \gen{B}}{iu_1}$, can be interpreted as the classical limit of a Reshetikhin twist \cite{Reshetikhin:1990ep}. This twist is necessary as the quantum $\psucentral$ generators act nontrivially on tensor products \cite{Gomez:2006va,Plefka:2006ze}, i.e. they have a deformed coproduct.
On the classical level investigated here, this twist implies that the cobrackets resulting from the classical r-matrix \eqref{eq:rclasspsucentral} get modified. They are defined in the usual way by

\[
 \cobra(\gen{J}^a) = \scomm{\gen{J}^a}{\crmat},
\]
as studied in section \ref{sec:quasitribi}. We have spelled them out explicitly in table \ref{tab:cobra}.

\begin{table}\centering
\<
\coprocl(\gen{C}_n)\eq 0
\nln
\coprocl(\gen{B}_n)\eq
+\sum_{k=0}^{n-1}
(\gen{Q}_k)^\alpha{}_b\wedge(\gen{S}_{n-1-k})^b{}_\alpha
\nl
+\sum_{k=1}^{n-1}
\alpha^{-1}\beta\varepsilon^{bd}\varepsilon_{\alpha\gamma}
(\gen{Q}_{k-1})^\alpha{}_b\wedge(\gen{Q}_{n-1-k})^\gamma{}_d
\nl
-\sum_{k=1}^{n-1}
\alpha\beta\varepsilon^{\beta\delta}\varepsilon_{ac}
(\gen{S}_{k-1})^a{}_\beta\wedge(\gen{S}_{n-1-k})^c{}_\delta
\nln
\coprocl(\gen{R}_n)^a{}_b\eq
+\sum_{k=0}^{n-1}
(\gen{R}_k)^a{}_c\wedge(\gen{R}_{n-1-k})^c{}_b
\nl
-\sum_{k=0}^{n-1}
\Bigl[
(\gen{S}_k)^a{}_\gamma\wedge(\gen{Q}_{n-1-k})^\gamma{}_b
-\half\delta^a_b\,(\gen{S}_k)^d{}_\gamma\wedge(\gen{Q}_{n-1-k})^\gamma{}_d
\Bigr]
\nln
\coprocl(\gen{L}_n)^\alpha{}_\beta\eq
-\sum_{k=0}^{n-1}
(\gen{L}_k)^\alpha{}_\gamma\wedge(\gen{L}_{n-1-k})^\gamma{}_\beta
\nl
+\sum_{k=0}^{n-1}
\Bigl[
(\gen{Q}_k)^\alpha{}_c\wedge(\gen{S}_{n-1-k})^c{}_\beta
-\half\delta^\alpha_\beta\,(\gen{Q}_k)^\delta{}_c\wedge(\gen{S}_{n-1-k})^c{}_\delta
\Bigr]
\nln
\coprocl(\gen{Q}_{n})^\alpha{}_b\eq
-\sum_{k=0}^{n-1}
(\gen{L}_k)^\alpha{}_\gamma\wedge(\gen{Q}_{n-1-k})^\gamma{}_b
-\sum_{k=0}^{n-1}
(\gen{R}_k)^c{}_b\wedge(\gen{Q}_{n-1-k})^\alpha{}_c
\nl
-\sum_{k=0}^{n}
\gen{C}_{k-1}\wedge(\gen{Q}_{n-k})^\alpha{}_b
+\sum_{k=0}^{n-1}
2\alpha\beta\varepsilon^{\alpha\gamma}\varepsilon_{bd}\gen{C}_{k-1}\wedge(\gen{S}_{n-1-k})^d{}_\gamma
\nl
\nln
\coprocl(\gen{S}_n)^a{}_\beta\eq
+\sum_{k=0}^{n-1}
(\gen{R}_k)^a{}_c\wedge(\gen{S}_{n-1-k})^c{}_\beta
+\sum_{k=0}^{n-1}
(\gen{L}_k)^\gamma{}_\beta\wedge(\gen{S}_{n-1-k})^a{}_\gamma
\nl
+\sum_{k=0}^{n}
\gen{C}_{k-1}\wedge(\gen{S}_{n-k})^a{}_\beta
+\sum_{k=0}^{n-1}
2\alpha^{-1}\beta\varepsilon^{ac}\varepsilon_{\beta\delta}\gen{C}_{k-1}\wedge(\gen{Q}_{n-1-k})^\delta{}_c
\nn
\>
\caption{Cobrackets of the Lie bialgebra generators.}
\label{tab:cobra}
\end{table}
These cobrackets differ from the standard cobrackets for the undeformed $\lutt$ algebra, as spelled out in table \ref{tab:cobrapslcentral} in section \ref{sec:glnmbialgebra}, exactly by terms generated by the twist. In particular, even the fermionic generators of degree zero have nontrivial cobrackets, in contrast to the standard cobrackets of degree zero generators in a loop algebra. \\

Having established the abstract, universal form of the classical r-matrix, we can straightforwardly evaluate it on the fundamental evaluation representation, as defined in equations \eqref{eq:fermigencentralrep},\eqref{eq:bosegencentralrep} in section \ref{sec:pslcentral}, using the parametrisation for the labels as in \eqref{eq:abcd}, and the spectral parameter \eqref{eq:uofx}. We get the result as listed in table \ref{tab:rCoeff}.

\begin{table}
\<
\crmat\state{\phi^a\phi^b}\eq
\half (A_{12}-B_{12})\state{\phi^a\phi^b}
+\half (A_{12}+B_{12})\state{\phi^b\phi^a}
+\half C_{12}\varepsilon^{ab}\varepsilon_{\alpha\beta}\state{\psi^\alpha\psi^\beta}
\nln
\crmat\state{\psi^\alpha\psi^\beta}\eq
-\half (D_{12}-E_{12})\state{\psi^\alpha\psi^\beta}
-\half (D_{12}+E_{12})\state{\psi^\beta\psi^\alpha}
-\half F_{12}\varepsilon^{\alpha\beta}\varepsilon_{ab}\state{\phi^a\phi^b}
\nln
\crmat\state{\phi^a\psi^\beta}\eq
G_{12}\state{\phi^a\psi^\beta}
+H_{12}\state{\psi^\beta\phi^a}
\nln
\crmat\state{\psi^\alpha\phi^b}\eq
K_{12}\state{\phi^b\psi^\alpha}
+L_{12}\state{\psi^\alpha\phi^b}
\nonumber
\>

\<
\half(A_{12}+B_{12})\eq\frac{1}{iu_1-iu_2}\nln
\half(A_{12}-B_{12})\eq\frac{(x_1-x_2)^2(x_1x_2+1)^2}{4x_1x_2(x_1^2-1)(x^2_2-1)(iu_1-iu_2)}
      =\frac{+\half-\alg{C}^{-1}_{1,-1}\alg{C}_{2,-1}-\alg{C}^{-1}_{2,-1}\alg{C}_{1,-1}}{iu_1-iu_2}\nln
\half C_{12}\eq \frac{i\gat{1}\gat{2}(x_1-x_2)}{\alpha x_1x_2(iu_1-iu_2)}
      =\frac{a_1c_2-c_1a_2}{iu_1-iu_2}\nln
-\half(D_{12}+E_{12})\eq-\frac{1}{iu_1-iu_2}\nln
-\half(D_{12}-E_{12})\eq-\frac{(x_1-x_2)^2(x_1x_2+1)^2}{4x_1x_2(x_1^2-1)(x^2_2-1)(iu_1-iu_2)}
      =\frac{-\half+\alg{C}^{-1}_{2,-1}\alg{C}_{1,-1}+\alg{C}^{-1}_{1,-1}\alg{C}_{2,-1}}{iu_1-iu_2}\nln
-\half F_{12}\eq -\frac{i\alpha x_1x_2(x_1-x_2)}{\gat{1}\gat{2}(x_1^2-1)(x^2_2-1)(iu_1-iu_2)}
      =\frac{d_1b_2-b_1d_2}{iu_1-iu_2}\nln
G_{12}\eq \frac{(x^2_1-x^2_2)(x^2_1x^2_2-1)}{4x_1x_2(x_1^2-1)(x_2^2-1)(iu_1-iu_2)}
      =\frac{\alg{C}^{-1}_{2,-1}\alg{C}_{1,-1}-\alg{C}^{-1}_{1,-1}\alg{C}_{2,-1}}{iu_1-iu_2}\nln
H_{12}\eq \frac{\gat{1} x_2(x_1x_2-1)}{\gat{2} x_1(x_2^2-1)(iu_1-iu_2)}
      =\frac{a_1d_2-c_1b_2}{iu_1-iu_2}\nln
K_{12}\eq \frac{\gat{2}x_1(x_1x_2-1)}{\gat{1}x_2(x_1^2-1)(iu_1-iu_2)}
      =\frac{d_1a_2-b_1c_2}{iu_1-iu_2}\nln
L_{12}\eq -\frac{(x^2_1-x^2_2)(x^2_1x^2_2-1)}{4x_1x_2(x_1^2-1)(x_2^2-1)(iu_1-iu_2)}
      =\frac{\alg{C}^{-1}_{1,-1}\alg{C}_{2,-1}-\alg{C}^{-1}_{2,-1}\alg{C}_{1,-1}}{iu_1-iu_2}\nonumber
\>
\caption{The classical (light cone) r-matrix of AdS/CFT.}
\label{tab:rCoeff}
\end{table}
This result was obtained by using the representation for the automorphism as in \eqref{eq:Bclassfund}. It coincides with the light-cone S-matrix of \cite{Frolov:2006cc}. To obtain a result compatible with the so-called AFS phase of the string S-matrix at strong coupling \cite{Arutyunov:2004vx}, we have to add a term

\[
 \gC_0 \wedge\gC_{-1}
\]
to the r-matrix. This combines with the twist to $\gC_{-1}\wedge(\gen{B}_0 -\gC_0)$, and can hence be seen as a shift of the automorphism by $-\gC_0$. As argued before, such shift does not modify the commutation relations. Furthermore, such shift is the only remaining freedom not completely fixed by the symmetries at the classical level. This is because the antipode acts trivially on the classical Lie bialgebra. Hence, there is no natural algebraic form of the crossing equation acting at this level. Hence, one might in principle add other terms compatible with the classical Yang-Baxter and unitarity equations. Certainly, terms of the form
\[
 \gC_m\wedge\gC_n
\]
satisfy this requirement. Otherwise, we have succeeded in reproducing the classical r-matrix purely from the symmetries.

\subsection{The Magnon S-Matrix of \sym}\label{sec:smatsym}

In section \ref{sec:integrabilitysym}, we have argued that \sym in the large $N$ limit is integrable, and this integrability is realised by regarding the action of the dilatation generator on single trace local operators as the action of an integrable Hamiltonian on a spin chain. In principle, we can use the techniques of diagonalisation of the Hamiltonian of the XXX Heisenberg spin chain introduced in section \ref{sec:xxxspinchain} to derive the spectrum of the dilatation operator. In section \ref{sec:xxxspinchain}, we were dealing with $\sun$ spin chains with nearest neighbour interaction. Now, the spin chain has $\sconf$ symmetry, which is noncompact and also long-ranged. The interaction range of the Hamiltonian depends on the order in perturbation theory we are considering for the gauge theory. At order $\lambda$, the complete Hamiltonian was derived in \cite{Beisert:2003jj}, and the Bethe equations were written down in \cite{Beisert:2003yb}. The full $\sconf$ Yangian has only been found at tree level \cite{Dolan:2003uh,Dolan:2004ps}. Assuming the Yangian to survive, we will reconstruct the one-loop S-matrix from the universal R-matrix in section \ref{sec:1loopS}.\\

We should note that when the interaction range of the Hamiltonian is longer than the length of the operator whose anomalous dimension we want to calculate, there will be wrapping interactions \cite{Ambjorn:2005wa}. This makes the spin chain picture unsuitable, and the asymptotic Bethe ansatz will not work in that case. Indeed, to calculate anomalous dimensions at wrapping order, one needs to take into account Luscher corrections \cite{Janik:2007wt}, and ultimately the TBA equations \cite{Gromov:2009tv,Gromov:2009bc,Arutyunov:2009ur,Bombardelli:2009ns}. Here, we will derive only the asymptotic S-matrix, whose diagonalisation leads to the asymptotic Bethe ansatz. However, as the goal is to find a universal R-matrix in terms of the abstract generators, as done for classical string theory in section \ref{sec:adsbialgebra}, such universal R-matrix should, upon choosing a suitable representation, yield the bound state S-matrices leading to the TBA equations.

\subsubsection{The One-Loop S-Matrix from Yangian Symmetry}\label{sec:1loopS}

In this section we would like to discuss the scattering matrix for magnons of the $\sconf$ spin chain at one-loop. We have seen in section \ref{sec:xxxspinchain}, deriving the scattering matrix for the $\sun$ spin chain requires choosing a vacuum state, which breaks the symmetry to $\alg{su}(n-1)$. The magnons transform under this residual symmetry, and the S-matrix is henceforth invariant under $\alg{su}(n-1)$. Likewise, we have to choose a vacuum for the $\sconf$ spin chain. A suitable choice is the BPS state 

\[
 Z^L,
\]
where $Z = \phi_5 + i \phi_6$ is part of the scalar sector of the theory.

\begin{center}
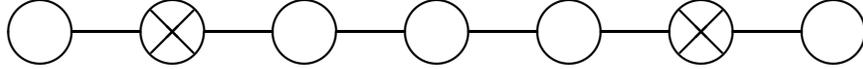
\begin{figure}
\setlength{\unitlength}{1pt}%
\small\thicklines%
\begin{center}
\begin{picture}(260,50)(0,-20)
\put(  0,00){\circle{25}}%
\put(  0,15){\makebox(0,0)[b]}%
\put(  0,-15){\makebox(0,0)[t]}
\put( 12,00){\line(1,0){26}}%
\put( 50,00){\circle{25}}%
\put( 50,15){\makebox(0,0)[b]}%
\put( 50,-15){\makebox(0,0)[t]}%
\put( 62,00){\line(1,0){26}}%
\put( 100,00){\circle{25}}%
\put( 100,15){\makebox(0,0)[b]}%
\put( 100,-15){\makebox(0,0)[t]}%
\put( 112,00){\line(1,0){26}}%
\put(150,00){\circle{25}}%
\put(150,15){\makebox(0,0)[b]}%
\put(150,-15){\makebox(0,0)[t]}%
\put(162,00){\line(1,0){26}}%
\put(200,00){\circle{25}}%
\put(200,15){\makebox(0,0)[b]}%
\put(200,-15){\makebox(0,0)[t]}%
\put(212,00){\line(1,0){26}}%
\put(250,00){\circle{25}}%
\put(250,15){\makebox(0,0)[b]}%
\put(250,-15){\makebox(0,0)[t]}%
\put(262,00){\line(1,0){26}}%
\put(300,00){\circle{25}}%
\put(300,15){\makebox(0,0)[b]}%
\put(300,-15){\makebox(0,0)[t]}%
\put( 42,-8){\line(1, 1){16}}%
\put( 42, 8){\line(1,-1){16}}%
\put(242,-8){\line(1, 1){16}}%
\put(242, 8){\line(1,-1){16}}%
\end{picture}
\end{center}
\caption{Dynkin diagram of $\sconf$ called ``Beauty``} \label{fig:Dynkinbeauty}
\end{figure}
\end{center}
The choice of the vacuum state breaks the symmetry to 

\[\label{def:residualsymm1l}
 \uone\ltimes\psu\oplus\psu\ltimes\uone.
\]
The two $\psu$'s correspond to the three nodes to the left/right of the middle node in figure \ref{fig:Dynkinbeauty}. The resulting S-matrix should hence be invariant under \eqref{def:residualsymm1l}. We can rewrite \eqref{def:residualsymm1l} as 

\[
 \utt\oplus\utt,
\]
such that the two $\uone$ charges in the decomposition $\utt=\uone\ltimes\psu\ltimes\uone$ are identified in both copies of $\utt$. Due to the direct sum structure, by integrability also the total S-matrix should factorise with respect to the algebra factorisation, i.e. one should have

\[
 \smat_{\utt\oplus\utt} =  \smat_{\utt} \otimes \smat_{\utt}. 
\]
We will henceforth assume this factorisation, and focus on deriving the $\utt$ invariant S-matrix. Note that the residual symmetry after symmetry breaking with the BPS vacuum is the same as the residual symmetry in the light-cone gauge, as studied in the previous section. The reason for this is that via the AdS/CFT correspondence the gauge theory operator $Z$ corresponds to excitations propagating along a big circle on the $S^5$ on the string side\cite{Berenstein:2002jq}. Note that for the moment, we have no reason to believe that the $\utt$'s undergo a central extension. 
\\

For the moment, let us assume that $\utt$ is enhanced to the Yangian, as we would expect from the existence of Yangian charges in the unbroken symmetry algebra. The fields of \sym transforming under the unbroken symmetry \eqref{def:residualsymm1l} are $8$ bosons and $8$ fermions, which carry two indices belonging to the fundamental representation of the two $\utt$'s. This is in precise analogy to the fields in the light-cone string theory, as studied in section \ref{sec:stringinlc}.\\

We now form scattering states in the same fashion as for $\sun$ spin chains considered section \ref{sec:xxxspinchain}. This means that the $\utt$ magnons look like

\[
 \state{X^M} = \sum_n e^{ipn}\state{Z \dots Z X^M Z \dots Z},
\]
with $M=1,\dots 4$, and $X^{1,2} = \phi^{1,2}$ are bosons, whereas $X^{3,4} = \psi^{1,2}$ are fermions. Hence, the problem of finding the S-matrix $\smat_{\utt\oplus\utt}$ from the $\sconf$ spin chain is equivalent to finding two copies of the S-matrix $\smat_{\utt}$, which can be though of as acting on excitations of a $\alg{su}(2|3)$ spin chain. \\

As the magnons transform in the fundamental representation of $\utt$, we expect the two-particle S-matrix $\smat_{\utt}$ to be related to the appropriate representation of the universal R-matrix of $\yutt$. The universal R-matrices for Yangians of $\slnm$ were studied in section \ref{sec:yangglnm}, so we will just briefly recall the results. We take the universal R-matrix from section \ref{sec:universalR}, and evaluate it as in section \ref{sec:funR} for $\glnn$, by simply putting $n=2$. We get the result

\[\label{eq:weakSmatrix}
 \smat = \frac{i/2 + u}{i/2 - u}\left(\frac{u}{u+i}\idm - \frac{i}{u+i}\perm\right).
\]
Here, we have used the Cartan matrix of the distinguished Dynkin diagram,

\[
A = \begin{pmatrix}
2&-1&0&0\\
-1&0&1&1\\
0&1&-2&0\\
0&1&0&0\\
\end{pmatrix},
\]
and chosen reality conditions as we deal with the real Lie algebra $\utt$. This requires the spectral parameter of section \ref{sec:funR} to be imaginary, i.e. we have put $u\rightarrow -i u$, such that the $u$ used in this section is real for physical particles.\\

Note that this S-matrix is obtained from a universal R-matrix in the same way as the relativistic $\sun$ S-matrices. Interestingly, as our method to construct universal R-matrices works well for all superalgebras $\slnm$, we note that for $n=m$ the complicated Gamma function prefactor disappears, and we remain with the simple rational factor $\frac{i/2 + u}{i/2 - u}$. To get the precise weak coupling limit of the spin chain S-matrix, the factor $\frac{i/2 + u}{i/2 - u}$ should reduce to $1$. This might be related to a twist of the S-matrix, which we are missing at the presence. Note that $\frac{i/2 + u}{i/2 - u}$ is just a phase of a freely moving particle and appears also in the Bethe equations, so we are at present not sure if the appearance of $\frac{i/2 + u}{i/2 - u}$ in the S-matrix has some deep physical meaning. We are convinced that it cannot be purely by chance that the Gamma functions cancel for $n=m$.\\

We should note that other choices for the outer automorphism, leading to different Dynkin diagrams, as in \ref{def:cartnnmula}, might restore Gamma functions. But for generic $\slnm$, $n\neq m$ and $n\neq m\pm 1$, there are always Gamma functions appearing in the S-matrix. Hence, we certainly believe it is no coincidence that the universal R-matrix for $\utt$ almost produces the correct result needed for physics. Also, the S-matrix in string and gauge theory looks slightly different, as noted in \cite{Arutyunov:2006yd}, and there is certainly a twist needed to modify the coproduct of the Lie generators \cite{Gomez:2006va,Plefka:2006ze}, which also involves a factor of the form $\frac{i/2 + u}{i/2 - u}$, even though the twist leading to this coproduct seems to not directly lead to the S-matrix \eqref{eq:weakSmatrix}.

\subsubsection{The All-Loop Magnon S-Matrix}\label{sec:allloopS}

The all-loop S-matrix was derived in \cite{Beisert:2005tm} from gauge theory arguments, up to its dressing factor. A crucial ingredient in the derivation is that the $\sconf$ spin chain has the property that some symmetry generators change the length. This was first realised in the $\alg{su}(3|2)$ subsector\footnote{We reverse $\alg{su}(2)$ and $\alg{su}(3)$ compared to \cite{Beisert:2003ys}, for compatibility with the conventions of chapter \ref{ch:yangians}.} \cite{Beisert:2003ys}. Let us consider this subsector, and the corresponding spin-chain. The fields transform in the fundamental representation of $\alg{su}(3|2)$ as defined in \eqref{def:funrepslnm}. We label the basis vectors as before by $\{\phi^1,\phi^2,\phi^3\equiv Z|\psi^1,\psi^2\}$. The third bosonic basis state $\phi^3\equiv Z$ will form the spin chain vacuum state $Z^L$, which corresponds to a BPS state in \sym. Hence, excitation states look like 

\[\label{def:asymptoticstate}
 \state{X_1,\dots X_M} = \sum_{n_1<<\dots << n_M} e^{i (p_1 n_1+\dots +p_M n_M)} \state{\dots Z,X_1,Z,\dots Z,X_M,Z,\dots}.
\]
As we mentioned before, the interaction range of the Hamiltonian depends on the order of the expansion in perturbation in $\lambda$. We now want to derive the all-loop S-matrix, which means the Hamiltonian has, in principle, infinite range of interaction. That is the reason why we should consider asymptotic states where the individual magnons are well separated, i.e. $n_1<<\dots << n_M$. This apparently requires us to work with an infinitely long chains. We emphasise that the Hamiltonian is not explicitly known to higher orders in perturbation theory. Nevertheless, one can fix the all-loop S-matrix purely by symmetry. This allows to extract the momenta via the Bethe equations, and, as one has knowledge of the all-loop dispersion relation \cite{Beisert:2004hm}, one can extract the exact all-loop energies of asymptotic states. Indeed, the structure of the dispersion relation is also fixed by the non-local Hopf symmetry resulting from the length changing of the underlying spin chain. \\

Let us consider the action of the symmetry generators which do not change the vacuum state $\state{Z}^L$, as in the previous sections where we discussed the perturbative string and gauge S-matrices. It turns out that some fermionic generators do change the length of the spin chain when acting on magnons, which means that they insert or remove vacuum states $Z$ into the chain. The generators form a $\psucentral$ algebra, i.e. the manifest $\sutt$ algebra is centrally extended \cite{Beisert:2005tm}. The action on one-particle states of the symmetry generators is given by 

\<
\gR{a}{b}\state{\phi^c}=\delta_b^c\state{\phi^a} -\frac{1}{2}\delta_b^a\state{\phi^c},\nln
\gL{\alpha}{\beta}\state{\psi^\gamma}=\delta_\beta^\gamma\state{\psi^\alpha}-\frac{1}{2}\delta_\beta^\gamma\state{\psi^\alpha},
\>
\[\label{eq:fermigenlengthch}
\begin{array}[b]{rclcrcl}
\gen{Q}^\alpha{}_a\state{\phi^b}\eq a\,\delta^b_a\state{\psi^\alpha},&&
\gen{Q}^\alpha{}_a\state{\psi^\beta}\eq b\,\varepsilon^{\alpha\beta}\varepsilon_{ab}\state{\phi^b Z^+},\\[3pt]
\gen{S}^a{}_\alpha\state{\phi^b}\eq c\,\varepsilon^{ab}\varepsilon_{\alpha\beta}\state{\psi^\beta Z^-},&&
\gen{S}^a{}_\alpha\state{\psi^\beta}\eq d\,\delta^\beta_\alpha\state{\phi^a},
\end{array}\]
with the action of the fermionic generators including the markers $Z^\pm$, which indicates that a vacuum field $Z$ is inserted or removed into the spin chain. The labels $a,b,c,d$ and full set of commutation relations was given in section \ref{sec:pslcentral}.

Indeed, on single magnon states, these extra markers have no physical effect, as single magnons are excitations over infinitely many $Z$'s. The crucial difference is that these markers modify the action on multi particle states. Consider a two magnon state $\state{XX}$. Then the action of e.g. a generator $\gen{Q}$ inserting a $Z$ would be 

\[
 \gen{Q}\state{XX} = \state{YZX} + \state{XYZ},
\]
where we have say $\gen{Q}\state{X} = \state{YZ}$. Recalling that we have asymptotic states \eqref{def:asymptoticstate}, we can shift the marker around the excitation $Y$, and redefining the indices $n_1, n_2$ we get an additional phase $e^{i p}$, where $p$ is the momentum corresponding to the magnon around we shifted the marker. These shifts mean that $\psucentral$ acts non-locally on tensor products, i.e. multi particle states, and the correct algebraic structure is a Hopf algebra \cite{Gomez:2006va,Plefka:2006ze}. The algebra relations of $\psucentral$ are not modified, but the coproduct is non-trivial, and spelled out in table \ref{tab:twistedcopro}\footnote{We choose the conventions of \cite{Beisert:2006qh}, upon formally identifying the marker $Y$ used in \cite{Beisert:2006qh} with $Z^{1/2}$. This leads to the S-matrix in the string frame \cite{Arutyunov:2006yd}.}.

\begin{table}\centering
\<
\copro\gen{R}^a{}_b\eq
\gen{R}^a{}_b\otimes 1
+1\otimes\gen{R}^a{}_b,
\nln
\copro\gen{L}^\alpha{}_\beta\eq
\gen{L}^\alpha{}_\beta\otimes 1
+1\otimes\gen{L}^\alpha{}_\beta,
\nln
\copro\gen{Q}^\alpha{}_b\eq
\gen{Q}^\alpha{}_b\otimes 1
+\eip^{+1}\otimes\gen{Q}^\alpha{}_b,
\nln
\copro\gen{S}^a{}_\beta\eq
\gen{S}^a{}_\beta\otimes 1
+\eip^{-1}\otimes\gen{S}^a{}_\beta,
\nln
\copro\gen{C}\eq
\gen{C}\otimes 1
+1\otimes\gen{C},
\nln
\copro\gen{P}\eq
\gen{P}\otimes 1
+\eip^{+2}\otimes\gen{P},
\nln
\copro\gen{K}\eq
\gen{K}\otimes 1
+\eip^{-2}\otimes\gen{K},
\nln
\copro\eip\eq
\eip\otimes \eip.
\nn
\>

\caption{The twisted coproduct of $\psucentral$.}
\label{tab:twistedcopro}
\end{table}
The generator $\eip$ appearing in the coproduct in table \ref{tab:twistedcopro} is a central generator of the enveloping algebra and has the eigenvalue $\eip = e^{i p /2}$ on the fundamental representation. Furthermore, the labels $a,b,c,d$ are fixed to be

\[\label{def:abcd}
a=\sqrt{g}\,\gamma,\quad
b=\sqrt{g}\,\frac{\alpha}{\gamma}\lrbrk{1-\frac{\xp{}}{\xm{}}},\quad
c=\sqrt{g}\,\frac{i\gamma}{\alpha \xp{}}\,,\quad
d=\sqrt{g}\,\frac{\xp{}}{i\gamma}\lrbrk{1-\frac{\xm{}}{\xp{}}},
\]
where $\xp{}, \xm{}$ are related to the momentum as

\[
 e^{ip} = \frac{x^+}{x^-}.
\]
They also satisfy the constraint
\[\label{eq:xpmconstraint}
\xp{}+\frac{1}{\xp{}}-\xm{}-\frac{1}{\xm{}}=\frac{i}{g}\,.
\]
$\gamma$ and $\alpha$ are two further generators related to the rescaling of the basis vectors. They do not appear in physical quantities. $g$ is related to the 't Hooft coupling as 

\[
 g = \frac{\sqrt{\lambda}}{4\pi}.
\]
The central charges $\gP, \gK$ of $\psucentral$ have to be cocommutative in order for the coproduct to be compatible with the existence of an S-matrix. This leads to the identifications

\[\label{eq:PKU}
\gen{P}=g\alpha\bigbrk{1-\eip^{+2}},
\quad
\gen{K}=g\alpha^{-1}\bigbrk{1-\eip^{-2}}.
\]
These identifications can be imposed abstractly, and are obviously compatible with the above representation labels. Furthermore, plugging these identifications into the shortening condition for the fundamental representation, the third central charge $\gC$ is given by 

\[
 \gC^2 = \frac{1}{4}-\gP \gK = \frac{1}{4} + 4 g^2 \sin^2(p/2).
\]
As the eigenvalue of $\gC$ corresponds to the energy of the underlying spin chain, this identification is identical to the dispersion relation of the theory. We stress that the derivation is purely in algebraic terms, and fixes the dispersion relation up to the choice of proportionality in front of $\sin^2(p/2)$\footnote{The S-matrix of the $AdS_4\times\mathbb{CP}^3$ correspondence is formally found by substituting the 't Hooft coupling by a more general function of $\lambda$ \cite{Ahn:2008aa}. Hence, the same algebraic structure seems to exist in this theory.}. \\

Let us now discuss the extension to the Yangian. We discussed the Yangian of $\ypslcentral$ in section \ref{sec:yangpslcentral}. Here, we only need to discuss the modifications due to the braiding generator $\eip$. The Yangian algebra relations remains unchanged, but the coproduct of the generators is now twisted in the same way as the Lie algebra generators, and the coproduct in the first realisation is explicitly given in table \ref{tab:coprotwistyang}.

\begin{table}\centering
\<
\copro\genY{R}^a{}_b\eq
\genY{R}^a{}_b\otimes 1
+1\otimes\genY{R}^a{}_b
\nl
+\half\gen{R}^a{}_c\otimes\gen{R}^c{}_b
-\half\gen{R}^c{}_b\otimes\gen{R}^a{}_c
\nl
-\half\gen{S}^a{}_\gamma\eip^{+1}\otimes\gen{Q}^\gamma{}_b
-\half\gen{Q}^\gamma{}_b\eip^{-1}\otimes\gen{S}^a{}_\gamma
\nl\qquad
+\quarter\delta^a_b\,\gen{S}^d{}_\gamma\eip^{+1}\otimes\gen{Q}^\gamma{}_d
+\quarter\delta^a_b\,\gen{Q}^\gamma{}_d\eip^{-1}\otimes\gen{S}^d{}_\gamma,
\nln
\copro\genY{L}^\alpha{}_\beta\eq
\genY{L}^\alpha{}_\beta\otimes 1
+1\otimes\genY{L}^\alpha{}_\beta
\nl
-\half\gen{L}^\alpha{}_\gamma\otimes\gen{L}^\gamma{}_\beta
+\half\gen{L}^\gamma{}_\beta\otimes\gen{L}^\alpha{}_\gamma
\nl
+\half\gen{Q}^\alpha{}_c\eip^{-1}\otimes\gen{S}^c{}_\beta
+\half\gen{S}^c{}_\beta\eip^{+1}\otimes\gen{Q}^\alpha{}_c
\nl\qquad
-\quarter\delta^\alpha_\beta\,\gen{Q}^\delta{}_c\eip^{-1}\otimes\gen{S}^c{}_\delta
-\quarter\delta^\alpha_\beta\,\gen{S}^c{}_\delta\eip^{+1}\otimes\gen{Q}^\delta{}_c,
\nln
\copro\genY{Q}^\alpha{}_b\eq
\genY{Q}^\alpha{}_b\otimes 1
+\eip^{+1}\otimes\genY{Q}^\alpha{}_b
\nl
-\half\gen{L}^\alpha{}_\gamma\eip^{+1}\otimes\gen{Q}^\gamma{}_b
+\half\gen{Q}^\gamma{}_b\otimes\gen{L}^\alpha{}_\gamma
\nl
-\half\gen{R}^c{}_b\eip^{+1}\otimes\gen{Q}^\alpha{}_c
+\half\gen{Q}^\alpha{}_c\otimes\gen{R}^c{}_b
\nl
-\half\gen{C}\eip^{+1}\otimes\gen{Q}^\alpha{}_b
+\half\gen{Q}^\alpha{}_b\otimes\gen{C}
\nl
+\half\varepsilon^{\alpha\gamma}\varepsilon_{bd}\gen{P}\eip^{-1}\otimes\gen{S}^d{}_\gamma
-\half\varepsilon^{\alpha\gamma}\varepsilon_{bd}\gen{S}^d{}_\gamma\eip^{+2}\otimes\gen{P},
\nln
\copro\genY{S}^a{}_\beta\eq
\genY{S}^a{}_\beta\otimes 1
+\eip^{-1}\otimes\genY{S}^a{}_\beta
\nl
+\half\gen{R}^a{}_c\eip^{-1}\otimes\gen{S}^c{}_\beta
-\half\gen{S}^c{}_\beta\otimes\gen{R}^a{}_c
\nl
+\half\gen{L}^\gamma{}_\beta\eip^{-1}\otimes\gen{S}^a{}_\gamma
-\half\gen{S}^a{}_\gamma\otimes\gen{L}^\gamma{}_\beta
\nl
+\half\gen{C}\eip^{-1}\otimes\gen{S}^a{}_\beta
-\half\gen{S}^a{}_\beta\otimes\gen{C}
\nl
-\half\varepsilon^{ac}\varepsilon_{\beta\delta}\gen{K}\eip^{+1}\otimes\gen{Q}^\delta{}_c
+\half\varepsilon^{ac}\varepsilon_{\beta\delta}\gen{Q}^\delta{}_c\eip^{-2}\otimes\gen{K},
\nln
\copro\genY{C}\eq
\genY{C}\otimes 1
+1\otimes\genY{C}
\nl
+\half\gen{P}\eip^{-2}\otimes\gen{K}
-\half\gen{K}\eip^{+2}\otimes\gen{P},
\nln
\copro\genY{P}\eq
\genY{P}\otimes 1
+\eip^{+2}\otimes\genY{P}
\nl
-\gen{C}\eip^{+2}\otimes\gen{P}
+\gen{P}\otimes\gen{C},
\nln
\copro\genY{K}\eq
\genY{K}\otimes 1
+\eip^{-2}\otimes\genY{K}
\nl
+\gen{C}\eip^{-2}\otimes\gen{K}
-\gen{K}\otimes\gen{C}.
\nn
\>
\caption{Coproduct of the twisted Yangian}
\label{tab:coprotwistyang}
\end{table}
If one would formally switch off the braiding generator $\eip$, one would recover the standard Yangian of $\psucentral$, as studied in section \ref{sec:yangpslcentral}. \\

We note that mathematically, we can generate $\eip$ by twisting the coproduct as

\[
 \copro\gen{X} = \mathcal F \copro_o \gen{X} \mathcal F^{-1},
\]
where $\mathcal F = \exp(i p/2\otimes \gen{B}_1^1)$ is a Reshetikhin twist. $p$ can be considered as an abstract central generator satisfying $\Delta p = p\otimes 1 + 1 \otimes p$. It can be considered as a formal logarithm of the generator $\eip$. $\gen{B}_1^1$ is one of the outer automorphisms of $\psucentral$, as defined in section \ref{sec:pslcentral}. Indeed, $\gen{B}_1^1$ acts only on the fermionic generators in a diagonal way, and can hence be easily seen to generate the coproduct \eqref{tab:twistedcopro}, \eqref{tab:coprotwistyang}.\\

We would now like to investigate the representation of the Yangian generators on the fundamental representation of $\psucentral$. We have already investigated the untwisted generators in the second realisation in section \ref{sec:drinf2central}. The generators of the first realisation can, in principle, be represented in the same way as for the unextended $\sutt$ algebra, namely as

\[
 \hat{\gen{X}} = u \gen{X},
\]
with some spectral parameter $u$. Similarly as in the case of the classical Lie bialgebra, as studied in section \ref{sec:adsbialgebra}, $u$ should be related to the momentum of the corresponding magnon, and hence be related to the generator $\eip$, or the related parameters $\xp{}, \xm{}$. The correct identification, found in \cite{Beisert:2007ds}, is given by

\[
 u = \frac{i g}{2}\frac{\xp{} + \xm{}}{1 + \frac{1}{\xp{}\xm{}}}.
\]
Up to an overall shift on all tensor factors, this is the only combination compatible with cocommutativity of the new central generators $\genY{P}$, $\genY{K}$. Hence, this identification is a necessary condition, should the Yangian be compatible with an S-matrix on two fundamental representations.\\

We are now in a position to fix the S-matrix on the fundamental representation. It is fully fixed up to a scalar factor by requiring

\[\label{eq:commSyang}
 \comm{S}{\Delta \gen{X}} = 0,
\]
for all generators $\gen{X}$ of the Yangian $\ypsucentral$. The answer is given in table \ref{tab:smatpsucentral}.

\begin{table}
\<
\smat_{12}\state{\phi_1^a\phi_2^b}
\eq
A_{12}\state{\phi_2^{\{a}\phi_1^{b\}}}
+B_{12}\state{\phi_2^{[a}\phi_1^{b]}}
+\half C_{12}\varepsilon^{ab}\varepsilon_{\alpha\beta}\state{\psi_2^\alpha\psi_1^\beta},
\nln
\smat_{12}\state{\psi_1^\alpha\psi_2^\beta}
\eq
D_{12}\state{\psi_2^{\{\alpha}\psi_1^{\beta\}}}
+E_{12}\state{\psi_2^{[\alpha}\psi_1^{\beta]}}
+\half F_{12}\varepsilon^{\alpha\beta}\varepsilon_{ab}\state{\phi_2^a\phi_1^b},
\nln
\smat_{12}\state{\phi_1^a\psi_2^\beta}
\eq
G_{12}\state{\psi_2^\beta\phi_1^{a}}
+H_{12}\state{\phi_2^{a}\psi_1^\beta},
\nln
\smat_{12}\state{\psi_1^\alpha\phi_2^b}
\eq
K_{12}\state{\psi_2^\alpha\phi_1^{b}}
+L_{12}\state{\phi_2^{b}\psi_1^\alpha}\nonumber.
\>
\<
A_{12}\eq S^0_{12}\,\frac{\xp{2}-\xm{1}}{\xm{2}-\xp{1}}\,,\nln
B_{12}\eq S^0_{12}\,\frac{\xp{2}-\xm{1}}{\xm{2}-\xp{1}}\lrbrk{1
          -2\,\frac{1-1/\xm{2}\xp{1}}{1-1/\xp{2}\xp{1}}\,\frac{\xm{2}-\xm{1}}{\xp{2}-\xm{1}}},\nln
C_{12}\eq S^0_{12}\,\frac{2\gamma_1\gamma_2}{\alpha \xp{1}\sqrt{\xp{2}\xm{2}}}\,
          \frac{1}{1-1/\xp{1}\xp{2}}\,\frac{\xm{2}-\xm{1}}{\xm{2}-\xp{1}}\,,\nln
D_{12}\eq -S^0_{12}\,\sqrt{\frac{\xm{1}\xp{2}}{\xp{1}\xm{2}}}\,,\nln
E_{12}\eq -S^0_{12}\,\sqrt{\frac{\xm{1}\xp{2}}{\xp{1}\xm{2}}}\lrbrk{1
          -2\,\frac{1-1/\xp{2}\xm{1}}{1-1/\xm{2}\xm{1}}\,\frac{\xp{2}-\xp{1}}{\xm{2}-\xp{1}}},\nln
F_{12}\eq -S^0_{12}\,\frac{2\alpha(\xp{1}-\xm{1})(\xp{2}-\xm{2})}{\gamma_1\gamma_2\sqrt{\xp{1}\xm{1}}\xm{2}}\,
          \frac{1}{1-1/\xm{1}\xm{2}}\,\frac{\xp{2}-\xp{1}}{\xm{2}-\xp{1}}\,,\nln
G_{12}\eq S^0_{12}\,\sqrt{\frac{\xm{1}}{\xp{1}}}\frac{\xp{2}-\xp{1}}{\xm{2}-\xp{1}}\,,\nln
H_{12}\eq S^0_{12}\,\sqrt{\frac{\xm{1}\xp{2}}{\xp{1}\xm{2}}}\frac{\gamma_1}{\gamma_2}\,\frac{\xp{2}-\xm{2}}{\xm{2}-\xp{1}}\,,\nln
K_{12}\eq S^0_{12}\,\frac{\gamma_2}{\gamma_1}\,\frac{\xp{1}-\xm{1}}{\xm{2}-\xp{1}}\,,\nln
L_{12}\eq S^0_{12}\,\sqrt{\frac{\xp{2}}{\xm{2}}}\,\frac{\xm{2}-\xm{1}}{\xm{2}-\xp{1}}\,.\nonumber
\>
\caption{All-loop S-matrix}
\label{tab:smatpsucentral}
\end{table}
Interestingly, this S-matrix is already fixed by invariance under the centrally extended Lie algebra $\psucentral$, without reference to the Yangian generators \cite{Beisert:2005tm}. The additional invariance under the Yangian generators was shown only later \cite{Beisert:2007ds}. This is in contrast to all other simple Lie superalgebras, where invariance under the Lie algebra generators constrains the S-matrix massively, but never fixes it. This fact is related to the distinguished properties of the representation theory of $\psucentral$, which was studied in \cite{Beisert:2006qh} (see also \cite{Gotz:2005ka} for the unextended case). Most notably, the tensor product of two fundamental representations is, for generic values of the central elements, irreducible. For simple Lie superalgebras, the tensor product usually decomposes into several irreducible components. However, in these cases, the representation of the corresponding Yangian will generically be irreducible. Hence, the complete S-matrix is usually fixed by imposing Yangian invariance on top of the usual Lie algebra invariance. Alternatively, one can solve the Yang-Baxter equation on top of the Lie algebra invariance. Note that the Yang-Baxter equation is defined on the triple tensor product, making the computation harder than using the Yangian invariance. The reason why solving the Yang-Baxter equation seems equivalent to solving the commutation relations with the Yangian is that the Yangian can be defined with the RTT relations.\\

As equation \eqref{eq:commSyang} is invariant under multiplication of $\smat$ with an arbitrary factor, the prefactor of $\smat$ needs to be fixed by crossing symmetry \cite{Janik:2006dc} and unitarity. This was done in \cite{Volin:2009uv}, confirming the earlier conjectured dressing factor of \cite{Beisert:2006ib,Beisert:2006ez}. We give the result of the dressing factor $\sigma^2$ in the integral form found in \cite{Dorey:2007xn}: 

\<
 \sigma^2 &=& \frac{R^2(\xp{1}\xp{2})R^2(\xm{1}\xm{2})}{R^2(\xp{1}\xm{2})R^2(\xm{1}\xp{2})}\nln
R^2 &=& \exp{2 i (\chi(x_1,x_2)-\chi(x_2,x_1))}\nln
\chi(x_1,x_2) &=& - i \oint \oint \frac{dz_1}{2\pi i}\frac{dz_2}{2\pi i}\frac{1}{x_1-z_1}\frac{1}{x_2-z_2}\log\Gamma(1 + i g (z_1+\frac{1}{z_1}-z_2-\frac{1}{z_2}))\nln
\>
The full $\psucentral$ invariant S-matrix is then completed by putting 

\[\label{def:factorofS}
 S^0_{12} = \sigma \sqrt{\frac{1-1/\xm{1}\xp{2}}{1-1/\xm{2}\xp{1}}\,\frac{\xm{2}-\xp{1}}{\xp{2}-\xm{1}}}.
\]
Recall that the full S-matrix of AdS/CFT is given by 

\<
\smat_{\psu\times\psucentral} = \smat_{\psucentral}\otimes\smat_{\psucentral},
\>
so the square root in \ref{def:factorofS} will not be troublesome.\\

{\bf Weak coupling Limit}

The S-matrix given in table \ref{tab:smatpsucentral} is a complete quantum S-matrix, depending on the coupling $g$. If we take $g\rightarrow 0$, we should recover the spin-chain S-matrix as studied in section \ref{sec:smatsym}. This S-matrix is not based on a classical r-matrix, but on the universal R-matrix of a full Yangian. To take this limit, it is important to keep in mind that the spectral parameters $\xp{},\xm{}$, and likewise the related momentum $p$, depend also on the coupling. To get the correct limit one has to rescale $\xp{},\xm{}$ with $1/g$, and then gets for the S-matrix\footnote{Here, we have also switched off the twist, working with the conventions of \cite{Beisert:2005tm}. This is the S-matrix in the spin-chain frame, see \cite{Arutyunov:2006yd,Arutyunov:2009ga} for discussions.}  

\[
 \smat_{12} = \frac{u}{u-i}\idm -\frac{i}{u-i}\perm
\]
in this limit, with $u=u_1-u_2$. Note that the S-matrix coming from the universal R-matrix of $\utt$ also has an extra prefactor, which, as discussed in section \ref{sec:smatsym}, might be related to some twists. \\

{\bf Strong coupling Limit}

To study the behaviour of the S-matrix in the strong coupling regime $g\rightarrow \infty$, it is useful to use the parametrisation 

\[
\xpm{}=x\sqrt{1-\frac{1}{4g^2(x-1/x)^2}}\pm \frac{i}{2g}\,\frac{x}{x-1/x},
\]
as found in \cite{Arutyunov:2006iu}. 

The parameter $\gamma$ appearing in the representation labels and the S-matrix scales as

\[
\gamma = \frac{1}{\sqrt{g}}\,\gat{},
\]
whereas $\alpha$ remains independent of $g$.\\

This limit corresponds to the near plane-wave regime \cite{Berenstein:2002jq,Parnachev:2002kk,Callan:2003xr,Minahan:2002ve}, or the classical limit of spinning strings \cite{Frolov:2003qc,Beisert:2003xu,Kruczenski:2003gt,Kazakov:2004qf}. The momentum $p$ scales as $\frac{1}{g}$ in this limit. \\

In this limit, the S-matrix behaves as \cite{Torrielli:2007mc}

\[
 \smat = \idm + \frac{1}{g}\crmat + O(\frac{1}{g^2}),
\]
with the classical r-matrix studied in section \ref{sec:adsbialgebra}. It can also be reproduced algebraically in a different way \cite{Moriyama:2007jt} than in section \ref{sec:adsbialgebra}. However, it was not shown that this proposal leads to a consistent quasi-triangular Lie bialgebra. Furthermore, only the classical r-matrix of section \ref{sec:adsbialgebra} leads to the correct classical limit of bound state S-matrices, as shown in \cite{deLeeuw:2008dp,Arutyunov:2009mi}. \\

In \cite{Arutyunov:2009mi}, also the full quantum S-matrix on higher representations was investigated. Indeed, the S-matrix for higher representations is not fixed by the Lie algebra symmetry alone, but only by Yangian invariance. This shows that for a complete physical picture one really needs the whole symmetry algebra, which is the Yangian. Note that the Yangian in this situation is the twisted Yangian $\ypsucentral$, where all three central elements are considered to be independent. On both classical limits $g\rightarrow\infty$, $g\rightarrow0$, effectively only one central charge survived, and the additional $\uone$ symmetry related to the external automorphism of the algebra appeared explicitly. As argued in \cite{Beisert:2005tm}, the two central charges $\gP$, $\gK$ are related to gauge transformations on the gauge side. However, it is at present not clear if one can consistently reduce them on the Yangian in an algebraically consistent way. The full quantum Yangian behind the all-loop S-matrix should be a mathematical quantisation of the Lie bialgebra of section \ref{sec:adsbialgebra}, and hence a deformed and twisted version of the Yangian $\mathcal{Y}(\utt)$. Another hint for this on the quantum level was found in \cite{Matsumoto:2007rh,Beisert:2007ty}, where it was shown that the Yangian generator corresponding to the external automorphism is also a symmetry of the all-loop S-matrix.\\

For these reasons, we believe that the all-loop S-matrix should be a representation of the universal R-matrix defined by 

\bea
\rmat &=& \fmat_{21}R_+ R_H R_-\fmat^{-1}\nln
R_+ &=& \prod_{i=1,\dots,6}^\rightarrow\prod_{k=0 }^\infty \exp(-(-1)^{|\beta|}\mathcal{F}^{|\beta|}\cwgp{\beta+k\delta}\otimes \cwgm{\beta-(k+1)\delta})\,,\nln
R_- &=& \prod_{i=1,\dots,6}^\leftarrow\prod_{k = 0}^\infty \exp(-\mathcal{F}^{|\beta|}\cwgm{\beta+k\delta}\otimes \cwgp{\beta-(k+1)\delta})\,,\nln
R_H &=& \prod_{i,j}^4\exp\left(\sum_{t=0}^\infty\left(\left(\frac{d}{d\lambda_1}\log(\gfhp{i}{1})\right)_{t}\otimes\left( D^{-1}_{ij}\log(\csgh{j}^-(\lambda_2)\right)_{-(t+1)}\right)\right)\,.\nln
\eea
The corresponding Chevalley-Serre basis is given in section \ref{sec:drinf2central}, and the operator $D_{ij}$ is related in the usual way to the Cartan matrix. Two important questions need to be answered before evaluating this formula for the universal R-matrix. First of all, the twist $\fmat$ should contain the part generating the braiding $\eip$, which is in principle done by putting

\[
 \fmat = e^{i p/2\otimes \gB{1}{1}}.
\]
It can also contain other combinations of the central charges, as discussed on the classical level in section \ref{sec:adsbialgebra}. $\gB{1}{1}$ has, a priori, no fundamental evaluation representation, so we propose to take the only $4\times 4$ matrix compatible with the inner product instead. This was given in \eqref{def:representauto}. Furthermore, the precise mechanism of inversion of the operator $D_{ij}$ needs to be given. There might be subtleties, as the analytic structure of the AdS/CFT S-matrix is far more complicated than the one of the S-matrices found in other integrable field theories, as discussed in chapter \ref{ch:integrablemodels}. As we observed, the construction of $R_H$ is similar to the explicit solution of the crossing equation, so we believe that the inversion of $D$ here should be done along the line of \cite{Volin:2009uv}, where an explicit solution of the AdS/CFT crossing equation was obtained.
\section{Conclusions and Outlook}

In this paper, we have investigated the application of Yangians to integrable models. In particular, we propose that the Yangian is a unified symmetry algebra for many integrable models, containing more information about physical models than widely appreciated. Most notably, the universal R-matrix of the Yangian was shown to lead to crossing invariant, unitary S-matrices for the known models with $\sun$ invariance. Likewise, the AdS/CFT S-matrix appearing in the Bethe Ansatz is invariant under the Yangian, so it is likely that it can also be completely reconstructed by purely algebraic means. The Yangian of the AdS/CFT S-matrix has some special features, as the underlying Lie superalgebra is centrally extended. Consequently, we studied the modifications necessary to generalise previous results concerning universal R-matrices of simple Lie algebras to superalgebras, and in particular, to the case of the centrally extended $\psucentral$ algebra. We reconstructed the AdS/CFT S-matrix in the strong and weak coupling regimes purely from the algebra, and also derived Drinfeld's second realisation of the full quantum Yangian. However, a rigorous derivation of the underlying quantum double remains an important task to be completed. The resulting universal R-matrix should also be related to the bound state S-matrices of \cite{deLeeuw:2008dp,Arutyunov:2009mi,Arutyunov:2009pw}. In this paper, we only studied the Yangian on the fundamental representation corresponding to fundamental magnons. As the Yangian is an abstract algebraic structure, choosing a different, suitable representation leads to the mirror theory, which is important for the derivation of the TBA equations. We refer the reader to \cite{Arutyunov:2009mi,Arutyunov:2009iq} for interesting discussions. Indeed, some subsectors of the quantum S-matrix were shown to be related to a universal R-matrix in \cite{Arutyunov:2009ce}, and in \cite{Torrielli:2008wi} the fundamental S-matrix was rewritten in a way resembling the structure of represented universal R-matrices, as well as the classical r-matrix of AdS/CFT discussed in this paper.\\

As the S-matrices of $\sun$ invariant models where reproduced purely from the Yangian, and we made progress towards such construction for the AdS/CFT S-matrix, we believe that similar constructions should work for most of the other integrable models as well. Indeed, many known integrable models are invariant under a Quantum Group, which is usually either a Yangian in the rational case, or a quantum affine algebra for trigonometric R-matrices. Another advantage of our approach is that as it is purely algebraic, we could generalise existing constructions straightforwardly e.g. to superalgebras. To our knowledge, a systematic study of crossing and unitary equations for superalgebras, in the line of thought of \cite{Berg:1977dp,Ogievetsky:1987vv} for ordinary Lie algebras, has not been done. One reason is that e.g. simply generalising theories like the principal chiral field from target space manifolds based on simple Lie groups to supergroups does not lead to theories with physical unitarity.

Integrable models with supergroup symmetry have been studied mainly in the condensed matter literature, see e.g. \cite{Saleur:1999cx}. In particular, in \cite{Bassi:1999ua,Saleur:2001cw,Saleur:2009bf} scattering matrices with $\alg{osp}(n|m)$ symmetry where investigated, and it was also found that the phase of the S-matrix just depends on the dual Coxeter number, similarly to the case of $\slnm$ R-matrices investigated in this paper. We believe that the universal R-matrices found in this paper are indeed in principle valid for any simple Lie superalgebra, so an important task would be to reproduce the $\alg{osp}(n|m)$ S-matrices from our universal R-matrix formula. However, the representation theory of Yangians based on Lie algebras other than $\slnm$ is considerably more involved, as there are no simple evaluation representations. Henceforth, even the evaluation of the rank one algebra $\alg{osp}(1|2)$ was quite involved \cite{Arnaudon:2003ab}. Our formalism for the derivation of the universal R-matrix was, however, different to what was known before in the literature \cite{Khoroshkin:1994uk,Arnaudon:2003ab}, as our resulting S-matrices are unitary. Furthermore, our formalism is valid for algebras with non-integer valued Cartan matrices. In particular, we believe that one can also find R-matrices for the exceptional Lie superalgebra $D(2,1;\alpha)$ from our formula. This algebra has some applications in relation with the AdS/CFT correspondence, as the centrally extended $\psucentral$ algebra can be obtained as a reduction from $D(2,1;\alpha)$ \cite{Serganova:1985aa, Beisert:2005tm}. In \cite{Matsumoto:2008ww}, an attempt has been made to also derive the Yangian and the classical r-matrix of $\psucentral$ from $D(2,1;\alpha)$.
Another situation where non-integer valued Cartan matrices can arise is the series of $\glnn$ algebras. These have a $\uone$ automorphism, which can be rescaled and shifted by the central charge. This seems to affect the dressing factor of the resulting S-matrices, as was observed in the case $\alg{gl}(1|1)$ studied in \cite{Rej:2010mu}.\\
As we mentioned, the S-matrix of AdS/CFT investigated in this paper is an internal scattering matrix scattering magnons of spin chains, which are related to single trace local operators. These operators are local in the four space-time dimensions of \symng. Proper gluon scattering amplitudes of \sym are, a priori, not related to our S-matrix. However, symmetry seems again a unifying scheme. Indeed, the original spin chain is $\sconf$ invariant, and this Lie superalgebra should be enhanced to the Yangian \cite{Dolan:2003uh,Dolan:2004ps}. Likewise, scattering amplitudes in \symng, which are by construction invariant under $\sconf$, were shown to be also invariant under a Yangian \cite{Drummond:2009fd,Drummond:2010qh,Drummond:2010uq,Beisert:2010gn}. This Yangian is related to the dual superconformal symmetry of the amplitudes \cite{Drummond:2008vq}. As some amplitudes are related to strings on twistor space \cite{Witten:2003nn}, it was recently confirmed that states in twistor string theory indeed form representations of a Yangian \cite{Corn:2010uj}. The gluon amplitudes seem structurally quite different to the scattering matrices consider in this paper. It would be very interesting to investigate if the Yangian symmetries can also restrict the amplitudes at higher loops. \\

The question of obtaining an all-loop result for amplitudes may hence be solved in a similar way as the S-matrix of the spectral problem\footnote{Fascinating progress in this direction has been made just when this paper was completed. In \cite{ArkaniHamed:2010kv}, all-loop recursion formulas were given for the integrand of scattering amplitudes.}. We have seen that the S-matrix for the spectral problem in AdS/CFT is a coupling dependent quantum S-matrix. 
It is invariant under the whole quantum Yangian, and the coupling dependence entered through the eigenvalues of the central charges. However, this does not at all constitute a proof of quantum integrability, but the derivation of the S-matrix is rather based on the assumption of quantum integrability. We note that even a rigorous derivation of classical integrability is not straightforward. The existence of infinitely many conserved charges follows from the existence of a flat current, as derived in \cite{Bena:2003wd}. Their Poisson commutativity was only shown in \cite{Magro:2008dv}, and the underlying classical symmetry algebra was studied in \cite{Vicedo:2009sn,Vicedo:2010qd}. The classical Poisson structure contains non-ultra local terms, i.e. derivatives of delta functions. Similar terms appear in the Poisson brackets of the principal chiral field or the $O(n)$ sigma model \cite{Maillet:1985fn,Maillet:1985ec,Maillet:1985ek}. This makes for instance the evaluation of commutation relations of the charges difficult \cite{Evans:1999mj,Magro:2008dv}. In the case of the principal chiral field, it was argued that the non-ultra local terms come from a bad choice of a classical limit of the well-defined quantum theory \cite{Faddeev:1985qu}. The existence of quantum conserved non-local charges was argued to exist in \cite{Luscher:1977uq} for the $O(n)$ model and in \cite{Abdalla:1986xb} for other nonlinear sigma models on symmetric spaces.\\

Recently, after convincing evidence that string theory on $AdS_4\times\mathbb{CP}^3$ is also integrable \cite{Minahan:2008hf,Arutyunov:2008if,Stefanski:2008ik} and exactly dual to $\mathcal N =6$ Chern-Simons theory in three dimensions \cite{Aharony:2008ug}, attempts have been made to classify space-time backgrounds on which string theory is integrable \cite{Zarembo:2010yz,Zarembo:2010sg}. Important for such integrability is the existence of a $\mathbb{Z}_4$ grading of the symmetry algebra. The arguments given so far for this classification are classical or based on one-loop computations. In the pure spinor formalism of the \ads superstring, quantum non-local charges were also argued to exist \cite{Berkovits:2004xu}, but no explicit computations were performed. Explicit calculations at one loop were done in \cite{Mikhailov:2007eg, Puletti:2008ym}, and confirm the integrability of the superstring at least in the pure spinor formalism. It is important to explicitly confirm the existence of infinitely many conserved charges in the full quantum theory, to prove integrability and henceforth rectify the use of the S-matrix formalism. Furthermore, one should, in principle, derive the same algebraic structures both from the Yang-Mills and the string side of the AdS/CFT correspondence. Henceforth, one should be able to fully verify that the TBA equations conjectured in \cite{Gromov:2009tv,Gromov:2009bc,Arutyunov:2009ur,Bombardelli:2009ns} correctly describe the spectrum of the AdS/CFT correspondence.\\

Another problem of the TBA equations, besides the lack of rigour in their derivation, is that they are hard to evaluate. This is because there are infinitely many coupled TBA equations, and the evaluation even in the case of the simplest operators is hard. In \cite{Gromov:2009tv}, the Y-system for the Konishi operator has been solved to four loops, confirming an earlier result obtained from the asymptotic Bethe ansatz complimented by Luscher corrections \cite{Bajnok:2008bm}. To this loop order, the result was also checked in perturbation theory \cite{Fiamberti:2007rj,Fiamberti:2008sh}. To five loops, Luscher corrections have been computed analytically \cite{Bajnok:2009vm}, and this result is in agreement with numerical investigations of the TBA equations \cite{Gromov:2009zb,Arutyunov:2010gb,Frolov:2010wt}. Furthermore, the Luscher corrections have been computed for the whole $\alg{sl}(2)$ sector \cite{Lukowski:2009ce}, and this result is compatible with expectations from the BFKL equation. Attempts have been made to find a dual string state for the Konishi operator \cite{Roiban:2009aa}. On the string side it is less clear when finite size corrections should become important. In \cite{Rej:2009dk} it was shown that already the next-to-leading order in the expansion of the dressing phase of the Bethe ansatz leads to inconsistencies in the analytic expansion. An important task for analytic studies as well as improved numerics is to find equations which are easier to handle. For other integrable models there are often integral equations which are equivalent to TBA equations, but have the property to be easier to handle. Most notably, for Sine-Gordon theory, there are the Destri-deVega equations \cite{Destri:1997yz}, or, more recently, integral equations for the principal chiral field with $\alg{su}(2)\times\alg{su}(2)$ symmetry have been found in \cite{Gromov:2008gj}. As the integral equations of the principal chiral field follow from the S-matrix, which we have shown follows directly from the Yangian, one might hope for a direct derivation of the integral equations from the Yangian. This is however currently far off. However, we note the appearance of q-deformed Cartan matrices in TBA equations \cite{Volin:2010cq}. We have shown that asymptotic S-matrices can be directly related to the Yangian. It would be interesting to investigate further the algebraic structures for finite size systems, and if they can also lead directly to interesting physical quantities.\\

Some of the difficulties of showing integrability of the AdS sigma model are related to the fact that in the light-cone gauge, the sigma model is not relativistically invariant. This makes it different to most other known integrable field theories. One solution to overcome this problem is to not use the light-cone gauge, but attempt some other covariant methods. One idea is to use the so-called Pohlmeyer reduction, which should reduce the original theory in such way that only physical degrees of freedom survive, and the theory remains relativistically invariant. The Pohlmeyer reduction was proposed in \cite{Grigoriev:2007bu} for the AdS superstring, see also \cite{Roiban:2009vh,Hoare:2009rq} for further progress in this direction. Interestingly, the leading S-matrix in the reduced theory \cite{Hoare:2009fs} looks quite similar to the classical limit \cite{Beisert:2010kk} of the q-deformed S-matrix \cite{Beisert:2008tw} of the spectral problem discussed in this paper. As this classical S-matrix is constructed from a Lie bialgebra, which can be considered as a deformation of the bialgebra discussed in this paper, again quantum groups or their classical limits seem to be responsible for the constraints on the S-matrix.\\

Besides seeking for a further closure of the gap between the mathematics of Yangians and their physical applications, an important task is to generalise the mathematics. Some important generalisations are of direct physical concern, such as a better understanding of the representation theory of Yangians of superalgebras, in particular the one based on the centrally extended $\psucentral$ algebra. The dependence of the central charges on each other in the quantum Yangian should have a similarly consistent interpretation as for the classical Lie bialgebra studied in this paper. Such identification might then change the representation theory of the Yangian. Furthermore, as we believe that the results concerning Yangians and their universal R-matrices are valid for all simple Lie superalgebras, one should explicitly show that they lead on representations to the scattering matrices found in \cite{Saleur:2001cw} for the case $\alg{osp}(n|m)$. One should also be able to construct R-and S-matrices based on the exceptional Lie superalgebras. Another generalisation is towards trigonometric solutions of the Yang-Baxter equation, whose symmetry is related to quantum affine algebras. Corresponding universal R-matrices have been constructed in \cite{Khoroshkin:1991ur}. Indeed, the quantum affine algebra is in several aspects easier to treat than the Yangian. In principle, one can also recover the Yangian in a special limit of the deformation parameter $q$. This is expected, as trigonometric S-matrices can be reduced to rational S-matrices. In some cases, such as for XXX spin chain, it is well understood what the q-deformation physically means. Basically, $q$ corresponds to an anisotropy parameter, which generalises the XXX spin chain to the XXZ spin chain. However, for other models, it is less clear what a q-deformation should mean. In the context of the AdS/CFT correspondence, q-deformation was related to Leigh-Strassler deformations \cite{Frolov:2005ty,Bundzik:2005zg,Mansson:2007sh,Mansson:2008xv}, where integrability is only preserved for special values of the deformation parameter. It is not clear if this is related to the above-mentioned q-deformed S-matrix of the spectral problem.\\ 

In this paper, we mainly discussed field theories defined on a cylinder, whose circumference is taken to infinity when considering the scattering problem. Yangian charges exist precisely for such decompactification limit. We note that there are also Yangian charges in theories with a boundary, such as the principal chiral model \cite{Delius:2001yi} or the AdS/CFT correspondence \cite{Ahn:2010xa} with inserted boundaries. The corresponding boundary Yang-Baxter equation yields solutions to the reflection matrices, which describe the reflection of a particle at the boundary. Hence, they act just on one representation space, and not on the tensor product, as the previously discussed two particle scattering matrices. The corresponding Yangian acting on the boundary is a twisted Yangian \cite{Mackay:2002bd}. It would be interesting if the boundary scattering matrices can also be reconstructed algebraically, in line of thought of this paper. As the universal R-matrix of quantum groups generically lives on the tensor product of two copies of the quantum group, one might have to consider a natural reduction of the R-matrix to one copy of the Yangian. There are natural ways involving the antipode as well as the multiplication map of the Hopf algebra, i.e. one might consider the element $\mu(S\otimes\idm)(\rmat)$. \\

The Yangian symmetries in the systems we were dealing with are well-defined in the infinite length limit of the underlying world-sheet or corresponding spin chain. The spin chain underlying \sym is long-ranged, as the interaction range grows in perturbation theory. We had to consider infinitely long chains to construct asymptotic scattering states, so for those states there is no problem of defining Yangian charges. Following the TBA logic, the Yangian of the mirror model should also be behind the finite size spectrum. It would be interesting to find out what algebraic structure is ultimately acting on finite size spectrum, and if it is ultimately a Yangian at all, or only becomes a Yangian for infinite length \cite{Vicedo:2010talk}. Note that curiously, for certain other long-ranged spin chains it was possible to define Yangian charges also for closed chains \cite{Haldane:1992sj,Hikami:1995yang}. Other long ranged spin chains with Yangian invariance were investigated in \cite{Beisert:2007jv}. It seems that as long as the long-ranged Hamiltonian preserves integrability, some Quantum Group symmetry also survives.\\ 

Symmetries have played a prominent role in physics in the last century, and it is appreciated that one needs to know the precise algebra relations. The algebra determines the representation theory, and hence the physics. In integrable models, many authors studied the underlying Lie symmetries and appreciated the importance of the existence of additional conserved charges. In this paper, we advocated that it is worth to know also the commutation relations of the additional conserved charges. In integrable models, this algebra is usually a Quantum Group. In particular, if the Quantum Group is a Yangian, as for the principal chiral field or the AdS/CFT correspondence at strong and weak coupling, we showed that one can reproduce the complete S-matrix just from the Yangian. Henceforth, we propose to study more deeply the symmetries involved in integrable models, as we argue they are more closely related to physics than is widely appreciated.

\section*{Acknowledgements}

The research reviewed in this paper was mainly done during my time as PhD student at Imperial College London, and this paper is based on the PhD thesis which I have submitted in August 2010. This research would not have been possible without the help and support of many people. In particular, I would like to thank my supervisor, Arkady Tseytlin for many discussions and the enjoyable and fruitful time while I was staying at Imperial College, and Jan Plefka for successful guidance during my undergraduate as well as early PhD stage. PArticular thanks also to Adam Rej for many fruitful discussions and collaboration during the later stage of my PhD. I also had very enjoyable collaborations with Niklas Beisert, Istvan Heckenberger, Peter Koroteev, Alessandro Torrielli and  Hiroyuki Yamane, and I am very grateful to all of them for the opportunity to work with them. \\

Furthermore, I would like to thank $\Omega$ Mekareeya, Adam Rej, Arkady Tseytlin and Benoit Vicedo for useful comments on the draft of this paper. I am also grateful to the referees of my PhD thesis, Patrick Dorey and Dan Waldram as well as the referees of Reviews in Mathematical Physics, for thoroughly reading the thesis, and making a large number of useful suggestions for improvements.\\

There are many more people with whom I had interesting discussions influencing the present work. It is impossible to list all of them, so I would like to thank the whole community for an enjoyable time in the last years.\\

I am also grateful to the Deutsche Telekom Stiftung for financial support for the research on which this paper is based.

\bibliographystyle{JHEP}
\bibliography{phdthesis}

\providecommand{\href}[2]{#2}\begingroup\raggedright\begin{thebibliography}{100}

\bibitem{Gross:1973id}
D.~J. Gross and F.~Wilczek, {\it {Ultraviolet behavior of non-abelian gauge
  theories}},  {\em Phys. Rev. Lett.} {\bf 30} (1973) 1343--1346.

\bibitem{Politzer:1973fx}
H.~D. Politzer, {\it {Reliable Perturbative Results for Strong Interactions?}},
   {\em Phys. Rev. Lett.} {\bf 30} (1973) 1346--1349.

\bibitem{Gross:1973ju}
D.~J. Gross and F.~Wilczek, {\it {Asymptotically Free Gauge Theories. 1}},
  {\em Phys. Rev.} {\bf D8} (1973) 3633--3652.

\bibitem{Gross:1974cs}
D.~J. Gross and F.~Wilczek, {\it Asymptotically free gauge theories. 2},  {\em
  Phys. Rev.} {\bf D9} (1974) 980--993.

\bibitem{Jansen:2008vs}
K.~Jansen, {\it {Lattice QCD: a critical status report}},  {\em PoS} {\bf
  LATTICE2008} (2008) 010, [\href{http://xxx.lanl.gov/abs/0810.5634}{{\tt
  arXiv:0810.5634}}].

\bibitem{Moch:2004pa}
S.~Moch, J.~A.~M. Vermaseren, and A.~Vogt, {\it {The three-loop splitting
  functions in QCD: The non-singlet case}},  {\em Nucl. Phys.} {\bf B688}
  (2004) 101--134, [\href{http://xxx.lanl.gov/abs/hep-ph/0403192}{{\tt
  hep-ph/0403192}}].

\bibitem{Vogt:2004mw}
A.~Vogt, S.~Moch, and J.~A.~M. Vermaseren, {\it {The three-loop splitting
  functions in QCD: The singlet case}},  {\em Nucl. Phys.} {\bf B691} (2004)
  129--181, [\href{http://xxx.lanl.gov/abs/hep-ph/0404111}{{\tt
  hep-ph/0404111}}].

\bibitem{'tHooft:1974jz}
G.~'t~Hooft, {\it A planar diagram theory for strong interactions},  {\em Nucl.
  Phys.} {\bf B72} (1974) 461.

\bibitem{Becker:2007zj}
K.~Becker, M.~Becker, and J.~H. Schwarz, {\it {String theory and M-theory: A
  modern introduction}}, . Cambridge, UK: Cambridge Univ. Pr. (2007) 739 p.

\bibitem{DiVecchia:2007we}
P.~Di~Vecchia and A.~Schwimmer, {\it {The beginning of string theory: a
  historical sketch}},  {\em Lect. Notes Phys.} {\bf 737} (2008) 119--136,
  [\href{http://xxx.lanl.gov/abs/0708.3940}{{\tt arXiv:0708.3940}}].

\bibitem{Schwarz:2007yc}
J.~H. Schwarz, {\it {The Early Years of String Theory: A Personal
  Perspective}},  \href{http://xxx.lanl.gov/abs/0708.1917}{{\tt
  arXiv:0708.1917}}.

\bibitem{Green:1984sg}
M.~B. Green and J.~H. Schwarz, {\it {Anomaly Cancellation in Supersymmetric
  D=10 Gauge Theory and Superstring Theory}},  {\em Phys. Lett.} {\bf B149}
  (1984) 117--122.

\bibitem{Maldacena:1998re}
J.~M. Maldacena, {\it The large n limit of superconformal field theories and
  supergravity},  {\em Adv. Theor. Math. Phys.} {\bf 2} (1998) 231--252,
  [\href{http://xxx.lanl.gov/abs/hep-th/9711200}{{\tt hep-th/9711200}}].

\bibitem{Gubser:1998bc}
S.~S. Gubser, I.~R. Klebanov, and A.~M. Polyakov, {\it Gauge theory correlators
  from non-critical string theory},  {\em Phys. Lett.} {\bf B428} (1998)
  105--114, [\href{http://xxx.lanl.gov/abs/hep-th/9802109}{{\tt
  hep-th/9802109}}].

\bibitem{Witten:1998qj}
E.~Witten, {\it Anti-de sitter space and holography},  {\em Adv. Theor. Math.
  Phys.} {\bf 2} (1998) 253--291,
  [\href{http://xxx.lanl.gov/abs/hep-th/9802150}{{\tt hep-th/9802150}}].

\bibitem{Metsaev:1998it}
R.~R. Metsaev and A.~A. Tseytlin, {\it {Type IIB superstring action in
  $AdS_5\times S^5$ background}},  {\em Nucl. Phys.} {\bf B533} (1998)
  109--126, [\href{http://xxx.lanl.gov/abs/hep-th/9805028}{{\tt
  hep-th/9805028}}].

\bibitem{Berenstein:2002jq}
D.~Berenstein, J.~M. Maldacena, and H.~Nastase, {\it Strings in flat space and
  pp waves from {$\mathcal{N}=\mathord{}$4} {Super} {Yang Mills}},  {\em JHEP}
  {\bf 0204} (2002) 013, [\href{http://xxx.lanl.gov/abs/hep-th/0202021}{{\tt
  hep-th/0202021}}].

\bibitem{Blau:2002dy}
M.~Blau, J.~Figueroa-O'Farrill, C.~Hull, and G.~Papadopoulos, {\it Penrose
  limits and maximal supersymmetry},  {\em Class. Quant. Grav.} {\bf 19} (2002)
  L87--L95, [\href{http://xxx.lanl.gov/abs/hep-th/0201081}{{\tt
  hep-th/0201081}}].

\bibitem{Metsaev:2001bj}
R.~R. Metsaev, {\it {Type IIB Green-Schwarz superstring in plane wave
  Ramond-Ramond background}},  {\em Nucl. Phys.} {\bf B625} (2002) 70--96,
  [\href{http://xxx.lanl.gov/abs/hep-th/0112044}{{\tt hep-th/0112044}}].

\bibitem{Metsaev:2002re}
R.~R. Metsaev and A.~A. Tseytlin, {\it Exactly solvable model of superstring in
  plane wave ramond-ramond background},  {\em Phys. Rev.} {\bf D65} (2002)
  126004, [\href{http://xxx.lanl.gov/abs/hep-th/0202109}{{\tt
  hep-th/0202109}}].

\bibitem{Gubser:2002tv}
S.~S. Gubser, I.~R. Klebanov, and A.~M. Polyakov, {\it {A semi-classical limit
  of the gauge/string correspondence}},  {\em Nucl. Phys.} {\bf B636} (2002)
  99--114, [\href{http://xxx.lanl.gov/abs/hep-th/0204051}{{\tt
  hep-th/0204051}}].

\bibitem{Frolov:2003qc}
S.~Frolov and A.~A. Tseytlin, {\it Multi-spin string solutions in {$AdS_5\times
  S^5$}},  {\em Nucl. Phys.} {\bf B668} (2003) 77--110,
  [\href{http://xxx.lanl.gov/abs/hep-th/0304255}{{\tt hep-th/0304255}}].

\bibitem{Tseytlin:2003ii}
A.~A. Tseytlin, {\it {Spinning strings and AdS/CFT duality}},
  \href{http://xxx.lanl.gov/abs/hep-th/0311139}{{\tt hep-th/0311139}}.

\bibitem{Plefka:2005bk}
J.~Plefka, {\it {Spinning strings and integrable spin chains in the AdS/CFT
  correspondence}},  {\em Living. Rev. Relativity} {\bf 8} (2005) 9,
  [\href{http://xxx.lanl.gov/abs/hep-th/0507136}{{\tt hep-th/0507136}}].

\bibitem{Beisert:2006ez}
N.~Beisert, B.~Eden, and M.~Staudacher, {\it Transcendentality and crossing},
  {\em J. Stat. Mech.} {\bf 07} (2007) P01021,
  [\href{http://xxx.lanl.gov/abs/hep-th/0610251}{{\tt hep-th/0610251}}].

\bibitem{Minahan:2002ve}
J.~A. Minahan and K.~Zarembo, {\it The bethe-ansatz for
  {$\mathcal{N}=\mathord{}$4} super yang-mills},  {\em JHEP} {\bf 0303} (2003)
  013, [\href{http://xxx.lanl.gov/abs/hep-th/0212208}{{\tt hep-th/0212208}}].

\bibitem{Beisert:2003yb}
N.~Beisert and M.~Staudacher, {\it The {$\mathcal{N}=\mathord{}$4} sym
  integrable super spin chain},  {\em Nucl. Phys.} {\bf B670} (2003) 439--463,
  [\href{http://xxx.lanl.gov/abs/hep-th/0307042}{{\tt hep-th/0307042}}].

\bibitem{Beisert:2003tq}
N.~Beisert, C.~Kristjansen, and M.~Staudacher, {\it The dilatation operator of
  {$\mathcal{N}=\mathord{}$4} conformal super yang-mills theory},  {\em Nucl.
  Phys.} {\bf B664} (2003) 131--184,
  [\href{http://xxx.lanl.gov/abs/hep-th/0303060}{{\tt hep-th/0303060}}].

\bibitem{Bena:2003wd}
I.~Bena, J.~Polchinski, and R.~Roiban, {\it Hidden symmetries of the
  {$AdS_5\times S^5$} superstring},  {\em Phys. Rev.} {\bf D69} (2004) 046002,
  [\href{http://xxx.lanl.gov/abs/hep-th/0305116}{{\tt hep-th/0305116}}].

\bibitem{Mandal:2002fs}
G.~Mandal, N.~V. Suryanarayana, and S.~R. Wadia, {\it {Aspects of semiclassical
  strings in AdS(5)}},  {\em Phys. Lett.} {\bf B543} (2002) 81--88,
  [\href{http://xxx.lanl.gov/abs/hep-th/0206103}{{\tt hep-th/0206103}}].

\bibitem{Dolan:2003uh}
L.~Dolan, C.~R. Nappi, and E.~Witten, {\it A relation between approaches to
  integrability in superconformal yang-mills theory},  {\em JHEP} {\bf 10}
  (2003) 017, [\href{http://xxx.lanl.gov/abs/hep-th/0308089}{{\tt
  hep-th/0308089}}].

\bibitem{Dolan:2004ps}
L.~Dolan, C.~R. Nappi, and E.~Witten, {\it Yangian symmetry in $d=$4
  superconformal yang-mills theory},
  \href{http://xxx.lanl.gov/abs/hep-th/0401243}{{\tt hep-th/0401243}}.

\bibitem{Arnold1995}
V.~I. {Arnol'd}, {\em {Mathematical methods of classical mechanics}}.
\newblock 1995.

\bibitem{Luscher:1977rq}
M.~L{\"u}scher and K.~Pohlmeyer, {\it Scattering of massless lumps and nonlocal
  charges in the two-dimensional classical nonlinear sigma model},  {\em Nucl.
  Phys.} {\bf B137} (1978) 46.

\bibitem{Luscher:1977uq}
M.~Luscher, {\it {Quantum Nonlocal Charges and Absence of Particle Production
  in the Two-Dimensional Nonlinear Sigma Model}},  {\em Nucl. Phys.} {\bf B135}
  (1978) 1--19.

\bibitem{Castillejo:1955ed}
L.~Castillejo, R.~H. Dalitz, and F.~J. Dyson, {\it Low's scattering equation
  for the charged and neutral scalar theories},  {\em Phys. Rev.} {\bf 101}
  (1956) 453--458.

\bibitem{Karowski:1977th}
M.~Karowski, H.~J. Thun, T.~T. Truong, and P.~H. Weisz, {\it {On the Uniqueness
  of a Purely Elastic s Matrix in (1+1)- Dimensions}},  {\em Phys. Lett.} {\bf
  B67} (1977) 321.

\bibitem{Zamolodchikov:1978xm}
A.~B. Zamolodchikov and A.~B. Zamolodchikov, {\it {Factorized S-matrices in two
  dimensions as the exact solutions of certain relativistic quantum field
  models}},  {\em Annals Phys.} {\bf 120} (1979) 253--291.

\bibitem{Iagolnitzer:1977sw}
D.~Iagolnitzer, {\it {Factorization of the Multiparticle s Matrix in Two-
  Dimensional Space-Time Models}},  {\em Phys. Rev.} {\bf D18} (1978) 1275.

\bibitem{Shankar:1977cm}
R.~Shankar and E.~Witten, {\it {The S Matrix of the Supersymmetric Nonlinear
  Sigma Model}},  {\em Phys. Rev.} {\bf D17} (1978) 2134.

\bibitem{Parke:1980ki}
S.~J. Parke, {\it {Absence of particle production and factorization of the
  S-matrix in (1+1) -dimensional models}},  {\em Nucl. Phys.} {\bf B174} (1980)
  166.

\bibitem{Polyakov:1983tt}
A.~M. Polyakov and P.~B. Wiegmann, {\it {Theory of nonabelian Goldstone bosons
  in two dimensions}},  {\em Phys. Lett.} {\bf B131} (1983) 121--126.

\bibitem{Wiegmann:1984pw}
P.~B. Wiegmann, {\it {On the theory of nonabelian Goldstone bosons in two
  dimensions: exact solution of the O(3) nonlinear sigma model}},  {\em Phys.
  Lett.} {\bf B141} (1984) 217.

\bibitem{Wiegmann:1984ec}
P.~Wiegmann, {\it {Exact factorized S-matrix of the chiral field in two
  dimensions}},  {\em Phys. Lett.} {\bf B142} (1984) 173--176.

\bibitem{MacKay:1992rc}
N.~J. MacKay, {\it {On the classical origins of Yangian symmetry in integrable
  field theory}},  {\em Phys. Lett.} {\bf B281} (1992) 90--97.

\bibitem{Abdalla:1986xb}
E.~Abdalla, M.~C.~B. Abdalla, and M.~Forger, {\it {Exact S-matrices for anomaly
  free nonlinear sigma models on symmetric spaces}},  {\em Nucl. Phys.} {\bf
  B297} (1988) 374.

\bibitem{Evans:1999mj}
J.~M. Evans, M.~Hassan, N.~J. MacKay, and A.~J. Mountain, {\it {Local conserved
  charges in principal chiral models}},  {\em Nucl. Phys.} {\bf B561} (1999)
  385--412, [\href{http://xxx.lanl.gov/abs/hep-th/9902008}{{\tt
  hep-th/9902008}}].

\bibitem{Drinfeld:1985rx}
V.~G. Drinfel'd, {\it Hopf algebras and the quantum yang-baxter equation},
  {\em Sov. Math. Dokl.} {\bf 32} (1985) 254--258.

\bibitem{Khoroshkin:1994uk}
S.~M. Khoroshkin and V.~N. Tolstoi, {\it {Yangian double and rational R
  matrix}},  \href{http://xxx.lanl.gov/abs/hep-th/9406194}{{\tt
  hep-th/9406194}}.

\bibitem{Faddeev:1990qg}
L.~D. {Faddeev}, {\it {Quantum Groups}},  in {\em Frontiers of Physics}
  ({E.~A.~Gotsman, Y.~Ne'Eman, \& A.~Voronel}, ed.), pp.~113--+, 1990.

\bibitem{Faddeev:1996iy}
L.~D. Faddeev, {\it {How Algebraic Bethe Ansatz works for integrable model}},
  \href{http://xxx.lanl.gov/abs/hep-th/9605187}{{\tt hep-th/9605187}}.

\bibitem{Jimbo:1985zk}
M.~Jimbo, {\it {A q difference analog of U(g) and the Yang-Baxter equation}},
  {\em Lett. Math. Phys.} {\bf 10} (1985) 63--69.

\bibitem{Agarwal:2004sz}
A.~Agarwal and S.~G. Rajeev, {\it Yangian symmetries of matrix models and spin
  chains: The dilatation operator of {$\mathcal{N}=\mathord{}$4} sym},  {\em
  Int. J. Mod. Phys.} {\bf A20} (2005) 5453--5490,
  [\href{http://xxx.lanl.gov/abs/hep-th/0409180}{{\tt hep-th/0409180}}].

\bibitem{Zwiebel:2006cb}
B.~I. Zwiebel, {\it {Yangian symmetry at two-loops for the $su(2|1)$ sector of
  N = 4 SYM}},  {\em J. Phys.} {\bf A40} (2007) 1141--1152,
  [\href{http://xxx.lanl.gov/abs/hep-th/0610283}{{\tt hep-th/0610283}}].

\bibitem{Staudacher:2004tk}
M.~Staudacher, {\it {The factorized S-matrix of CFT/AdS}},  {\em JHEP} {\bf 05}
  (2005) 054, [\href{http://xxx.lanl.gov/abs/hep-th/0412188}{{\tt
  hep-th/0412188}}].

\bibitem{Beisert:2005fw}
N.~Beisert and M.~Staudacher, {\it Long-range psu(2,2$/$4) bethe ansaetze for
  gauge theory and strings},  {\em Nucl. Phys.} {\bf B727} (2005) 1--62,
  [\href{http://xxx.lanl.gov/abs/hep-th/0504190}{{\tt hep-th/0504190}}].

\bibitem{Beisert:2005tm}
N.~Beisert, {\it {The $su(2|2)$ dynamic S-matrix}},  {\em Adv. Theor. Math.
  Phys.} {\bf 12} (2008) 945,
  [\href{http://xxx.lanl.gov/abs/hep-th/0511082}{{\tt hep-th/0511082}}].

\bibitem{Klose:2006zd}
T.~Klose, T.~McLoughlin, R.~Roiban, and K.~Zarembo, {\it {Worldsheet scattering
  in $AdS_5\times S^5$}},  {\em JHEP} {\bf 03} (2007) 094,
  [\href{http://xxx.lanl.gov/abs/hep-th/0611169}{{\tt hep-th/0611169}}].

\bibitem{Beisert:2007ds}
N.~Beisert, {\it {The S-Matrix of AdS/CFT and Yangian Symmetry}},  {\em PoS}
  {\bf Solvay} (2007) 002, [\href{http://xxx.lanl.gov/abs/0704.0400}{{\tt
  0704.0400}}].

\bibitem{Janik:2006dc}
R.~A. Janik, {\it {The $AdS_5\times S^5$ superstring worldsheet S-matrix and
  crossing symmetry}},  {\em Phys. Rev.} {\bf D73} (2006) 086006,
  [\href{http://xxx.lanl.gov/abs/hep-th/0603038}{{\tt hep-th/0603038}}].

\bibitem{Arutyunov:2004vx}
G.~Arutyunov, S.~Frolov, and M.~Staudacher, {\it Bethe ansatz for quantum
  strings},  {\em JHEP} {\bf 10} (2004) 016,
  [\href{http://xxx.lanl.gov/abs/hep-th/0406256}{{\tt hep-th/0406256}}].

\bibitem{Kazakov:2004qf}
V.~A. Kazakov, A.~Marshakov, J.~A. Minahan, and K.~Zarembo, {\it
  {Classical/quantum integrability in AdS/CFT}},  {\em JHEP} {\bf 05} (2004)
  024, [\href{http://xxx.lanl.gov/abs/hep-th/0402207}{{\tt hep-th/0402207}}].

\bibitem{Beisert:2005cw}
N.~Beisert and A.~A. Tseytlin, {\it {On Quantum Corrections to Spinning Strings
  and Bethe Equations}},  {\em Phys. Lett.} {\bf B629} (2005) 102--110,
  [\href{http://xxx.lanl.gov/abs/hep-th/0509084}{{\tt hep-th/0509084}}].

\bibitem{Hernandez:2006tk}
R.~Hern{\'a}ndez and E.~L{\'o}pez, {\it Quantum corrections to the string bethe
  ansatz},  {\em JHEP} {\bf 07} (2006) 004,
  [\href{http://xxx.lanl.gov/abs/hep-th/0603204}{{\tt hep-th/0603204}}].

\bibitem{Beisert:2006ib}
N.~Beisert, R.~Hern\'andez, and E.~L\'opez, {\it {A Crossing-Symmetric Phase
  for $AdS_5 \times S^5$ Strings}},  {\em JHEP} {\bf 11} (2006) 070,
  [\href{http://xxx.lanl.gov/abs/hep-th/0609044}{{\tt hep-th/0609044}}].

\bibitem{Volin:2009uv}
D.~Volin, {\it {Minimal solution of the AdS/CFT crossing equation}},  {\em J.
  Phys.} {\bf A42} (2009) 372001,
  [\href{http://xxx.lanl.gov/abs/0904.4929}{{\tt arXiv:0904.4929}}].

\bibitem{Frolov:2002av}
S.~Frolov and A.~A. Tseytlin, {\it Semiclassical quantization of rotating
  superstring in {$AdS_5 \times S^5$}},  {\em JHEP} {\bf 0206} (2002) 007,
  [\href{http://xxx.lanl.gov/abs/hep-th/0204226}{{\tt hep-th/0204226}}].

\bibitem{Kotikov:2007cy}
A.~V. Kotikov, L.~N. Lipatov, A.~Rej, M.~Staudacher, and V.~N. Velizhanin, {\it
  {Dressing and Wrapping}},  {\em J. Stat. Mech.} {\bf 0710} (2007) P10003,
  [\href{http://xxx.lanl.gov/abs/0704.3586}{{\tt arXiv:0704.3586}}].

\bibitem{Bajnok:2008bm}
Z.~Bajnok and R.~A. Janik, {\it {Four-loop perturbative Konishi from strings
  and finite size effects for multiparticle states}},  {\em Nucl. Phys.} {\bf
  B807} (2009) 625--650, [\href{http://xxx.lanl.gov/abs/0807.0399}{{\tt
  arXiv:0807.0399}}].

\bibitem{Janik:2007wt}
R.~A. Janik and T.~Lukowski, {\it {Wrapping interactions at strong coupling --
  the giant magnon}},  {\em Phys. Rev.} {\bf D76} (2007) 126008,
  [\href{http://xxx.lanl.gov/abs/0708.2208}{{\tt arXiv:0708.2208}}].

\bibitem{Hofman:2006xt}
D.~M. Hofman and J.~M. Maldacena, {\it Giant magnons},  {\em J. Phys.} {\bf
  A39} (2006) 13095--13118, [\href{http://xxx.lanl.gov/abs/hep-th/0604135}{{\tt
  hep-th/0604135}}].

\bibitem{Rej:2009dk}
A.~Rej and F.~Spill, {\it {Konishi at strong coupling from ABE}},  {\em J.
  Phys.} {\bf A42} (2009) 442003,
  [\href{http://xxx.lanl.gov/abs/0907.1919}{{\tt arXiv:0907.1919}}].

\bibitem{Zamolodchikov:1989cf}
A.~B. Zamolodchikov, {\it {Thermodynamic Bethe ansatz in relativistic models.
  scaling three state Potts and Lee-Yang models}},  {\em Nucl. Phys.} {\bf
  B342} (1990) 695--720.

\bibitem{Arutyunov:2007tc}
G.~Arutyunov and S.~Frolov, {\it {On String S-matrix, Bound States and TBA}},
  {\em JHEP} {\bf 12} (2007) 024,
  [\href{http://xxx.lanl.gov/abs/0710.1568}{{\tt arXiv:0710.1568}}].

\bibitem{Arutyunov:2009zu}
G.~Arutyunov and S.~Frolov, {\it {String hypothesis for the $AdS_5 x S^5$
  mirror}},  {\em JHEP} {\bf 03} (2009) 152,
  [\href{http://xxx.lanl.gov/abs/0901.1417}{{\tt arXiv:0901.1417}}].

\bibitem{Gromov:2009tv}
N.~Gromov, V.~Kazakov, and P.~Vieira, {\it {Exact Spectrum of Anomalous
  Dimensions of Planar N=4 Supersymmetric Yang-Mills Theory}},  {\em Phys. Rev.
  Lett.} {\bf 103} (2009) 131601,
  [\href{http://xxx.lanl.gov/abs/0901.3753}{{\tt arXiv:0901.3753}}].

\bibitem{Gromov:2009bc}
N.~Gromov, V.~Kazakov, A.~Kozak, and P.~Vieira, {\it {Exact Spectrum of
  Anomalous Dimensions of Planar N = 4 Supersymmetric Yang-Mills Theory: TBA
  and excited states}},  {\em Lett. Math. Phys.} {\bf 91} (2010) 265--287,
  [\href{http://xxx.lanl.gov/abs/0902.4458}{{\tt arXiv:0902.4458}}].

\bibitem{Arutyunov:2009ur}
G.~Arutyunov and S.~Frolov, {\it {Thermodynamic Bethe Ansatz for the ${AdS_5 x
  S^5}$ Mirror Model}},  {\em JHEP} {\bf 05} (2009) 068,
  [\href{http://xxx.lanl.gov/abs/0903.0141}{{\tt arXiv:0903.0141}}].

\bibitem{Bombardelli:2009ns}
D.~Bombardelli, D.~Fioravanti, and R.~Tateo, {\it {Thermodynamic Bethe Ansatz
  for planar AdS/CFT: a proposal}},  {\em J. Phys.} {\bf A42} (2009) 375401,
  [\href{http://xxx.lanl.gov/abs/0902.3930}{{\tt arXiv:0902.3930}}].

\bibitem{Dorey:1996re}
P.~Dorey and R.~Tateo, {\it {Excited states by analytic continuation of TBA
  equations}},  {\em Nucl. Phys.} {\bf B482} (1996) 639--659,
  [\href{http://xxx.lanl.gov/abs/hep-th/9607167}{{\tt hep-th/9607167}}].

\bibitem{Destri:1997yz}
C.~Destri and H.~J. de~Vega, {\it {Non-linear integral equation and
  excited-states scaling functions in the sine-Gordon model}},  {\em Nucl.
  Phys.} {\bf B504} (1997) 621--664,
  [\href{http://xxx.lanl.gov/abs/hep-th/9701107}{{\tt hep-th/9701107}}].

\bibitem{Gromov:2008gj}
N.~Gromov, V.~Kazakov, and P.~Vieira, {\it {Finite Volume Spectrum of 2D Field
  Theories from Hirota Dynamics}},  {\em JHEP} {\bf 12} (2009) 060,
  [\href{http://xxx.lanl.gov/abs/0812.5091}{{\tt arXiv:0812.5091}}].

\bibitem{Arutyunov:2009mi}
G.~Arutyunov, M.~de~Leeuw, and A.~Torrielli, {\it {The Bound State S-Matrix for
  $AdS_5 \times S^5$ Superstring}},  {\em Nucl. Phys.} {\bf B819} (2009)
  319--350, [\href{http://xxx.lanl.gov/abs/0902.0183}{{\tt arXiv:0902.0183}}].

\bibitem{Drummond:2009fd}
J.~M. Drummond, J.~M. Henn, and J.~Plefka, {\it {Yangian symmetry of scattering
  amplitudes in N=4 super Yang-Mills theory}},  {\em JHEP} {\bf 05} (2009) 046,
  [\href{http://xxx.lanl.gov/abs/0902.2987}{{\tt arXiv:0902.2987}}].

\bibitem{Drummond:2010qh}
J.~M. Drummond and L.~Ferro, {\it {Yangians, Grassmannians and T-duality}},
  {\em JHEP} {\bf 07} (2010) 027,
  [\href{http://xxx.lanl.gov/abs/1001.3348}{{\tt arXiv:1001.3348}}].

\bibitem{Drummond:2010uq}
J.~M. Drummond and L.~Ferro, {\it {The Yangian origin of the Grassmannian
  integral}},  \href{http://xxx.lanl.gov/abs/1002.4622}{{\tt arXiv:1002.4622}}.

\bibitem{Beisert:2010gn}
N.~Beisert, J.~Henn, T.~McLoughlin, and J.~Plefka, {\it {One-Loop
  Superconformal and Yangian Symmetries of Scattering Amplitudes in N=4 Super
  Yang-Mills}},  {\em JHEP} {\bf 04} (2010) 085,
  [\href{http://xxx.lanl.gov/abs/1002.1733}{{\tt arXiv:1002.1733}}].

\bibitem{Beisert:2010jq}
N.~Beisert, {\it {On Yangian Symmetry in Planar N=4 SYM}},
  \href{http://xxx.lanl.gov/abs/1004.5423}{{\tt arXiv:1004.5423}}.

\bibitem{Alday:2008yw}
L.~F. Alday and R.~Roiban, {\it {Scattering Amplitudes, Wilson Loops and the
  String/Gauge Theory Correspondence}},  {\em Phys. Rept.} {\bf 468} (2008)
  153--211, [\href{http://xxx.lanl.gov/abs/0807.1889}{{\tt arXiv:0807.1889}}].

\bibitem{Alday:2009zza}
L.~F. Alday, {\it {Scattering amplitudes and the AdS/CFT correspondence}},
  {\em J. Phys.} {\bf A42} (2009) 254006.

\bibitem{Rej:2010mu}
A.~Rej and F.~Spill, {\it {The Yangian of $sl(n|m)$ and the universal
  R-matrix}},  \href{http://xxx.lanl.gov/abs/1008.0872}{{\tt arXiv:1008.0872}}.

\bibitem{Spill:2008tp}
F.~Spill and A.~Torrielli, {\it {On Drinfeld's second realization of the
  AdS/CFT $su(2|2)$ Yangian}},  {\em J. Geom. Phys.} {\bf 59} (2009) 489--502,
  [\href{http://xxx.lanl.gov/abs/0803.3194}{{\tt arXiv:0803.3194}}].

\bibitem{Beisert:2007ty}
N.~Beisert and F.~Spill, {\it {The Classical r-matrix of AdS/CFT and its Lie
  Bialgebra Structure}},  {\em Commun. Math. Phys.} {\bf 285} (2009) 537--565,
  [\href{http://xxx.lanl.gov/abs/0708.1762}{{\tt arXiv:0708.1762}}].

\bibitem{Spill:2008yr}
F.~Spill, {\it {Weakly coupled $N=4$ Super Yang-Mills and N=6 Chern-Simons
  theories from $u(2|2)$ Yangian symmetry}},  {\em JHEP} {\bf 03} (2009) 014,
  [\href{http://xxx.lanl.gov/abs/0810.3897}{{\tt arXiv:0810.3897}}].

\bibitem{Molev:2002}
A.~Molev, {\it Yangians and their applications},  vol.~3 of {\em Handbook of
  Algebra}, pp.~907 -- 959.
\newblock North-Holland, 2003.
\newblock \href{http://xxx.lanl.gov/abs/math/0211288}{{\tt math/0211288}}.

\bibitem{Arutyunov:2009pw}
G.~Arutyunov, M.~de~Leeuw, and A.~Torrielli, {\it {On Yangian and Long
  Representations of the Centrally Extended $su(2|2)$ Superalgebra}},  {\em
  JHEP} {\bf 06} (2010) 033, [\href{http://xxx.lanl.gov/abs/0912.0209}{{\tt
  arXiv:0912.0209}}].

\bibitem{Fuchs:1997jv}
J.~Fuchs and C.~Schweigert, {\it {Symmetries, Lie algebras and representations:
  A graduate course for physicists}}, . Cambridge, UK: Univ. Pr. (1997) 438 p.

\bibitem{Kac:1977em}
V.~G. Kac, {\it {Lie Superalgebras}},  {\em Adv. Math.} {\bf 26} (1977) 8--96.

\bibitem{Kac:1977qb}
V.~G. Kac, {\it {A Sketch of Lie Superalgebra Theory}},  {\em Commun. Math.
  Phys.} {\bf 53} (1977) 31--64.

\bibitem{Chari:1994pz}
V.~Chari and A.~Pressley, {\em A guide to quantum groups}.
\newblock Cambridge University Press, Cambridge, UK, 1994.

\bibitem{Belavin:1982}
A.~A. Belavin and V.~G. Drinfeld, {\it Solutions of the classical
  {Y}ang-{B}axter equation for simple {L}ie algebras},  {\em Funktsional. Anal.
  i Prilozhen.} {\bf 16} (1982), no.~3 1--29, 96.

\bibitem{Kassel:1995xr}
C.~Kassel, {\it {Quantum groups}}, . New York, USA: Springer (1995) 531 p.
  (Graduate text in mathematics, 155).

\bibitem{Drinfeld:1987sy}
V.~G. Drinfel'd, {\it A new realization of yangians and quantized affine
  algebras},  {\em Sov. Math. Dokl.} {\bf 36} (1988) 212--216.

\bibitem{Gow:2007th}
L.~Gow, {\it {Yangians of Lie Superalgebras}},  {\em
  http://www.aei.mpg.de/~lucy/thesis.pdf} (2007).

\bibitem{Gow:2007aa}
L.~Gow, {\it {Gauss decomposition of the Yangian $Y(gl_{m|n})$}},  {\em Comm.
  Math. Phys. 276} (2007).

\bibitem{Stukopin:2005aa}
V.~Stukopin, {\it {Superalgebra A(m,n) and computation of Universal R-matrix}},
   \href{http://xxx.lanl.gov/abs/math/0504302}{{\tt math/0504302}}.

\bibitem{Beisert:2006qh}
N.~Beisert, {\it {The Analytic Bethe Ansatz for a Chain with Centrally Extended
  su(2$/$2) Symmetry}},  {\em J. Stat. Mech.} {\bf 07} (2007) P01017,
  [\href{http://xxx.lanl.gov/abs/nlin.SI/0}{{\tt arXiv:nlin.SI/0}}].

\bibitem{Berg:1977dp}
B.~Berg, M.~Karowski, P.~Weisz, and V.~Kurak, {\it {Factorized U(n) Symmetric s
  Matrices in Two-Dimensions}},  {\em Nucl. Phys.} {\bf B134} (1978) 125.

\bibitem{Bernard:1990jw}
D.~Bernard, {\it {Hidden Yangians in 2-D massive current algebras}},  {\em
  Commun. Math. Phys.} {\bf 137} (1991) 191--208.

\bibitem{Bernard:1992mu}
D.~Bernard and A.~Leclair, {\it {The Quantum double in integrable quantum field
  theory}},  {\em Nucl. Phys.} {\bf B399} (1993) 709--748,
  [\href{http://xxx.lanl.gov/abs/hep-th/9205064}{{\tt hep-th/9205064}}].

\bibitem{Volin:2010cq}
D.~Volin, {\it {Quantum integrability and functional equations}},
  \href{http://xxx.lanl.gov/abs/1003.4725}{{\tt arXiv:1003.4725}}.

\bibitem{Bethe:1931hc}
H.~Bethe, {\it Zur theorie der metalle i. eigenwerte und eigenfunktionen der
  linearen atomkette},  {\em Z. Phys.} {\bf 71} (1931) 205--226.

\bibitem{Yang:1967bm}
C.-N. Yang, {\it {Some exact results for the many body problems in one
  dimension with repulsive delta function interaction}},  {\em Phys. Rev.
  Lett.} {\bf 19} (1967) 1312--1314.

\bibitem{Sutherland:1975vr}
B.~Sutherland, {\it Model for a multicomponent quantum system},  {\em Phys.
  Rev.} {\bf B12} (1975) 3795--3805.

\bibitem{Nepomechie:1998jf}
R.~I. Nepomechie, {\it {A Spin Chain Primer}},  {\em Int. J. Mod. Phys.} {\bf
  B13} (1999) 2973--2986, [\href{http://xxx.lanl.gov/abs/hep-th/9810032}{{\tt
  hep-th/9810032}}].

\bibitem{Beisert:2004ry}
N.~Beisert, {\it The dilatation operator of {$\mathcal{N}=\mathord{}$4} super
  yang-mills theory and integrability},  {\em Phys. Rept.} {\bf 405} (2004)
  1--202, [\href{http://xxx.lanl.gov/abs/hep-th/0407277}{{\tt
  hep-th/0407277}}].

\bibitem{Dorey:1996gd}
P.~Dorey, {\it {Exact S matrices}},
  \href{http://xxx.lanl.gov/abs/hep-th/9810026}{{\tt hep-th/9810026}}.

\bibitem{MacKay:2004tc}
N.~J. MacKay, {\it {Introduction to Yangian symmetry in integrable field
  theory}},  {\em Int. J. Mod. Phys.} {\bf A20} (2005) 7189--7218,
  [\href{http://xxx.lanl.gov/abs/hep-th/0409183}{{\tt hep-th/0409183}}].

\bibitem{Babelon:2003book}
O.~{Babelon}, D.~{Bernard}, and M.~{Talon}, {\em {Introduction to Classical
  Integrable Systems}}.
\newblock June, 2003.

\bibitem{Goldschmidt:1980wq}
Y.~Y. Goldschmidt and E.~Witten, {\it {Conservation laws in some
  two-dimensional models}},  {\em Phys. Lett.} {\bf B91} (1980) 392.

\bibitem{Maillet:1985fn}
J.~M. Maillet, {\it {Kac-Moody algebra and extended Yang-Baxter relations in
  the O(N) nonlinear sigma model}},  {\em Phys. Lett.} {\bf B162} (1985) 137.

\bibitem{Maillet:1985ec}
J.~M. Maillet, {\it {Hamiltonian structures for integrable classical theories
  from graded Kac-Moody algebras}},  {\em Phys. Lett.} {\bf B167} (1986) 401.

\bibitem{Maillet:1985ek}
J.~M. Maillet, {\it {New integrable canonical structures in two-dimensional
  models}},  {\em Nucl. Phys.} {\bf B269} (1986) 54.

\bibitem{Abdalla:1984iq}
E.~Abdalla, M.~C.~B. Abdalla, and A.~Lima-Santos, {\it {On the exact solution
  of the principal chiral model}},  {\em Phys. Lett.} {\bf B140} (1984) 71--75.

\bibitem{Fateev:1994ai}
V.~A. Fateev, V.~A. Kazakov, and P.~B. Wiegmann, {\it {Principal chiral field
  at large N}},  {\em Nucl. Phys.} {\bf B424} (1994) 505--520,
  [\href{http://xxx.lanl.gov/abs/hep-th/9403099}{{\tt hep-th/9403099}}].

\bibitem{Fateev:1994dp}
V.~A. Fateev, P.~B. Wiegmann, and V.~A. Kazakov, {\it {Large N chiral field in
  two-dimensions}},  {\em Phys. Rev. Lett.} {\bf 73} (1994) 1750--1753.

\bibitem{Karbach:1998ba1}
M.~{Karbach} and G.~{M{\"u}ller}, {\it {Introduction to the Bethe ansatz I}},
  {\em ArXiv Condensed Matter e-prints} (Sept., 1998)
  [\href{http://xxx.lanl.gov/abs/cond-mat/}{{\tt cond-mat/}}].

\bibitem{Karbach:1998ba2}
M.~{Karbach}, K.~{Hu}, and G.~{M{\"u}ller}, {\it {Introduction to the Bethe
  Ansatz II}},  {\em Computers in Physics} {\bf 12} (Nov., 1998) 565--573,
  [\href{http://xxx.lanl.gov/abs/cond-mat/}{{\tt cond-mat/}}].

\bibitem{Kirillov:1985ab}
A.~Kirillov, {\it Combinatorial identities, and completeness of eigenstates of
  the heisenberg magnet},  {\em Journal of Mathematical Sciences} {\bf 30}
  (1985) 2298--2310. 10.1007/BF02105347.

\bibitem{Faddeev:1984aa}
L.~D. Faddeev and L.~A. Takhtadzhyan, {\it Spectrum and scattering of
  excitations in the one-dimensional isotropic heisenberg model},  {\em Journal
  of Mathematical Sciences} {\bf 24} (1984) 241--267. 10.1007/BF01087245.

\bibitem{'tHooft:1993gx}
G.~'t~Hooft, {\it {Dimensional reduction in quantum gravity}},
  \href{http://xxx.lanl.gov/abs/gr-qc/9310026}{{\tt gr-qc/9310026}}.

\bibitem{Susskind:1994vu}
L.~Susskind, {\it {The World as a hologram}},  {\em J. Math. Phys.} {\bf 36}
  (1995) 6377--6396, [\href{http://xxx.lanl.gov/abs/hep-th/9409089}{{\tt
  hep-th/9409089}}].

\bibitem{Aharony:2008ug}
O.~Aharony, O.~Bergman, D.~L. Jafferis, and J.~Maldacena, {\it {N=6
  superconformal Chern-Simons-matter theories, M2-branes and their gravity
  duals}},  \href{http://xxx.lanl.gov/abs/0806.1218}{{\tt arXiv:0806.1218}}.

\bibitem{Minahan:2008hf}
J.~A. Minahan and K.~Zarembo, {\it {The Bethe ansatz for superconformal
  Chern-Simons}},  {\em JHEP} {\bf 09} (2008) 040,
  [\href{http://xxx.lanl.gov/abs/0806.3951}{{\tt arXiv:0806.3951}}].

\bibitem{Arutyunov:2008if}
G.~Arutyunov and S.~Frolov, {\it {Superstrings on $AdS_4 x CP^3$ as a Coset
  Sigma-model}},  {\em JHEP} {\bf 09} (2008) 129,
  [\href{http://xxx.lanl.gov/abs/0806.4940}{{\tt arXiv:0806.4940}}].

\bibitem{Stefanski:2008ik}
B.~Stefanski, jr, {\it {Green-Schwarz action for Type IIA strings on
  $AdS_4\times CP^3$}},  {\em Nucl. Phys.} {\bf B808} (2009) 80--87,
  [\href{http://xxx.lanl.gov/abs/0806.4948}{{\tt arXiv:0806.4948}}].

\bibitem{Aharony:1999ti}
O.~Aharony, S.~S. Gubser, J.~M. Maldacena, H.~Ooguri, and Y.~Oz, {\it Large n
  field theories, string theory and gravity},  {\em Phys. Rept.} {\bf 323}
  (2000) 183--386, [\href{http://xxx.lanl.gov/abs/hep-th/9905111}{{\tt
  hep-th/9905111}}].

\bibitem{Green:1987sp}
M.~B. Green, J.~H. Schwarz, and E.~Witten, {\it {Superstring Theory. Vol. 1:
  Introduction}}, . Cambridge, Uk: Univ. Pr. ( 1987) 469 P. ( Cambridge
  Monographs On Mathematical Physics).

\bibitem{Polchinski:1998rq}
J.~Polchinski, {\it {String theory. Vol. 1: An introduction to the bosonic
  string}}, . Cambridge, UK: Univ. Pr. (1998) 402 p.

\bibitem{Polchinski:1998rr}
J.~Polchinski, {\it {String theory. Vol. 2: Superstring theory and beyond}}, .
  Cambridge, UK: Univ. Pr. (1998) 531 p.

\bibitem{Dorey:2009zz}
N.~Dorey, {\it {Notes on integrability in gauge theory and string theory}},
  {\em J. Phys.} {\bf A42} (2009) 254001.

\bibitem{Arutyunov:2009ga}
G.~Arutyunov and S.~Frolov, {\it {Foundations of the ${AdS_5 x S^5}$
  Superstring. Part I}},  {\em J. Phys.} {\bf A42} (2009) 254003,
  [\href{http://xxx.lanl.gov/abs/0901.4937}{{\tt arXiv:0901.4937}}].

\bibitem{Rej:2009je}
A.~Rej, {\it {Integrability and the AdS/CFT correspondence}},  {\em J. Phys.}
  {\bf A42} (2009) 254002, [\href{http://xxx.lanl.gov/abs/0907.3468}{{\tt
  arXiv:0907.3468}}].

\bibitem{Sohnius:1981sn}
M.~F. Sohnius and P.~C. West, {\it Conformal invariance in
  {$\mathcal{N}=\mathord{}$4} supersymmetric yang-mills theory},  {\em Phys.
  Lett.} {\bf B100} (1981) 245.

\bibitem{Magro:2008dv}
M.~Magro, {\it {The Classical Exchange Algebra of $AdS_5 \times S^5$}},  {\em
  JHEP} {\bf 01} (2009) 021, [\href{http://xxx.lanl.gov/abs/0810.4136}{{\tt
  arXiv:0810.4136}}].

\bibitem{Vicedo:2009sn}
B.~Vicedo, {\it {Hamiltonian dynamics and the hidden symmetries of the $AdS_5 x
  S^5$ superstring}},  {\em JHEP} {\bf 01} (2010) 102,
  [\href{http://xxx.lanl.gov/abs/0910.0221}{{\tt arXiv:0910.0221}}].

\bibitem{Vicedo:2010qd}
B.~Vicedo, {\it {The classical R-matrix of AdS/CFT and its Lie dialgebra
  structure}},  \href{http://xxx.lanl.gov/abs/1003.1192}{{\tt
  arXiv:1003.1192}}.

\bibitem{Lipatov:1993yb}
L.~N. Lipatov, {\it {High-energy asymptotics of multicolor QCD and exactly
  solvable lattice models}},
  \href{http://xxx.lanl.gov/abs/hep-th/9311037}{{\tt hep-th/9311037}}.

\bibitem{Faddeev:1994zg}
L.~D. Faddeev and G.~P. Korchemsky, {\it {High-energy QCD as a completely
  integrable model}},  {\em Phys. Lett.} {\bf B342} (1995) 311--322,
  [\href{http://xxx.lanl.gov/abs/hep-th/9404173}{{\tt hep-th/9404173}}].

\bibitem{Beisert:2003jj}
N.~Beisert, {\it The complete one-loop dilatation operator of
  {$\mathcal{N}=\mathord{}$4} super yang-mills theory},  {\em Nucl. Phys.} {\bf
  B676} (2004) 3, [\href{http://xxx.lanl.gov/abs/hep-th/0307015}{{\tt
  hep-th/0307015}}].

\bibitem{Dorey:2007xn}
N.~Dorey, D.~M. Hofman, and J.~Maldacena, {\it On the singularities of the
  magnon s-matrix},  {\em Phys. Rev.} {\bf D76} (2007) 025011,
  [\href{http://xxx.lanl.gov/abs/hep-th/0703104}{{\tt hep-th/0703104}}].

\bibitem{Arutyunov:2009kf}
G.~Arutyunov and S.~Frolov, {\it {The Dressing Factor and Crossing Equations}},
   {\em J.Phys.A} {\bf A42} (2009) 425401,
  [\href{http://xxx.lanl.gov/abs/0904.4575}{{\tt arXiv:0904.4575}}].

\bibitem{Gomez:2006va}
C.~Gomez and R.~Hern\'andez, {\it {The magnon kinematics of the AdS/CFT
  correspondence}},  {\em JHEP} {\bf 11} (2006) 021,
  [\href{http://xxx.lanl.gov/abs/hep-th/0608029}{{\tt hep-th/0608029}}].

\bibitem{Plefka:2006ze}
J.~Plefka, F.~Spill, and A.~Torrielli, {\it {On the Hopf algebra structure of
  the AdS/CFT S-matrix}},  {\em Phys. Rev.} {\bf D74} (2006) 066008,
  [\href{http://xxx.lanl.gov/abs/hep-th/0608038}{{\tt hep-th/0608038}}].

\bibitem{Gomez:2007zr}
C.~Gomez and R.~Hern\'andez, {\it {Quantum deformed magnon kinematics}},  {\em
  JHEP} {\bf 03} (2007) 108,
  [\href{http://xxx.lanl.gov/abs/hep-th/0701200}{{\tt hep-th/0701200}}].

\bibitem{Young:2007wd}
C.~A.~S. Young, {\it q-deformed supersymmetry and dynamic magnon
  representations},  {\em J. Phys.} {\bf A40} (2007) 9165--9176,
  [\href{http://xxx.lanl.gov/abs/0704.2069}{{\tt 0704.2069}}].

\bibitem{Ambjorn:2005wa}
J.~Ambj{\o}rn, R.~A. Janik, and C.~Kristjansen, {\it Wrapping interactions and
  a new source of corrections to the spin-chain / string duality},  {\em Nucl.
  Phys.} {\bf B736} (2006) 288--301,
  [\href{http://xxx.lanl.gov/abs/hep-th/0510171}{{\tt hep-th/0510171}}].

\bibitem{Arutyunov:2006ak}
G.~Arutyunov, S.~Frolov, J.~Plefka, and M.~Zamaklar, {\it {The Off-shell
  Symmetry Algebra of the Light-cone $AdS_5\times S^5$ Superstring}},  {\em J.
  Phys.} {\bf A40} (2007) 3583--3606,
  [\href{http://xxx.lanl.gov/abs/hep-th/0609157}{{\tt hep-th/0609157}}].

\bibitem{Klose:2007rz}
T.~Klose, T.~McLoughlin, J.~A. Minahan, and K.~Zarembo, {\it {World-sheet
  scattering in $AdS_5\times S^5$ at two loops}},  {\em JHEP} {\bf 08} (2007)
  051, [\href{http://xxx.lanl.gov/abs/0704.3891}{{\tt 0704.3891}}].

\bibitem{Berkovits:2004jw}
N.~Berkovits, {\it Brst cohomology and nonlocal conserved charges},  {\em JHEP}
  {\bf 02} (2005) 060, [\href{http://xxx.lanl.gov/abs/hep-th/0409159}{{\tt
  hep-th/0409159}}].

\bibitem{Beisert:2004hm}
N.~Beisert, V.~Dippel, and M.~Staudacher, {\it A novel long range spin chain
  and planar {$\mathcal{N}=\mathord{}$4} super yang-mills},  {\em JHEP} {\bf
  07} (2004) 075, [\href{http://xxx.lanl.gov/abs/hep-th/0405001}{{\tt
  hep-th/0405001}}].

\bibitem{Arutyunov:2006iu}
G.~Arutyunov and S.~Frolov, {\it {On $AdS_5\times S^5$ string S-matrix}},  {\em
  Phys. Lett.} {\bf B639} (2006) 378--382,
  [\href{http://xxx.lanl.gov/abs/hep-th/0604043}{{\tt hep-th/0604043}}].

\bibitem{Reshetikhin:1990ep}
N.~Reshetikhin, {\it Multiparameter quantum groups and twisted quasitriangular
  hopf algebras},  {\em Lett. Math. Phys.} {\bf 20} (1990) 331--335.

\bibitem{Frolov:2006cc}
S.~Frolov, J.~Plefka, and M.~Zamaklar, {\it {The $AdS_5\times S^5$ superstring
  in light-cone gauge and its Bethe equations}},  {\em J. Phys.} {\bf A39}
  (2006) 13037--13082, [\href{http://xxx.lanl.gov/abs/hep-th/0603008}{{\tt
  hep-th/0603008}}].

\bibitem{Arutyunov:2006yd}
G.~Arutyunov, S.~Frolov, and M.~Zamaklar, {\it {The Zamolodchikov-Faddeev
  algebra for $AdS_5\times S^5$ superstring}},  {\em JHEP} {\bf 04} (2007) 002,
  [\href{http://xxx.lanl.gov/abs/hep-th/0612229}{{\tt hep-th/0612229}}].

\bibitem{Beisert:2003ys}
N.~Beisert, {\it The $su(2|3)$ dynamic spin chain},  {\em Nucl. Phys.} {\bf
  B682} (2004) 487--520, [\href{http://xxx.lanl.gov/abs/hep-th/0310252}{{\tt
  hep-th/0310252}}].

\bibitem{Ahn:2008aa}
C.~Ahn and R.~I. Nepomechie, {\it {N=6 super Chern-Simons theory S-matrix and
  all-loop Bethe ansatz equations}},  {\em JHEP} {\bf 09} (2008) 010,
  [\href{http://xxx.lanl.gov/abs/0807.1924}{{\tt arXiv:0807.1924}}].

\bibitem{Gotz:2005ka}
G.~G{\"o}tz, T.~Quella, and V.~Schomerus, {\it Tensor products of psl(2$/$2)
  representations},  \href{http://xxx.lanl.gov/abs/hep-th/0506072}{{\tt
  hep-th/0506072}}.

\bibitem{Parnachev:2002kk}
A.~Parnachev and A.~V. Ryzhov, {\it {Strings in the near plane wave background
  and AdS/CFT}},  {\em JHEP} {\bf 10} (2002) 066,
  [\href{http://xxx.lanl.gov/abs/hep-th/0208010}{{\tt hep-th/0208010}}].

\bibitem{Callan:2003xr}
C.~G. Callan, Jr., H.~K. Lee, T.~McLoughlin, J.~H. Schwarz, I.~Swanson, and
  X.~Wu, {\it {Quantizing string theory in $AdS_5\times S^5$: Beyond the
  pp-wave}},  {\em Nucl. Phys.} {\bf B673} (2003) 3--40,
  [\href{http://xxx.lanl.gov/abs/hep-th/0307032}{{\tt hep-th/0307032}}].

\bibitem{Beisert:2003xu}
N.~Beisert, J.~A. Minahan, M.~Staudacher, and K.~Zarembo, {\it Stringing spins
  and spinning strings},  {\em JHEP} {\bf 09} (2003) 010,
  [\href{http://xxx.lanl.gov/abs/hep-th/0306139}{{\tt hep-th/0306139}}].

\bibitem{Kruczenski:2003gt}
M.~Kruczenski, {\it Spin chains and string theory},  {\em Phys. Rev. Lett.}
  {\bf 93} (2004) 161602, [\href{http://xxx.lanl.gov/abs/hep-th/0311203}{{\tt
  hep-th/0311203}}].

\bibitem{Torrielli:2007mc}
A.~Torrielli, {\it Classical r-matrix of the su(2$/$2) sym spin-chain},  {\em
  Phys. Rev.} {\bf D75} (2007) 105020,
  [\href{http://xxx.lanl.gov/abs/hep-th/0701281}{{\tt hep-th/0701281}}].

\bibitem{Moriyama:2007jt}
S.~Moriyama and A.~Torrielli, {\it {A Yangian Double for the AdS/CFT Classical
  r-matrix}},  {\em JHEP} {\bf 06} (2007) 083,
  [\href{http://xxx.lanl.gov/abs/0706.0884}{{\tt 0706.0884}}].

\bibitem{deLeeuw:2008dp}
M.~de~Leeuw, {\it {Bound States, Yangian Symmetry and Classical r-matrix for
  the AdS5 x S5 Superstring}},  {\em JHEP} {\bf 06} (2008) 085,
  [\href{http://xxx.lanl.gov/abs/0804.1047}{{\tt arXiv:0804.1047}}].

\bibitem{Matsumoto:2007rh}
T.~Matsumoto, S.~Moriyama, and A.~Torrielli, {\it {A Secret Symmetry of the
  AdS/CFT S-matrix}},  {\em JHEP} {\bf 09} (2007) 099,
  [\href{http://xxx.lanl.gov/abs/0708.1285}{{\tt 0708.1285}}].

\bibitem{Arutyunov:2009iq}
G.~Arutyunov, M.~de~Leeuw, R.~Suzuki, and A.~Torrielli, {\it {Bound State
  Transfer Matrix for AdS5 x S5 Superstring}},  {\em JHEP} {\bf 10} (2009) 025,
  [\href{http://xxx.lanl.gov/abs/0906.4783}{{\tt arXiv:0906.4783}}].

\bibitem{Arutyunov:2009ce}
G.~Arutyunov, M.~de~Leeuw, and A.~Torrielli, {\it {Universal blocks of the
  AdS/CFT Scattering Matrix}},  {\em JHEP} {\bf 05} (2009) 086,
  [\href{http://xxx.lanl.gov/abs/0903.1833}{{\tt arXiv:0903.1833}}].

\bibitem{Torrielli:2008wi}
A.~Torrielli, {\it {Structure of the string R-matrix}},  {\em J. Phys.} {\bf
  A42} (2009) 055204, [\href{http://xxx.lanl.gov/abs/0806.1299}{{\tt
  arXiv:0806.1299}}].

\bibitem{Ogievetsky:1987vv}
E.~Ogievetsky, P.~Wiegmann, and N.~Reshetikhin, {\it {The principal chiral
  field in two-dimensions on classical Lie algebras: The Bethe ansatz solution
  and factorized theory of scattering}},  {\em Nucl. Phys.} {\bf B280} (1987)
  45--96.

\bibitem{Saleur:1999cx}
H.~Saleur, {\it The continuum limit of $sl(n/k)$ integrable super spin chains},
   {\em Nucl. Phys.} {\bf B578} (2000) 552--576,
  [\href{http://xxx.lanl.gov/abs/solv-int/9905007}{{\tt solv-int/9905007}}].

\bibitem{Bassi:1999ua}
Z.~S. Bassi and A.~LeClair, {\it {The exact S-matrix for an $osp(2|2)$
  disordered system}},  {\em Nucl. Phys.} {\bf B578} (2000) 577--627,
  [\href{http://xxx.lanl.gov/abs/hep-th/9911105}{{\tt hep-th/9911105}}].

\bibitem{Saleur:2001cw}
H.~Saleur and B.~Wehefritz-Kaufmann, {\it {Integrable quantum field theories
  with OSP(m/2n) symmetries}},  {\em Nucl. Phys.} {\bf B628} (2002) 407--441,
  [\href{http://xxx.lanl.gov/abs/hep-th/0112095}{{\tt hep-th/0112095}}].

\bibitem{Saleur:2009bf}
H.~Saleur and B.~Pozsgay, {\it {Scattering and duality in the 2 dimensional
  $OSP(2|2)$ Gross Neveu and sigma models}},  {\em JHEP} {\bf 02} (2010) 008,
  [\href{http://xxx.lanl.gov/abs/0910.0637}{{\tt arXiv:0910.0637}}].

\bibitem{Arnaudon:2003ab}
D.~{Arnaudon}, N.~{Cramp{\'e}}, L.~{Frappat}, and E.~{Ragoucy}, {\it {Super
  Yangian $Y(osp(1|2))$ and the Universal R-matrix of Its Quantum Double}},
  {\em Communications in Mathematical Physics} {\bf 240} (2003) 31--51,
  [\href{http://xxx.lanl.gov/abs/math/0209167}{{\tt math/0209167}}].

\bibitem{Serganova:1985aa}
V.~V. Serganova, {\it Automorphisms of simple lie superalgebras},  {\em Math.
  USSR Izv.} {\bf 24} (1985) 539--551.

\bibitem{Matsumoto:2008ww}
T.~Matsumoto and S.~Moriyama, {\it {An Exceptional Algebraic Origin of the
  AdS/CFT Yangian Symmetry}},  {\em JHEP} {\bf 04} (2008) 022,
  [\href{http://xxx.lanl.gov/abs/0803.1212}{{\tt arXiv:0803.1212}}].

\bibitem{Drummond:2008vq}
J.~M. Drummond, J.~Henn, G.~P. Korchemsky, and E.~Sokatchev, {\it {Dual
  superconformal symmetry of scattering amplitudes in N=4 super-Yang-Mills
  theory}},  {\em Nucl. Phys.} {\bf B828} (2010) 317--374,
  [\href{http://xxx.lanl.gov/abs/0807.1095}{{\tt arXiv:0807.1095}}].

\bibitem{Witten:2003nn}
E.~Witten, {\it {Perturbative gauge theory as a string theory in twistor
  space}},  {\em Commun. Math. Phys.} {\bf 252} (2004) 189--258,
  [\href{http://xxx.lanl.gov/abs/hep-th/0312171}{{\tt hep-th/0312171}}].

\bibitem{Corn:2010uj}
J.~Corn, T.~Creutzig, and L.~Dolan, {\it {Yangian in the Twistor String}},
  \href{http://xxx.lanl.gov/abs/1008.0302}{{\tt arXiv:1008.0302}}.

\bibitem{ArkaniHamed:2010kv}
N.~Arkani-Hamed, J.~L. Bourjaily, F.~Cachazo, S.~Caron-Huot, and J.~Trnka, {\it
  {The All-Loop Integrand For Scattering Amplitudes in Planar N=4 SYM}},
  \href{http://xxx.lanl.gov/abs/1008.2958}{{\tt arXiv:1008.2958}}.

\bibitem{Faddeev:1985qu}
L.~D. Faddeev and N.~Y. Reshetikhin, {\it {Integrability of the Principal
  Chiral Model in (1+1)-Dimension}},  {\em Ann. Phys.} {\bf 167} (1986) 227.

\bibitem{Zarembo:2010yz}
K.~Zarembo, {\it {Algebraic Curves for Integrable String Backgrounds}},
  \href{http://xxx.lanl.gov/abs/1005.1342}{{\tt arXiv:1005.1342}}.

\bibitem{Zarembo:2010sg}
K.~Zarembo, {\it {Strings on Semisymmetric Superspaces}},  {\em JHEP} {\bf 05}
  (2010) 002, [\href{http://xxx.lanl.gov/abs/1003.0465}{{\tt
  arXiv:1003.0465}}].

\bibitem{Berkovits:2004xu}
N.~Berkovits, {\it {Quantum consistency of the superstring in $AdS_5\times S^5$
  background}},  {\em JHEP} {\bf 03} (2005) 041,
  [\href{http://xxx.lanl.gov/abs/hep-th/0411170}{{\tt hep-th/0411170}}].

\bibitem{Mikhailov:2007eg}
A.~Mikhailov and S.~Schafer-Nameki, {\it {Algebra of transfer-matrices and
  Yang-Baxter equations on the string worldsheet in AdS(5) x S(5)}},  {\em
  Nucl. Phys.} {\bf B802} (2008) 1--39,
  [\href{http://xxx.lanl.gov/abs/0712.4278}{{\tt arXiv:0712.4278}}].

\bibitem{Puletti:2008ym}
V.~G.~M. Puletti, {\it {Aspects of quantum integrability for pure spinor
  superstring in $AdS_5\times S^5$}},  {\em JHEP} {\bf 09} (2008) 070,
  [\href{http://xxx.lanl.gov/abs/0808.0282}{{\tt arXiv:0808.0282}}].

\bibitem{Fiamberti:2007rj}
F.~Fiamberti, A.~Santambrogio, C.~Sieg, and D.~Zanon, {\it {Wrapping at four
  loops in $N=4$ SYM}},  {\em Phys. Lett.} {\bf B666} (2008) 100--105,
  [\href{http://xxx.lanl.gov/abs/0712.3522}{{\tt arXiv:0712.3522}}].

\bibitem{Fiamberti:2008sh}
F.~Fiamberti, A.~Santambrogio, C.~Sieg, and D.~Zanon, {\it {Anomalous dimension
  with wrapping at four loops in N=4 SYM}},  {\em Nucl. Phys.} {\bf B805}
  (2008) 231--266, [\href{http://xxx.lanl.gov/abs/0806.2095}{{\tt
  arXiv:0806.2095}}].

\bibitem{Bajnok:2009vm}
Z.~Bajnok, A.~Hegedus, R.~A. Janik, and T.~Lukowski, {\it {Five loop Konishi
  from AdS/CFT}},  {\em Nucl. Phys.} {\bf B827} (2010) 426--456,
  [\href{http://xxx.lanl.gov/abs/0906.4062}{{\tt arXiv:0906.4062}}].

\bibitem{Gromov:2009zb}
N.~Gromov, V.~Kazakov, and P.~Vieira, {\it {Exact Spectrum of Planar ${\cal
  N}=4$ Supersymmetric Yang- Mills Theory: Konishi Dimension at Any Coupling}},
   {\em Phys. Rev. Lett.} {\bf 104} (2010) 211601,
  [\href{http://xxx.lanl.gov/abs/0906.4240}{{\tt arXiv:0906.4240}}].

\bibitem{Arutyunov:2010gb}
G.~Arutyunov, S.~Frolov, and R.~Suzuki, {\it {Five-loop Konishi from the Mirror
  TBA}},  {\em JHEP} {\bf 04} (2010) 069,
  [\href{http://xxx.lanl.gov/abs/1002.1711}{{\tt arXiv:1002.1711}}].

\bibitem{Frolov:2010wt}
S.~Frolov, {\it {Konishi operator at intermediate coupling}},
  \href{http://xxx.lanl.gov/abs/1006.5032}{{\tt arXiv:1006.5032}}.

\bibitem{Lukowski:2009ce}
T.~Lukowski, A.~Rej, and V.~N. Velizhanin, {\it {Five-Loop Anomalous Dimension
  of Twist-Two Operators}},  {\em Nucl. Phys.} {\bf B831} (2010) 105--132,
  [\href{http://xxx.lanl.gov/abs/0912.1624}{{\tt arXiv:0912.1624}}].

\bibitem{Roiban:2009aa}
R.~Roiban and A.~A. Tseytlin, {\it {Quantum strings in $AdS_5 \times S^5$:
  strong-coupling corrections to dimension of Konishi operator}},  {\em JHEP}
  {\bf 11} (2009) 013, [\href{http://xxx.lanl.gov/abs/0906.4294}{{\tt
  arXiv:0906.4294}}].

\bibitem{Grigoriev:2007bu}
M.~Grigoriev and A.~A. Tseytlin, {\it {Pohlmeyer reduction of $AdS_5 x S^5$
  superstring sigma model}},  {\em Nucl. Phys.} {\bf B800} (2008) 450--501,
  [\href{http://xxx.lanl.gov/abs/0711.0155}{{\tt arXiv:0711.0155}}].

\bibitem{Roiban:2009vh}
R.~Roiban and A.~A. Tseytlin, {\it {UV finiteness of Pohlmeyer-reduced form of
  the $AdS_5xS^5$ superstring theory}},  {\em JHEP} {\bf 04} (2009) 078,
  [\href{http://xxx.lanl.gov/abs/0902.2489}{{\tt arXiv:0902.2489}}].

\bibitem{Hoare:2009rq}
B.~Hoare, Y.~Iwashita, and A.~A. Tseytlin, {\it {Pohlmeyer-reduced form of
  string theory in $AdS_5 x S^5$: semiclassical expansion}},  {\em J. Phys.}
  {\bf A42} (2009) 375204, [\href{http://xxx.lanl.gov/abs/0906.3800}{{\tt
  arXiv:0906.3800}}].

\bibitem{Hoare:2009fs}
B.~Hoare and A.~A. Tseytlin, {\it {Tree-level S-matrix of Pohlmeyer reduced
  form of $AdS_5 x S^5$ superstring theory}},  {\em JHEP} {\bf 02} (2010) 094,
  [\href{http://xxx.lanl.gov/abs/0912.2958}{{\tt arXiv:0912.2958}}].

\bibitem{Beisert:2010kk}
N.~Beisert, {\it {The Classical Trigonometric r-Matrix for the Quantum-
  Deformed Hubbard Chain}},  \href{http://xxx.lanl.gov/abs/1002.1097}{{\tt
  arXiv:1002.1097}}.

\bibitem{Beisert:2008tw}
N.~Beisert and P.~Koroteev, {\it {Quantum Deformations of the One-Dimensional
  Hubbard Model}},  {\em J. Phys.} {\bf A41} (2008) 255204,
  [\href{http://xxx.lanl.gov/abs/0802.0777}{{\tt arXiv:0802.0777}}].

\bibitem{Khoroshkin:1991ur}
S.~M. {Khoroshkin} and V.~N. {Tolstoy}, {\it {Universal R-matrix for quantized
  (super)algebras}},  {\em Communications in Mathematical Physics} {\bf 141}
  (Nov., 1991) 599--617.

\bibitem{Frolov:2005ty}
S.~A. Frolov, R.~Roiban, and A.~A. Tseytlin, {\it {Gauge - string duality for
  superconformal deformations of N = 4 super Yang-Mills theory}},  {\em JHEP}
  {\bf 07} (2005) 045, [\href{http://xxx.lanl.gov/abs/hep-th/0503192}{{\tt
  hep-th/0503192}}].

\bibitem{Bundzik:2005zg}
D.~Bundzik and T.~Mansson, {\it {The general Leigh-Strassler deformation and
  integrability}},  {\em JHEP} {\bf 01} (2006) 116,
  [\href{http://xxx.lanl.gov/abs/hep-th/0512093}{{\tt hep-th/0512093}}].

\bibitem{Mansson:2007sh}
T.~Mansson, {\it {The Leigh-Strassler Deformation and the Quest for
  Integrability}},  {\em JHEP} {\bf 06} (2007) 010,
  [\href{http://xxx.lanl.gov/abs/hep-th/0703150}{{\tt hep-th/0703150}}].

\bibitem{Mansson:2008xv}
T.~Mansson and K.~Zoubos, {\it {Quantum Symmetries and Marginal Deformations}},
   \href{http://xxx.lanl.gov/abs/0811.3755}{{\tt arXiv:0811.3755}}.

\bibitem{Delius:2001yi}
G.~W. Delius, N.~J. MacKay, and B.~J. Short, {\it {Boundary remnant of Yangian
  symmetry and the structure of rational reflection matrices}},  {\em Phys.
  Lett.} {\bf B522} (2001) 335--344,
  [\href{http://xxx.lanl.gov/abs/hep-th/0109115}{{\tt hep-th/0109115}}].

\bibitem{Ahn:2010xa}
C.~Ahn and R.~I. Nepomechie, {\it {Yangian symmetry and bound states in AdS/CFT
  boundary scattering}},  {\em JHEP} {\bf 05} (2010) 016,
  [\href{http://xxx.lanl.gov/abs/1003.3361}{{\tt arXiv:1003.3361}}].

\bibitem{Mackay:2002bd}
N.~J. {MacKay}, {\it {Rational K-matrices and representations of twisted
  Yangians}},  {\em Journal of Physics A Mathematical General} {\bf 35} (Sept.,
  2002) 7865--7876, [\href{http://xxx.lanl.gov/abs/math/0205}{{\tt
  math/0205}}].

\bibitem{Vicedo:2010talk}
B.~Vicedo, {\it {The classical R-matrix of AdS/CFT}},  {\em Talk at IGST2010
  NORDITA, Stockholm} (2010).

\bibitem{Haldane:1992sj}
F.~D.~M. Haldane, Z.~N.~C. Ha, J.~C. Talstra, D.~Bernard, and V.~Pasquier, {\it
  {Yangian symmetry of integrable quantum chains with long range interactions
  and a new description of states in conformal field theory}},  {\em Phys. Rev.
  Lett.} {\bf 69} (1992) 2021--2025.

\bibitem{Hikami:1995yang}
K.~{Hikami}, {\it {Yangian symmetry and Virasoro character in a lattice spin
  system with long-range interactions}},  {\em Nuclear Physics B} {\bf 441}
  (Feb., 1995) 530--548.

\bibitem{Beisert:2007jv}
N.~Beisert and D.~Erkal, {\it {Yangian Symmetry of Long-Range gl(N) Integrable
  Spin Chains}},  {\em J. Stat. Mech.} {\bf 0803} (2008) P03001,
  [\href{http://xxx.lanl.gov/abs/0711.4813}{{\tt arXiv:0711.4813}}].

\end{thebibliography}\endgroup

\end{document}